\documentclass[a4paper,fleqn,usenatbib]{mnras}

\usepackage[T1]{fontenc}
\usepackage{ae,aecompl}
\usepackage{natbib}
\usepackage{lscape}
\usepackage{lastpage}

\usepackage{graphicx}	
\usepackage{amsmath}	
\usepackage{amssymb}	
  \usepackage{multicol}        
   \usepackage{bm}		
   \usepackage{pdflscape}	
    \usepackage{subfig}
 \usepackage{enumerate}
 \usepackage{rotating}
\usepackage{float}

\newcommand{\Msol}{M$_{\odot}$}                           
\newcommand{\Zsol}{Z$_{\odot}$} 				
\newcommand{\hii}{H\,{\sc ii}\rm}

\newcommand{\ho}{$\itl H_{0}$}

\def\kmsmpc {\rm {km~s^{-1}}~Mpc^{-1}}
\newcommand{\lsig}{$L(\mathrm{H}\beta) - \sigma$}
\def\lsigma{$L-\sigma$~}
\newcommand{\hb}{\mathrm{H}\beta}
\def\hbeta {H$\beta$}
\def\halpha {H$\alpha$}

\newcounter{mycount}
\newcommand\lbl[1]{\refstepcounter{mycount}\label{my:#1}}

\author[D. Fern\'andez Arenas et al.] {David Fern\'andez Arenas $^{1}$,
Elena Terlevich$^{1}$, 
Roberto Terlevich$^{1,2}$, 
Jorge Melnick$^{3,7}$,
\newauthor Ricardo Ch\'avez$^{4,5}$, 
Fabio Bresolin$^{6}$,
Eduardo Telles$^{7}$,
Manolis Plionis$^{8,9}$ and \newauthor
Spyros Basilakos$^{10}$\\ \\
$^{1}$ Instituto Nacional de Astrof\'\i sica, \'Optica y Electr\'onica,Tonantzintla, Puebla, Mexico \\
$^{2}$ Institute of Astronomy, University of Cambrige, Cambridge, CB3 0HA, UK\\
$^{3}$ European Southern Observatory, Santiago de Chile, Chile \\
$^{4}$ Cavendish Laboratory, University of Cambridge, 19 J. J. Thomson Ave, Cambridge CB3 0HE, UK\\
$^{5}$ Kavli Institute for Cosmology, University of Cambridge, Madingley Road, Cambridge CB3 0HA, UK\\
$^{6}$ Institute for Astronomy, University of Hawaii, 2680 Woodlawn Drive, 96822 Honolulu,HI USA \\ 
$^{7}$ Observatorio Nacional, Rua Jos\'e Cristino 77, 20921-400 Rio de Janeiro, Brasil\\
$^{8}$ National Observatory of Athens, P.Pendeli, Athens, Greece \\
$^{9}$ Physics Dept., Aristotle Univ. of Thessaloniki, Thessaloniki 54124, Greece \\
$^{10}$ Academy of Athens Research Center for Astronomy \& Applied Mathematics, Soranou Efessiou 4, 11-527 Athens, Greece \\
}

\title[Determination of  \ho\ ]{An independent determination of the local Hubble constant.}
\pubyear{2017}
\begin{document}
\setcounter{mycount}{0}

\label{firstpage}
\pagerange{\pageref{firstpage}--\pageref{lastpage}}
\maketitle

\begin{abstract}
The relationship between the integrated H$\beta$ line luminosity and the velocity dispersion of the ionized gas of \hii\ galaxies and giant \hii\ regions represents an exciting standard candle that presently can be used up to redshifts $z \sim 4$. Locally it is used to obtain precise  measurements of  the  Hubble constant  by combining the slope of the relation obtained from nearby ($z \leq $ 0.2) \hii\ galaxies with the zero point determined from giant \hii\ regions belonging to an  `anchor sample' of  galaxies for which accurate redshift-independent distance moduli are available.   We present new data for 36 giant \hii\ regions in 13 galaxies of the anchor sample that includes the megamaser galaxy NGC~4258.  Our data is the result of the first four years of observation of our primary sample of 130 giant \hii\ regions in 73 galaxies with Cepheid determined distances.  Our best estimate of the Hubble parameter is  $71.0\pm2.8(random)\pm2.1(systematic)$ $\kmsmpc$. This result is the product of an independent approach and, although at present  less precise than the latest SNIa results, it is  amenable to substantial improvement.


\end{abstract}

\begin{keywords}
Hubble constant, \hii\ galaxies  , \hii\ regions, cosmology
\end{keywords}


\section{Introduction}

{In the past two decades a combination of different distance indicators \cite[Cepheids, SNIa, surface brightness fluctuations, etc. see][]{Freedman2010} has been used to vastly improve the accuracy in the determination of the Hubble constant \ho. The Hubble Space Telescope (HST) Key Project and Carnegie Hubble Program \citep{Freedman2001, freedman2012} among others, have obtained an accuracy of  3\% on the measurement of \ho\ reporting values of $73.8\pm2.4$, and $74.3\pm2.1$, $\kmsmpc$. Subsequently  \citet{Humphreys2013} using the megamaser galaxy NGC~4258 (for which very precise `geometric'  distance measurements are available) reported $72.7\pm2.4$ $\kmsmpc$. 

On the other hand, Planck  observations of the Cosmic Microwave Background (CMB) combined with a flat $\Lambda$CDM cosmology derived a value of  \ho\ = $67.3\pm1.2$ $\kmsmpc$ ~\citep{Planck132014} that indicates a 2.5$\sigma$  tension with the direct estimate reported by \cite{riess2011} suggesting the possible need of new Physics to solve the problem.

\cite{Efstathiou2014} re-examined the Cepheids analysis of \cite{riess2011} and  found  \ho\  = $72.5\pm2.5$ $\kmsmpc$; while using as the central calibrator the  NGC~ 4258 megamaser and the SNIa database he obtained a value of \ho\ = $70.6\pm2.4$ $\kmsmpc$ concluding that there is no evidence for a need to postulate new Physics.
\cite{Riess2016} addressed \cite{Efstathiou2014}  result suggesting that a change in the colour selection of the Cepheids removes the difference  in the \ho\ values.
The  new results from SNIa \citep{Riess2016} of \ho\ =  $73.24\pm1.74$ $\kmsmpc$ have reinstated the ``tension", now at the 3.1$\sigma$ level with the value obtained  by 
\cite{Planck152016}  of \ho\ = $67.8\pm0.9$ $\kmsmpc$.

This ``tension" has prompted us to explore using new  and improved data,  our local estimate of \ho\ based on the standard candle provided by the correlation that exists in  \hii\ galaxies (HIIGs) and giant \hii\ regions (GHIIRs) between the turbulent emission line velocity dispersion $(\sigma)$ of the Balmer lines and its integrated  luminosity  \citep{terlevich1981,melnick1987,melnick1988,Chavez2012}. 

HIIGs are compact and massive systems experiencing luminous bursts of star formation generated by the formation of young super stellar clusters (SSCs) with a high 
luminosity per unit mass and with properties similar, if not identical, to GHIIRs. The potential of GHIIRs as distance indicators was originally realized 
 from the existence of a correlation between the GHIIR diameter and the parent galaxy luminosity \citep{Sersic1960,Sandage1962} see also \cite{Kennicutt1979}. A different approach was proposed by \cite{mel77,melnick1978}, who found that the turbulent width of the nebular emission lines is correlated with the GHIIR diameters. 
 
  \cite{terlevich1981} (hereinafter TM81) found a tight correlation between the turbulent emission lines velocity dispersion  and their integrated luminosity: the \lsigma  relation. This correlation, 
 valid for HIIGs and GHIIRs,  links a distance dependent parameter, the  integrated H$\beta$ line luminosity, with a parameter that is independent of distance, the velocity dispersion of the ionized gas,  therefore defining a redshift independent  distance estimator.

The \lsigma  relation represents a rather interesting distance indicator that with present instrumentation can be utilized out to $z \sim 4$ \citep{melnick2000,sie05,pli11,terlevich2015, Chavez2016}.

\cite{Chavez2012} confirmed that the  \lsigma relation does provide a reliable independent method to measure the Hubble constant.
To determine the value of the local Hubble constant,  the   \lsigma  relation for HIIGs  is anchored to a  sample of GHIIRs in nearby galaxies having  accurate distances determined  using primary distance indicators. 

Although the scatter of the \lsigma  distance indicator is about a factor of two larger than the one based on SNIa \citep{Chavez2014}, this is partially compensated by the larger number of local calibrators available for the \lsigma method, i.e.~galaxies with distance determination independent of redshift, compared to those available for SNIa, plus the fact that the number of GHIIRs per galaxy is usually more than one, thus reducing the uncertainty per anchor galaxy. 

A fundamental problem with the determination of the Hubble constant using SNIa is related to the low expected rate of SNIa inside the 30 Mpc reach of the HST for accurate Cepheid studies \citep{Riess2016}. The present sample of SNIa in galaxies with accurate distance estimates is 19   and it would not substantially increase over the remaining lifetime of the HST given that their average rate is only about one SNIa per year \citep{Riess2016}.
On the other hand the number of anchor galaxies with GHIIRs and accurate Cepheid distances is presently 73 in our primary sample, with a total of 130 GHIIRs. Moreover, GHIIRs in  special galaxies like the LMC, the SMC, and NGC~~4258  with very accurate redshift-independent distance  determinations are also included in our sample of anchor galaxies.}

The  \lsigma distance indicator assumes a linear relation between the logarithm of the $H\beta$ emission-line luminosity L(H$\beta$) (proportional to the number of ionizing photons) and the logarithm of the width of the emission lines $\sigma$, proportional to the total mass of the system.   Although there is a solid framework for understanding the underlying physics of the \lsigma relation \citep{Chavez2014}, it remains empirical in the sense that we are not yet able to predict accurately  the coefficients of the relation starting from basic principles. 

Thus, the application of the \lsigma relation as a distance estimator requires care when determining the slope of the relation, especially because standard least-squares techniques are usually not adequate for data with observational errors in the independent coordinate. Additionally,  a good understanding of the random and systematic errors of the data is needed.  For example,   \cite{Chavez2014}  found that the size of the system, albeit difficult to measure,  is a strong second parameter that reduces the scatter of the \lsigma relation by about 40\%{.} 

We also know that in very young starburst clusters,  capable of ionizing the surrounding gas, the intensity of the emission lines fades rapidly as the massive stars evolve, while the velocity dispersions remain roughly constant for much longer, which may introduce a  systematic effect as discussed by \cite{Melnick2017}. 

Furthermore, both the presence of dust, ubiquitous in young star forming regions, and the possible escape of Lyman continuum photons may also introduce systematic effects that are difficult to remove.  Potential systematic effects  regarding the line profiles are the broad wings  associated with the stellar winds of the most massive stars and the presence of multiple cores inside the spectrograph aperture.
Systematic effects are also the main limitation for the SNIa distance estimator so an important sub-product of our technique is to provide a comparative method to study the systematics of both empirical methods.

The paper is organized as follows, in \S \ref{data}, we describe our  new GHIIR data 
for the ``anchor sample". In \S \ref{extabs} we present the corrections to the observed fluxes due to  extinction and underlying absorption and \S \ref{age} deals with evolutionary corrections. Section \S \ref{distlum} discusses distances and luminosities.
  In \S \ref{calH0}, we present our  method for determining \ho. \S \ref{explore} is a detailed study of the systematic errors that may affect the application of the \lsigma\ relation to measure distances and to determine \ho . \S \ref{SN} presents a comparison with previous results for \ho\ in particular those from SNIa and the Planck collaboration. The conclusions are given in \S \ref{conclusions}.

\begin{table*}
	\caption{Adopted distance moduli for the new anchor sample.}
	\begin{tabular}{lccl}\\\cline{1-4}
		\hline
		Object &  Distance Modulus (mag)   & Distance (Mpc)  & Reference\\  \hline
		
		IC10  &  24.22  $\pm$  0.13 & 0.70   $\pm$ 0.04 & \ref{my:Sakai1999} \\ 

		M101  &  29.15  $\pm$  0.10 & 6.76   $\pm$ 0.32 & \ref{my:Paturel2002b}, \ref{my:Saha2006}, \ref{my:Freedman2001}, \ref{my:Willick2001}, \ref{my:Sakai2004}, \ref{my:Shappee2011}, \ref{my:Mager2013} \\ 

		M33  &  24.58  $\pm$  0.10 & 0.82   $\pm$ 0.03 & \ref{my:Paturel2002b}, \ref{my:Saha2006}, \ref{my:Freedman2001}, \ref{my:Willick2001}, \ref{my:Sakai2004}, \ref{my:Lee2002}, \ref{my:Scowcroft2009}, \ref{my:An2007}, \ref{my:Bhardwaj2016}, \ref{my:Gieren2013} \\ 

		M81  &  27.80  $\pm$  0.10 & 3.63   $\pm$ 0.17 & \ref{my:Paturel2002b}, \ref{my:Saha2006}, \ref{my:Freedman2001}, \ref{my:Sakai2004}, \ref{my:McCommas2009}, \ref{my:Gerke2011}, \ref{my:Kanbur2003} \\  

		MRK116  &  31.35  $\pm$  0.22 & 18.62  $\pm$ 1.98 & \ref{my:Fiorentino2010}, \ref{my:Aloisi2007}, \ref{my:Marconi2010} \\ 

		N2366  &  27.63  $\pm$  0.14 & 3.36   $\pm$ 0.22 & \ref{my:Ferrarese2000} \\ 

		N2403 &  27.49  $\pm$  0.23 & 3.15   $\pm$ 0.35 &  \ref{my:Saha2006}, \ref{my:Freedman2001}, \ref{my:Madore1991}, \ref{my:Freedman1988} \\ 

		N4258 &  29.37  $\pm$  0.06 & 7.48   $\pm$ 0.03 & \ref{my:Paturel2002b}, \ref{my:Saha2006}, \ref{my:Freedman2001}, \ref{my:Willick2001}, \ref{my:An2007}, \ref{my:Gerke2011}, \ref{my:Kanbur2003}, \ref{my:Hoffmann2015}, \ref{my:van-Leeuwen2007},  \ref{my:Di-Benedetto2013}, \ref{my:Macri2006}, \ref{my:Efstathiou2014}, \ref{my:Ngeow2003}, \ref{my:Fausnaugh2015}, \ref{my:Mager2008}, \ref{my:Caputo2002}, \ref{my:Newman2001}, \ref{my:Maoz1999}
		 $\star$\\ 

		N4395 &  28.22  $\pm$  0.12 & 4.41   $\pm$ 0.25 & \ref{my:Thim2004} \\ 

		N0925 &  29.80  $\pm$  0.10 & 9.12  $\pm$ 0.43 & \ref{my:Paturel2002b}, \ref{my:Saha2006}, \ref{my:Freedman2001}, \ref{my:Willick2001}, \ref{my:Kanbur2003} \\     

		N2541 & 30.35  $\pm$  0.12 & 11.75  $\pm$ 0.67 & \ref{my:Paturel2002b}, \ref{my:Saha2006}, \ref{my:Freedman2001}, \ref{my:Willick2001}, \ref{my:Kanbur2003} \\

		N3319 & 30.65  $\pm$  0.14 & 13.49  $\pm$ 0.90 &  \ref{my:Paturel2002b}, \ref{my:Saha2006}, \ref{my:Freedman2001}, \ref{my:Willick2001}, \ref{my:Kanbur2003} \\

		N3198 & 30.75  $\pm$  0.13 & 14.13  $\pm$ 0.87 &  \ref{my:Paturel2002b}, \ref{my:Saha2006}, \ref{my:Freedman2001}, \ref{my:Willick2001}, \ref{my:Kanbur2003} \\          \hline
	\end{tabular}

\begin{flushleft}
1:~\protect\cite{Sakai1999}\lbl{Sakai1999}
2:~\protect\cite{Paturel2002b}\lbl{Paturel2002b} 
3:~\protect\cite{Saha2006}\lbl{Saha2006} 
4:~\protect\cite{Freedman2001}\lbl{Freedman2001}
5:~\protect\cite{Willick2001}\lbl{Willick2001}
6:~\protect\cite{Sakai2004}\lbl{Sakai2004}
7:~\protect\cite{Shappee2011}\lbl{Shappee2011}
8:~\protect\cite{Mager2013}\lbl{Mager2013}
9:~\protect\cite{Lee2002}\lbl{Lee2002}
10:~\protect\cite{Scowcroft2009}\lbl{Scowcroft2009}
11:~\protect\cite{An2007}\lbl{An2007}
12:~\protect\cite{Bhardwaj2016}\lbl{Bhardwaj2016}
13:~\protect\cite{Gieren2013}\lbl{Gieren2013}
14:~\protect\cite{McCommas2009}\lbl{McCommas2009}
15:~\protect\cite{Gerke2011}\lbl{Gerke2011}
16:~\protect\cite{Kanbur2003}\lbl{Kanbur2003}
17:~\protect\cite{Fiorentino2010}\lbl{Fiorentino2010}
18:~\protect\cite{Aloisi2007}\lbl{Aloisi2007}
19:~\protect\cite{Marconi2010}\lbl{Marconi2010}
20:~\protect\cite{Ferrarese2000}\lbl{Ferrarese2000}
21:~\protect\cite{Madore1991}\lbl{Madore1991}
22:~\protect\cite{Freedman1988}\lbl{Freedman1988}
23:~\protect\cite{Hoffmann2015}\lbl{Hoffmann2015}
24:~\protect\cite{van-Leeuwen2007}\lbl{van-Leeuwen2007}
25:~\protect\cite{Di-Benedetto2013}\lbl{Di-Benedetto2013}
26:~\protect\cite{Macri2006}\lbl{Macri2006}
27:~\protect\cite{Efstathiou2014}\lbl{Efstathiou2014}
28:~\protect\cite{Ngeow2003}\lbl{Ngeow2003}
29:~\protect\cite{Fausnaugh2015}\lbl{Fausnaugh2015}
30:~\protect\cite{Mager2008}\lbl{Mager2008}
31:~\protect\cite{Caputo2002}\lbl{Caputo2002}
32:~\protect\cite{Newman2001}\lbl{Newman2001}
33:~\protect\cite{Maoz1999}\lbl{Maoz1999}
34:~\protect\cite{Thim2004}\lbl{Thim2004} \\

$\star$ The geometric maser distance for NGC 4258 is 7.60 $\pm$ 0.32 Mpc  \protect\citep{Humphreys2013}. 

\end{flushleft}
\label{tableDistan}
\end{table*}

\section{The data}\label{data}

The use of the \lsigma relation as a distance indicator and as a tool to derive the Hubble constant, requires accurate determination of  both the luminosity and the FWHM or velocity dispersion of the emission lines in GHIIRs and  HIIGs.  In this section we discuss the observations and the quality of the obtained data in our new  sample of GHIIRs in nearby galaxies.

\subsection{The New Anchor Sample}\label{appa}

To improve the early work on GHIIRs and to obtain a fiducial anchor sample we started in 2012  a long term project to acquire  integrated $\hb$ fluxes and velocity dispersions of a new sample of 130 GHIIRs in 73 galaxies for which  accurate  distances  have been determined using primary distance indicators. 
Here we present  the results of the observations of 36 GHIIRs hosted by 13 such nearby galaxies representing about $1/4$th of our primary sample of GHIIRs. 

Much of the variance in the value of \ho\  is related to the choice of  distance to the galaxies in the anchor sample which in turn is intimately linked to the choice of calibration of the Cepheids period-luminosity (PL) relation. A thorough discussion of this aspect can be found in \cite{Riess2016}.


\subsubsection{Adopted distances}

The Cepheid distances to our sample galaxies were obtained from NASA/IPAC Extragalactic Database\footnote{This research has made use of the NASA/IPAC Extragalactic Database (NED) which is operated by the Jet Propulsion Laboratory, California Institute of Technology, under contract with the National Aeronautics and Space Administration.}.
Our adopted distance for each galaxy is the average value provided by the references in Table \ref{tableDistan}, weighted by the reciprocal of the quoted distance modulus error.
We only considered distances based on CCD photometry, that have been obtained, almost entirely, from determinations published more recently than the year 2000. Where necessary  the published distance moduli were adapted using as reference an LMC value  of $(m-M)_{LMC}$=18.50.

In addition to the references in Table \ref{tableDistan}, we provide the following specific comments:

1) From the Hubble Space Telescope Key Project team papers we only used the result published by \cite{Freedman2001}, adopting their metallicity-corrected distance values.

2) From \cite{Kanbur2003} we adopted the metallicity-corrected distances obtained from the LMC Cepheid PL relation.

3) In the case of  \cite{Paturel2002b}, where they used the period-luminosity relation for Galactic Cepheids with HIPPARCOS distances, we used their adopted distance moduli, given in their Table 4 (Column 8).

We note that only one Cepheid distance was available for the galaxies IC 10, NGC~ 2366 and NGC~ 4395; for these three galaxies the adopted distance is the average of the Cepheid value and the mean of the  Tip of the Red Giant Branch (TRGB) values. For MRK116 (I Zw 18), the distance values reported in the literature rely on theoretical models, because of the very low metallicity of this system (1/40th \Zsol), which prevents the  use of empirical Period-Luminosity relations. 

\begin{figure}
 \includegraphics[width=0.45\textwidth]{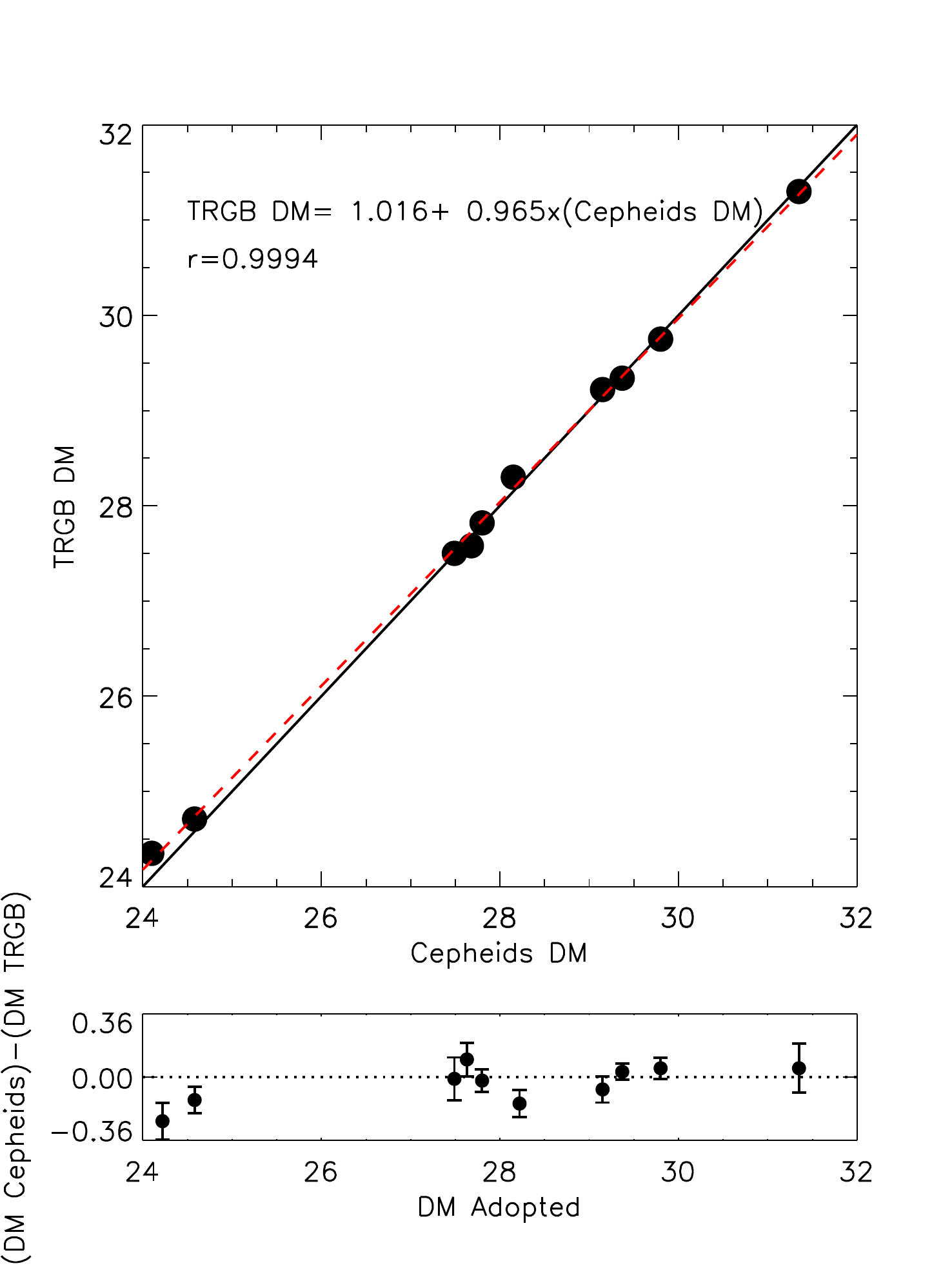}
    \caption{Comparison of the distance modulii for our anchor sample obtained using the TRGB or the Cepheids  PL relation. The solid line is the 1-to-1 relation and the red dashed line is the fit to the points, r is the correlation coefficient. The inset shows the coefficients of the fit. The residuals are shown in the bottom panel against the adopted distance modulii from Table \ref{dist}.}
    \label{dist}
\end{figure}

The distances obtained from the TRGB provided an important sanity check. The good agreement between the two sources of distance is shown in Figure \ref{dist}.

\subsubsection{The GHIIR sample }

In this section we present the results of the observations of  36 GHIIRs hosted by 13 nearby galaxies with redshift-independent distances.
The targets are listed in Table \ref{regions} and a journal of observations is given in Table \ref{Journal}. 
Table~\ref{tablaPar} presents the relevant data for the new sample that we use in this paper to determine the zero-point of the \lsigma relation and thus to derive the value of the Hubble constant. 

As in our previous work, further selection conditions are  \\

\noindent (i) a lower limit for the equivalent width, $EW(H\beta)>$ 50\AA{} to exclude highly evolved regions, diminish contamination by an underlying older stellar  population  \citep[cf.][]{melnick2000} and avoid objects with high rate of escape of  ionizing photons, plus \\

\noindent (ii) an upper limit to the Balmer line widths,  $\log\sigma<1.8$ km s$^{-1}$ to minimize the possibility of including systems supported by rotation  or with multiple young ionizing clusters, as discussed in \citet{melnick1988}.




 \begin{table}
 \caption{Regions observed}
  \label{regions}
  \small
 \begin{tabular}{|l|l|l|l|l|}    \\\cline{1-4}
Index &          GHIIR& $\alpha$(J2000) & $\delta$(J2000)\\\hline
1 &           IC~10-111 & 00 20 27.0   & +59 17 29      \\
2 &           IC~10-C01 & 00 20 17.0 & +59 18 34      \\
3 &       M101-NGC~5447 & 14 02 28.0 & +54 16 33  \\
4 &       M101-NGC~5455 & 14 03 01.2 & +54 14 29   \\
5 &       M101-NGC~5461 & 14 03 41.0 & +54 19 02  \\
6 &       M101-NGC~5462 & 14 03 53.1 & +54 22 06   \\
7 &       M101-NGC~5471 & 14 04 28.6 & +54 23 53  \\
8 &           M33-NGC~588 & 01 32 45.9 & +30 38 51      \\
9  &           M33-NGC~592 & 01 33 11.7 & +30 38 42     \\
10 &           M33-NGC~595 & 01 33 33.8 &  +30 41 30    \\
11 &           M33-NGC~604 & 01 34 33.2 & +30 47 06     \\
12 &          M81-HK268 & 09 55 52.8 & +68 59 03 \\
13 &          M81-HK652 & 09 54 57.0 & +69 08 48 \\
14 &            MRK~116 & 09 34 02.0 & +55 14 28     \\
15 &        NGC~2366-HK110 & 07 28 30.1 & +69 11 37     \\
16 &         NGC~2366-HK54 & 07 28 46.6 & +69 11 27	    \\
17 &         NGC~2366-HK72 & 07 28 43.0 & +69 11 23     \\
18 &             NGC~2366 & 07 28 54.6 & +69 12 57     \\
19 &         NGC~2403-VS24 & 07 36 45.5 & +65 37 01     \\
20 &          NGC~2403-VS3 & 07 36 20.0 & +65 37 04       \\
21 &         NGC~2403-VS44 & 07 37 07.0 & +65 36 39       \\
22 &          NGC~925-120 & 02 27 01.6 & +33 34 28     \\
23 &          NGC~925-128 & 02 26 58.6 & +33 34 40     \\
24 &           NGC~925-42 & 02 27 21.6 & +33 33 31     \\
25 &       NGC~4258-RC01 & 12 18 55.3 & +47 16 46  \\
26 &       NGC~4258-RC02 & 12 19 01.4 & +47 15 25 \\
27 &    NGC~4395-NGC~4399 & 12 25 42.9 & +33 30 57  \\
28 &   NGC~4395-NGC~4400  & 12 25 56.0 & +33 30 54   \\
29 &    NGC~4395-NGC~4401 & 12 25 57.6 & +33 31 42 \\
30 &    NGC~2541-A         & 08 14 47.6 &	+49 03 59  \\
31 &    NGC~2541-B         & 08 14 37.3 &	+49 02 59   \\
32 &    NGC~2541-C         & 08 14 37.2 &	+49 03 53 \\
33 &    NGC~3319-A         & 10 39 03.9 &	+41 39 41 \\
34 &    NGC~3319-B         & 10 39 00.3 &	+41 40 08  \\
35 &    NGC~3319-C         & 10 39 17.7 &	+41 42 07   \\
36 &    NGC~3198-A         & 10 19 46.1 & 	+45 31 03 \\\hline
\end{tabular} 
\end{table}


\begin{table*}
  \caption{Journal of observations.}
 \label{Journal}
 \newcommand{\mc}[3]{\multicolumn{#1}{#2}{#3}}
   \begin{turn}{90}
   \begin{tabular}{|c|c|c|c|c|c|c|c|c|c|c|c|c|}    \\\cline{1-13}
\hline
 &  \mc{6}{c}{Low-resolution spectroscopy  }  &  \mc{6}{c}{High-resolution spectroscopy  } \\  \cline{2-13}
 &  \mc{3}{c}{ OAN-2.1m-B\&C} &\mc{3}{c}{ OAGH-2.1m-B\&C}& \mc{3}{c}{ OAN-2.1m-MEZCAL} &\mc{3}{c}{ OAGH-2.1m-CanHiS}\\ 
 &  Exp. time & Number & Date & Exp. time & Number & Date& Exp. time & Number & Date & Exp. time & Number & Date \\
INDEX &    (s)& of Exp. &   &  (s)& of Exp. &    &  (s) &of Exp.   & &  (s) &  of Exp.  \\ \hline 
1   &360 &4 & Oct 13,2012  & & &                  &900 &3 &Sep 23,2013 & & &   \\ 
2   &360 &4 & Oct 13,2012  & & &                  & & & &1200 &3 &Oct 11,2013  \\  
3  &    &  &              &900  &4 &Apr 14,2013  & & & &1200 &3 &Mar 21,2015 \\ 
4   &    &  &              &900  &3 &Apr 15,2013  & & & &1500 &3 &Mar 22,2015  \\ 
5   &    &  &              &900  &3 &Apr 13,2013  & & & &1200 &3 &Mar 21,2015  \\ 
6   &    &  &              &600  &3 &Apr 11,2013  & & & &1800 &3 &Mar 22,2015   \\ 
7   &    &  &              &900  &3 &Apr 11,2013  &1200 &3 &Mar 23,2014 &900  &3 &Mar 21,2015   \\
8   &360 &4 &Oct 13,2012  &600  &3 &Nov 13,2012   &1200 &3 &Sep 25,2013 &1500 &3 &Oct 9,2013 \\
9   &360 &4 &Oct 12,2012   &600  &3 &Nov 13,2012  & & & &1200 &2 &Oct 15,2014   \\ 
10  &360 &4 &Oct 12,2012   &300  &5 &Nov 13,2012  &600  &3 &Sep 25,2013  &900 &2 &Oct 17,2014    \\
11  &360 &4 &Oct 14,2012   &900  &2 &Nov 12,2012  &1200 &3 &Aug 27, 2013 &900 &2 &Oct 17,2014    \\ 
12  &    &  &              &1200 &3 &Apr 11,2013  &900 &3 &Mar 20,2014   &1600 &3 &Mar 29,2014    \\ 
13  &    &  &              &1200 &3 &Apr 13,2013  &1800 &2 &Mar 15,2015 & & &    \\ 
14  &    &  &              &900  &3 &Apr 11,2013  & & & &1200 &3 &Mar 26,2015    \\ 
15  &360 &4 &Oct 13,2012   & & &                  & & & &1200 &2 &Mar 23,2015    \\ 
16	    &360 &4 &Oct 13,2012   & & &          &1800 &2 &Mar 12,2015 &1200 &3 &Mar 23,2015     \\ 
17  &360 &4 &Oct 13,2012   & & &                  &900 &2 &Mar 13,2015 &1800 &3 & Mar 23,2015     \\ 
18  &300 &3 &Oct 12,2012   &600  &3 &Nov 14, 2012 &2400 &2 &Mar 12,2015 & & &       \\ 
19  &360 &4 &Oct 14,2012   &900  &3 &Nov 14, 2012 & & & &900 &3 &Oct 17,2014      \\  
20  &360 &4 &Oct 14,2012   &420  &4 &Nov 13, 2012 & & & &1000 &3 &Oct 12,2013       \\ 
21  &360 &4 &Oct 14,2012   &600  &3 &Nov 14, 2012 & & & &1200 &3 &Oct 10,2013      \\  
22  &360 &4 &Oct 12,2012   &1200 &3 &Nov 14, 2012 &1200 &3 &Sep 26,2013 & & &       \\  
23  &480 &5 &Oct 12,2012   &1200 &3 &Nov 14, 2012 &1200 &3 &Sep 24,2013 & & &       \\ 
24  &360 &4 &Oct 12,2012   &900  &4 & Nov 14, 2012&1200 &3 &Sep 24,2013 &1200 &3 &Oct 10,2013       \\  
25  &    &  &              &1200 &3 &Apr 15,2013  &1800 &3 &Mar 21,2014 & & &        \\  
26  &    &  &              &1200 &3 &Apr 15,2013  &1200 &3 &Mar 22,2014 &1800 &4 &Mar 23,2015         \\ 
27  &    &  &              &1200 &3 &Apr 16,2013  &1200 &2 &Mar 15,2015 & & &         \\ 
28  &    &  &              &1200 &3 &Apr 16,2013  &1200 &4 &Mar 30,2014 & & &          \\ 
29  &    &  &              &1500 &3 &Apr 12,2013  &1800 &2 &Mar 15,2015 & & &           \\
30  &    &  &              &900  &3 &Mar 13,2016  &1800 &3 &Mar 1, 2016  & &  &         \\
31  &    &  &              &900  &3 &Mar 13,2016  &1200 &3 &Feb 27,2016 &1500 &3  &Mar 10,2016  \\
32  &    &  &              &900  &3 &Mar 13,2016  & & & &1800 &3 &Mar 12,2016     \\
33  &    &  &              &1200 &3 &Mar 15,2016  &1800 &3 &Feb 27,2016 & &  &    \\
34  &    &  &              &1200 &3 &Mar 15,2016  &1800 &3 &Feb 27,2016 & &  &    \\
35  &    &  &              &1200 &3 &Mar 16,2016  &1800 &3 &Feb 27,2016 & &  &    \\
36  &    &  &              &1200 &4 &Mar 14,2016  &1800 &3 &Feb 29,2016 & &  &    \\\\ \hline\cline{1-13}
\end{tabular}
 \end{turn}
\end{table*}


\begin{center}
\begin{table*}
 \caption{New anchor sample of Giant HII regions. Luminosities are corrected using \citet{Gordon2003} extinction law.}
\begin{tabular}{|l|c|r|c|c|c|c|c|}\\\cline{1-8}
\hline
Name           & Flux (H$\beta$)             & EW (H$\beta$)             & $A_{V}$ & $Q$x100    & FWHM(H$\alpha$)           & $\log L$(H$\beta$) & $\log\sigma$(H$\alpha$) \\
               & (10$^{-14}$erg s$^{-1}$cm$^{-2}$)             & \AA{}             &  &     & \AA{}            &  (erg s$^{-1}$)      &  (km s$^{-1}$)\\ \hline
IC10-111        &     5297.      $\pm$     626.  &   118.      $\pm$  7.4    &  0.057   $\pm$  0.053     &  0.182    &  1.048   $\pm$  0.030       & 39.52    $\pm$    0.08 & 1.180 $\pm$ 0.013 \\
IC10-C01        &     3297.      $\pm$     380. &   221.      $\pm$  11.0   &  0.013   $\pm$  0.078     &  0.000    &  0.817    $\pm$  0.015       & 39.29    $\pm$    0.08 & 1.114 $\pm$ 0.012 \\
M101-NGC~5447    &     169.      $\pm$     8.3    &   169.      $\pm$  3.8    &  1.490   $\pm$  0.113     &  0.022    &  1.433   $\pm$  0.018      & 40.68    $\pm$    0.07 & 1.414 $\pm$ 0.023 \\
M101-NGC~5455    &     73.5      $\pm$     6.6    &   112.      $\pm$  6.5    &  0.443   $\pm$  0.075     &  0.007    &  1.353   $\pm$  0.021      & 39.82    $\pm$    0.07 & 1.369 $\pm$ 0.021 \\
M101-NGC~5461    &     82.2      $\pm$     9.8    &   135.      $\pm$  1.7    &  0.464   $\pm$  0.063     &  0.087    &  1.120   $\pm$  0.021      &  39.88    $\pm$    0.07 & 1.295 $\pm$ 0.018 \\
M101-NGC~5462    &     29.2      $\pm$     1.5    &   180.      $\pm$  4.4    &  0.908   $\pm$  0.070     &  1.828    &  1.130   $\pm$  0.012      &  39.65    $\pm$    0.06 & 1.304 $\pm$ 0.018 \\
M101-NGC~5471    &     98.0      $\pm$     11.   &   256.      $\pm$  0.7    &  0.371   $\pm$  0.072     &  0.027    &  1.268   $\pm$  0.014       & 39.91    $\pm$    0.07 & 1.302 $\pm$ 0.018 \\
M33-NGC~588        &     271.      $\pm$     24.   &   75.  $\pm$  0.8    &  0.513   $\pm$  0.184     &  0.012    &  0.876   $\pm$  0.012          & 38.59    $\pm$    0.11 & 1.105 $\pm$ 0.011 \\
M33-NGC~592        &     173.      $\pm$     15.   &   50.      $\pm$  0.4    &  0.261   $\pm$  0.149     &  14.62    &  0.797   $\pm$  0.017      & 38.34    $\pm$    0.09 & 1.010 $\pm$ 0.010 \\
M33-NGC~595        &     652.      $\pm$     74.   &   68.      $\pm$  0.5    &  0.549   $\pm$  0.137     &  11.65    &  0.868   $\pm$  0.013      & 39.15    $\pm$    0.09 & 1.245 $\pm$ 0.016 \\
M33-NGC~604        &     1508.      $\pm$     141.  &   95.      $\pm$  1.0    &  0.266   $\pm$  0.136     &  0.007    &  1.132   $\pm$  0.013     & 39.22    $\pm$    0.09 & 1.269 $\pm$ 0.017 \\
M81-HK268       &     33.0      $\pm$     3.0    &   64.      $\pm$  4.7    &  0.973   $\pm$  0.245     &  0.199    &  0.877   $\pm$  0.021        & 39.19    $\pm$    0.13 & 1.130 $\pm$ 0.012 \\
M81-HK652       &     188.3      $\pm$     25.   &   67.      $\pm$  1.8    &  1.004   $\pm$  0.247     &  0.181    &  1.113   $\pm$  0.018        & 39.96    $\pm$    0.15 & 1.275 $\pm$ 0.017 \\
MRK116         &     16.4      $\pm$     1.2    &   88.      $\pm$  2.5    &  0.011   $\pm$  0.259     &  15.28    &  1.205   $\pm$  0.036         & 39.91    $\pm$    0.16 & 1.307 $\pm$ 0.018 \\
NGC~2366-HK110     &     22.0      $\pm$     3.9    &   114.      $\pm$  7.5    &  0.011   $\pm$  0.273     &  0.013    &  0.748   $\pm$ 0.018     &  38.48    $\pm$    0.17 & 0.968 $\pm$ 0.008 \\
NGC~2366-HK54      &     127.  $\pm$     14.   &   215.      $\pm$  3.1    &  0.589   $\pm$  0.073     &  0.061    &  0.970   $\pm$  0.022         &  39.27    $\pm$    0.08 & 1.177 $\pm$ 0.014 \\
NGC~2366-HK72      &     184.      $\pm$     16.   &   265.      $\pm$  7.7    &  0.014   $\pm$  0.072     &  0.080    &  0.953   $\pm$  0.019     &  39.40    $\pm$    0.08 & 1.196 $\pm$ 0.014 \\
NGC~2366          &     164.      $\pm$     15.   &   84.      $\pm$  7.2    &  0.016   $\pm$  0.090     &  0.064    &  0.929   $\pm$  0.054       & 39.35    $\pm$    0.08 & 1.202 $\pm$ 0.014 \\
NGC~2403-VS24      &     55.3      $\pm$     6.6    &   149.      $\pm$  3.9    &  0.421   $\pm$  0.159     &  0.100    &  0.960   $\pm$ 0.013     & 39.02    $\pm$    0.13 & 1.176 $\pm$ 0.014 \\
NGC~2403-VS3       &     75.7      $\pm$     8.7    &   87.      $\pm$  0.9    &  0.442   $\pm$  0.030     &  0.000    &  1.105   $\pm$  0.021     & 39.17    $\pm$    0.11 & 1.234 $\pm$ 0.015 \\
NGC~2403-VS44      &     46.2      $\pm$     6.7    &   125.      $\pm$  1.2    &  1.375   $\pm$  0.062     &  0.002    &  1.234   $\pm$  0.011    & 39.40    $\pm$    0.12 & 1.291 $\pm$ 0.018 \\
NGC~925-120        &     2.1      $\pm$     0.5    &   102.      $\pm$  4.4    &  0.631   $\pm$  0.127     &  3.289    &  0.950   $\pm$  0.048     & 38.63    $\pm$    0.12 & 1.030 $\pm$ 0.010 \\
NGC~925-128        &     8.0      $\pm$     1.0    &   118.      $\pm$  4.0    &  0.012   $\pm$  0.058     &  0.001    &  0.964   $\pm$  0.035     & 38.91    $\pm$    0.08 & 1.178 $\pm$ 0.013 \\
NGC~925-42         &     3.8      $\pm$     0.8    &   116.      $\pm$  3.7    &  0.011   $\pm$  0.015     &  0.011    &  0.901   $\pm$  0.017     & 38.58    $\pm$    0.11 & 1.126 $\pm$ 0.012 \\
NGC~4258-RC01    &     35.0      $\pm$     4.3    &   98.      $\pm$  6.6    &  0.679   $\pm$  0.254     &  2.437    &  0.913   $\pm$  0.046       & 39.71    $\pm$    0.13 & 1.210 $\pm$ 0.015 \\
NGC~4258-RC02    &     147.      $\pm$     32.   &   69.      $\pm$  0.7    &  0.295   $\pm$  0.096     &  0.591    &  1.218   $\pm$  0.032        & 40.14    $\pm$    0.11 & 1.329 $\pm$ 0.019 \\
NGC~4395-NGC~4399 &     9.8      $\pm$     2.5    &   47.      $\pm$  6.0    &  0.327   $\pm$  0.089     &  2.053    &  0.879   $\pm$  0.025       &  38.52    $\pm$    0.13 & 1.101 $\pm$ 0.012 \\
NGC~4395-NGC~4400 &     147.      $\pm$     41.   &   82.      $\pm$  1.0    &  0.014   $\pm$  0.060     &  7.096    &  1.061   $\pm$  0.045      &  39.62    $\pm$    0.13 & 1.297 $\pm$ 0.018 \\
NGC~4395-NGC~4401 &     11.1      $\pm$     1.9    &   73.      $\pm$  1.9    &  0.019   $\pm$  0.043     &  0.115    &  0.785   $\pm$  0.045     &  38.42    $\pm$    0.09 & 1.022 $\pm$ 0.010 \\
NGC~2541-A           &      7.1      $\pm$     0.5    &   94.      $\pm$  14.0   &  0.001   $\pm$  0.111     &  0.329    & 0.860    $\pm$  0.009   & 39.25    $\pm$    0.07 & 1.225 $\pm$ 0.034 \\
NGC~2541-B           &      15.5      $\pm$     0.1    &   98.      $\pm$  10.0   &  0.002   $\pm$  0.080     &  0.015    & 0.980    $\pm$  0.015  & 39.42    $\pm$    0.06 & 1.243 $\pm$ 0.054 \\
NGC~2541-C           &      13.9      $\pm$     0.1    &   73.      $\pm$  6.7    &  0.001   $\pm$  0.030     &  0.278    & 0.940    $\pm$  0.017  & 39.50    $\pm$    0.05 & 1.261 $\pm$ 0.062 \\
NGC~3319-A           &      21.6      $\pm$     0.1    &   114.      $\pm$  4.7    &  0.396   $\pm$  0.221     &  0.329    & 1.090    $\pm$  0.090 & 39.93    $\pm$    0.13 & 1.326 $\pm$ 0.030 \\
NGC~3319-B           &      17.5      $\pm$     0.1    &   80.      $\pm$  4.9    &  0.000   $\pm$  0.106     &  0.153    & 0.990    $\pm$  0.031  & 39.66    $\pm$    0.08 & 1.286 $\pm$ 0.011 \\
NGC~3319-C           &      3.9      $\pm$     0.2    &   123.      $\pm$  17.0   &  1.440    $\pm$  0.251     &  0.132    & 0.910    $\pm$  0.085 & 39.68    $\pm$    0.26 & 1.221 $\pm$ 0.070 \\
NGC~3198-A           &      15.0      $\pm$     1.2    &   90.      $\pm$  11.0   &  0.015   $\pm$  0.141     &  0.327    & 0.950    $\pm$  0.077  & 39.73    $\pm$    0.09 & 1.238 $\pm$ 0.028 \\ \hline
\end{tabular}
 \label{tablaPar}
 \end{table*}
\end{center}
 \begin{figure}

\includegraphics[scale=0.375]{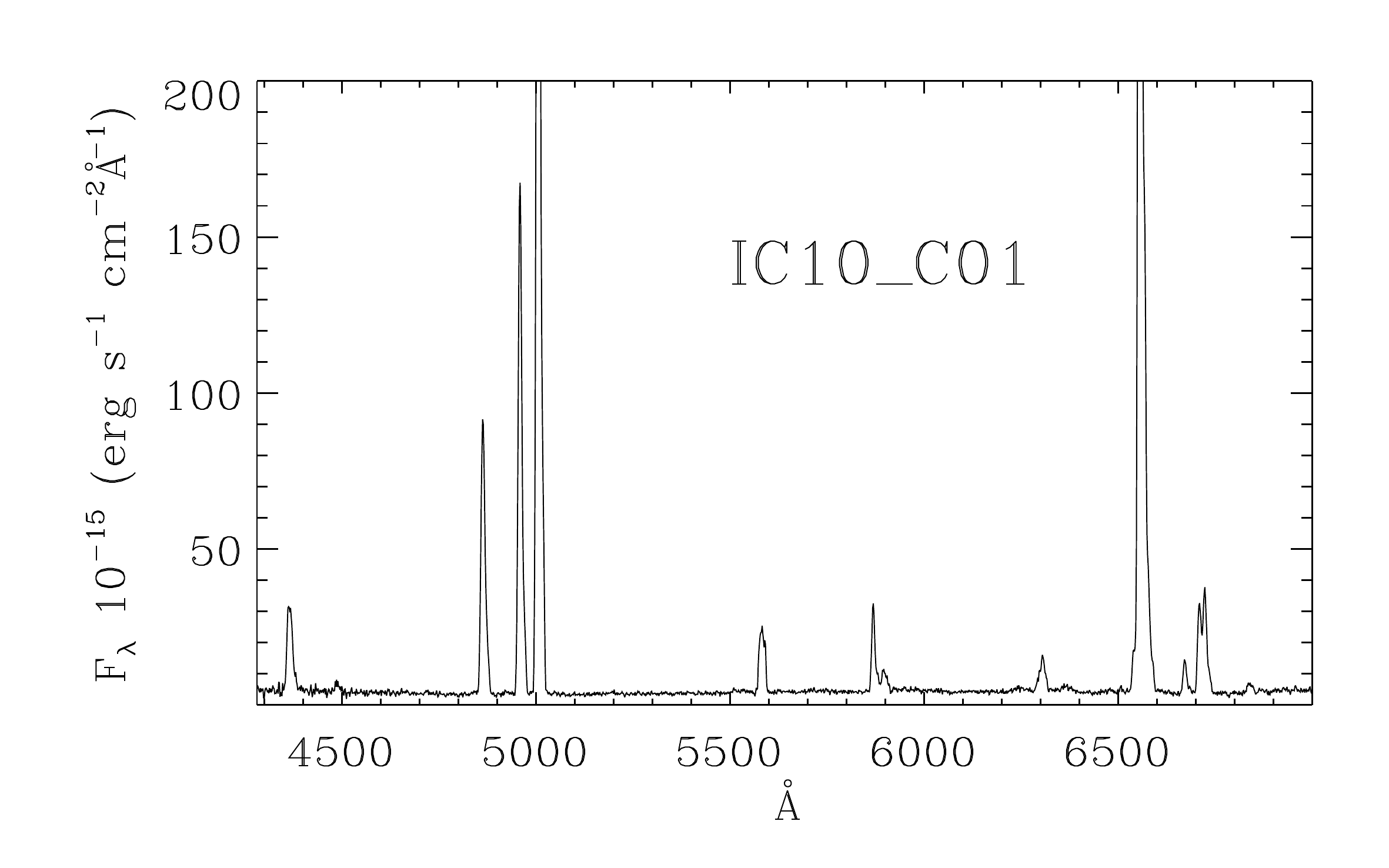}
  \caption{IC10 C01 low-resolution spectrum obtained at  OAN-SPM.}
   \label{LRSpec}
  \end{figure}

 \begin{figure*}
  \centering
  \includegraphics[width=0.4\textwidth]{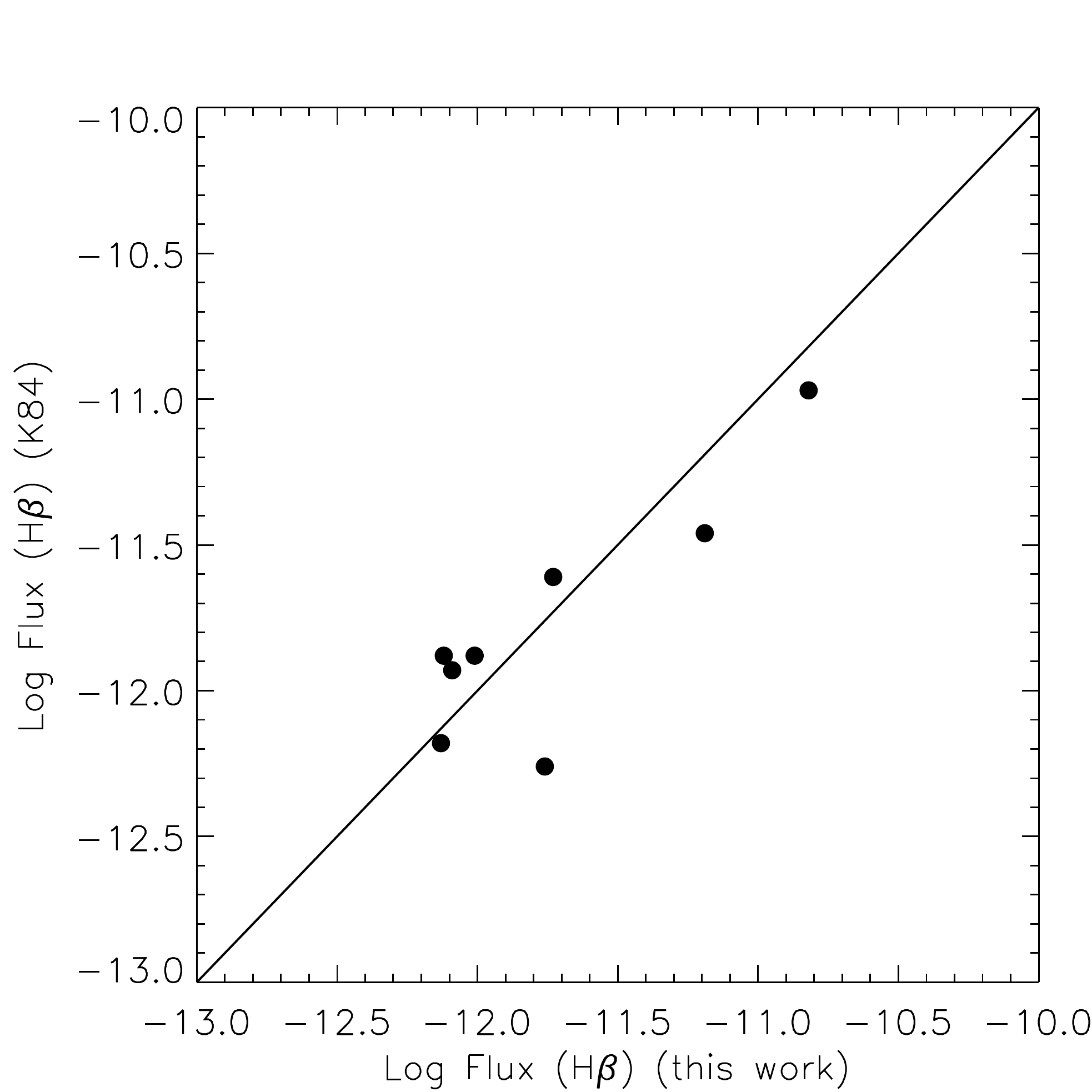}
  \includegraphics[width=0.4\textwidth]{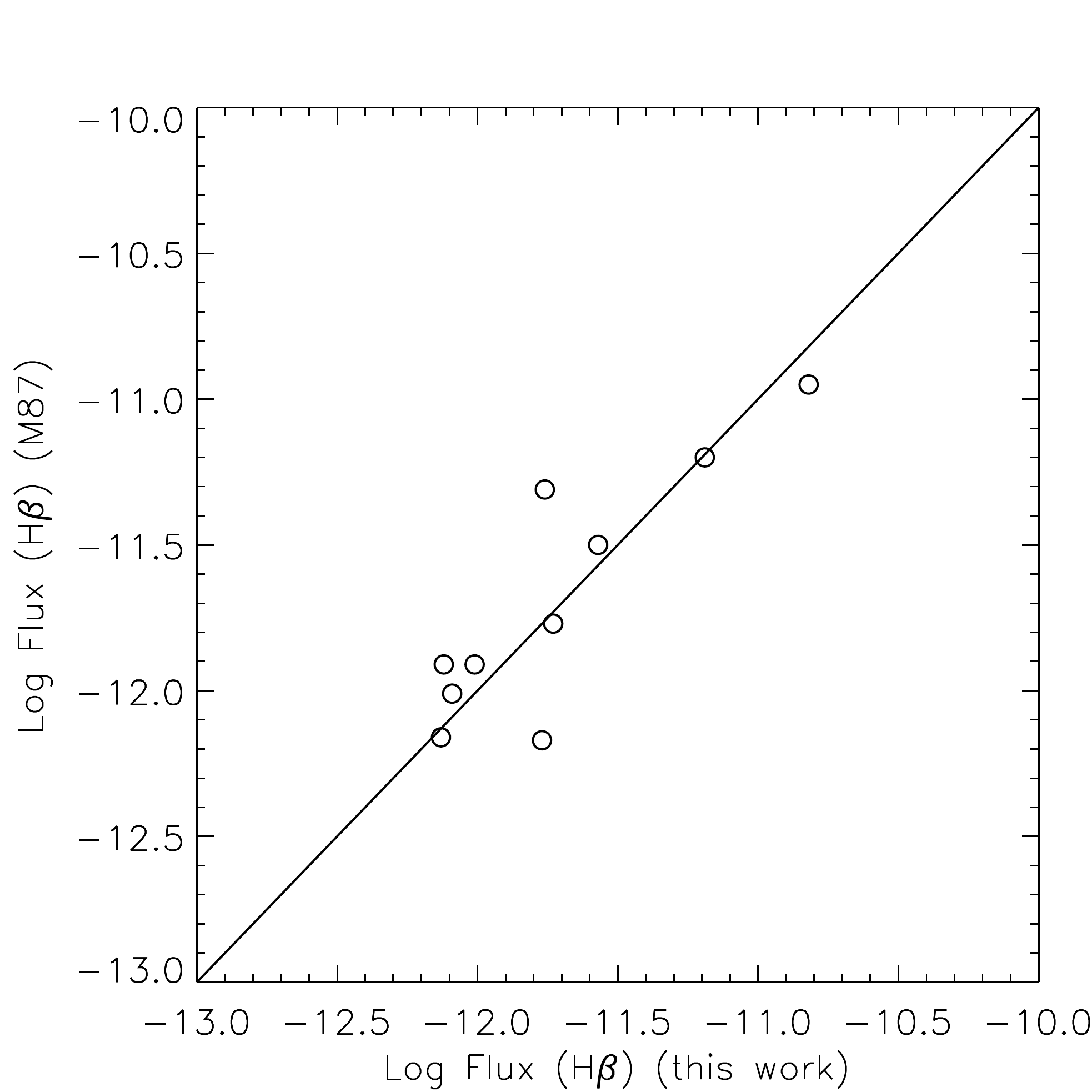}
   \includegraphics[width=0.4\textwidth]{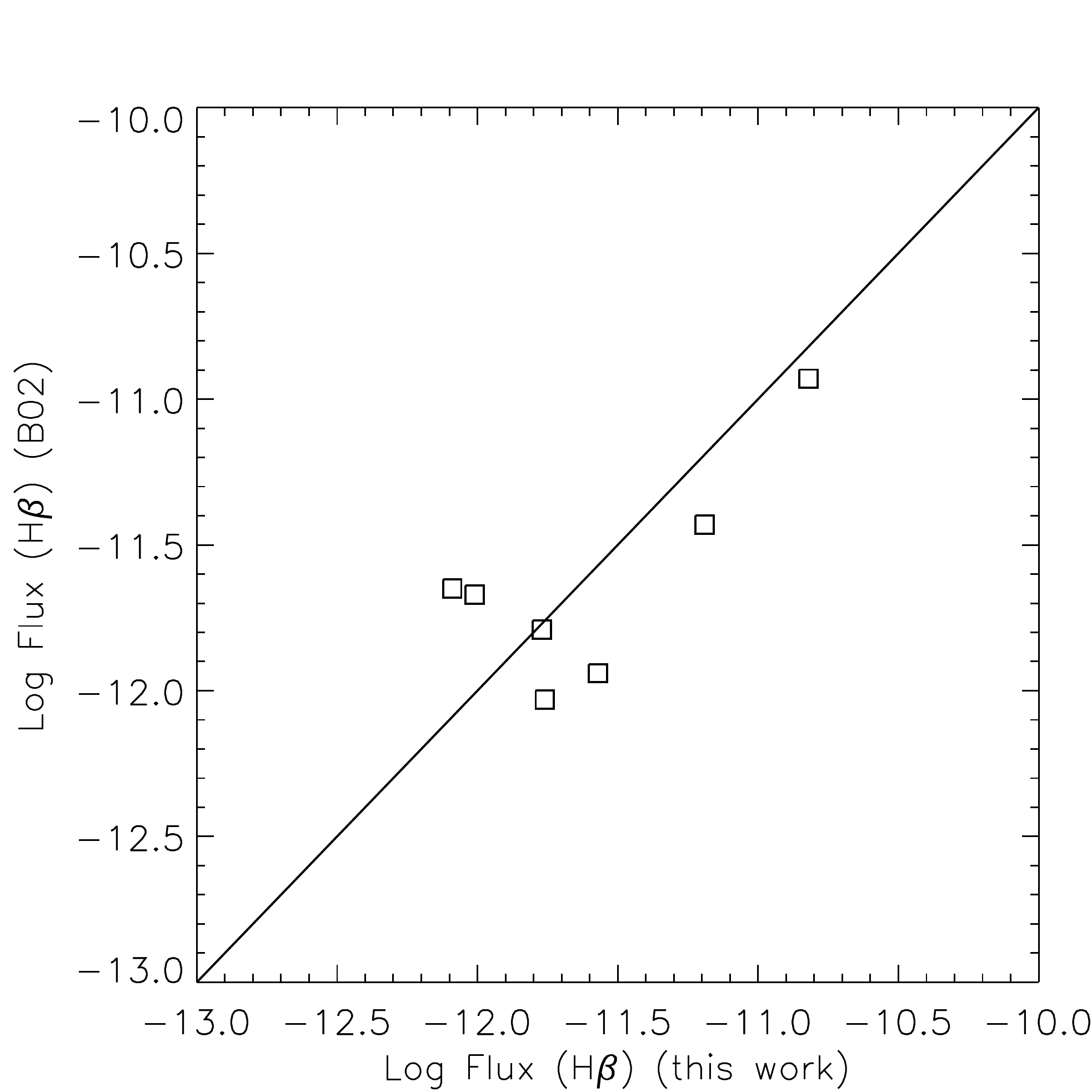}
  \includegraphics[width=0.4\textwidth]{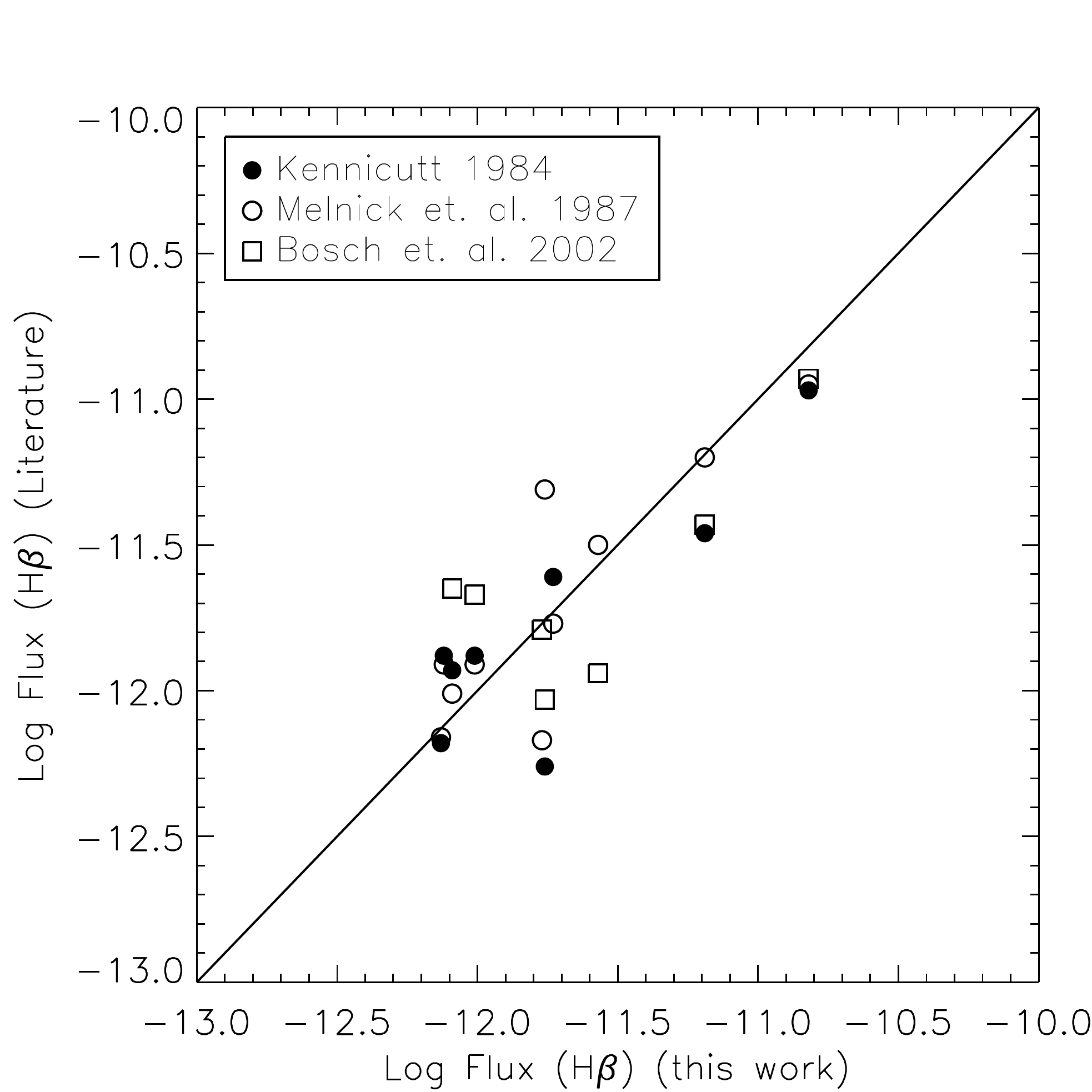}
  \caption{Comparison between our photometry and  \protect\cite{Kennicutt1984} (top left) , \protect\cite{melnick1987} (top right)  and  \protect\cite{bosch2002} (bottom left). . All together are shown at the bottom right panel.  }
  \label{photF15M77}
 \end{figure*}

\subsubsection{Low-resolution spectrophotometry}

We obtained the integrated emission-line fluxes from wide slit low resolution spectrophotometry. When necessary, we converted the fluxes of H$\beta$ to H$\alpha$ using the theoretical ratio for case B recombination.

The low-resolution spectroscopy, used for measuring the emission lines flux, was obtained using similar Boller \& Chivens Cassegrain spectrographs at two 2.1-m  telescopes with similar configurations in long-slit mode. The telescopes are located at the Observatorio Astron{\'o}mico Nacional (OAN-SPM) in San Pedro M{\'a}rtir (Baja California) and at the Observario Astrof{\'i}sico Guillermo Haro (OAGH)  in Cananea (Sonora), both situated in northern M{\'e}xico.
The observations at OAGH were performed using a 150 gr mm$^{-1}$ grating with a blaze angle of 3$^{\circ}$25$\arcmin$\ centred at $\lambda \sim 5000$\AA{} and a slit-width of 9 arcsec. The data from OAN were obtained using a 400 gr mm$^{-1}$ grating with a blaze angle of 6$^{\circ}$30$\arcmin$, the grating was centred at  $\lambda \sim 5850$\AA{} and the slit-width  was 13 arcsec.
At least three spectrophotometric standard stars were observed each night, and  at least one GHIIR  was repeated every night in order to concatenate the different observing runs. The objects were observed at small zenith distances. All nights reported here were  photometric; 
the seeing in  most nights varied between 1.1 and 1.4 arcsec.

The spectra  were reduced using the standard procedure in IRAF\footnote{IRAF is distributed by the National Optical Astronomy Observatories, which are operated by the Association of Universities for Research     in Astronomy, Inc., under cooperative agreement with the National Science Foundation.}.
The spectrophotometric  standard stars observed were  Feige 110, G191-B2B, BD+28, G158-100, Hz4, Feige 34, Feige 66 and BD+33. A typical low resolution spectrum obtained at OAN-SPM is shown as an example in  Figure \ref{LRSpec}.

 \subsubsection{Comparison with previous  work}\label{mh0data}
 
We have compared our wide aperture spectrophotometry with published aperture photometry  from \cite{melnick1987}, \cite{Kennicutt1984} and \cite{bosch2002} hereinafter M87, K84, B02 respectively, for the giant HII regions  NGC~ 588, NGC~ 592, NGC~ 595 and NGC~ 604 in M33 and NGC~ 5447, NGC~ 5461, NGC~ 5462 and NGC~ 5471 in M101.

K84 obtained photoelectric \halpha\ photometry using single channel photometers on the Kitt Peak 0.9m, CTIO 0.6m and the Manastash Ridge 0.8m telescopes with  20 \AA\  FWHM
interference filters for line and continuum. The observations were performed through apertures large enough to include the outer edge of the \hii\ region.  As K84 provides no estimate of the extinction  we have assumed negligible  extinction at \halpha\ and transformed K84  fluxes to \hbeta\  scaling them by the theoretical Balmer decrement (2.86).

Observations by M87  were made with the 1.52-m
telescope of the Observatorio Astron�mico Nacional at Calar Alto, Spain, using 100 \AA\  FWHM
interference filters to define the line and continuum photometric bands. M87 used an RCA C31034 Ga-As photomultiplier and at least three concentric apertures, of  which the biggest was always larger  than the halo of the \hii\ region
With these apertures a curve of growth was constructed to estimate the contamination by diffuse emission in the host galaxy and the total emission line flux.

B02  measured the total fluxes  of \hbeta\ and \halpha\ emission lines from CCD narrow band  images  obtained at the 1.0-m Jacobus Kapteyn Telescope at the Observatorio del Roque de los Muchachos in the Canary Islands. The GHIIRs and the flux standard stars, were
observed using four narrow-band filters (FWHM$\sim $50\AA ) centred at  \hbeta\ and \halpha\  and their adjacent continua.  B02 \hbeta\ fluxes are not as reliable as their  \halpha\ ones (see the original paper), so we have compared our photometry with B02 \halpha\ fluxes scaled by the theoretical Balmer decrement.

Figure \ref{photF15M77} shows the comparison of our photometry with that of K84 (top left), M87 (top right), B02 (bottom left); using the same symbols all are plotted together  in the bottom right panel. 

Regarding  the comparison of our fluxes with those from K84,  except for NGC~ 592 in M33 there is a reasonable agreement inside $\pm$ 0.25 dex.

The comparison of the \hbeta\ fluxes shows a  $\pm$ 0.2 dex concordance with  M87 except for NGC~592 in M33 and  NGC~ 5447 in M101. These  regions are known to have  multiple `knots'  so the discrepancy could be related to differences in the pointing. The case for NGC~5447 is reinforced by the fact that also the emission line width observed by M87 in this GHIIR shows a discrepancy with our value. In both parameters, emission line flux and line width M87 show lower values than our present observations.

It is important to note that for NGC~592 the average  between K84 and M87 measurement differs only 0.05 dex from our measurement.
The comparison with B02 shows slightly higher scatter.

We can conclude that the comparison illustrated in Figure \ref{photF15M77} shows no systematic trends with the data from the literature and illustrates the difficulty in performing integrated photometry in extended \hii\ regions.

  
 \subsubsection{High-resolution spectroscopy}
 
 Given that typical values of the velocity dispersion  of GHIIRs are in the region of 10-30 km/s high resolution spectrographs are needed to accurately measure their emission line widths. To this end high resolution spectra were obtained using echelle spectrographs at the observatories  OAN and OAGH.

At the  OAN we used the  Manchester echelle spectrometer  \citep[][MEZCAL]{meaburn2003}  a
long-slit nebular  echelle high resolution spectrograph built to obtain spatially-resolved profiles of individual emission lines from faint extended sources. MEZCAL operates in the wavelength range 3900\textendash9000 \AA{} with a spectral resolving power $\mathcal{R}\sim100,000$. This echelle spectrograph has no cross-disperser so it isolates single orders using interference filters. We used a 90\AA{} bandwidth filter to isolate the 87th order containing the H$\alpha$ and [N II] nebular emission lines with $\lambda_{c}=$ 6575\AA{}.
The observations were performed with a 70$\mu$m (0.95 arsec) slit corresponding to a velocity resolution of $\sigma_{inst}=6.0$~km s$^{-1}$. Two pixel binning was applied in both the spatial and spectral directions. 

The  Cananea High-Resolution Spectrograph (CanHis) is a high spatial and very high spectral resolution echelle spectrograph $\mathcal{R}\sim140,000$ at the 2.1m telescope at OAGH. Like MEZCAL, CanHis utilizes medium-band interference filters to isolate individual  orders \citep{hunten1991}. We used the filter centred in  H$\alpha$, $\lambda_{c}=$ 6563\AA{}, covering a bandwidth of 90\AA{}. The observations were performed with a slit width of 50.7$\mu$m (0.45 arsec) resulting in a velocity resolution 
$\sigma_{inst}=3.0$~km s$^{-1}$.

Not having a cross disperser, both MEZCAL and CanHiS are very  efficient instruments.

The data were reduced  using standard IRAF tasks. The wavelength calibration and the instrumental resolutions were obtained using  an internal U-Ne lamp in  CanHis, and a Th-Ar lamp for  MEZCAL.

\begin{figure*}
  \includegraphics[width=0.45\textwidth]{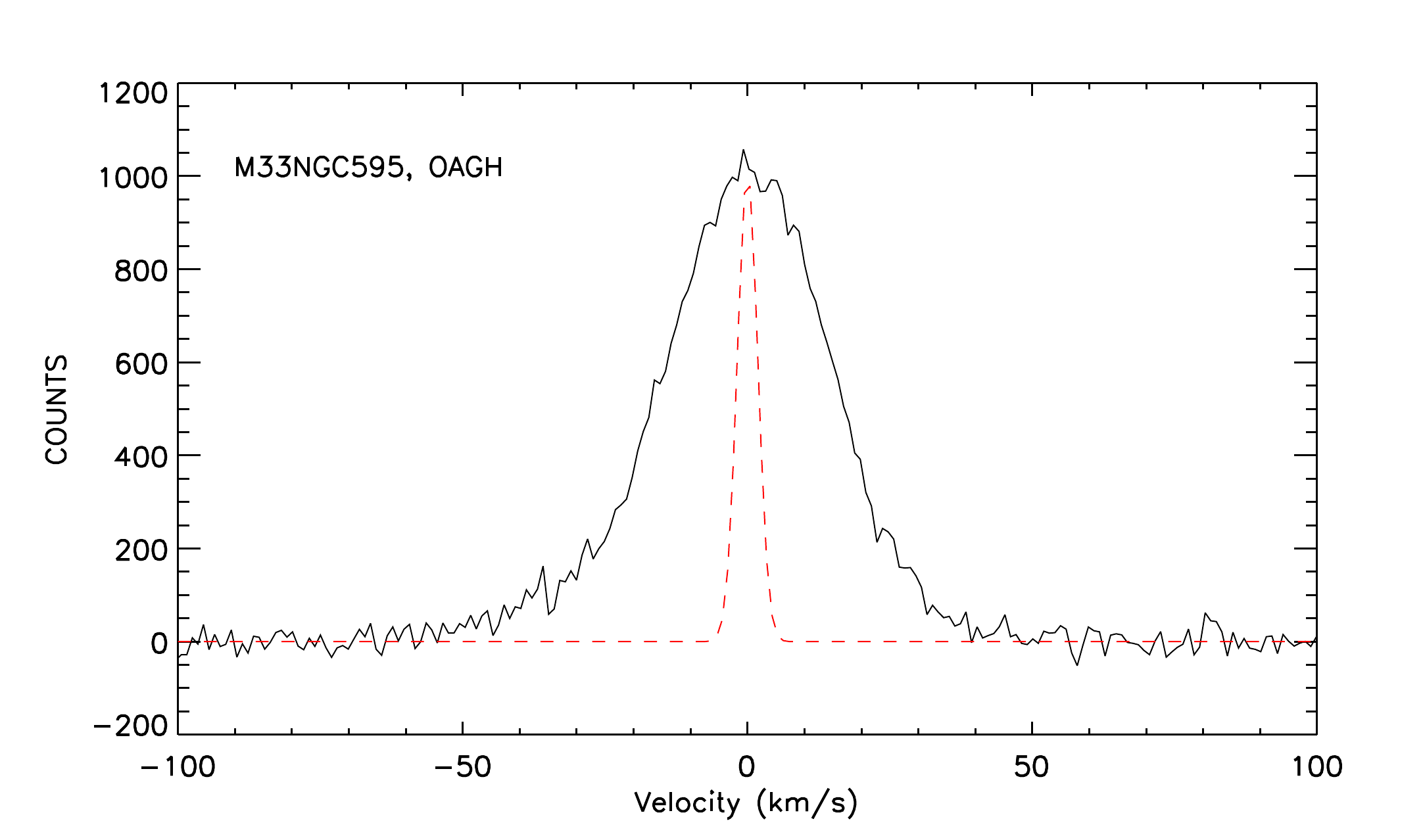}
  \includegraphics[width=0.45\textwidth]{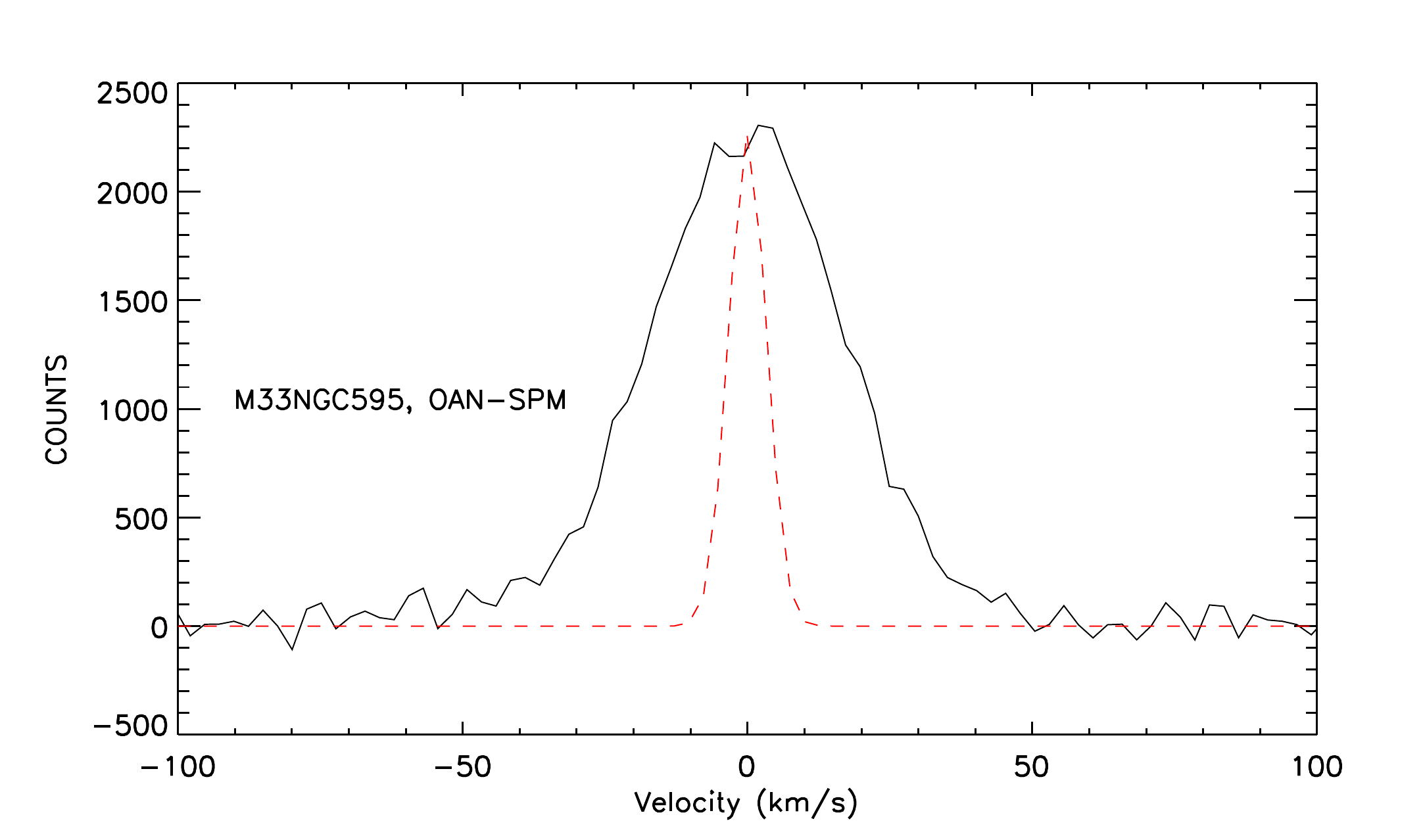}
  \caption{Spectrum of NGC~ 595 using OAGH CanHis (left) and OAN-SPM MEZCAL (right) in H$\alpha$. The  dashed (red) line is the fit to the instrumental profile obtained for the  calibration lamps at 6585.3\AA{}{} and 6583.9\AA{}  at OAGH and OAN-SPM  respectively.}
   \label{HRSpec}
  \end{figure*}

\begin{figure}
\includegraphics[width=0.48\textwidth]{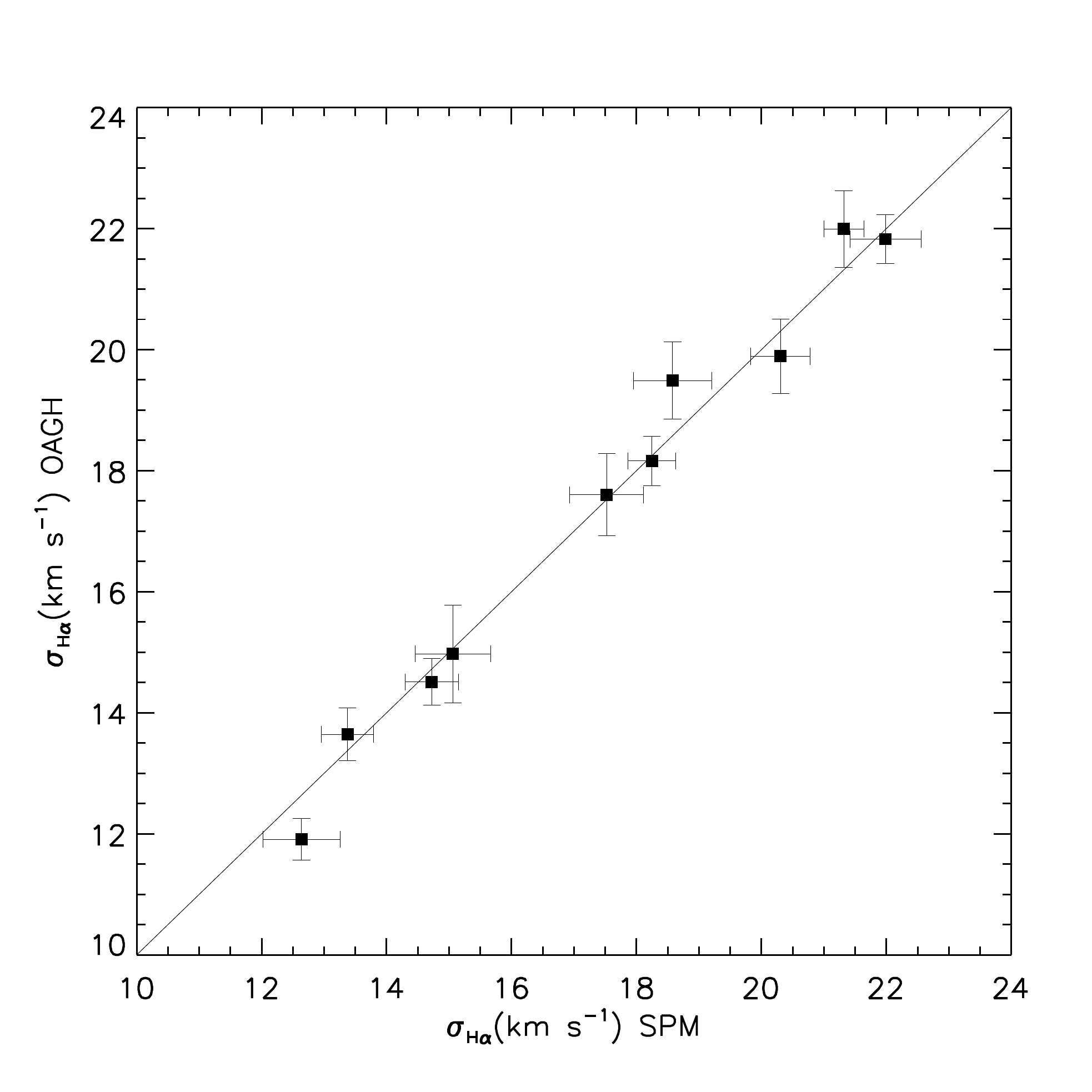}
   \caption{Velocity dispersion for the 10 objects observed with both OAGH CanHis  and OAN-SPM MEZCAL.  }
    \label{CanMez}
  \end{figure}

Repeated observations of ten targets  were obtained with MEZCAL and CanHiS in order to estimate observational errors and night-to-night variations  and to compare the performance of both instruments. NGC~~595 (a GHIIR in M~33)  high resolution spectra  are shown as an example in  Figure \ref{HRSpec}.  Figure \ref{CanMez}  shows a comparison of the $\sigma$ values obtained with both instruments. 

\subsubsection{Line profiles }

A commonly used  method  to study the distribution of gas and its kinematic properties is based on  fitting the profile of the emission lines with Gaussians, although asymmetric non-Gaussian profiles in the emission lines are frequently found in the literature (e.g. \citealt{Bordalo2011, hagele2013, Chavez2014}).
To compensate,  multiple Gaussian curves or Lorentzian functions are fitted, in  which case the information contained in the wings might be lost or  strongly dependent on the adopted  initial values.

The  presence of weak extended non-Gaussian wings  in the profiles of the emission lines may introduce a
small systematic effect in the determination of the FWHM. This effect may be  associated with stellar winds and multiple simultaneous starbursts. This can equally affect the determination of the FWHM of HIIGs and GHIIRs and hence, the distance estimator used to calculate \ho .

The alternative strategy of eliminating objects that have multiple profiles or extended wings in the profiles seems risky because multiplicity will sometimes appear as a structure in the velocity dispersion and sometimes as brighter spots depending on the relative radial velocity of the regions that are superposed along the line of sight. 
So for the distance indicator it seems safer to  include all the objects in the sample, paying the price of  a larger scatter in the correlation as in \cite{Bordalo2011} and in \cite{Chavez2014}.

 Our approach for the present work is to obtain the FWHM of  H$\alpha$ from the high-resolution spectra and to fit  both a single Gaussian and  a Gauss-H\'ermite series to the line profile. 
 
\begin{figure}
 \includegraphics[width=0.48\textwidth]{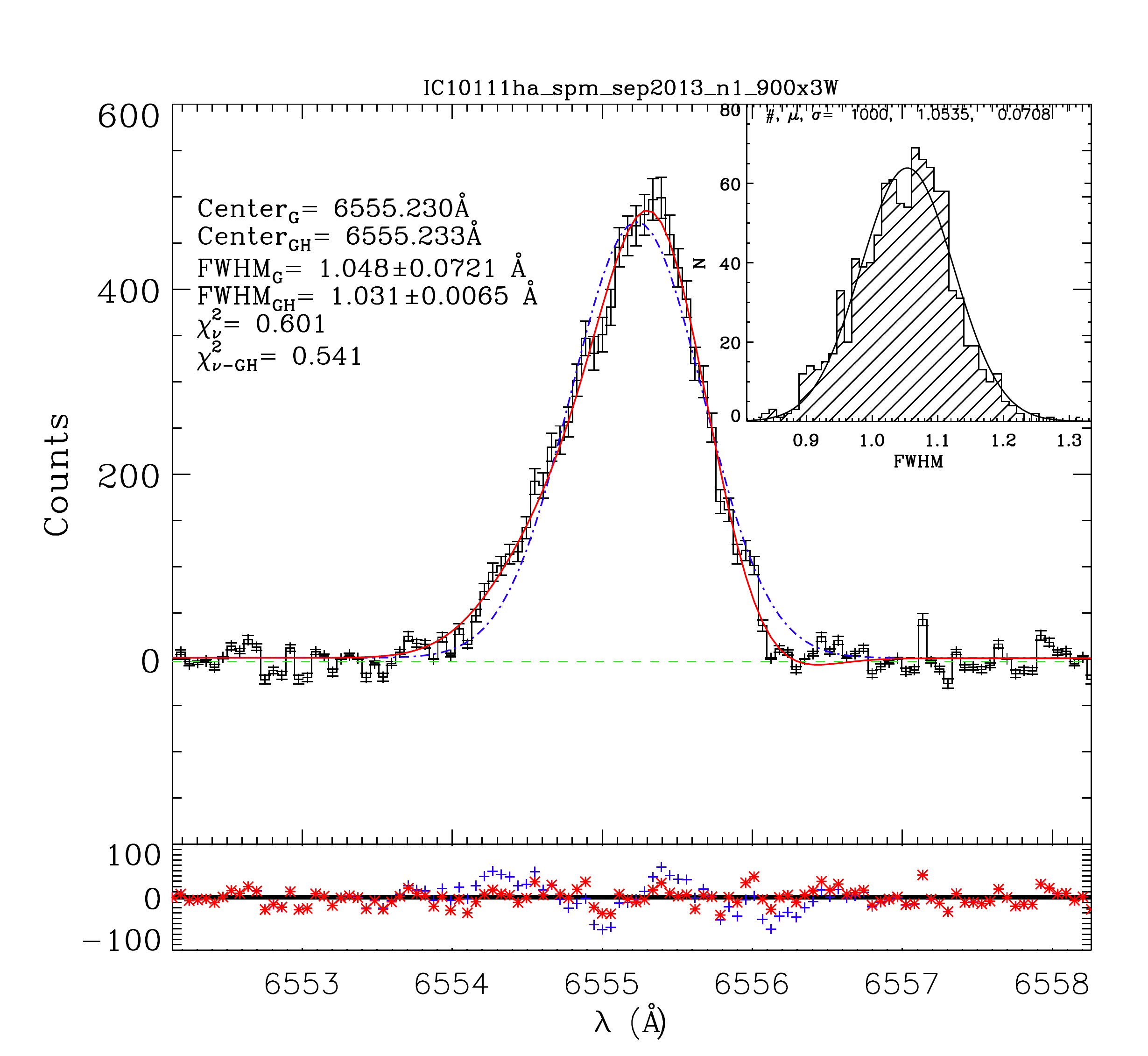}
    \caption{Fit of the H$\alpha$ profile using a Gaussian (blue dashed line) and a Gauss-H\'ermite fit (continuous red).
   The residuals are shown in the lower panel with the same colour code. The inset shows the results of a Monte Carlo simulation to estimate the errors in the parameters of the best fit. }
   \label{HRSpecFit}
   \end{figure}

\begin{figure}
 \includegraphics[scale=0.34]{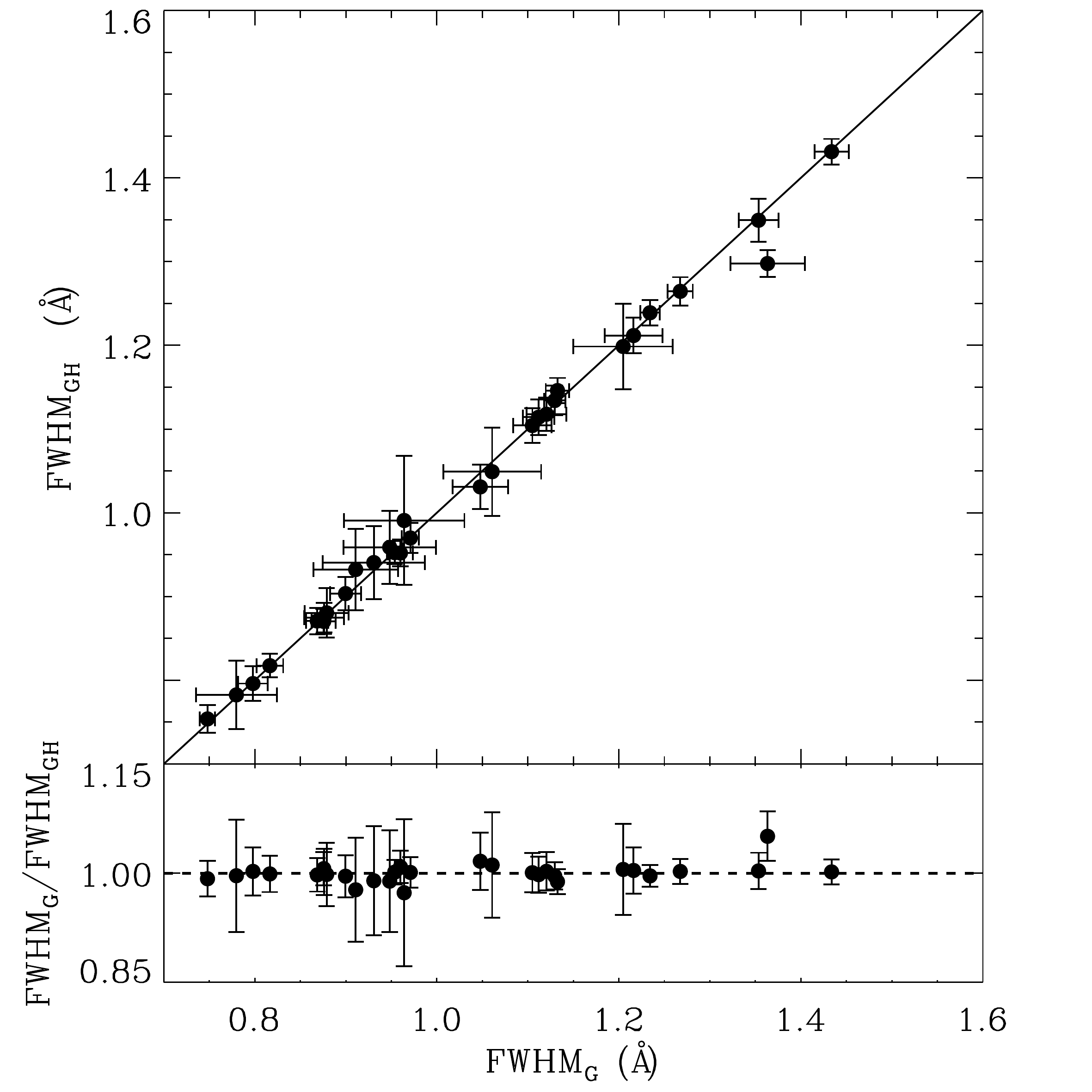}
    \caption{FWHM obtained applying  Gauss vs.~Gauss-H\'ermite fits. Lower panel: the ratio of the FWHM obtained using these fits. }
     \label{GussHer}
   \end{figure}

 The Gauss-H\'ermite function  preserves the information of the velocity of the gas  by fitting the wings of the emission-line profiles. It has the added  advantage that this fitting can be implemented in a hands-off  routine, by varying the moments of the function automatically \citep{riffel2010}. An example profile comparison using both  methods  is shown in  Figure \ref{HRSpecFit}.  For most of our data the profiles are not far from  Gaussian and therefore the estimates of the FWHM by both methods  are not very different. In  Figure \ref{GussHer}  we show the comparison of the FWHM obtained by fitting Gauss or Gauss-H\'ermite functions to the line profiles. Both methods are equivalent within the errors indicating that the presence of wings or slight asymmetries in the sample does not affect the measurement of the FWHM of the emission lines.

 The observed velocity dispersions  ($\sigma_{obs}$)  are then  corrected by thermal ($\sigma_{t}$) and instrumental ($\sigma_{i}$) broadening, thus the intrinsic velocity dispersion is given by: 
 \begin{equation}
  \sigma=\sqrt{\sigma_{obs}^{2}-\sigma_{i}^{2}-\sigma_{t}^{2}}
 \end{equation}
 where the thermal broadening was calculated assuming a Maxwellian velocity distribution of the hydrogen ion, from the equation:
 \begin{equation}
  \sigma_{t}=\sqrt[]{\frac{kT_{e}}{m}}
 \end{equation}
 where $k$ is the Boltzmann constant, $m$ is the mass of the ion and $T_{e}$ is the electron temperature in  Kelvin.   The instrumental broadening is $\sigma_{i}$= 3 and 6  km s$^{-1}$ for data obtained with CanHiS and MEZCAL respectively.


  \subsubsection{Comparison with previous  work}

\begin{figure}
\includegraphics[width=0.48\textwidth]{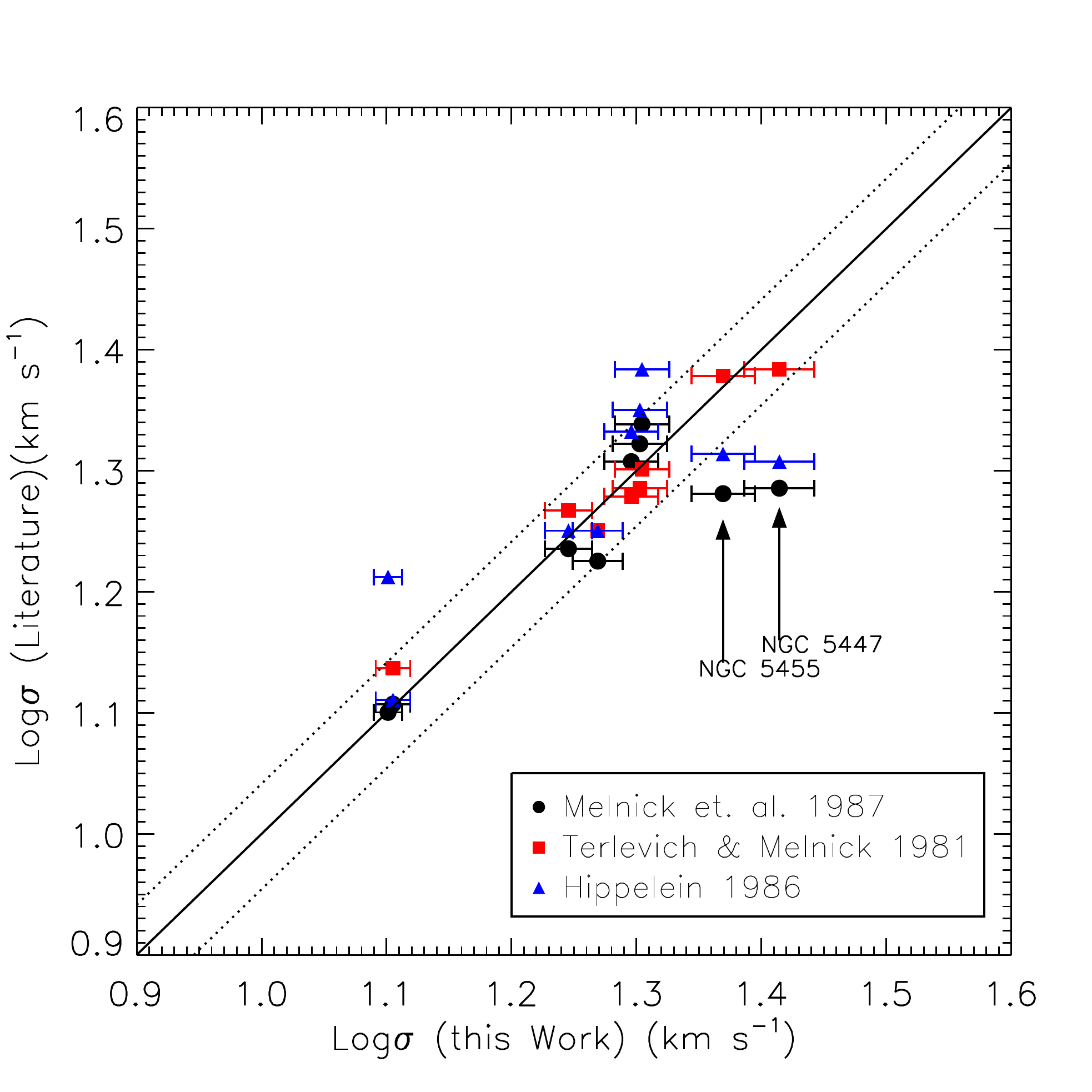}
   \caption{Comparison of the velocity dispersion for the eight GHIIRs in common with M87, H86 and TM81. The dotted lines indicate $\pm\ $10\%.}
    \label{sigmalite}
  \end{figure}

Eight of our GHIIRs have published  determinations of their velocity dispersion. Figure \ref{sigmalite} shows the comparison of our measurements with M87,  TM81 and \cite{hip86} (hereinafter H86). The dotted lines indicate concordance within  10\%. The points with the greatest discrepancy correspond to  NGC~~5447 and NGC~~5455 in M~101. NGC~~5447 has multiple `knots' that could  have lead to a different pointing when the M87 data was obtained as discused in Section \ref {mh0data}. 
On the other hand our values for NGC~~5447 and NGC~~5455  agree within 10\%\ with those reported by TM81. It is interesting to note  that while TM81(as this work) uses only the Balmer line FWHM measurements, M87 uses the average between the [OIII] and Balmer FWHM. Because the [OIII] lines are systematically narrower than the Balmer emission lines, see \cite{Melnick2017},   it is expected that M87 FWHM values will be  systematically  smaller than the pure \halpha\ ones.

 From this comparison we can conclude that there is  agreement in the values of the FWHM inside $\pm $10\%\ with TM81 and also with M87 except  for NGC~ 5447 and NGC~ 5455 in M101. The comparison with H86 however shows a larger scatter with five GHIIRs outside the $\pm $10\%\  band out of eight GHIIRs in common.

 
\subsection{\hii\ galaxies} 

We use the data for 107  nearby (z$\leq$0.16) HIIGs defined by \cite{Chavez2014}  as their `best dataset';  we will refer to it as the Ch14 sample obtained, as the GHIIRs in the present work, both with low and high spectral resolution in order to measure the total fluxes and the emission line profiles respectively. The high resolution observations were performed with the HDS on Subaru, and UVES on the VLT and the low resolution ones with the Mexican 2m telescopes at OAN-SPM and OAGH.

The line fluxes in \cite{Chavez2014} were measured  through a very wide (typically $8''$) entrance slit.  As the HIIGs  in the sample are quite compact (typical diameters of less than $5''$)  the use of $8''$ wide slits guarantees a complete sampling of the line emitting region. These large aperture observations have a systematic effect with respect to the SDSS spectrophotometry for those objects with sizes larger than the $3''$  fibre aperture of SDSS.
Here we use  the \cite{Chavez2014}  measurements to compute the H$\beta$ luminosities  and  compare them with  those resulting from the SDSS photometry.
 
The \cite{Chavez2014} HDS observations were taken through a $4''$ slit, which nominally corresponds to an instrumental velocity dispersion $\sigma_{inst}=12.3~$km/s, although that results in an instrumental profile that is flat-top and not Gaussian. The UVES observations were obtained through a $2''$ slit corresponding to a nominal instrumental resolution of $\sigma_{inst}=4.65$~km/s, and again the instrumental profiles are  flat-top. Although the seeing during the observations was substantially better than $2''$, the sizes of most objects are larger than the slit width, so no elaborate procedure was required for the instrumental corrections as in the case of HDS.  Notice, however, that both for UVES and HDS the box-shaped instrumental profiles imply that the emission-line profiles have non-Gaussian cores that can be  appreciated in the residuals of the Gaussian fits (\citealt{Chavez2014}).

As the  SDSS Petrosian diameters of most objects in the Ch14 sample are about $3''$ to $5''$, the instrumental resolution in the HDS spectra under normal seeing conditions  is biased by the surface brightness profile of the objects. Although this effect is difficult to  quantify with the available data,  \cite{Chavez2014} corrected their observations for instrumental broadening using an ``equivalent" SDSS Petrosian diameter, with reasonably good results.

\section{Extinction and underlying absorption}\label{extabs}

Massive bursts of star formation are embedded in large amounts of gas and dust and this dust is responsible for the extinction of light in the line of sight due to absorption and scattering. The amount of extinction can be estimated using hydrogen recombination lines  through  the Balmer decrement, although contamination by  underlying stellar Balmer absorption lines changes the ratio of observed emission lines such that the internal extinction can be overestimated.
 
We have  derived the `true' visual extinction and determined the underlying Balmer absorption using two different extinction laws, the one by  \cite{calzetti2000} which has been widely used for massive starburst galaxies and the one by \cite{Gordon2003} that corresponds to the LMC2 supershell near the prototypical GHIIR 30 Doradus in the LMC. Notice that since the photometric errors in the \hbeta\ fluxes of the HIIGs  in \cite{Chavez2014} are  small, the errors in their luminosity are dominated by the errors in the extinction correction. As in our previous papers \citep[see e.g.][]{Chavez2014}, the uncertainties in the fluxes and equivalent widths have been estimated using the  expressions from  \citet{tresse1999}. When more than one measurement was available the fluxes were calculated using a mean weighted by the errors. 
  
\begin{figure}
  \centering
  \includegraphics[width=0.48\textwidth]{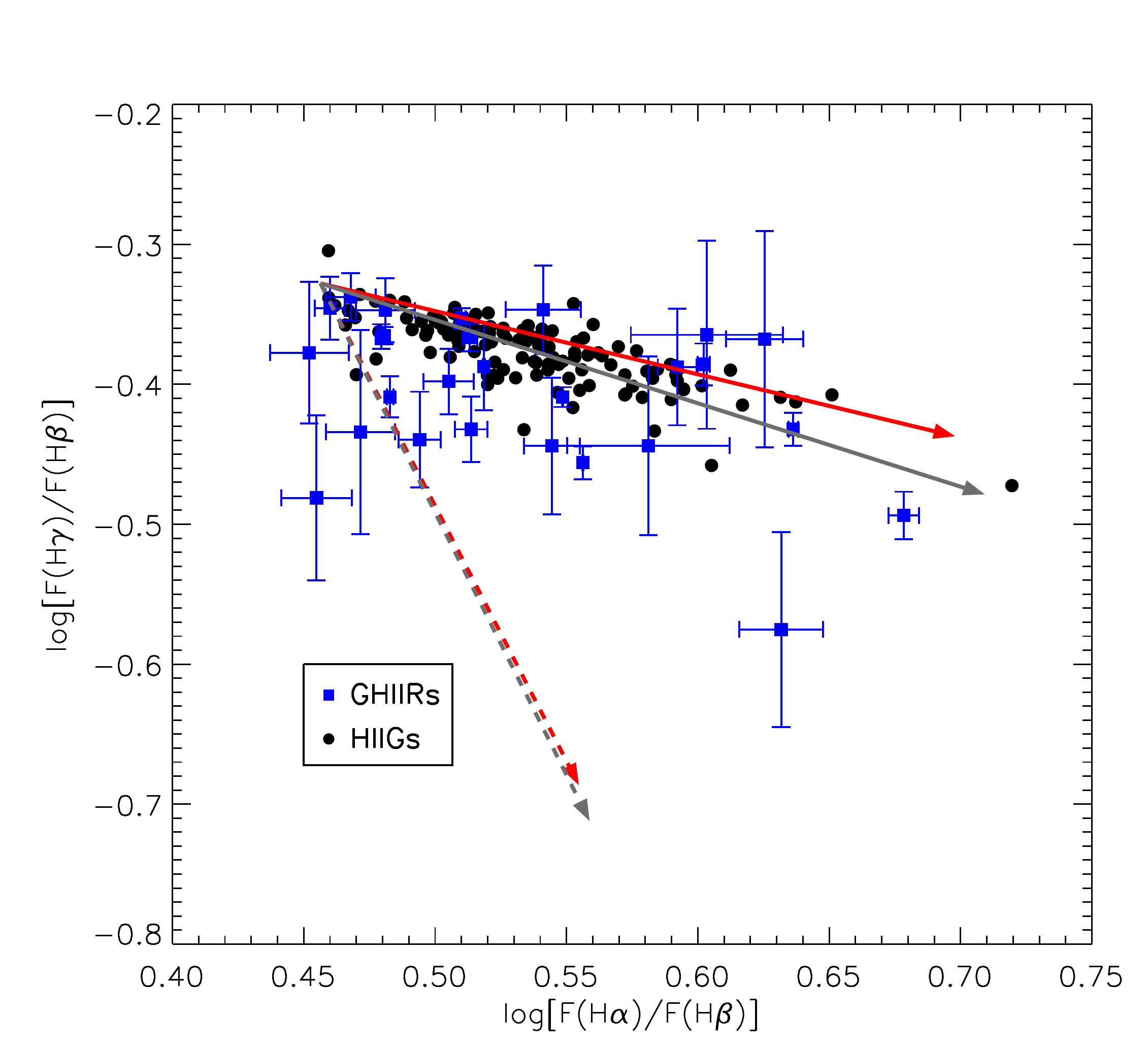}
  \caption{ Balmer decrement plane $\log (F(H\alpha)/F(H\beta))$ vs. $\log (F(H\gamma)/ F(H\beta))$. The vectors indicate the direction of shifts due to extinction [solid lines for two different extinction laws (red: \protect\citet{Gordon2003}, gray: \protect\citet{calzetti2000})] or to underlying absorption (dashed line) from the intrinsic values given by recombination.}
  \label{Extin}
 \end{figure}

To correct for extinction we used a modification of  the Balmer decrement method. We corrected the Balmer line emissions for the effect of stellar absorption lines  using the technique proposed by \citet[][see equations therein]{rosa2002}. This method allows us to obtain simultaneously the values of $Q$ and $A_{V}$ that correspond to the underlying absorption and the visual extinction respectively. In  Figure \ref{Extin} we represent the Balmer decrement plane $\log (F(H\alpha)/F(H\beta))$ vs. $\log (F(H\gamma)/ F(H\beta))$. In the absence of underlying absorption all points should be distributed along the extinction vector, while in the absence of extinction all points should be distributed along the underlying absorption line. We can see from the figure that  most of our objects fall, within the errors, in the region close to the reddening vector and to the right of the underlying absorption vector.

We calculated $A_{V}$ and $Q$ using the theoretical ratios for case B recombination $F(\mathrm{H}\alpha) / F(\mathrm{H}\beta) = 2.86$ and $F(\mathrm{H}\gamma) / F(\mathrm{H}\beta) = 0.47$ \citep{ost89}. We measured the Balmer lines from the SDSS spectra, and propagate the uncertainties by a Monte Carlo procedure.  Errors in the luminosities are dominated by uncertainties in the correction for extinction. The dereddened fluxes were obtained from the expression:
 \begin{equation}
  F_{0}(\lambda)=F_{obs}(\lambda)10^{0.4A_{V}k(\lambda)/R_{V}}
 \end{equation}
 where $k(\lambda)=A(\lambda)/E(B-V)$ is given by the extinction law either  \citet{calzetti2000} or \citet{Gordon2003}, and $R_{V}=A_{V}/(B-V)$ is the optical total-to-selective extinction. 
 We adopted  $R_{V}=4.05 $  and $2.77 $ from 
  \citet{calzetti2000} and \citet{Gordon2003} respectively.
 
 Finally the dereddened fluxes were corrected by the underlying absorption \citep[equation from][]{rosa2002}:
 \begin{equation}
  F(\lambda)=\frac{F_{0}(\lambda)}{1-Q}
 \end{equation}

\section{Evolution correction}\label{age}

In young SSCs, as the ones in our sample, the UV luminosity and therefore their emission line luminosity decays rapidly in less than 7 million years while the  optical luminosity remains relatively constant or increases due to the evolution of the cluster main sequence to lower temperatures and smaller bolometric correction.

Even in our sample of HIIGs and GHIIRs, chosen to be the youngest systems,  it is  crucial  to verify that the rapid luminosity evolution of the stellar cluster does not introduce a systematic bias in the distance indicator. This would happen, for example, if the average age of the GHIIRs is different from that of the HIIGs or if luminous and faint HIIGs have different  average ages.  In general, if velocity dispersion measures mass, younger clusters will be more luminous than older ones  for a given $\sigma$. 

The evolution effect can be  scrambled by the superposition of bursts of different ages along the line of sight. Nevertheless, even small systematics can have a sizeable effect in the value of \ho\ so it is important to remove the evolution  effect  from the data in a similar fashion as for the dust extinction.  The equivalent width of \hbeta\ [EW(\hbeta)]  is a useful age estimator \citep{dottori1981,sta96,mar08} or at least it provides an upper limit of the age of the burst \citep{terlevich2004}. Indeed there is some empirical evidence for this as discussed by  \cite{melnick2000}  and \cite{Bordalo2011}.
\citet{Chavez2014} explored the posibility that  the age of the burst  is a second parameter, using the EW(\hbeta) as an age estimator  in the  \lsigma relation for HIIGs, and found  a rather weak dependence.

\subsection{Determining the evolution  correction with stellar population synthesis models}\label{agecorrection}

\begin{figure}
\includegraphics[width=0.45\textwidth]{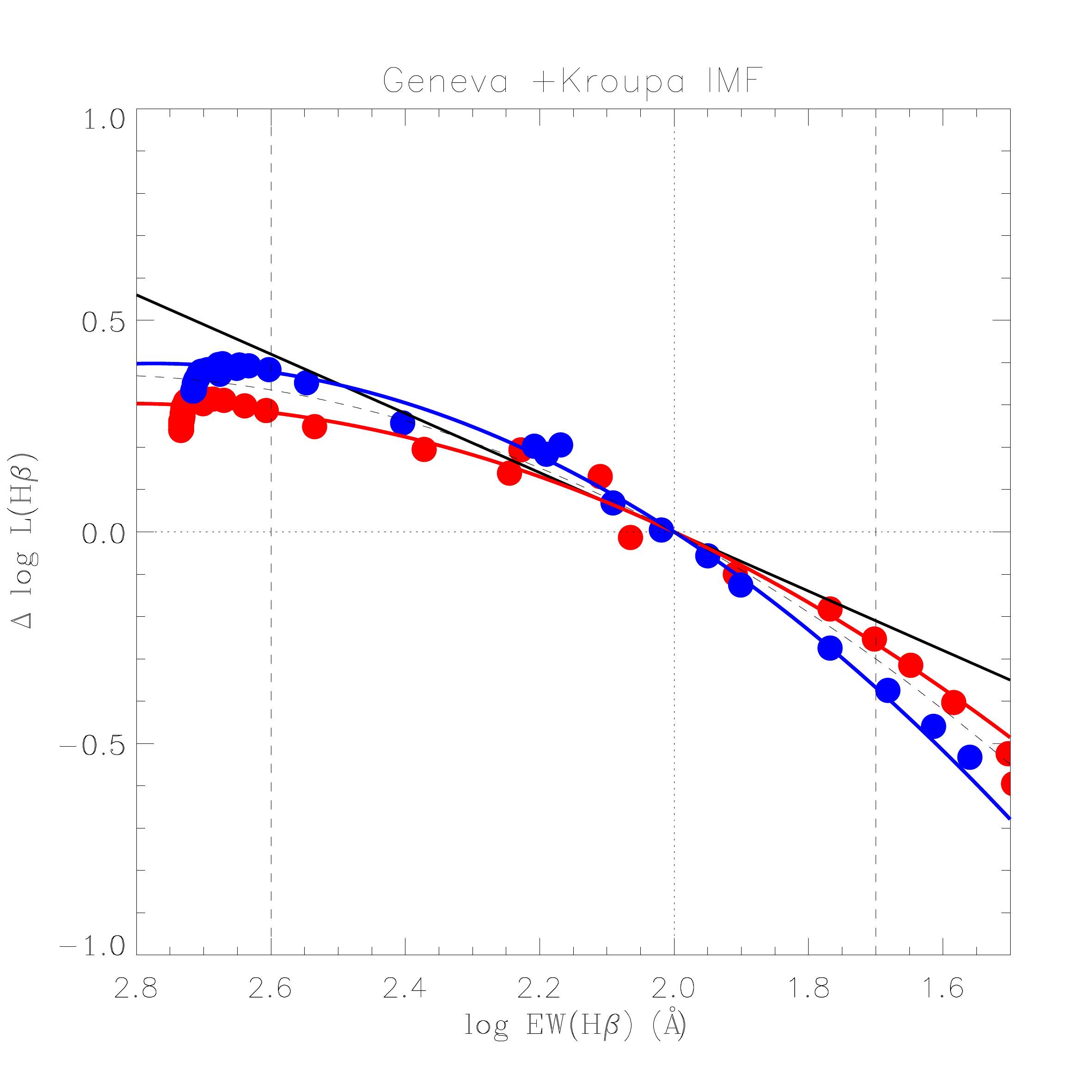}
\caption{ Differential change in the  L(\hbeta) vs.  EW(\hbeta)  for a  Kroupa IMF with upper mass  limit of 120 \Msol\ and Geneva tracks  for metallicities Z=0.004 and Z=0.008 computed using SB99. 
The vertical dotted line shows the median of the equivalent width of the sample. The vertical dashed lines indicate the extremes of the EW in the samples. The  solid black line is the linear fit to the models, in colour the quadratic fits: blue and red for the different metallicities Z=0.004 and 0.008 respectively, and the dashed line is the total fit.}
\label{hbetacorrected}
\end{figure}

The ionizing luminosity of young stellar clusters and consequently the Balmer line luminosity of the associated \hii\ region remain almost constant during the first $10^6$ yrs of evolution and then decay rapidly after  the first 3~Myrs while the continuum luminosity remains approximately constant during the first 8 Myrs.
This combined effect is illustrated in Figure \ref{hbetacorrected}  where we can see the  change in the  L(\hbeta) vs.  EW(\hbeta) for instant bursts with a  Kroupa IMF and the  Geneva tracks  for metallicity  Z=0.004 and Z=0.008 computed using Starburst 99 \citep[][SB99]{leitherer1999}.

We corrected the luminosities  to the value they would have when the EW is equal to the median EW of the GHIIR of the anchor sample. 
The median EW ({\it w}$_{med}$) defines therefore a  `median-age' for the corrected sample.\footnote{ In \cite{Melnick2017} we erroneously
concluded that the choice of $w_{med}$ strongly influences
the value of \ho. This mistake was due to 
a numerical error that has now been corrected.
For any value of $w_{med}$, the correction
to the fluxes of the HIIGs is almost exactly offset by the 
change in zero point resulting from the correction
of the GHIIRs luminosities.}

\begin{figure}
 \centering
 \includegraphics[scale=0.33]{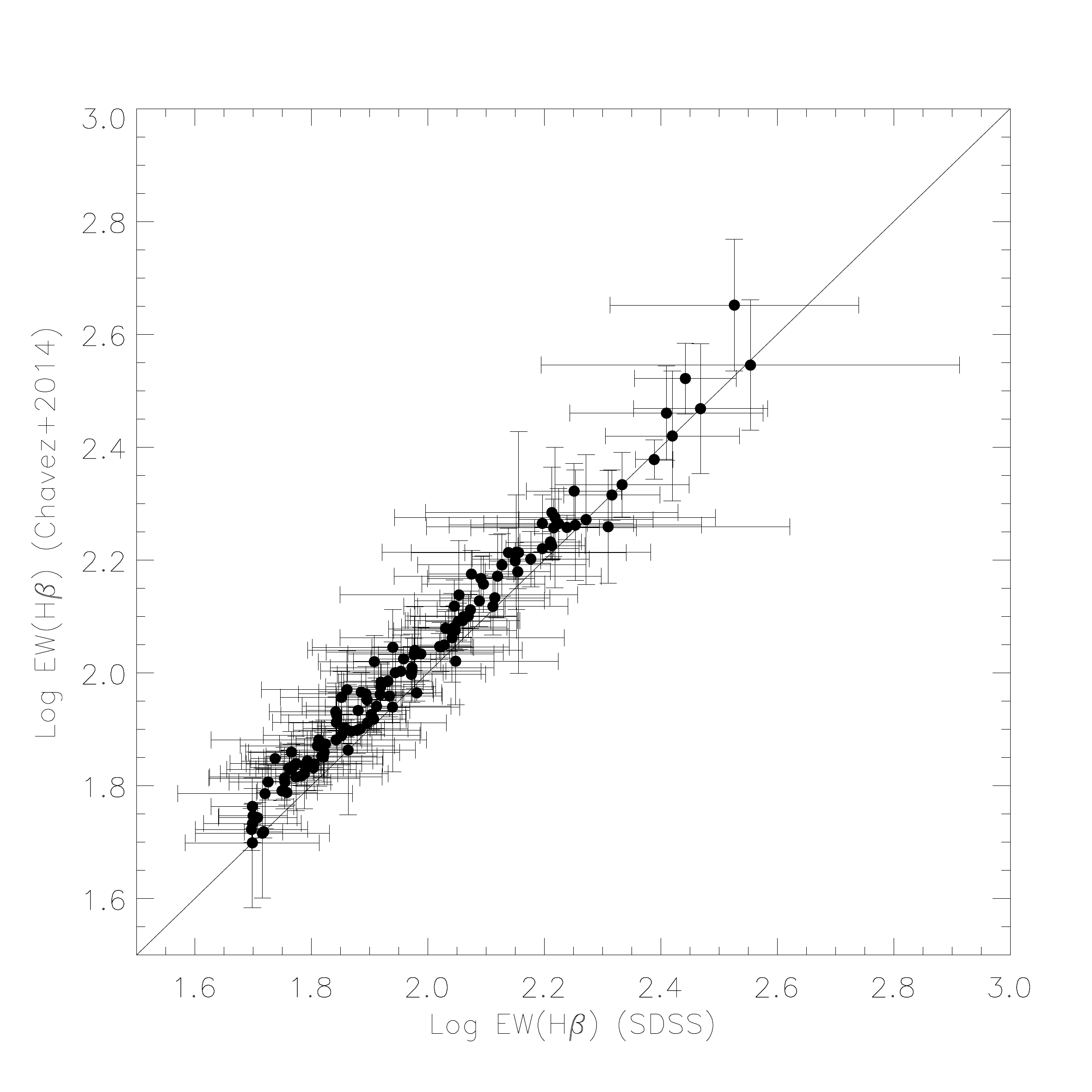}
  \caption{Comparison between the equivalent widths of \hbeta\ from \protect\cite{Chavez2014} and those from  SDSS spectra. The 1:1 line is shown.}
 \label{comEW}
\end{figure}

The changes can be quantified through the relation $\Delta L(H\beta) = c\times\Delta EW(H\beta)$ \citep[equation 3 in ][]{hagele2013}, where $c$ varies between 0.5 and 0.9 for  different stellar synthesis models. Clearly, however, as shown in Figure \ref{hbetacorrected} the relation between log(L) and log(EW) is not linear, but for the range of ages of interest it can be approximated by a parabola of the form
\begin{equation}
\log L(H\beta)= constant +a_{1}[\log EW({H\beta})]+a_{2}[\log EW({H\beta})]^2
\end{equation}
Thus we corrected the observed flux as,
\begin{equation}
\Delta \log F(H\beta) = (w_{i}-w_{med})[a_{1}+a_{2}(w_{i}+w_{med})]
\end{equation}
where $w_{med}$ and $w_i$ are the $ \log$ median[$EW({H\beta})$] and the individual $\log EW({H\beta})$ respectively. 

The fit to the average of the models gives values of the coefficients  of $a_1$=2.55$\pm$0.08  and $a_2$=-0.44$\pm$0.02.

The associated error is:
\begin{equation}
 \begin{split}
  \delta \Delta \log  F(H\beta)^{2}=(w_{i}-w_{med})^2(\delta a_{1})^2+(w_{i}^2-w_{med}^2)^2(\delta a_{2})^2  \\
  +(a_{1}+2a_{2}w_{i})^2(\delta w_{i})^2
 \end{split}
\end{equation}
with $\delta a_i$  being the error in the fit coefficients. 

The EW values in \citet{Chavez2014} are systematically larger than those from the SDSS spectra. This result is as expected  for the larger spectral apertures used in  \citet{Chavez2014} if the nebular emission is significantly more extended than the continuum as it is probably the case in these systems. This is illustrated in Figure \ref{comEW} where we  show the comparison between the equivalent width measured from   \citet{Chavez2014}  and those from SDSS including the observational error bars.


\subsubsection{Caveats}

There are four main uncertainties associated with our approach to  evolution correction:  firstly, as discussed in \cite{calzetti2000}, in local star-forming galaxies there is a differential extinction between stars and ionized gas being the nebular lines  more attenuated  than the stellar continuum by a factor of about two thus also biasing the observed EW to lower values.

Secondly, as discussed in \cite{ terlevich2004,Melnick2017}, and Telles \& Melnick (2017, in preparation)  the observed EW of HIIGs is biased to lower values due to the contribution  of an underlying older stellar population to the observed continuum.  This effect can be seen in the images of the GHIIRs of Appendix A with about half of the objects showing a complex morphology. 
The fact that the distribution of EW in GHIIRs is similar to that of HIIGs, see    Figure \ref{histograms}, supports the view that the underlying continuum and differential reddening effects in the measured EW are similar in the two samples.

Thirdly, there is a dearth of  models of the evolution of SSCs that include self-consistently the photoionization of the interstellar gas. SB99 includes Balmer emission line EW estimates, but these estimates are simply based in the total UV ionizing flux without  taking into account \hii\ region parameters such as metallicity, density, or ionization parameter. 

Finally, SB99 models do not include massive interacting binaries nor the contribution of stars more massive than 120M$_{\odot}$, all of which are expected to be important in massive SSCs. In fact, studies of local SSCs like 30-Doradus indicate that the binary fraction among the most massive stars could be as high as 100\%\ \citep{Bosch2009}, and that a substantial fraction of these could be interacting \citep{Sana2013}. These effects imply that our  corrections are still rather tentative and should  be  considered as indicative of the effect of evolution.  

In \S \ref{EvoCorr} we discuss the effect  of a  contribution  to the observed EW  of an underlying older stellar population and of the differential extinction on the value of \ho .


\section{Distances and luminosities}\label{distlum}

We computed the luminosities L  for the HIIG from the observed fluxes F as  $L = 4\pi F D_{L}^2$  where  the luminosity distance D$_L$ was derived   using either the linear Hubble law  $D_L=c z/H_{0}$, or the complete set of cosmological parameters.   In general the former $D_{L}$  is smaller than the latter.

For  systems with $z<0.1$ using the  linear Hubble law to determine $D_{L}$ underestimates the luminosity distance $D_{L}$ by less than 8\% relative to the distances computed using standard cosmology. Thus, using the linear relation for objects with $z>0.1$ will  result in \ho\ being overestimated. On the other hand estimating $D_L$ using the cosmological parameters makes the method sensitive to the choice of cosmological parameters such as $\Omega_\Lambda$  that will  slightly  change  $D_{L}$ and therefore  the derived \ho .  To estimate the size of this effect we built from our primary sample (hereinafter S1) a subsample (S2) constrained  to HIIGs with $z <0.1$ and use for it the linear relationship to estimate the distance independently of cosmology. 

\begin{figure}
  \includegraphics[scale=0.18]{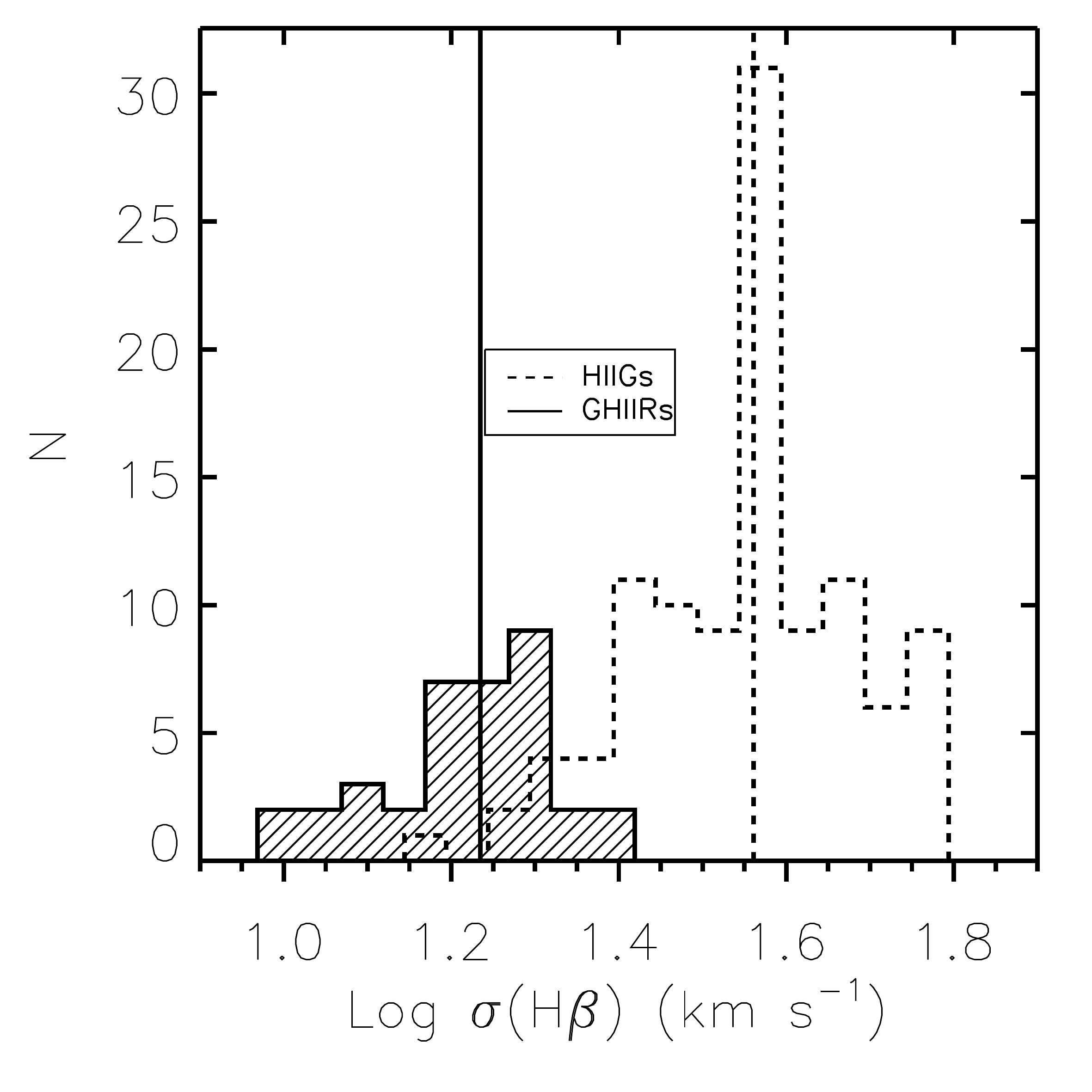}\includegraphics[scale=0.18]{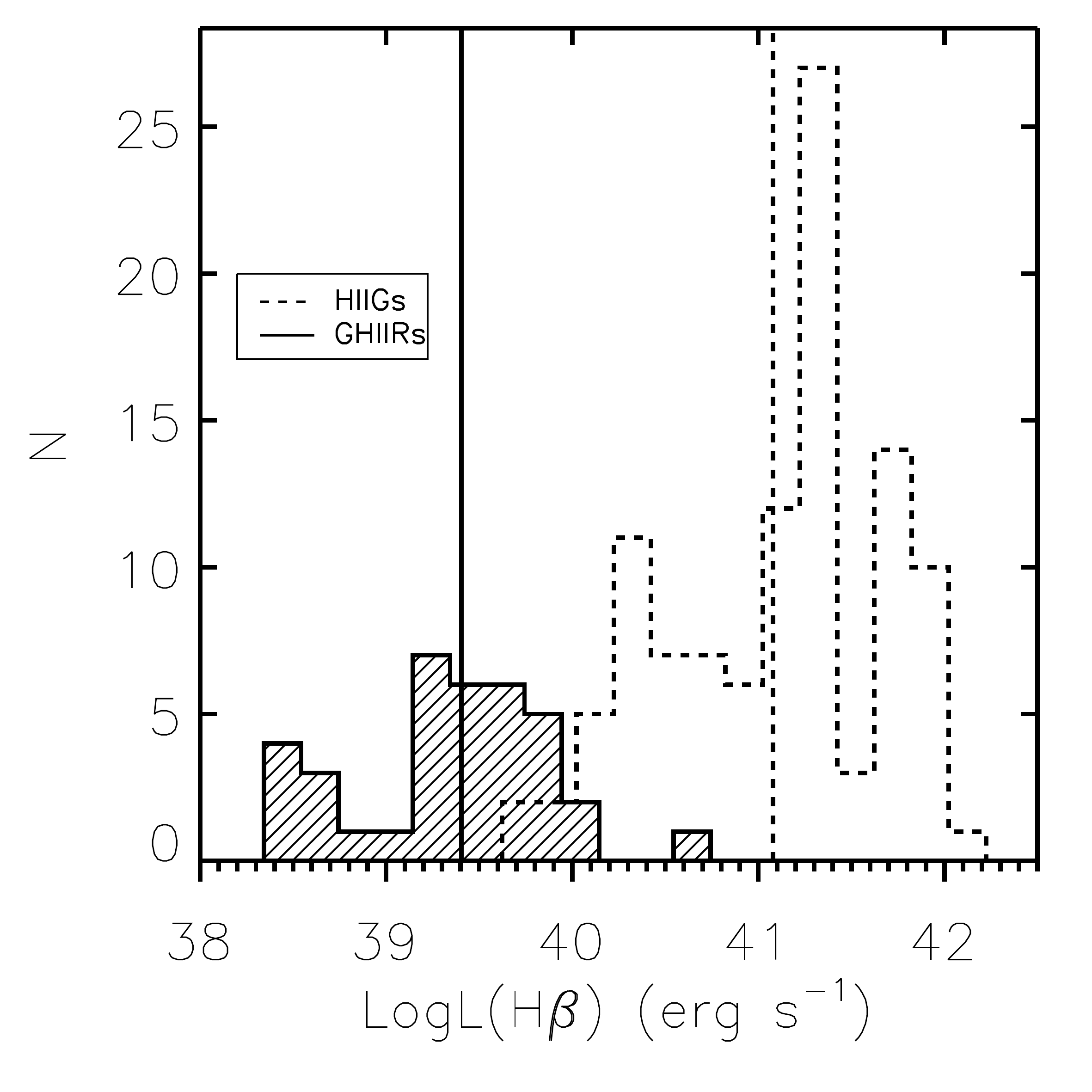}
  \includegraphics[scale=0.18]{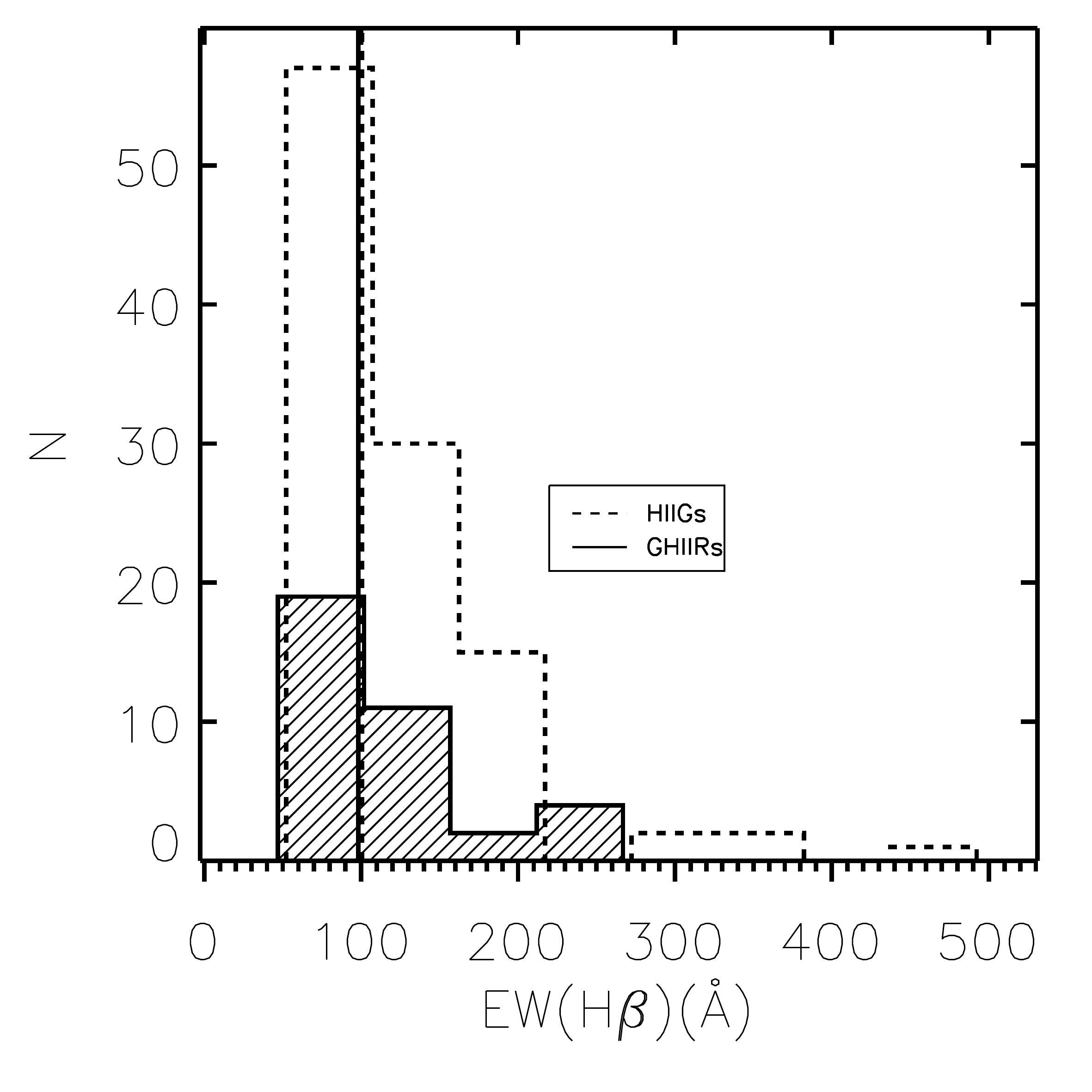}\includegraphics[scale=0.18]{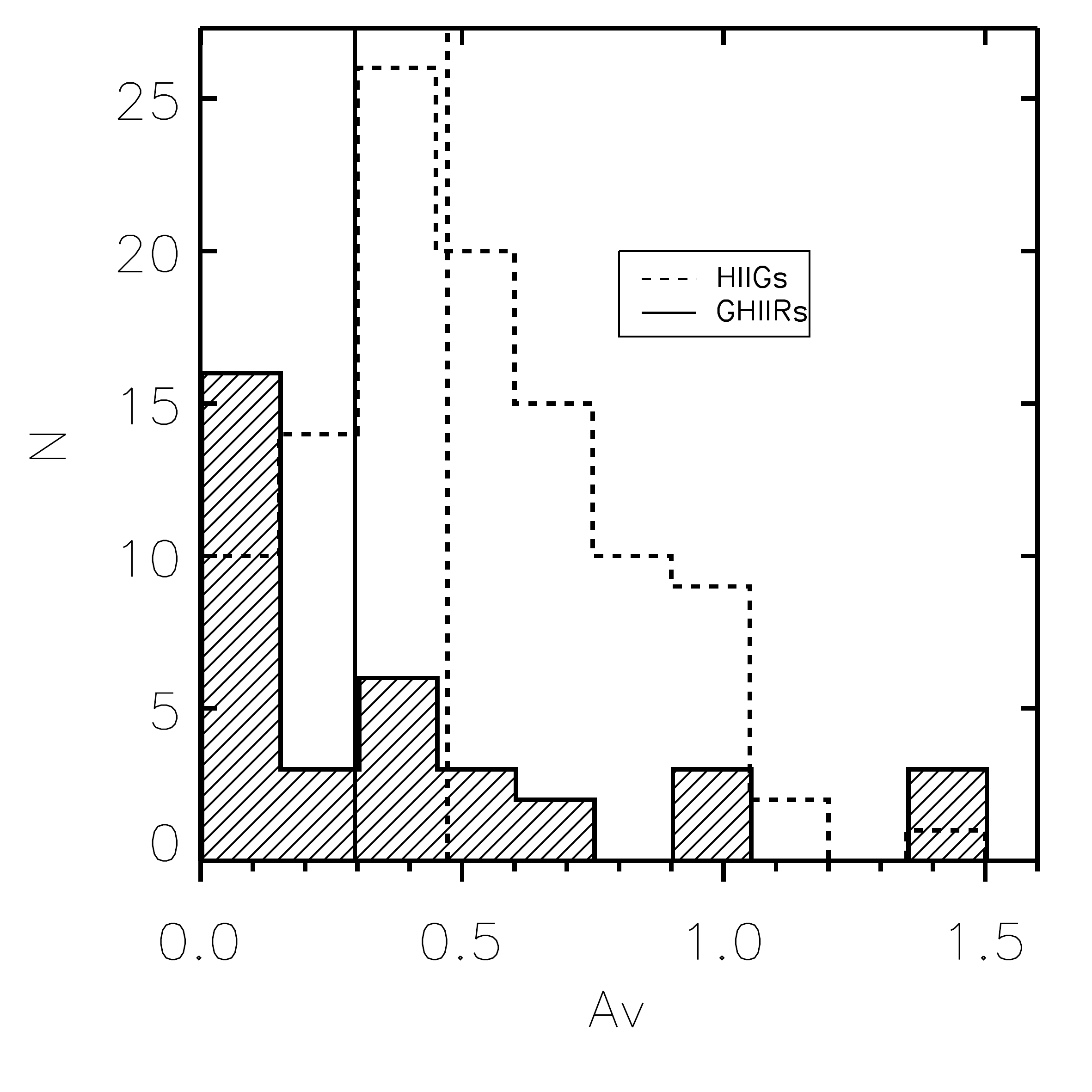}
    \caption{ Comparison of the distribution of properties of the GHIIRs (shaded area)  and the \protect\cite{Chavez2014} HIIGs sample
   (dashed line); {\bf upper left}:  H$\beta$ velocity dispersion,  {\bf down left}: EW(H$\beta$), {\bf upper right}: H$\beta$ luminosity, {\bf down right}: $A_{V}$ extinction parameter obtained using  \protect\cite{Gordon2003} extinction law. The vertical lines represent the median value for each sample. }
   \label{histograms}
   \end{figure}

In summary, the extinction corrected fluxes were used to calculate the luminosity of the HIIGs, $(L = 4\pi$$D$$_{L}^2$) where D$_L$ was derived (depending on z) using either the linear relationship of the Hubble law ($D=cz/H_{0}$) or  a flat cosmology with  $\Omega_m$~=~0.29. 
The distribution of the parameters used for the distance estimator are shown in Figure \ref{histograms}.

\subsubsection{Aperture effects}

From the very beginning of this project our conventional wisdom has been to measure the velocity dispersions through relatively narrow slits to preserve the spectral resolution and the luminosities through wide apertures to include all the flux. The underlying assumption is that the turbulence is isotropic and the internal extinction modest, so that even through a narrow slit we still sample the full turbulent cascade. This is a pretty good assumption for single objects, but a  small fraction of our objects have complex profiles, a sign of complex structure. Thus, even narrow slits may encompass more than one starburst along the line of sight. Therefore, using different entrance apertures for luminosities and velocity dispersions may introduce systematic effects that need to be quantified (see  \citet{Melnick2017} for further discussion).

Notice also that while the fluxes have been measured through wide slits, the extinction correction is derived using SDSS fluxes measured through a $3''$ fibre aperture. The evolution correction on the other hand is determined using  \cite{Chavez2014} spectrophotometry for the Ch14  data and SDSS spectrophotometry for the SDSS data. This is relevant because the luminosity errors are dominated by the uncertainties in the extinction and evolution corrections.

\section{The Hubble constant}\label{calH0}

\begin{figure*}
 \includegraphics[width=0.75\textwidth]{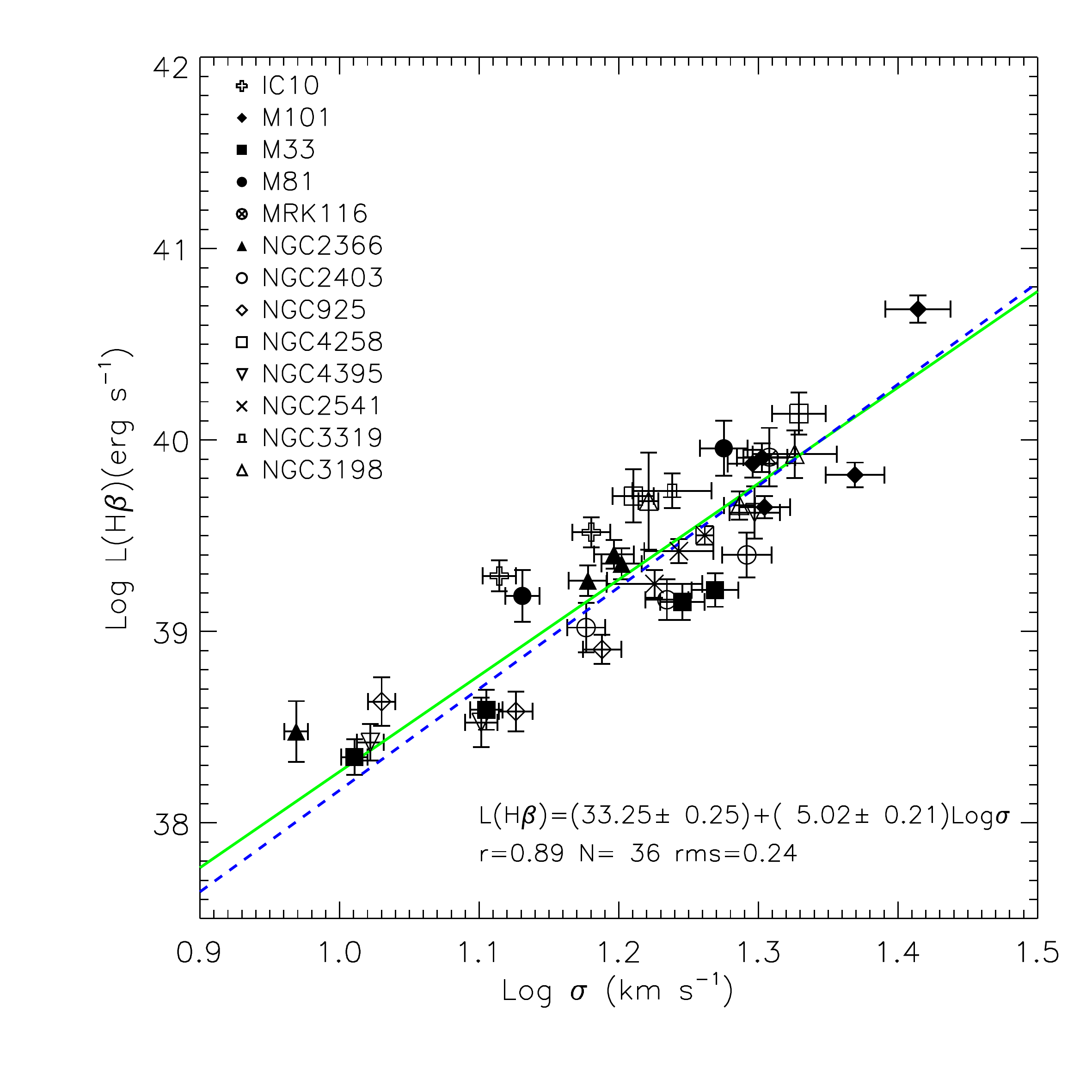}
    \caption{L(H$\beta)$-$\sigma$ relation for the GHIIRs. The adopted distances to the parent galaxies (see Table  \ref{tableDistan}) are derived from primary distance estimators.  The  green solid line is the fit to the data given in the inset and the dashed line is the fit to the anchor sample of \protect\cite{Chavez2012}.}
   \label{lsigmaGRHII}
   \end{figure*}

 The Hubble constant is determined as follows: first we fix the slope of the \lsigma relation using the velocity dispersions and luminosities of the HIIGs. The slope is independent of the actual value of \ho. The said slope is then used to determine the zero-point of the relation using our new data of 36 GHIIRs whose \lsigma relation is shown in Figure \ref{lsigmaGRHII}. The slope and zero-point define the distance indicator that is then applied to the sample of HIIGs to determine \ho. 


\subsection{Methodology}

The method used for the determination of the Hubble constant using the $L-\sigma$ relation and  the analysis of the propagation of errors follows the formalism presented in \cite{Melnick2017}.

The error propagation includes the observational errors plus the covariance of the two variables which must be included even when the observational errors are uncorrelated.

The error $\delta Y_i$ in the prediction of a linear correlation of the form
$y=a+bx$
at a given value of $x=x_i$,  when the parameters $a$ and $b$ are determined using least-squares techniques and including the
experimental errors in both variables ($\delta x_i,\delta y_i$) is,
\begin{equation}\label{errp3}
(\delta Y_i)^2 =(\delta y_i)^2 + (b\times\delta x_i)^2 + (\delta a)^2 + (\delta b)^2 (x_i - <x>)^2
\end{equation}

The standard least-squares solution is in general biased when the independent variable is subject to error. In such cases Eq.\ref{errp3} is a very good approximation, but is not exact.

To estimate the Hubble constant we use the slope ($\alpha$) of the  \lsigma relation of the HIIGs and our anchor sample to calibrate the zero point ($Z_p$) of the distance indicator  as follows

\begin{equation}\label{one}
Z_p=\frac{\sum_{i=1}^{36}{ {\it W_i}(\log L_{\rm GHR,i} - \alpha\times \log\sigma_{\rm GHR,i}) }} { \rm \sum_{i=1}^{36}{{\it W_i} } }
\end{equation}
where $L_{GHR,i}$ is the H$\beta$ luminosity of each GHIIR and $\sigma_{\rm GHR,i}$ the corresponding velocity dispersion. The statistical weights $W_i$ are calculated as:

\begin{equation}\label{zero}
\begin{split}
W_i^{-1}= \Bigl(0.4343\frac{\delta L_{\rm GHR,i}}{L_{\rm GHR,i}}\Bigr)^2 + \Bigl(0.4343\alpha\frac{\delta\sigma_{\rm GHR,i}}{\sigma_{\rm GHR,i}}\Bigr)^2 \\
+ (\delta\alpha)^2(\sigma_{\rm GHR,i}-<\sigma_{\rm HIIG}>)^2
\end{split}
\end{equation} 
where $<\sigma_{\rm HIIG}>$ is the average velocity dispersion of the HIIGs that define the slope of the relation. Thus, the calibrated \lsigma relation or distance estimator is: $ \log L(H\beta) = \alpha \log\sigma+Z_p$. 
To calculate the Hubble constant we minimise the function, 

\begin{equation}\label{mins}
\chi^2( \itl H_{0}  ) = \sum_{i=1}^{N}[{W_i(\mu_i - \mu_{\itl H_{0},i})^2 - ln (W_i) ]}
\end{equation}
where $\mu_i$ is the logarithmic distance  modulus to each HIIG calculated using the distance indicator and the 
H$\beta$ flux F(H$\beta$) as  

\begin{equation}\label{mods}
\mu_i=2.5[ Z_p + \alpha\times \log\sigma_i - \log F_i(H\beta) - \log 4\pi]
\end{equation}
and $\mu_{\itl H_{0},i}$ is the distance modulus calculated from the redshift using either the linear relation $D_L=zc/\itl H_{0}$ or the full cosmological prescription with $\Omega_{\Lambda}=0.71$.

The best value of $\itl H_{0}$ is then obtained minimising $\chi^2$ with statistical weights $W_i^{-1}=\delta\mu_i^2+\delta\mu_{\itl H_{0},i}^2$ calculated as,

\begin{equation}\label{weights}
\begin{split}
W_i^{-1}= 6.25 [(\delta Z_p)^2 + \Bigl(0.4343\frac{\delta F_i}{F_i}\Bigr)^2 + \Bigr(0.4343\alpha\frac{\delta\sigma_i}{\sigma_i}\Bigl)^2  \\
     +  (\delta \alpha)^2(\sigma_i-<\sigma>)^2]
\end{split}
\end{equation}




 

\section{Systematics}\label{explore}
\subsection{Exploring  the parameter space}

In previous sections we have discussed the known statistical uncertainties associated with our sample. Genuine systematic errors are difficult to estimate, so  in this Section we include a range of parameters to quantify at least part of the systematic error component.  In particular, we explore alternative parametrizations that can not be easily included in the error scheme, as follows:

\itemize{
	\item  Two samples: S1 with 107 galaxies or S2 with $z<0.1$ and 92 galaxies; \smallskip
	\item  Two different sources for the \hbeta\ photometry: \cite{Chavez2014} (Ch14) or SDSS; \smallskip
	\item  Two formulations for the luminosity distance for the HIIGs:  $D_L=H_0/cz$ (LR) or full $\Lambda CDM$ cosmology with $\Omega_{\Lambda}=0.71$; \smallskip
	\item  For these three cases we use two different extinction laws: \cite{calzetti2000} (C00) or \cite{Gordon2003} (G03).
}
\smallskip

This yields 16 different combinations (models; the preferred one is model 13 as we will discuss below) see Table \ref{hnotiA2}, which we compute for the two evolutionary scenarios discussed above: no evolution, and evolution corrected using quadratic fits to SB99 models as detailed in Section \ref{age}. The resulting parameters of the \lsigma relation for each case are listed in Tables~\ref{hnotiA3} and \ref{hnotiA7}, and shown in graphical form  in Figure~\ref{modelosH} that illustrates  the sensitivity of the value of \ho\ to the adopted combination.

   \begin{table}\footnotesize
 \centering
    \caption{{\bf Model parameters for \ho\ computation} \\
    Column 1: Model code (* denotes the preferred model). Column 2: Sample. Column 3: Number of HIIGs. Column 4:  Extinction law. Column 5: Source of \hbeta\ photometry. Column 6: Distance estimate of HIIGs.}
    \newcommand{\mc}[3]{\multicolumn{#1}{#2}{#3}}
    \begin{tabular}{ l c c c c c c l }   \\\hline
             M&\mc{1}{c}{Sample}& \mc{1}{c}{N} &\mc{1}{c}{Extinction}& \mc{1}{c}{Photometry}&\mc{1}{c}{Distance}\\\hline
      1 &  S1 & 107&C00 &Ch14&    LR\\
     2  &  S2 & ~92 &C00 &Ch14&    LR\\
     3  &  S1 & 107&C00 &SDSS&    LR\\
     4  &  S2 & ~92 &C00 &SDSS&    LR\\
     5  &  S1 & 107&C00 &Ch14&   $\Omega_{\Lambda}=0.71$\\
     6  &  S2 & ~92 &C00 &Ch14&   $\Omega_{\Lambda}=0.71$\\
     7  &  S1 & 107&C00 &SDSS&   $\Omega_{\Lambda}=0.71$\\
     8  &  S2 & ~92 &C00 &SDSS&   $\Omega_{\Lambda}=0.71$\\
     9  &  S1 & 107&G03 &Ch14&    LR\\
     10 &  S2 &  ~92 &G03 &Ch14&    LR\\
     11 &  S1 & 107&G03 &SDSS&    LR\\
     12 &  S2 & ~92 &G03 &SDSS&    LR\\
     13\ $\star$ &  S1 & 107&G03 &Ch14&   $\Omega_{\Lambda}=0.71$\\
     14 &  S2 & ~92 &G03 &Ch14&   $\Omega_{\Lambda}=0.71$\\
     15 &  S1 & 107&G03 &SDSS&   $\Omega_{\Lambda}=0.71$\\
     16 &  S2 & ~92 &G03 &SDSS&   $\Omega_{\Lambda}=0.71$\\\\\hline
          \end{tabular}
\label{hnotiA2}
\end{table}


 \begin{table*}\footnotesize
  \centering                             
     \caption{\bf The $L-\sigma$ relation without correction for evolution}       
     \newcommand{\mc}[3]{\multicolumn{#1}{#2}{#3}}
     \begin{tabular}{l  l  c  c  c  c  c   c } \\\hline
 N & \mc{1}{c}{Sample} &  $H_0$   &  a    &  b  & rms(HIIG) & rms(DI)  & $\chi^2_{min}/dof$\\ \hline
 1 &  S1-C00-Chavez14-LIN                      & 77.1 $^{+  3.1  }_{- 3.0 }$   &  33.49  $\pm$  0.25  & 4.81   $\pm$    0.14   &  0.343  &   0.343  &   0.931 \\
 2 &  S2-C00-Chavez14-LIN                      & 75.4 $^{+  3.3  }_{- 3.1 }$   &  33.39  $\pm$  0.25  & 4.89   $\pm$    0.16   &  0.365  &   0.365  &   1.048 \\
 3 &  S1-C00-SDSS-LIN                          & 72.8 $^{+  2.9  }_{- 2.8 }$   &  33.20  $\pm$  0.26  & 5.05   $\pm$    0.14   &  0.372  &   0.372  &   1.075 \\
 4 &  S2-C00-SDSS-LIN                          & 70.3 $^{+  3.1  }_{- 2.9 }$   &  33.04  $\pm$  0.26  & 5.18   $\pm$    0.16   &  0.398  &   0.398  &   1.220 \\
 5 &  S1-C00-Chavez14-$\Omega_{\Lambda}=0.71$  & 75.4 $^{+  3.0  }_{- 2.9 }$   &  33.29  $\pm$  0.25  & 4.97   $\pm$    0.14   &  0.343  &   0.353  &   0.970 \\
 6 &  S2-C00-Chavez14-$\Omega_{\Lambda}=0.71$  & 74.3 $^{+  3.2  }_{- 3.1 }$   &  33.23  $\pm$  0.25  & 5.02   $\pm$    0.16   &  0.365  &   0.373  &   1.088 \\
 7 &  S1-C00-SDSS-$\Omega_{\Lambda}=0.71$      & 71.1 $^{+  2.9  }_{- 2.8 }$   &  32.99  $\pm$  0.26  & 5.22   $\pm$    0.14   &  0.372  &   0.384  &   1.119 \\
 8 &  S2-C00-SDSS-$\Omega_{\Lambda}=0.71$      & 69.1 $^{+  3.0  }_{- 2.9 }$   &  32.87  $\pm$  0.26  & 5.32   $\pm$    0.17   &  0.398  &   0.408  &   1.262 \\
 9 &  S1-G03-Chavez14-LIN                      & 76.3 $^{+  3.0  }_{- 2.9 }$   &  33.46  $\pm$  0.23  & 4.84   $\pm$    0.14   &  0.345  &   0.345  &   0.976 \\
10 & S2-G03-Chavez14-LIN                       & 74.3 $^{+  3.1  }_{- 3.0 }$   &  33.35  $\pm$  0.23  & 4.93   $\pm$    0.16   &  0.367  &   0.367  &   1.105 \\
11 & S1-G03-SDSS-LIN                           & 71.7 $^{+  2.8  }_{- 2.7 }$   &  33.16  $\pm$  0.24  & 5.09   $\pm$    0.14   &  0.374  &   0.374  &   1.136 \\
12 & S2-G03-SDSS-LIN                           & 68.8 $^{+  2.9  }_{- 2.8 }$   &  32.99  $\pm$  0.24  & 5.23   $\pm$    0.17   &  0.401  &   0.401  &   1.294 \\
13* & S1-G03-Chavez14-$\Omega_{\Lambda}=0.71$  & 74.6 $^{+  2.9  }_{- 2.8 }$   &  33.27  $\pm$  0.23  & 5.00   $\pm$    0.14   &  0.345  &   0.355  &   1.020 \\
14 & S2-G03-Chavez14-$\Omega_{\Lambda}=0.71$   & 73.2 $^{+  3.1  }_{- 3.0 }$   &  33.19  $\pm$  0.24  & 5.06   $\pm$    0.16   &  0.367  &   0.376  &   1.147 \\
15 &  S1-G03-SDSS-$\Omega_{\Lambda}=0.71$      & 70.0 $^{+  2.8  }_{- 2.7 }$   &  32.96  $\pm$  0.24  & 5.25   $\pm$    0.14   &  0.373  &   0.385  &   1.182 \\
16 & S2-G03-SDSS-$\Omega_{\Lambda}=0.71$       & 67.6 $^{+  2.9  }_{- 2.8 }$   &  32.82  $\pm$  0.24  & 5.37   $\pm$    0.17   &  0.401  &   0.410  &   1.338  \\ \hline
 \end{tabular}
 \label{hnotiA3}
 \end{table*}

\begin{table*}\footnotesize
  \centering 
     \caption{\bf The $L-\sigma$ relation corrected for evolution using EW(\hbeta)}   
     \newcommand{\mc}[3]{\multicolumn{#1}{#2}{#3}}
     \begin{tabular}{l l c c c c c  c} \\\hline
  N &  \mc{1}{c}{Sample} & {H$_{0}$ }          &     {a}    &  {b} & rms (HIIG) & rms(DI) &$\chi^2_{min}/dof$\\\hline
 1 & S1-C00-Chavez14-LIN                       & 73.9  $^{+ 2.9  }_{- 2.8}$ & 33.33  $\pm$  0.24  &  4.93  $\pm$  0.14 &  0.355  &  0.354  &   1.014 \\
 2 & S2-C00-Chavez14-LIN                       & 71.7  $^{+ 3.1  }_{- 3.0}$ & 33.18  $\pm$  0.24  &  5.05  $\pm$  0.17 &  0.375  &  0.374  &   1.124 \\
 3 & S1-C00-SDSS-LIN                           & 73.0  $^{+ 2.9  }_{- 2.8}$ & 33.07  $\pm$  0.25  &  5.14  $\pm$  0.14 &  0.388  &  0.383  &   1.179 \\
 4 & S2-C00-SDSS-LIN                           & 70.0  $^{+ 3.0  }_{- 2.9}$ & 32.88  $\pm$  0.25  &  5.30  $\pm$  0.17 &  0.411  &  0.407  &   1.307 \\
 5 & S1-C00-Chavez14-$\Omega_{\Lambda}=0.71$   & 72.3  $^{+ 2.9  }_{- 2.8}$ & 33.13  $\pm$  0.24  &  5.09  $\pm$  0.14 &  0.355  &  0.364  &   1.046 \\
 6 & S2-C00-Chavez14-$\Omega_{\Lambda}=0.71$   & 70.6  $^{+ 3.1  }_{- 2.9}$ & 33.03  $\pm$  0.25  &  5.18  $\pm$  0.17 &  0.375  &  0.382  &   1.159 \\
 7 & S1-C00-SDSS-$\Omega_{\Lambda}=0.71$       & 71.2  $^{+ 2.9  }_{- 2.7}$ & 32.87  $\pm$  0.25  &  5.30  $\pm$  0.15 &  0.388  &  0.395  &   1.215 \\
 8 & S2-C00-SDSS-$\Omega_{\Lambda}=0.71$       & 68.8  $^{+ 3.0  }_{- 2.9}$ & 32.71  $\pm$  0.26  &  5.44  $\pm$  0.18 &  0.411  &  0.416  &   1.346 \\
 9 & S1-G03-Chavez14-LIN                       & 72.6  $^{+ 2.9  }_{- 2.8}$ & 33.28  $\pm$  0.24  &  4.98  $\pm$  0.14 &  0.360  &  0.359  &   1.033 \\
10 & S2-G03-Chavez14-LIN                       & 70.2  $^{+ 3.0  }_{- 2.9}$ & 33.13  $\pm$  0.25  &  5.11  $\pm$  0.17 &  0.381  &  0.380  &   1.147 \\
11 & S1-G03-SDSS-LIN                           & 71.1  $^{+ 2.8  }_{- 2.7}$ & 33.01  $\pm$  0.25  &  5.21  $\pm$  0.14 &  0.394  &  0.389  &   1.203 \\
12 & S2-G03-SDSS-LIN                           & 67.8  $^{+ 3.0  }_{- 2.8}$ & 32.79  $\pm$  0.25  &  5.38  $\pm$  0.18 &  0.418  &  0.414  &   1.337 \\
13*& S1-G03-Chavez14-$\Omega_{\Lambda}=0.71$   & 71.0  $^{+ 2.8  }_{- 2.7}$ & 33.09  $\pm$  0.25  &  5.14  $\pm$  0.14 &  0.360  &  0.369  &   1.064 \\
14 & S2-G03-Chavez14-$\Omega_{\Lambda}=0.71$   & 69.1  $^{+ 3.0  }_{- 2.9}$ & 32.97  $\pm$  0.25  &  5.24  $\pm$  0.17 &  0.381  &  0.388  &   1.180 \\
15 & S1-G03-SDSS-$\Omega_{\Lambda}=0.71$       & 69.4  $^{+ 2.8  }_{- 2.7}$ & 32.81  $\pm$  0.25  &  5.37  $\pm$  0.15 &  0.394  &  0.401  &   1.237 \\
16 & S2-G03-SDSS-$\Omega_{\Lambda}=0.71$       & 66.6  $^{+ 2.9  }_{- 2.8}$ & 32.63  $\pm$  0.26  &  5.52  $\pm$  0.18 &  0.418  &  0.424  &   1.374 \\\hline
 \end{tabular}
 \label{hnotiA7}
 \end{table*}\subsubsection{Sensitivity to $\hb\ $ photometry}\label{photom}

The sensitivity to $\hb\ $ photometry was explored by comparing  SDSS and  \cite{Chavez2014} photometric measurements. A cursory inspection of Figure \ref{modelosH} and  Table \ref{hnotiA3} and \ref{hnotiA7} reveals that the values of \ho\ are  on average systematically lower by about 4.9~$\kmsmpc$ for SDSS fluxes and no evolutionary corrections, this systematic difference is reduced to 1.7~$\kmsmpc$ when evolutionary corrections are included.  The smaller value of \ho\ for the SDSS photometry is related to the systematically steeper slope of the \lsigma relation compared with that obtained when using  \cite{Chavez2014} photometry, both with and without evolution correction.  Since the lower luminosity HIIGs are also the closer ones, the steepening is a consequence of the smaller SDSS aperture,  compared with that of \cite{Chavez2014}, underestimating the line fluxes of the nearest galaxies. We therefore favour the results with the larger aperture photometry from \cite{Chavez2014}.

\subsubsection{Sensitivity to extinction laws}

The effect of using different extinction laws [\cite[][Models 1 to 8]{calzetti2000} or \cite[][Models 9 to 16]{Gordon2003}] is explored next. In general  \cite{calzetti2000} law yields larger values of \ho\  than \cite{Gordon2003} typically by about 1.5~$\kmsmpc$.  The \cite{calzetti2000} law was derived from a sample of eight heterogeneous starburst galaxies where only two, Tol 1924-416 and UGCS410  are bonafide  HIIGs and the rest are evolved high metallicity starburst galaxies, while the \cite{Gordon2003} extinction curve  corresponds to the LMC2 supershell near the 30 Doradus starforming region, the prototypical GHIIR, in the Large Magellanic Cloud. Therefore, we are inclined to prefer the results using the  \cite{Gordon2003} extinction law.

\subsubsection{Sensitivity to evolution corrections}\label{EvoCorr}

\begin{figure}
  \includegraphics[scale=0.45]{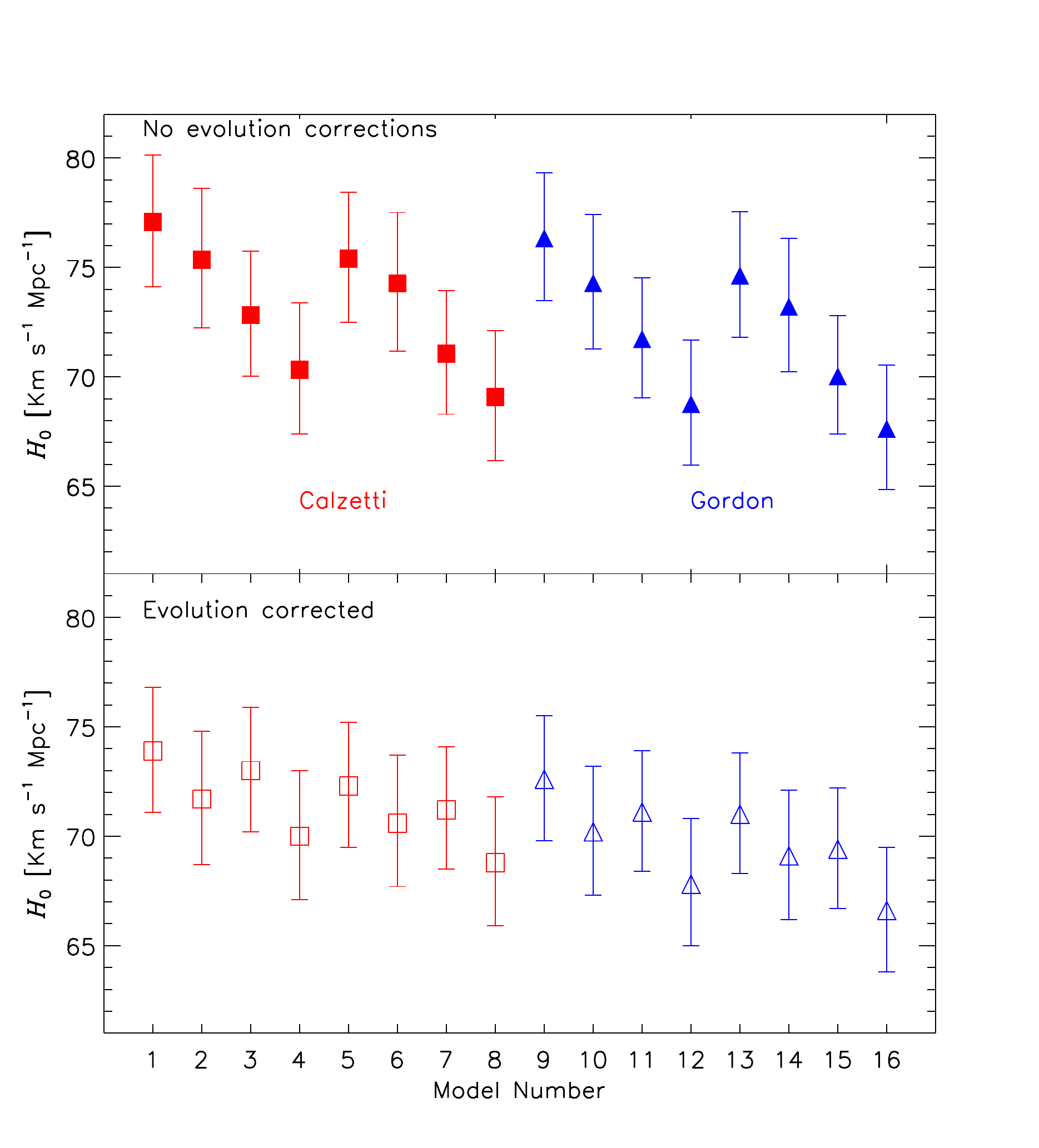} 
  \caption[]{  \small  Graphic representation of the values of \ho\ for the various parameters listed in Tables~\ref{hnotiA3} and \ref{hnotiA7}. {\bf Top}. Resulting \ho\ without correcting the luminosities for evolution. {\bf Bottom}. \small Same as the top panel, but using the fluxes  corrected for evolution, see Section~\ref{age}. As discussed in the text, the difference between models 1-8 (in red) and models 9-16 (in blue) is the adopted extinction law as indicated by the figure legends. }
\label{modelosH}
\end{figure}

Figure~\ref{modelosH} shows the resulting values of \ho\ for the models without  (upper panel) and with evolution correction (lower panel).
Comparing the upper and lower panels we clearly see that we obtain  lower values for $H_0$ when we apply the evolutionary corrections particularly in those models using \cite{Chavez2014} photometry.  The difference between the SDSS and \cite{Chavez2014} correction is  probably linked to the systematic difference in the masured EW  as discussed above and is an indication of the systematic errors involved in this correction. The values of \ho\  obtained after applying the evolution correction show as a family   less scatter  (r.m.s.=1.5 $\kmsmpc$) than the uncorrected results (r.m.s.=2.5 $\kmsmpc$) indicating that the evolution corrected results  are  more self consistent.

The evolution correction should be taken with care because, as we have already mentioned (see Section \ref {agecorrection}), even for pure starbursts the SB99 models provide only indicative values. Furthermore the observed EW are  contaminated by underlying populations of older stars and the nebular lines are a factor of two  more attenuated than the continuum regions. These two effects act in the same direction and as a consequence the observed EW are smaller than the intrinsic EW.

We  have included in the estimate of the evolution correction the contribution  of an underlying older stellar population and of  the differential extinction. To this end we have computed models where the observed EW was increased by a factor that represents the change in the EW after the removal of the presumed older stellar population continuum and the correction due to the differential extinction. Telles \& Melnick (2017, in preparation) have estimated from SED fits to the observed spectrum of HIIGs that the contribution of the underlying older population is  on average less than $50\%$ of the intrinsic continuum. Regarding the differential extinction effect, given that in general nebular extinction is small in our sample we can assume that such effect will be smaller than that of the older population. We have therefore assumed  for our evolution corrected estimates that the intrinsic EW of the ionizing SSC is on average over the whole sample about  $33\%$ larger than the observed EW. In all 16 models increasing the EW of GHIIRs and HIIGs  by $33\%$ results in a decrease of the slope of the distance indicator that translates in an  increase of \ho\ of less than 1 $\kmsmpc$.

\begin{figure}
 \includegraphics[width=0.45\textwidth]{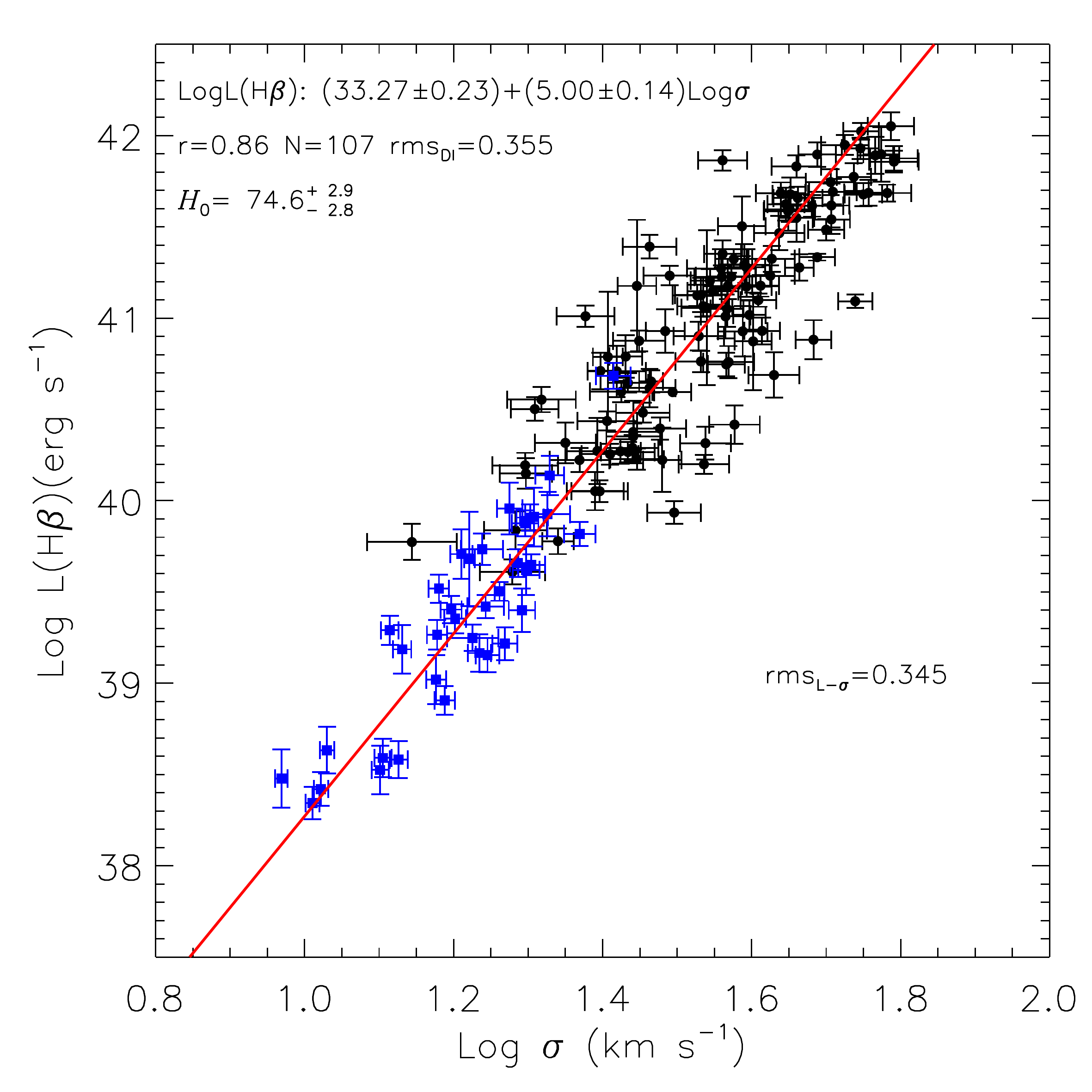}
 \vspace*{-0.0cm}
 \caption{\small  The \lsigma relation for the  \protect\cite{Chavez2014} sample using the velocity dispersions in the original paper; the fluxes have been corrected using \protect\cite{Gordon2003} extinction law. The solid line is the fit to the HIIGs points. The inset equation is the distance indicator where the slope is obtained from the fit to the HIIGs and the Zp determined following the procedure described in the text.}
 \label{lsChis1}
 \end{figure}

\begin{figure}
\includegraphics[width=0.45\textwidth]{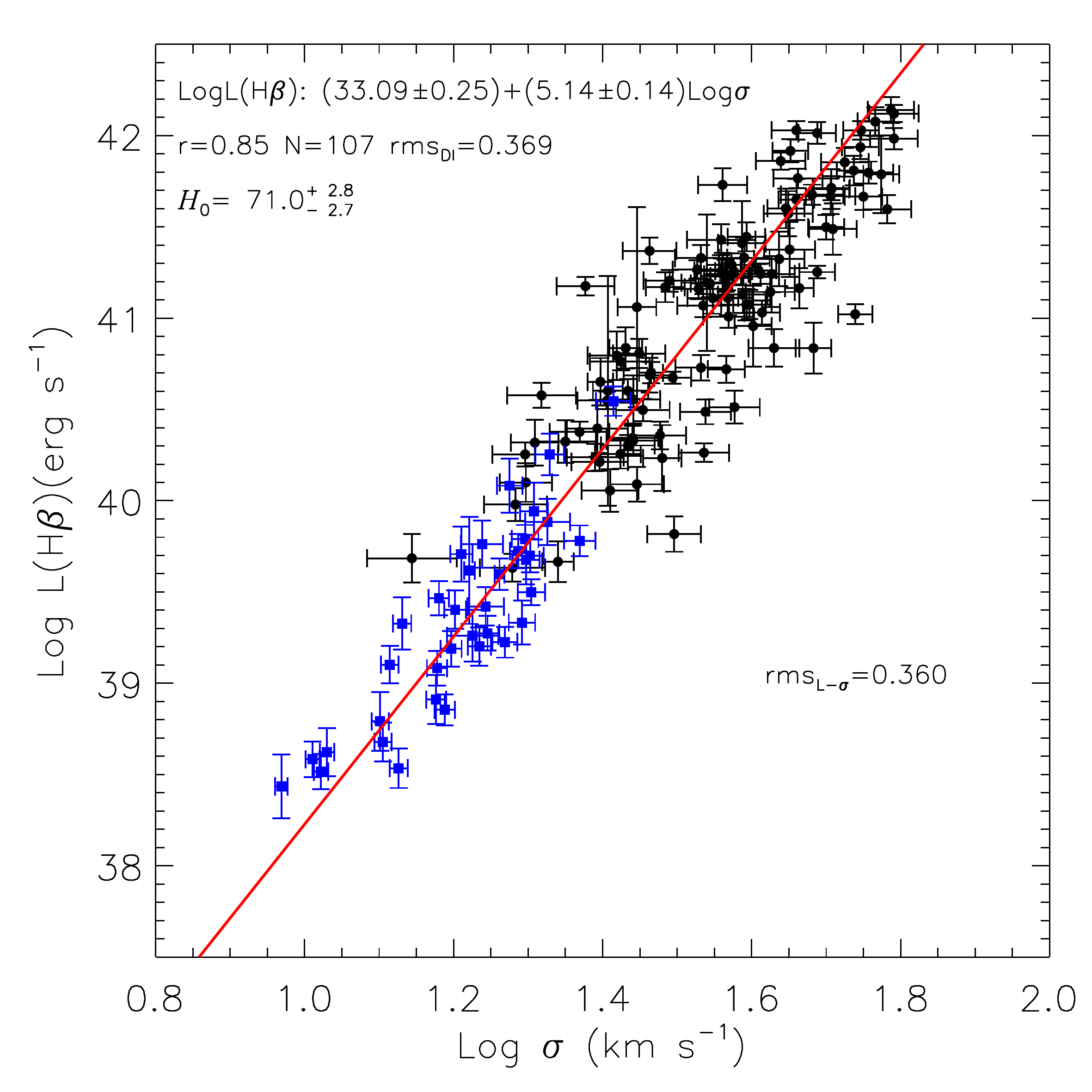}
\vspace*{-0.0cm}
\caption{\small  The \lsigma relation for the \protect\cite{Chavez2014} sample using the velocity dispersions of $\textmd{\hbeta}$ and the luminosities corrected to the median of EW(\hbeta) as discussed in the text. The solid line is the fit to the HIIGs. The inset equation is the distance indicator where the slope is obtained from the fit to the HIIGs  and the Zp determined following the procedure described in the text.   }
\label{lsighbetacorrected}
\end{figure}

Figures \ref{lsChis1}  and  \ref{lsighbetacorrected} show the \lsigma relation using the data   for model N=13 before and after correcting for evolution.

\subsubsection{Robustness of the slope}\label{robust}

Variations in the sample used to derive the \lsigma  relation can affect the values of the slope and $Z_p$. If by removing  or replacing a few data points the slope and  $Z_p$ change, and therefore also does the value of \ho , then the \lsigma relation is not robust. 

To address this issue we applied a  bootstrap sample  test \citep{Simpson1986} selecting a subset of samples of our primary sample of HIIGs, by random resampling with replacement  for  10,000 trials. The statistic of interest is calculated for each bootstrap sample and the frequency distribution of the statistic over all the bootstrap samples is taken to represent our best information on the probability distribution of the parameters, in this case the slope. For this test we choose the combination of parameters given by model 13.

\begin{figure}
\includegraphics[scale=0.3]{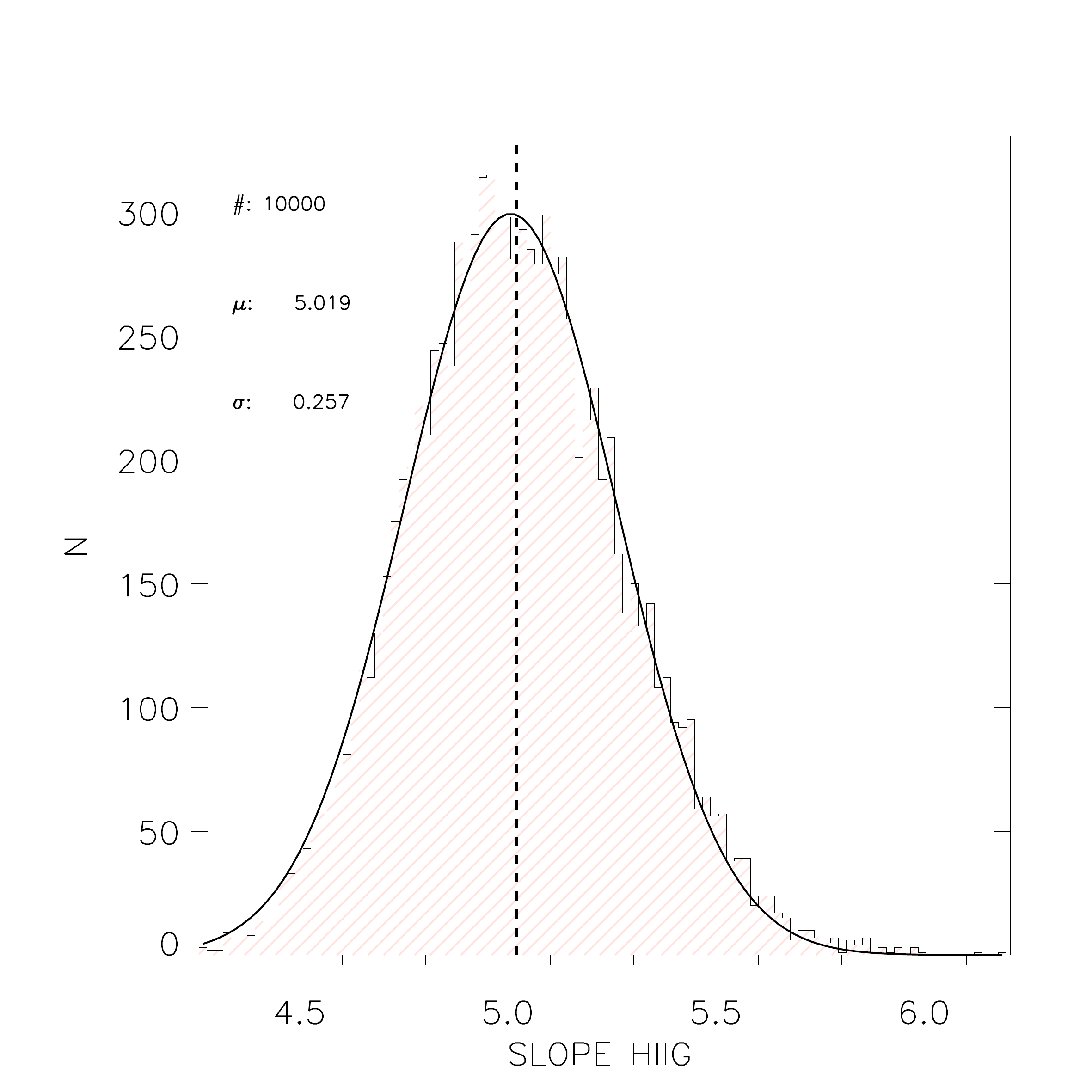}
\caption{Frequency distributions obtained using the bootstrap method for the slope of the HIIGs sample for model 13 of Table~\ref{hnotiA3}.}
\label{boots}
\end{figure}

The resulting frequency distribution is shown in Figure \ref{boots}.  The bootstrap gives $5.019\pm0.257$ while for the single solution we obtain $5.00\pm0.14$.  Since the results of the bootstrap  and of the single solution are similar we can conclude that the slope of the \lsigma relation  is  robust to random changes in the sample. 


\subsubsection{Sensitivity of \ho\ to changes in the sample.}

As already mentioned the sub-sample S2 has an upper redshift cutoff of z = 0.1 instead of the z = 0.16 limit for  our primary sample S1. Comparing S1 (in which the distances were computed using a flat cosmology with $\Omega_m$=0.29) with sample S2 (using the linear Hubble relation for the distances)  Figure~\ref{modelosH} shows that using S2 reduces  the value of \ho\ typically by about 2 $\kmsmpc$ with a range from 1.1 to 3.3 $\kmsmpc$; the results for  S2 give a slightly larger uncertainty than for S1, which is to be expected  given the smaller size of the sample.
The sensitivity of \ho\ to the actual value of $\Omega_m$ is low, amounting in our case to an uncertainty of about  0.1\% in \ho\ for an uncertainty in $\Omega_m$ of 0.02 \citep[see][]{Betoule2014}. This, together with the larger uncertainty when using  S2 drives us to use  S1 with the distance determined using a flat cosmology with  $\Omega_m$=0.29. 

Putting together these points with  the aspects discussed in the previous sections led us to choose model 13 (S1-G03-Ch14-$\Omega_{\Lambda}=0.71$) as our preferred one.  Model 13 gives \ho\ values of   $74.6 \pm2.9 $  and  $71.0 \pm2.8 $ $\kmsmpc$ for the uncorrected and evolution corrected cases respectively.

It is important to notice that the \ho\ results for model 10 (S2-G03-Ch14-LR) are  close to the results for model 13; they are  \ho\ =  $74.3 \pm3.1 $ and  $70.9 \pm3.0 $ $\kmsmpc$  for the uncorrected and evolution corrected cases respectively. Since in model 10 we are using the linear Hubble relation (restricting the sample to objects with z < 0.1), this reinforces the point  that we are not biasing the results by using additional Cosmological parameters. 

\subsubsection{Additional checks}

\begin{figure}
 \includegraphics[scale=0.33]{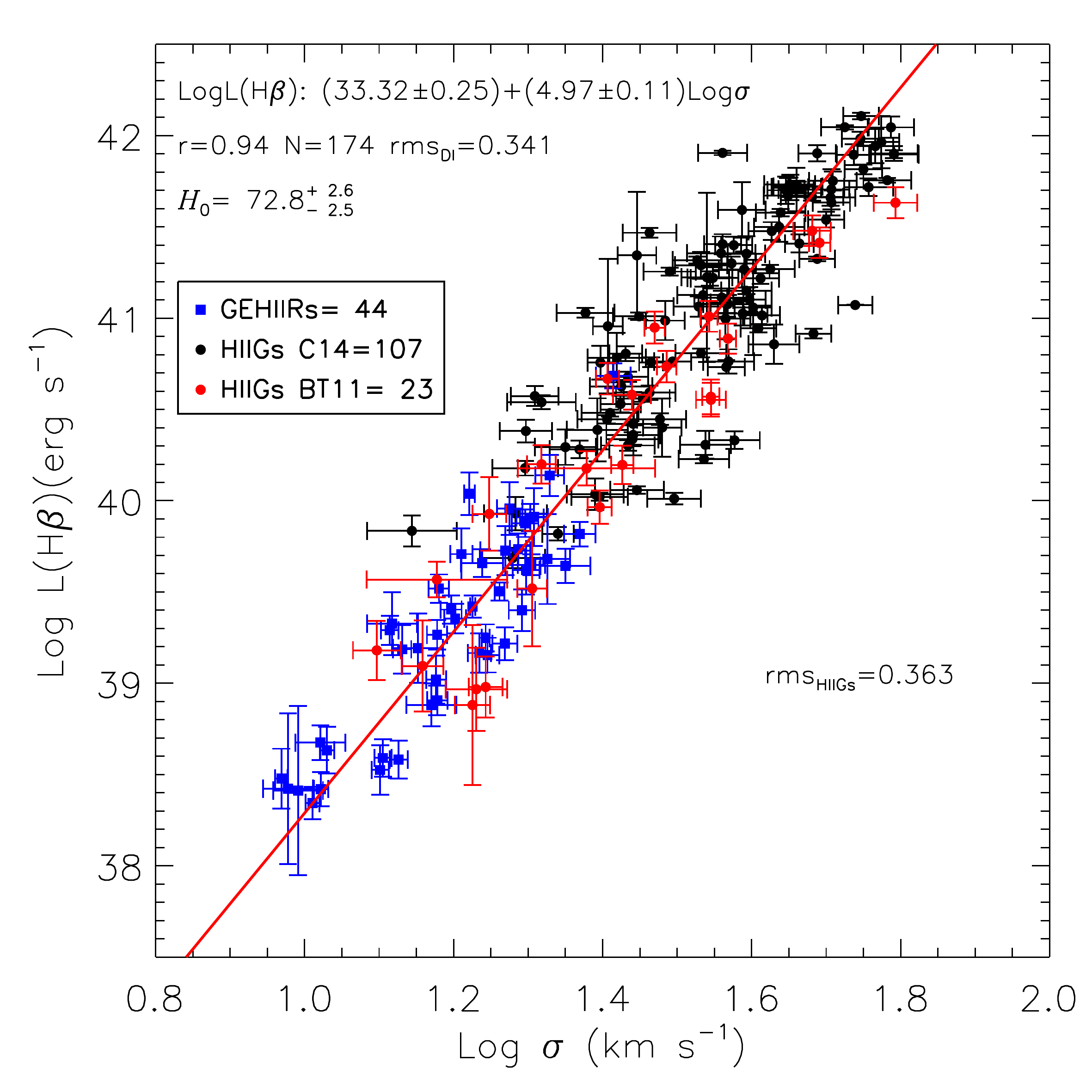}
 \caption{\lsigma relation for  \citet{Chavez2014}  and \citet{Bordalo2011} combined samples. The solid line is the fit to the HIIGs. The inset equation is the distance indicator where the slope is obtained from the fit to the HIIGs  and the Zp determined following the procedure described in the text.  }
 \label{C14BT11}
 \end{figure}

Following \cite{Melnick2017}, we have also analyzed the effect of expanding the sample by incorporating the data of HIIGs published in \cite{Bordalo2011} and GHIIRs from \cite{Chavez2012}. The resulting sample includes a total of 130 HIIGs and 44 GHIIRs.
To homogenize the enlarged sample we were forced to use the SDSS photometry for the 107 HIIGs \citep{Chavez2014} in  order to make it compatible with \citet{Bordalo2011}. Extinction correction was performed using \cite{Gordon2003} law and the HIIGs luminosities were computed using a flat cosmology with $\Omega_{\Lambda}=0.71$. The \lsigma relation using the data of \citet{Chavez2014}  plus  \citet{Bordalo2011} can be seen in Figure \ref{C14BT11}. Using this relation as the distance estimator we obtain a value of \ho\ =  $72.8 \pm2.6 $ $\kmsmpc$  that should be compared with the result  of \ho\ =  $70.0\pm2.8 $ $\kmsmpc$  obtained for model $N=15$ in Table \ref{hnotiA3}.

\begin{figure}
 \includegraphics[scale=0.33]{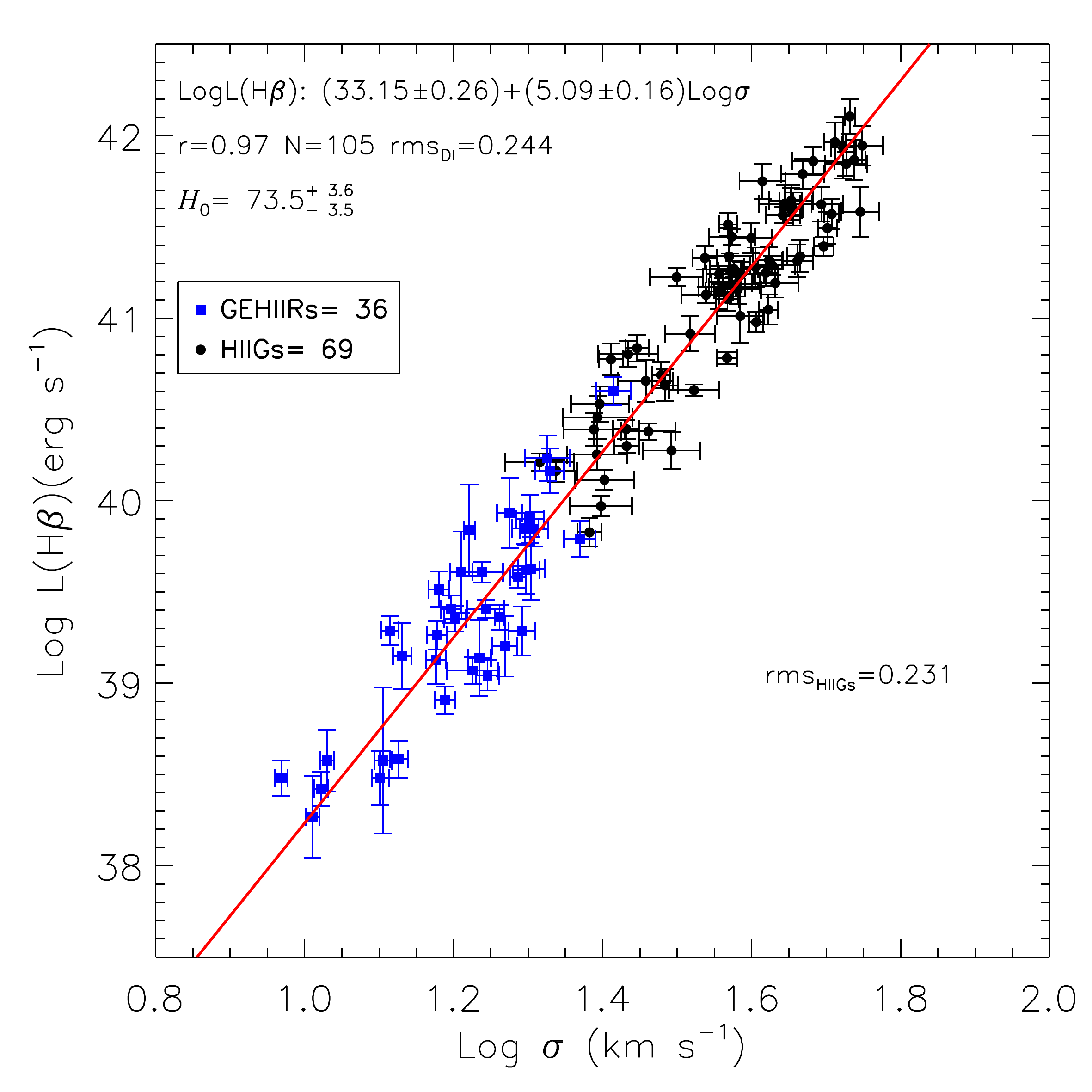}
 \caption{\small  The \lsig\ relation and \ho\ determination using the sample of 69 HIIGs of  \protect\cite{Chavez2012}. The solid line is the fit to the HIIGs. The inset equation is the distance indicator where the slope is obtained from the fit to the HIIGs  and the Zp determined following the procedure described in the text.  }
 \label{69C12}
 \end{figure}

A final check was done by including the more restrictive sample of HIIGs from \cite{Chavez2012} where galaxies with  asymmetric or multiple line profiles in either $\hb\ $ or [OIII]~5007\AA\, and galaxies with large photometric errors or uncertain extinction corrections were removed, thus reducing the  sample to 69 HIIGs. We computed \ho\ using these 69 galaxies from \cite{Chavez2012} and our new anchor sample of 36 GHIIRs as shown in Figure~\ref{69C12}) and obtain \ho\ =  $73.5\pm3.6$ $\kmsmpc$, in good agreement with the value of \ho\ =  $74.6 \pm2.9 $ from model 13  in  Table~\ref{hnotiA3}.


\subsection{Summary of systematics effects}

From the results of the different models in Tables  \ref{hnotiA3}  and  \ref{hnotiA7} and the discussion in this section we can infer that the systematic effects cause changes in the  \lsigma relation that translate into  r.m.s. variations  in the value of \ho\ of 1.5 $\kmsmpc$  and 2.5 $\kmsmpc$ for the solutions with and without evolution correction respectively and of 2.1 $\kmsmpc$ for the 32 solutions.

 \section{Comparison with SNI{a} and Planck CMB}\label{SN}\label{comparison}

The direct determination of the value of \ho\ derived from the HST Key Project and Carnegie Hubble Program \citep{Freedman2001, freedman2012} reached an accuracy of  3\% reporting values of  \ho\ = $73.8\pm2.4$ and \ho\ =  $74.3\pm2.1$, $\kmsmpc$ respectively.
\citet{Humphreys2013} using the megamaser galaxy NGC~4258  estimated \ho\ =  $72.7\pm2.4$ $\kmsmpc$.
 These determinations together  with the estimate by \cite{riess2011} of \ho\ =73.8$\pm2.4 \kmsmpc$ showed a 2.5$\sigma\ $ tension with the \cite{Planck132014} derived value of  \ho .
  \cite{Efstathiou2014} re-analysed   \cite{riess2011} Cepheid data and concluded that there is no evidence for a need to postulate new Physics.
 
  \cite{Rigault2015}, using the Nearby Supernova Factory sample, found that SNIa  are dimmer in  star forming than in passive environments. 
  As the majority of Cepheid based distances are for late type and star forming galaxies this can lead to a bias in cosmological measurements. 
  Correcting for this bias, they find a value of \ho= $70.6\pm 2.6$ $\kmsmpc$ when using the Large Magellanic Cloud distance, Milky Way parallaxes, and the NGC~ 4258 measurements as the Cepheid zero point, and \ho=$68.8\pm3.3$ $\kmsmpc$ when using only NGC~ 4258. This last value is within 1$\sigma$ of the Planck collaboration result. 
 It has to be mentioned that the \cite{Rigault2015} result was reanalized by \cite{Jones2015}  who found no evidence that SNIa in starforming environments are significantly fainter than in locally passive environments.

\cite{Riess2016} presented a comprehensive and thorough analysis  of an enlarged sample of SNIa and improved distances to the anchor sample. Interestingly while addressing the result from   \cite{Efstathiou2014}  found that a change in the colour cut removes the problem.  This most recent result from \citet{Riess2016} has reinstated the tension, now at the 3.1$\sigma$ level, between the value obtained  by  \cite{Planck152016}  of \ho\ = $67.8\pm0.9$ $\kmsmpc$ and \citet{Riess2016} of \ho\ =  $73.2\pm1.8$ $\kmsmpc$.

 
Our main result incorporating the evolution correction,  is\\ \ho\ =$71.0\pm3.5$ $\kmsmpc$(random+systematic) a value that is half way between    the  \cite{Planck152016} estimate and \cite{Riess2016} determination.
 
 

  \begin{figure*}
 \includegraphics[scale=0.8]{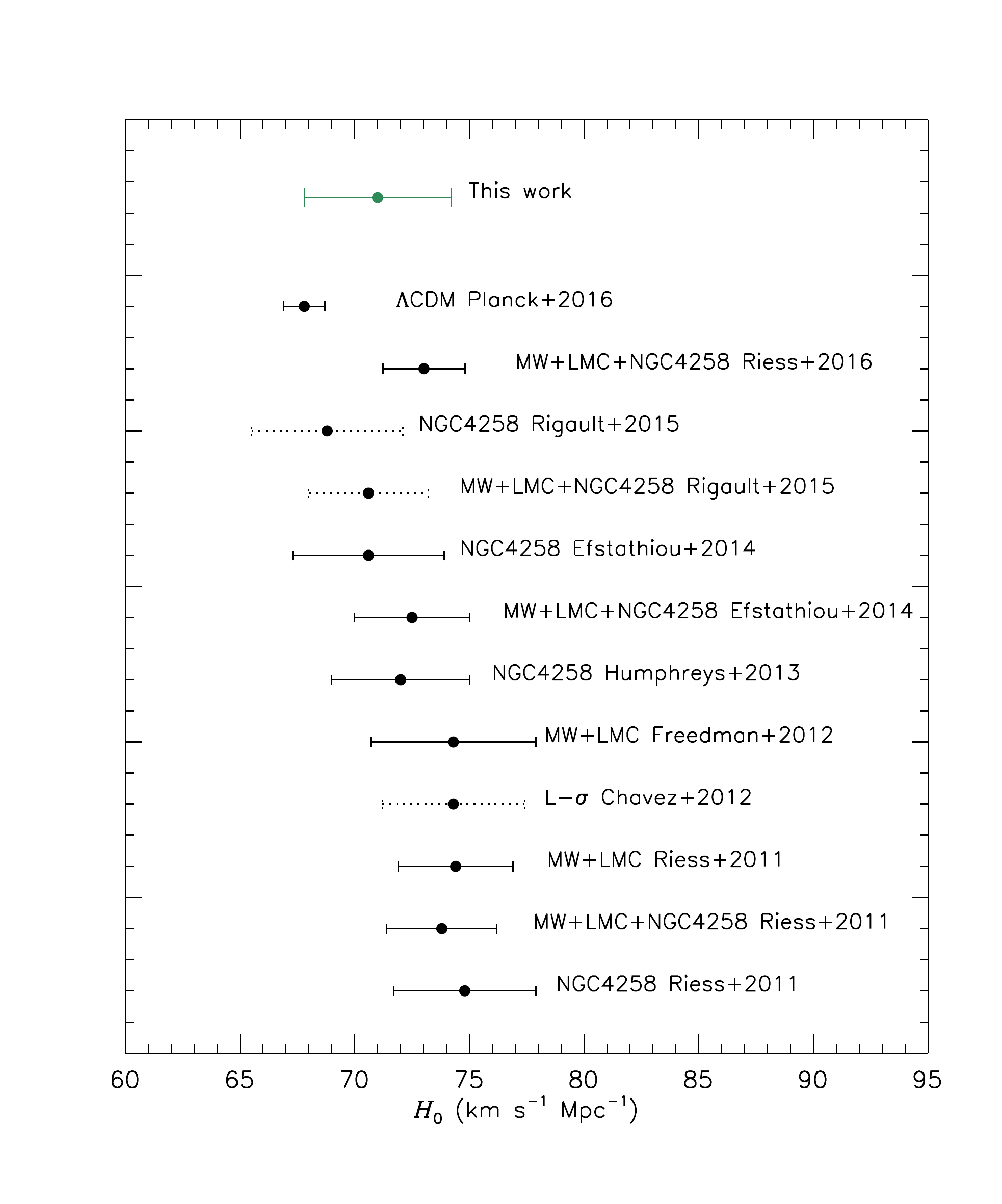}
 \caption{Most recent estimates of \ho\ from the literature. The top line shows our evolution corrected result. Dotted error bars indicate only random errors while continuous bars indicate random plus systematic errors. See discussion in \S \ref{comparison}.}
 \label{h0resul}
 \end{figure*}

\section{Conclusions}\label{conclusions}

 We have used the \lsigma distance indicator to derive an independent local value of the Hubble parameter \ho{.} To this end we have combined new data for 36 GHIIRs in 13 galaxies of the `anchor sample' that includes the megamaser galaxy NGC~4258, with the data for 107 HIIGs from \cite{Chavez2014}. Our new data is the result of the first four years of observation of our primary sample of 130 GHIIRs  in 73 galaxies with Cepheid distances.  
 
 The determination of \ho\ is a rather delicate undertaking and that is reflected in the range of values that we obtain for it.
Figure \ref{modelosH}  shows  the values obtained with different combinations of parameters such as: redshift, extinction laws, luminosities derived using either the linear relation for the distance or  cosmological parameters, the photometry derived from SDSS or \citet{Chavez2014} and the correction using a second parameter related to the age of the burst as parametrized by the EW($\hb$).

The results from  stellar population synthesis models, such as SB99, allow to estimate a theoretical evolutionary correction for all GHIIRs and HIIGs. One major problem with this approach is that different sets of isochrones give somewhat different correction coefficients, and also that the observed EW($\hb$) is affected by the presence of an underlying older population and differential extinction in a degree that is not simple to estimate. Furthermore, present day evolutionary models do not include the effect of massive binaries nor the photoionization is fully modelled.
All the aforementioned effects  add  to the uncertainty of the result.

Using the  SDSS photometry gives values of \ho\ slightly lower  than those calculated using the  photometry of \cite{Chavez2014}. This is probably related to the fact that the small  aperture of the SDSS  spectroscopy 
underestimates  the emission line fluxes in the nearest objects, that happen to be also the lowest luminosity ones. The result is a steeper slope  in the  \lsigma relation  leading to a smaller value of \ho .  

We estimated the effect of varying the extinction law on the derived value of \ho . For this we have used two different extinction laws:  \cite{calzetti2000} which has been widely used for starburst galaxies, and  the  one for 30-Doradus given by \cite{Gordon2003}. 
Using  the extinction law from \cite{calzetti2000} tends to produce values of \ho\ slightly larger than when using the extinction law from \cite{Gordon2003} but with a systematic difference  inside the \ho\ errors. Given that Gordon extinction law is derived from the prototypical massive star forming region 30-Doradus, while Calzetti extinction law is derived from global properties of mostly massive star forming galaxies and therefore includes aspects related to the parent galaxy, we chose to use the former.

A very small change in the Hubble constant is obtained when we take into account the effect of the underlying absorptions on Balmer emission lines.  
This effect is larger for the high order Balmer lines  and \hbeta\  is only weakly affected;  moreover the selection criterion with EW(\hbeta)> 50\AA{} minimizes this effect.

We  also investigated the stability of the solutions using an expanded sample that included the data from \cite{Bordalo2011} and GHIIRs from \cite{Chavez2012} leading to a total of 137 HIIGs and 45 giant GHIIRs.  Using in this case SDSS photometry we obtain a value of \ho\ =  $72.8\pm2.8$ $\kmsmpc$.

From our determinations of \ho\ we estimate that the systematic errors are  2.1 $\kmsmpc$ including the error associated with the evolution correction;
the way we propagate errors is not completely rigorous, however we consider it is an  appropriate statistical tool to investigate the systematic effects of the \lsigma relation to determine distances and  \ho.   

In sum our preferred model incorporates the  reddennig law from \cite{Gordon2003}, \hbeta\ photometry from  \cite{Chavez2014} and luminosity distances  with complete cosmology using the whole sample S1 (for full details see Section \ref{explore}). 
Under these conditions we obtain \ho\ =$74.52\pm2.85$ $\kmsmpc$, almost identical to the value reported by \cite {Chavez2012} of    \ho\  = $74.3\pm3.1$ and consistent within errors with the new results from SNIa \citep{Riess2016} of \ho\ =  $73.24\pm1.74$ $\kmsmpc$ (see model 13 marked with an asterisk in Table~\ref{hnotiA3}).

Including an evolution correction leads to our best estimate,\\  \ho\ = $71.0\pm2.8(random)\pm2.1(systematic)$ $\kmsmpc$   \\\
(see model 13 marked with an asterisk in Table~\ref{hnotiA7}) a value that is between the two best results so far, i.e. \cite{Planck152016} estimate of \ho\ = $67.8\pm0.9$ $\kmsmpc$, and \cite{Riess2016} determination of \ho\ =  $73.2\pm1.8$ $\kmsmpc$.




Regarding future improvement of the \lsigma\  distance indicator, our first priority is to increase the anchor sample from the present 13 galaxies to the 43 galaxies of our primary sample. Much of the error in the value of \ho\ is related to the uncertainty in the value of the slope of the \lsigma relation, thus  it will be important to include low luminosity HIIG, i.e. those with luminosities similar to the luminosity of GHIIR, and also  GHIIR in more distant galaxies.
The addition of a second parameter in the \lsigma\ relation can lead to important improvements in the distance indicator.  In particular the size of the starforming region has proven to be a real possibility potentially reducing the scatter  by about 40\%.  We  also plan to expand the analysis to include TRGB distances to the galaxies in the primary sample. Finally the evolution correction needs a  quantitative approach that takes into account the underlying stellar continuum and differential reddening effect in the measurement of the EW of the emission lines.

\section*{Acknowledgments}
The authors are grateful to an anonymous referee whose comments helped to greatly improve the clarity and accuracy of the paper.

David Fernandez Arenas is grateful to the Mexican Research Council (CONACYT)  for suporting this research under studentship 262132.
Elena Terlevich, Roberto Terlevich and Ricardo Ch{\'a}vez,  acknowledge CONACYT for supporting their research under grants: CB-2008-103365,  CB-2010-155046 and 277187.
The hospitality of 
Cananea and San Pedro M{\'a}rtir staff during the observing runs is gratefully acknowledged. 
Jorge Melnick acknowledges support from a CNPq {\it Ci{\^encia} sem Fronteiras}
grant at the Observatorio Nacional in Rio de Janeiro, the
hospitality of ON as a PVE visitor and the
hospitality of INAOE in Puebla during which parts of this research
were conducted. Eduardo Telles enjoyed useful discussions with Jailson Alcaniz.

\bibliographystyle{mnras}
\bibliography{biblio}

\clearpage

\appendix

\section{Results for individual objects}

In this Appendix we show for each GHIIR:

On the left panel the slit positions  over the H$\alpha$ image obtained from NASA/IPAC Extragalactic Database (NED).  The wider slit oriented E-W corresponds to the low dispersion spectrophotometry observations. The much narrower slit corresponds to the high dispersion spectroscopy.

The central panel shows the  high-resolution H$\alpha$ profile and two different fits: Gaussian(blue dashed line) and Gauss-H\'ermite (continuous red).
The residuals from the fits are shown in the lower panel with the same colour code. The inset shows the results of a Monte Carlo simulation to estimate the errors in the parameters of the best fit.

The right panel shows the  low-resolution spectrum and the name of the GHIIR.

 \begin{figure*}
\includegraphics[scale=0.2]{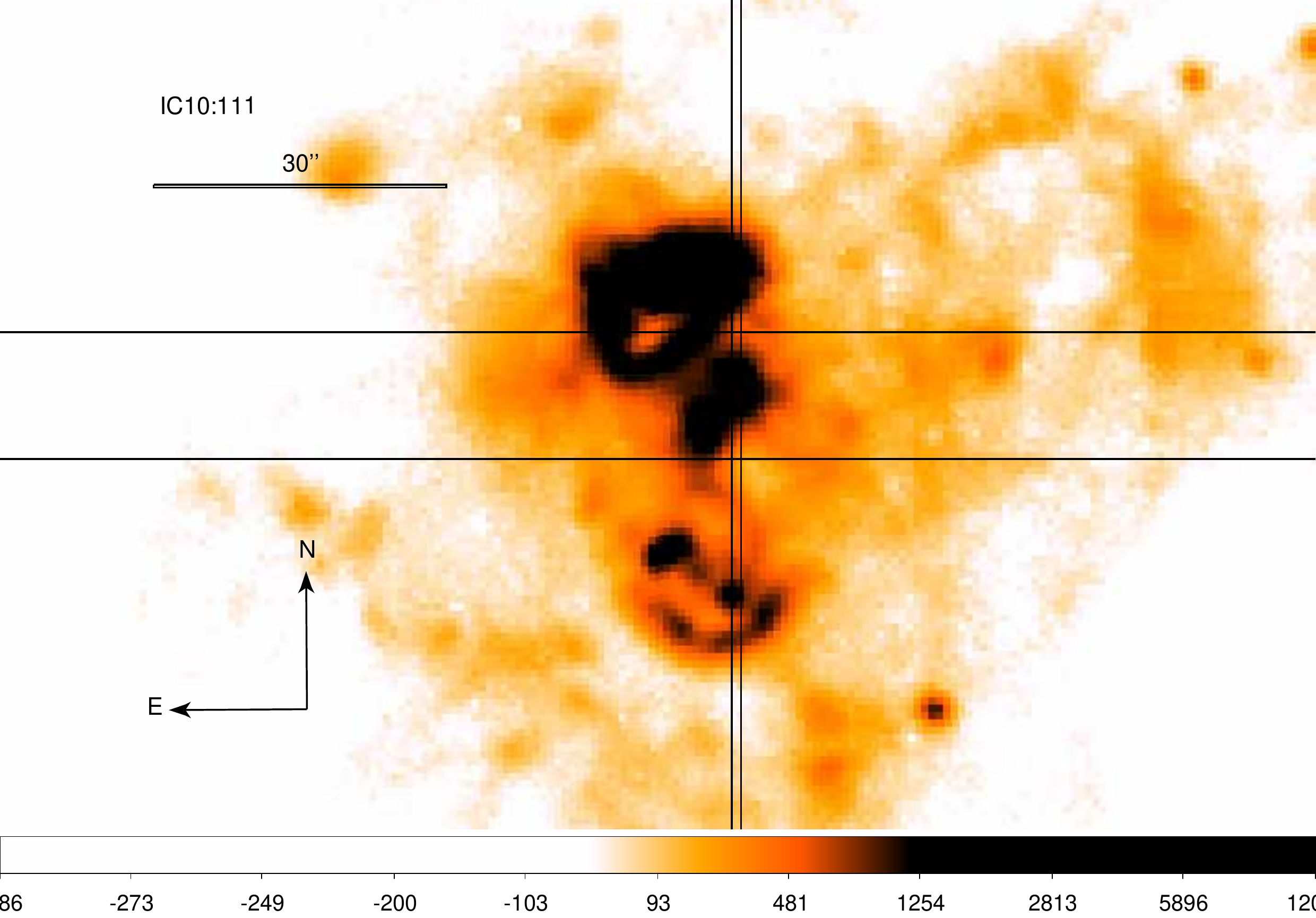} 
 \includegraphics[scale=0.18]{IC10111ha_spm_sep2013_n1_900x3W2017-01-24.pdf}
\includegraphics[scale=0.2]{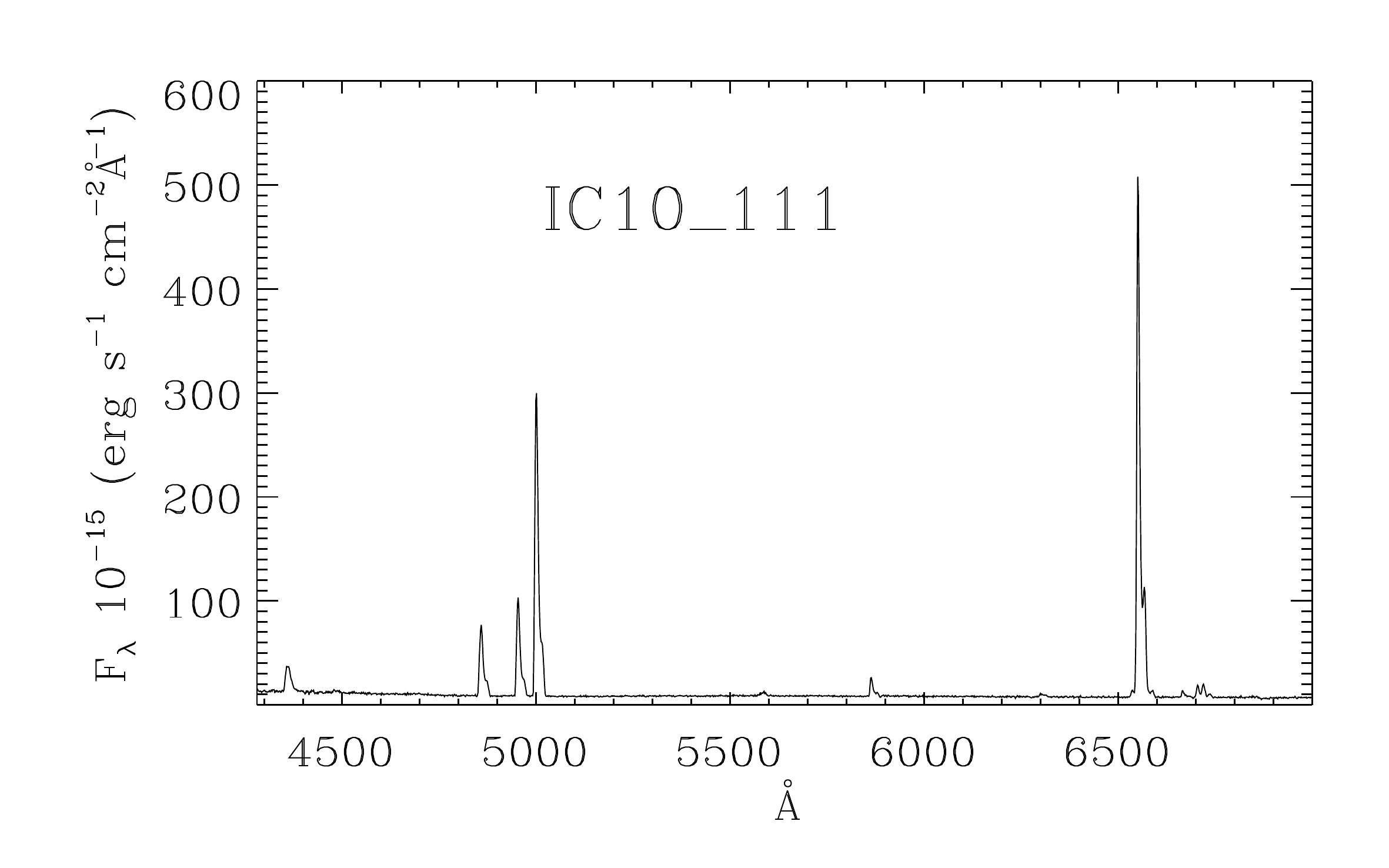}

\includegraphics[scale=0.2]{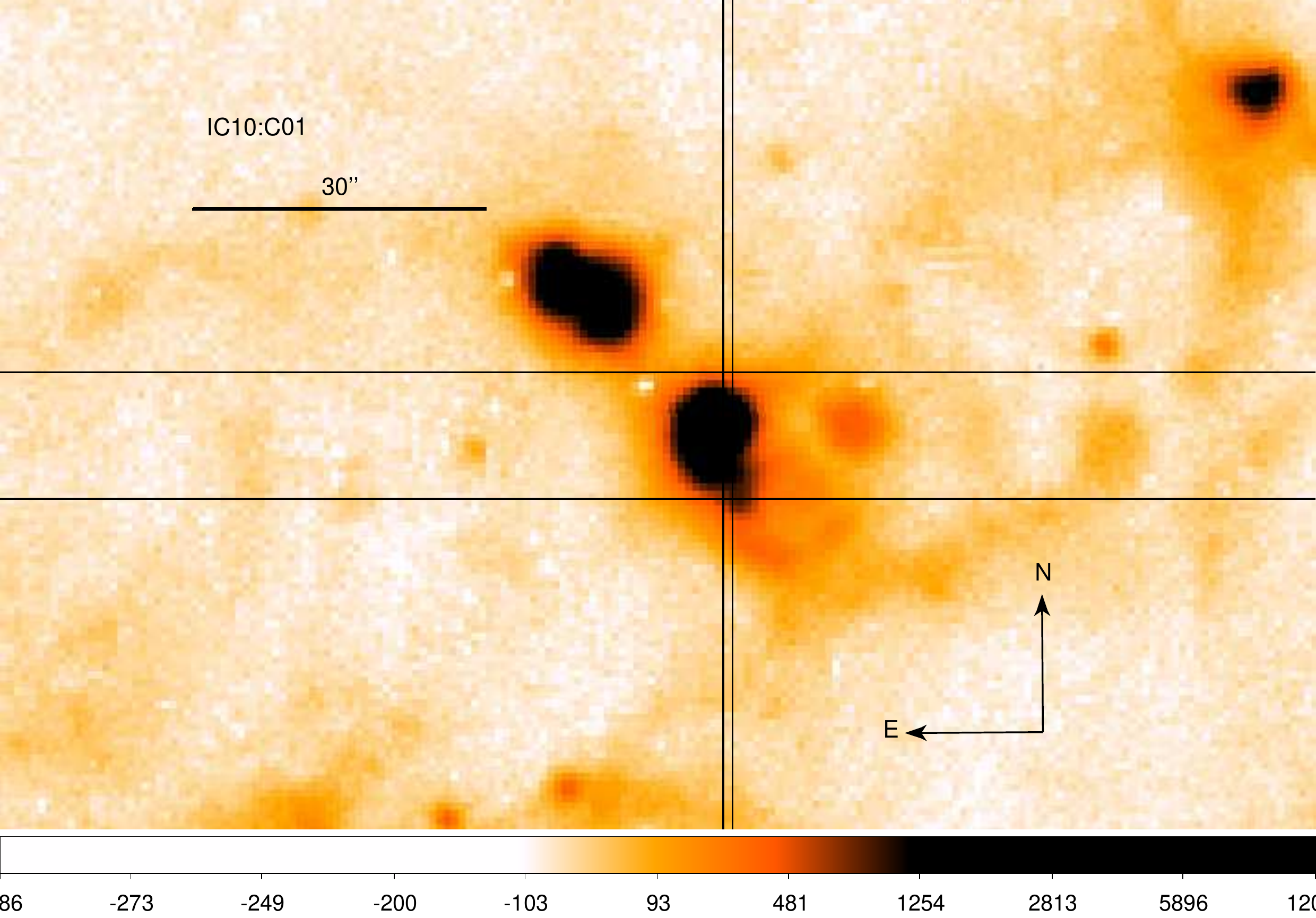}
\includegraphics[scale=0.18]{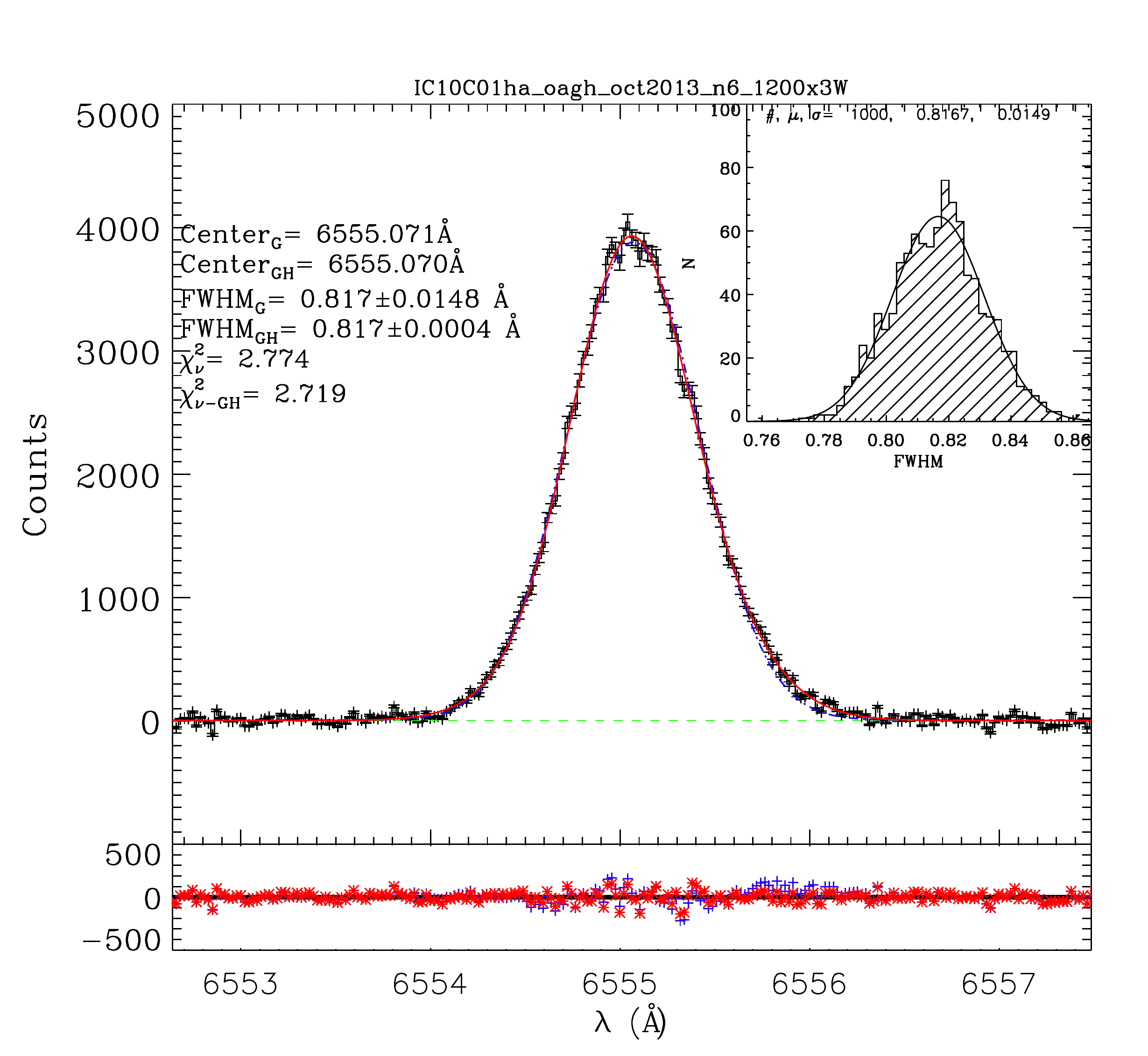}
\includegraphics[scale=0.2]{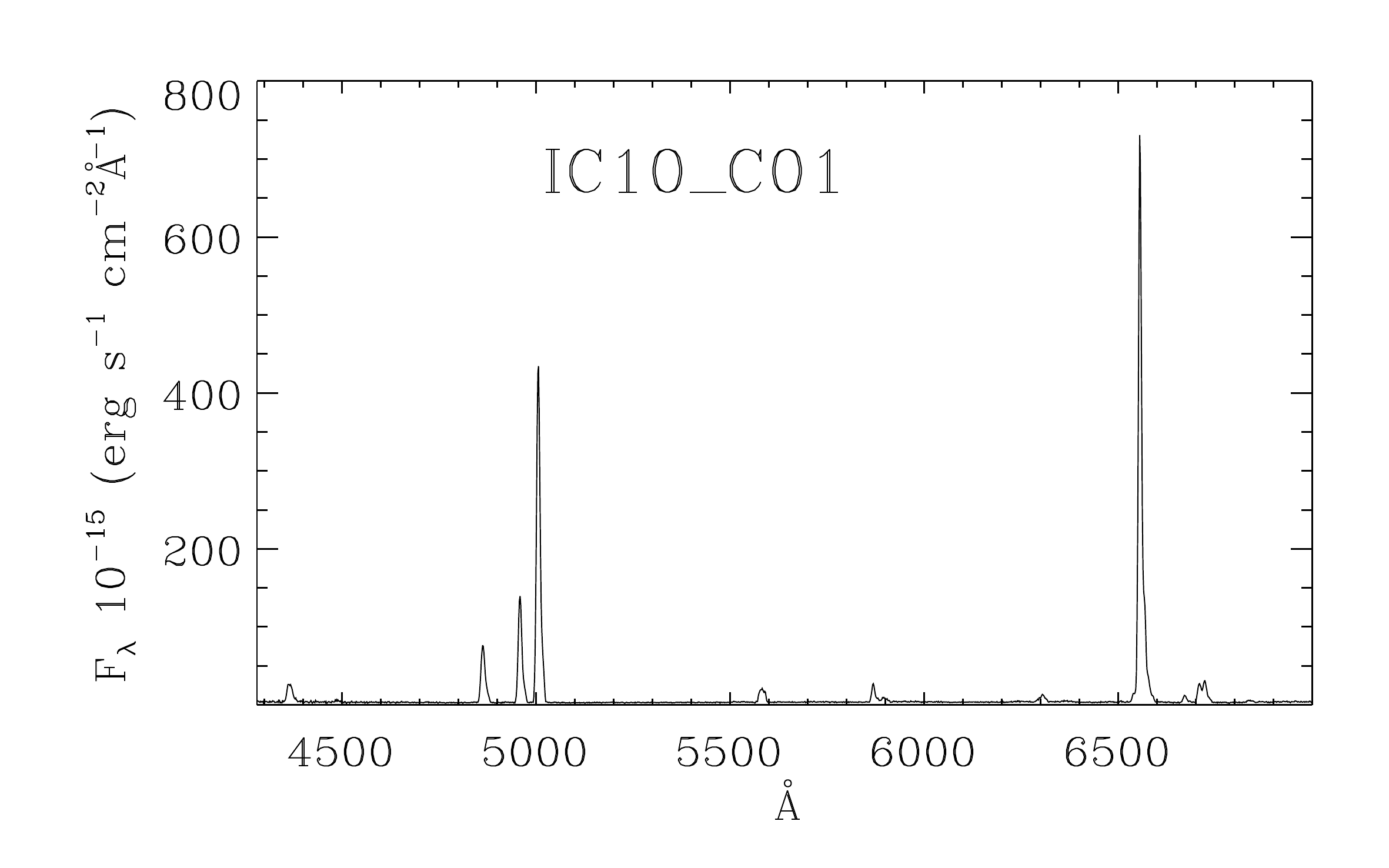}

\includegraphics[scale=0.2]{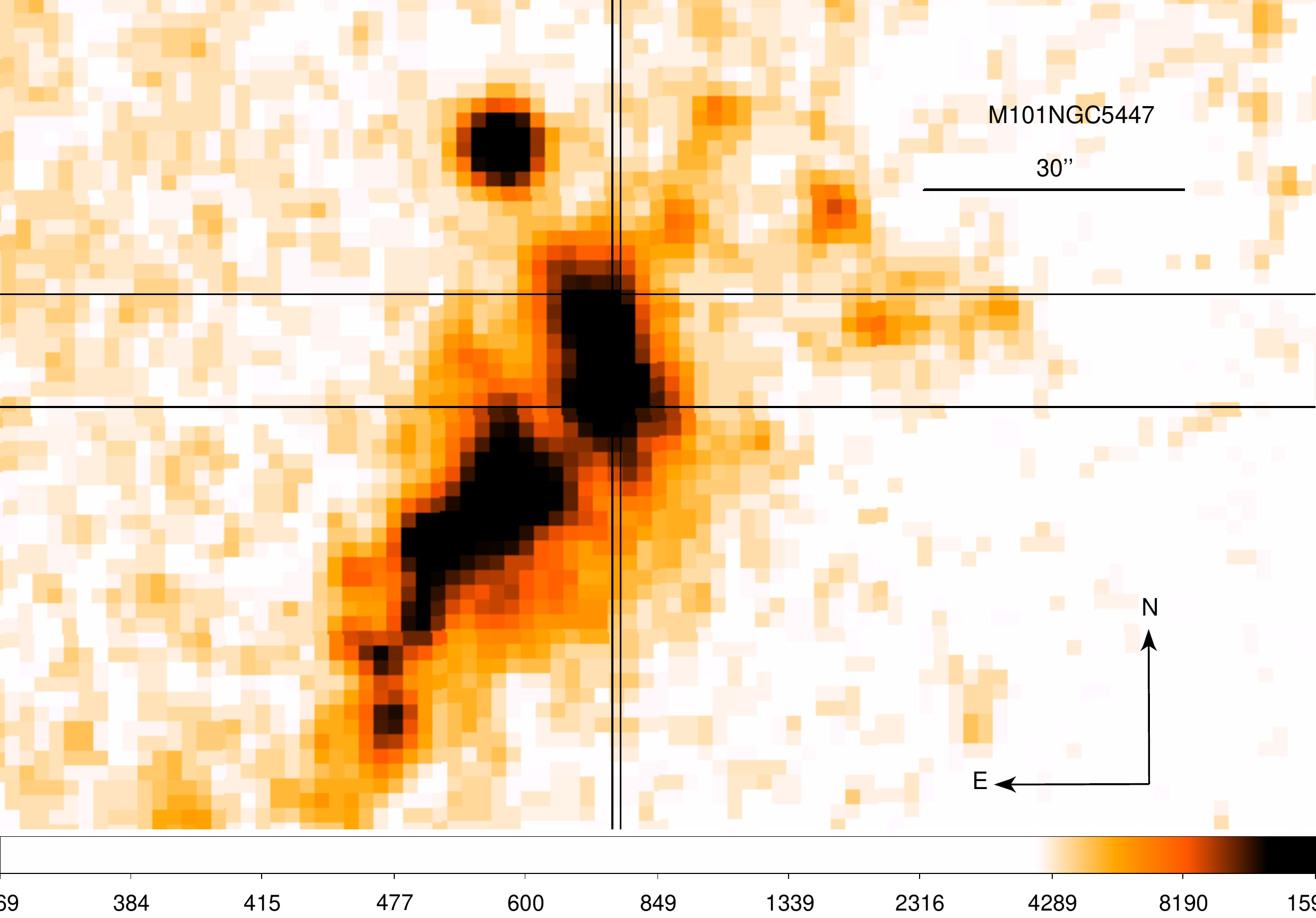}
\includegraphics[scale=0.18]{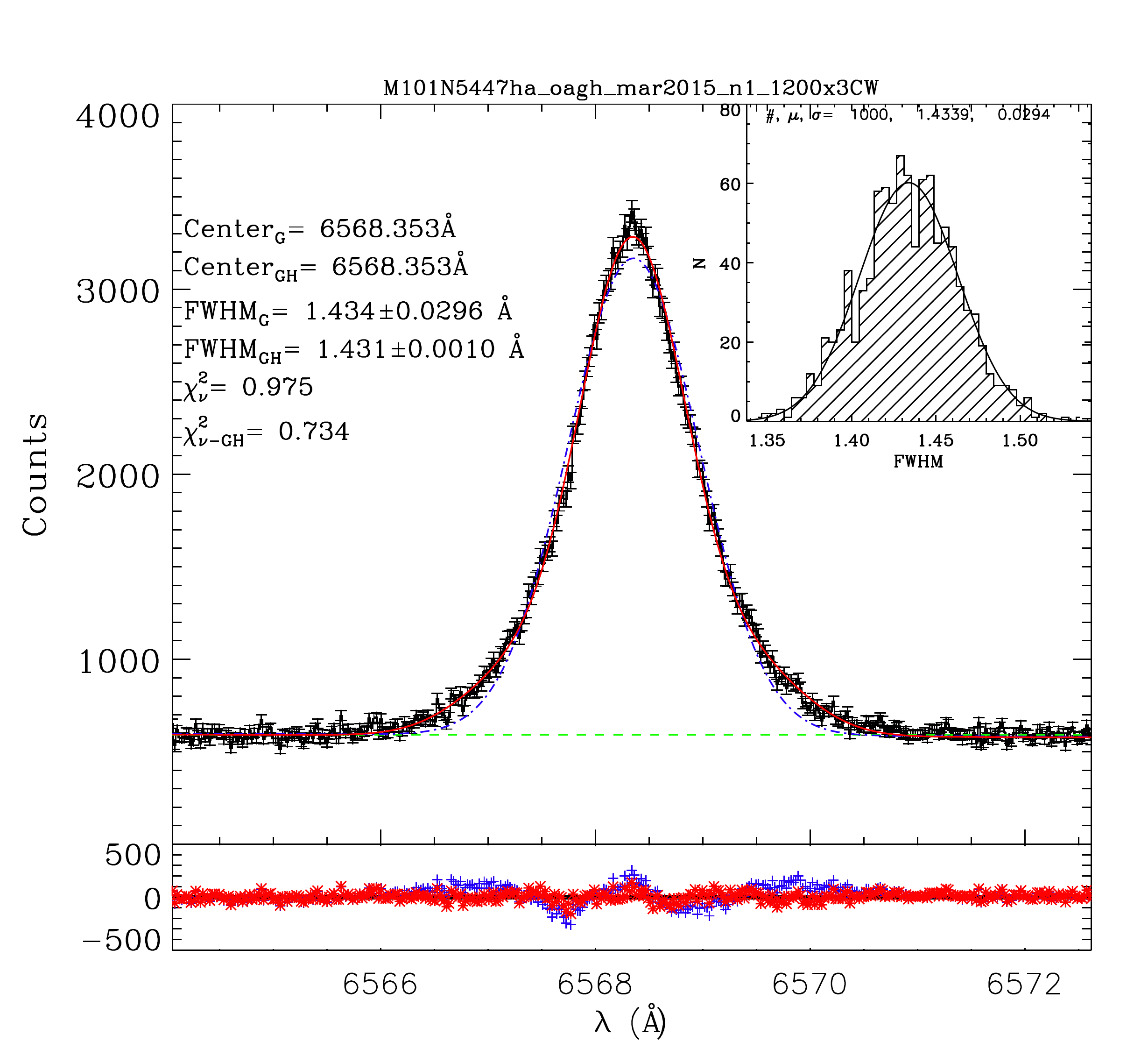}
\includegraphics[scale=0.2]{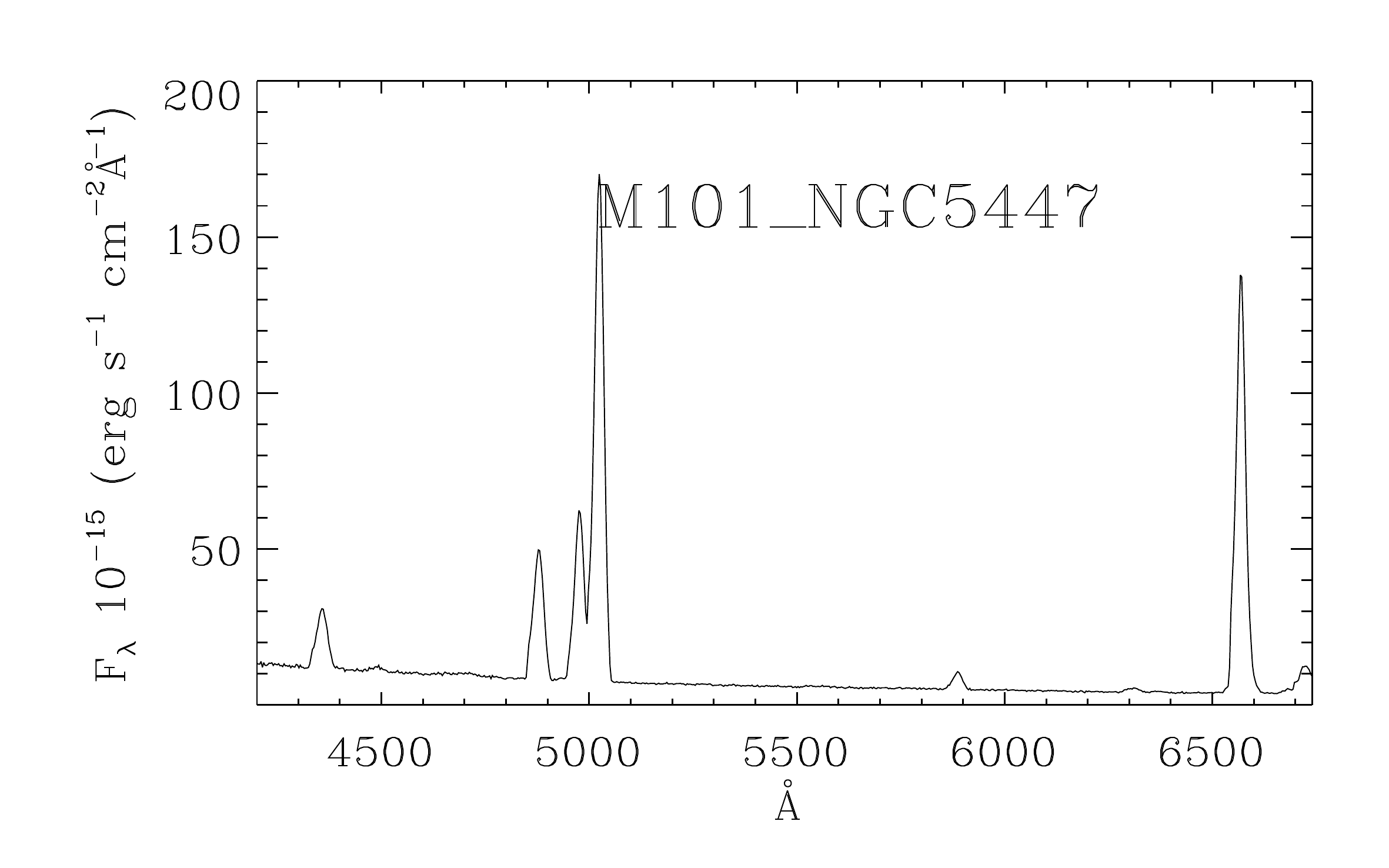}

 \includegraphics[scale=0.2]{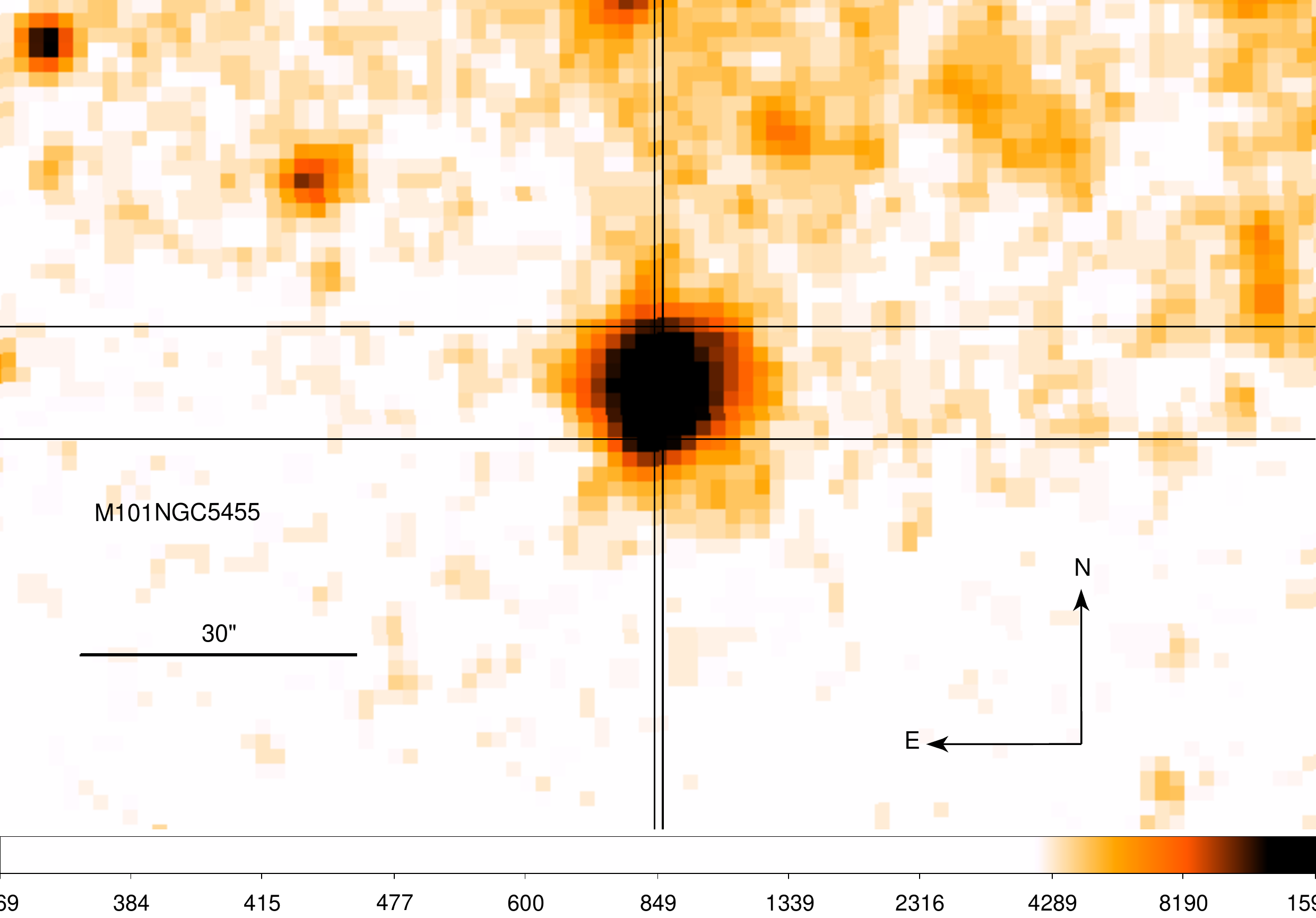}
  \includegraphics[scale=0.18]{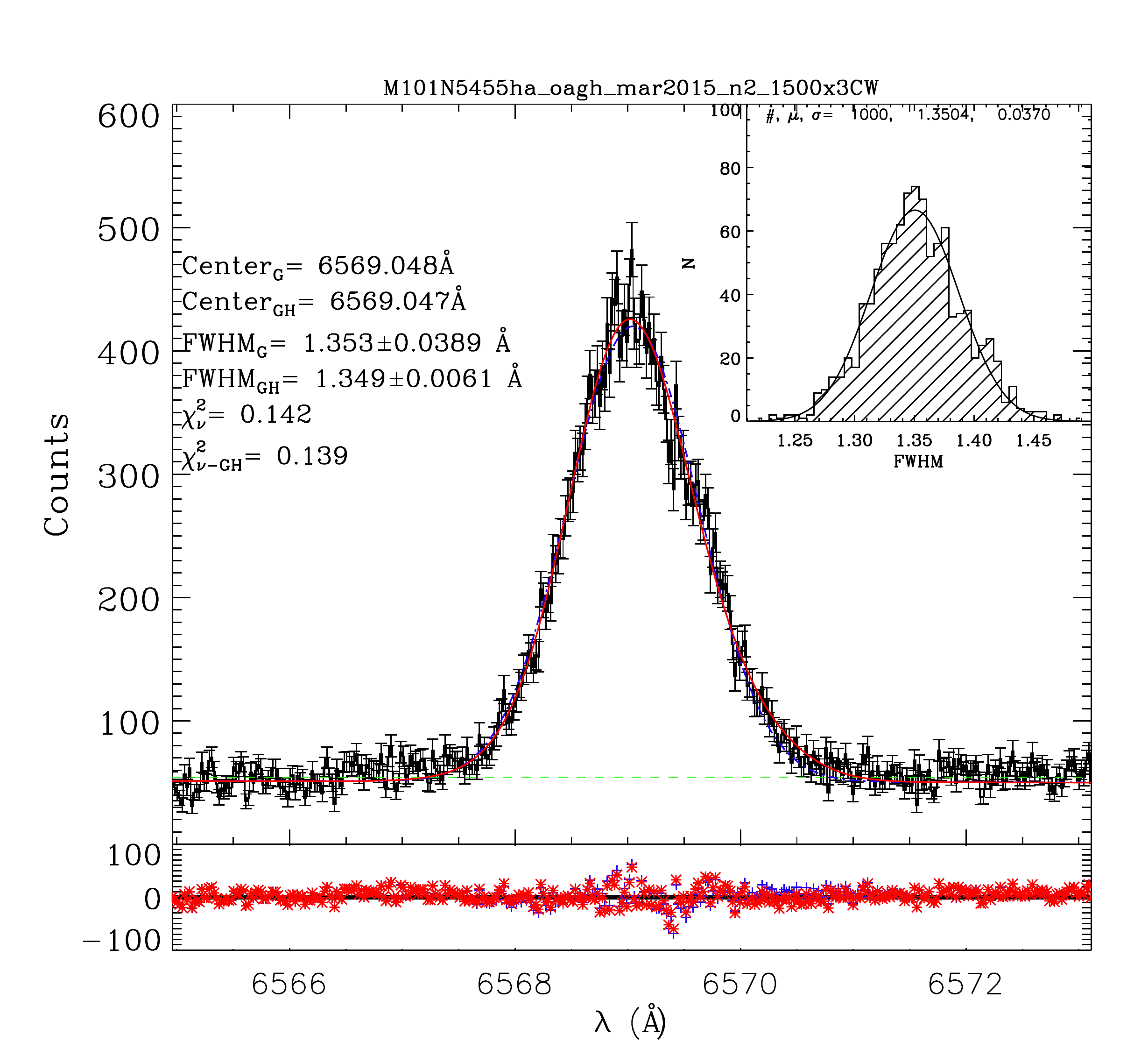}
  \includegraphics[scale=0.2]{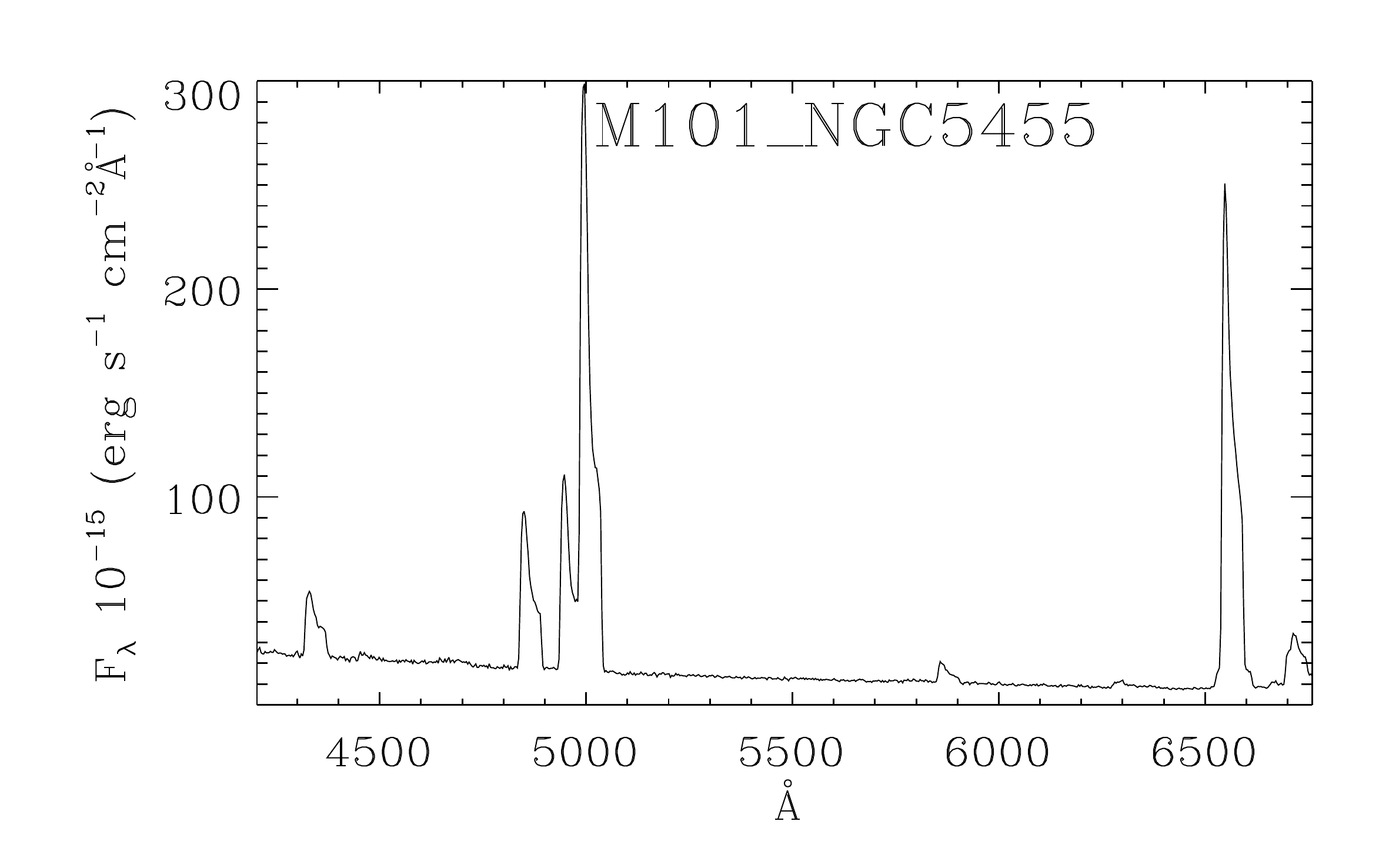}

 \caption{H$\alpha$ image obtained from NASA/IPAC Extragalactic Database (NED), high-resolution profile for the GHIIR and low-resolution spectrum.}

 \end{figure*}

 \begin{figure*}
 \includegraphics[scale=0.2]{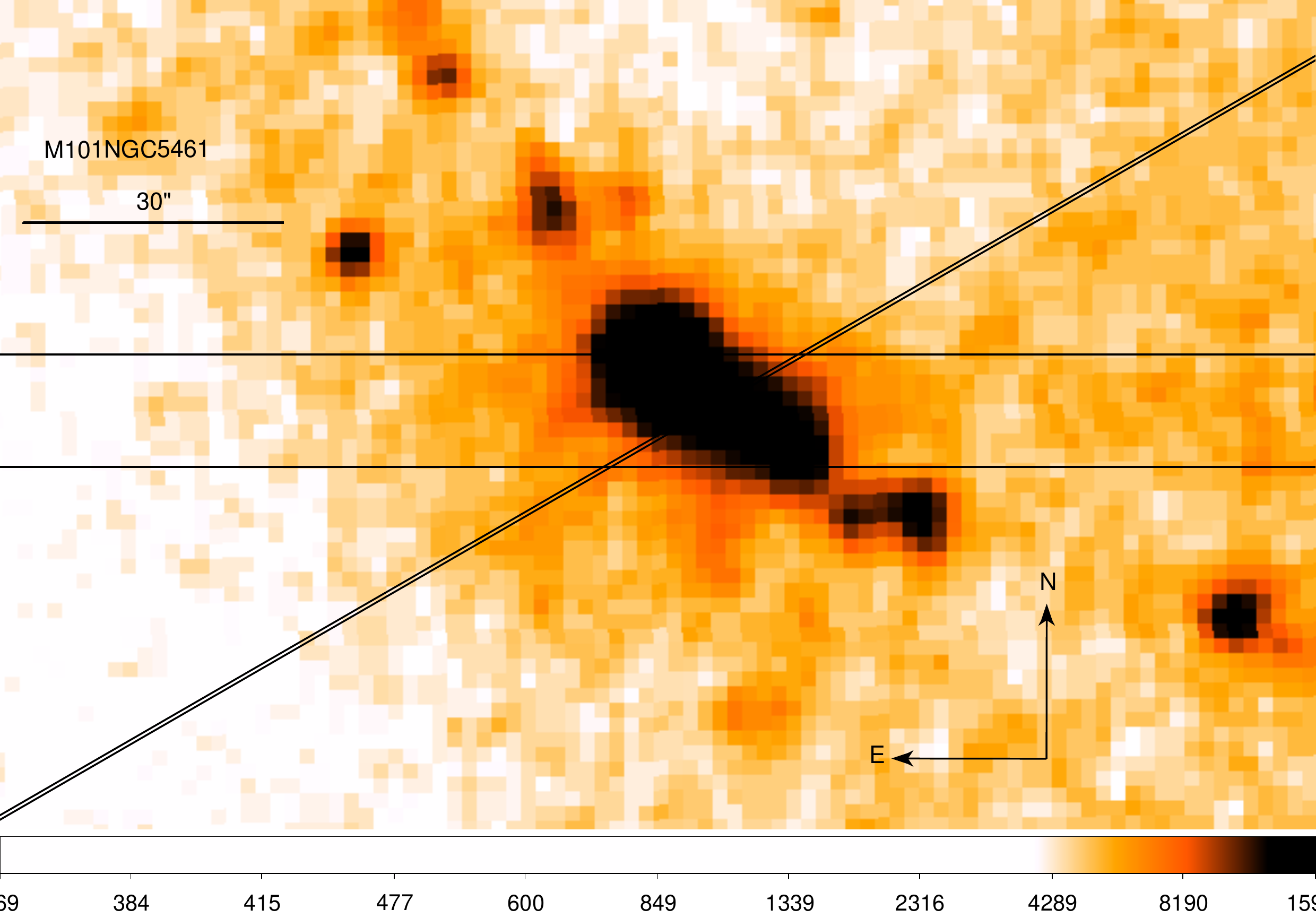}
 \includegraphics[scale=0.18]{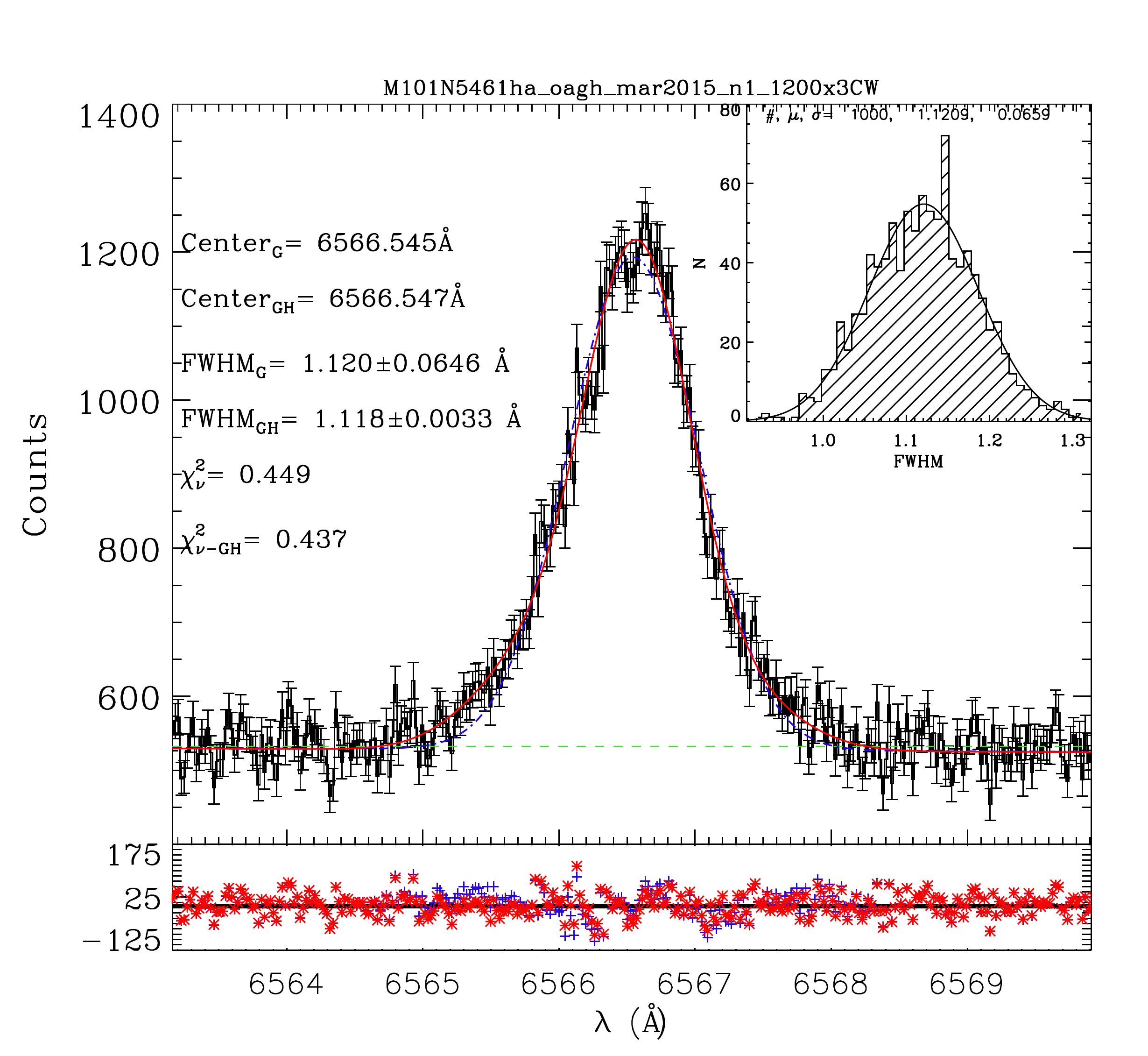}
 \includegraphics[scale=0.2]{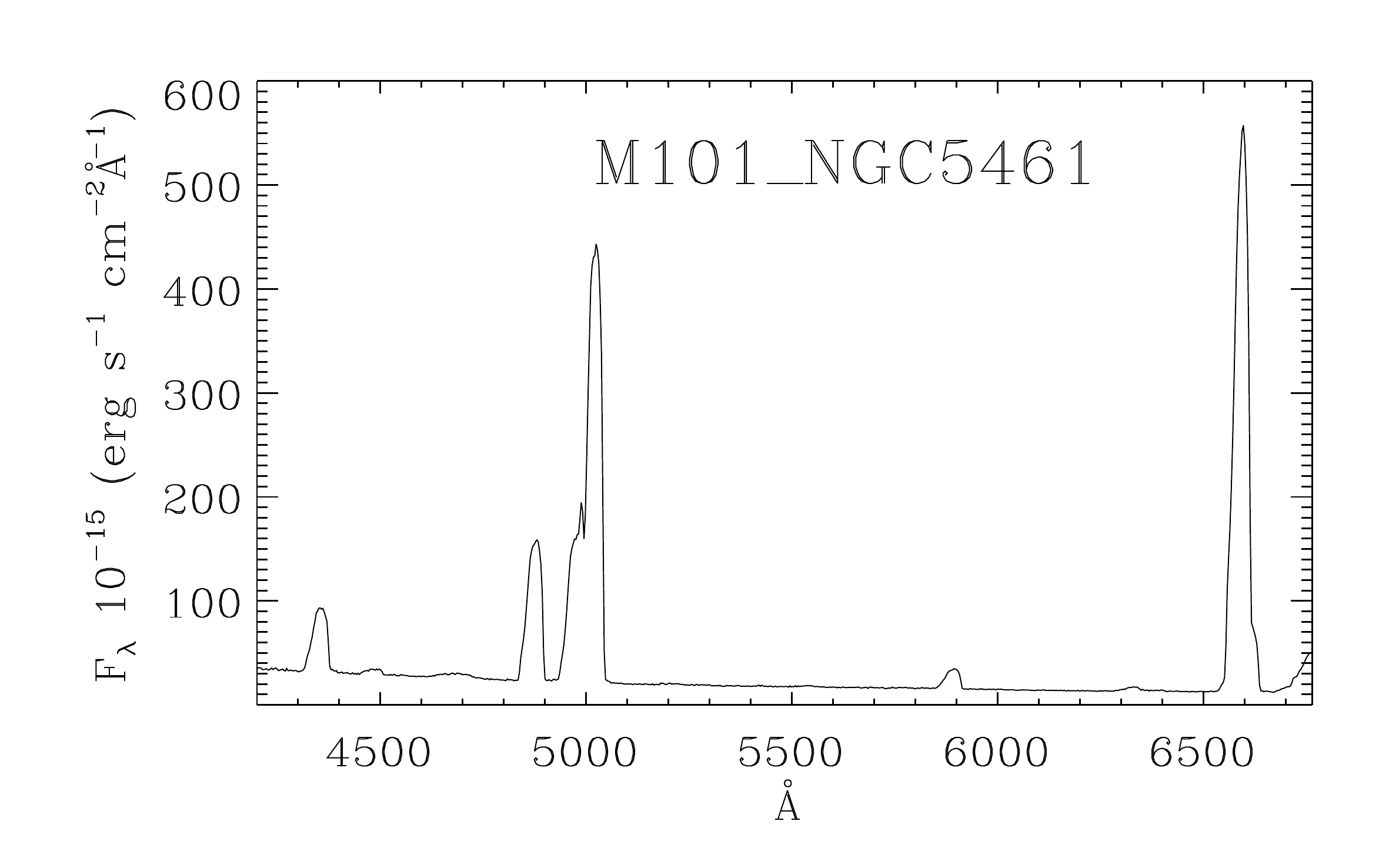}
 
 \includegraphics[scale=0.2]{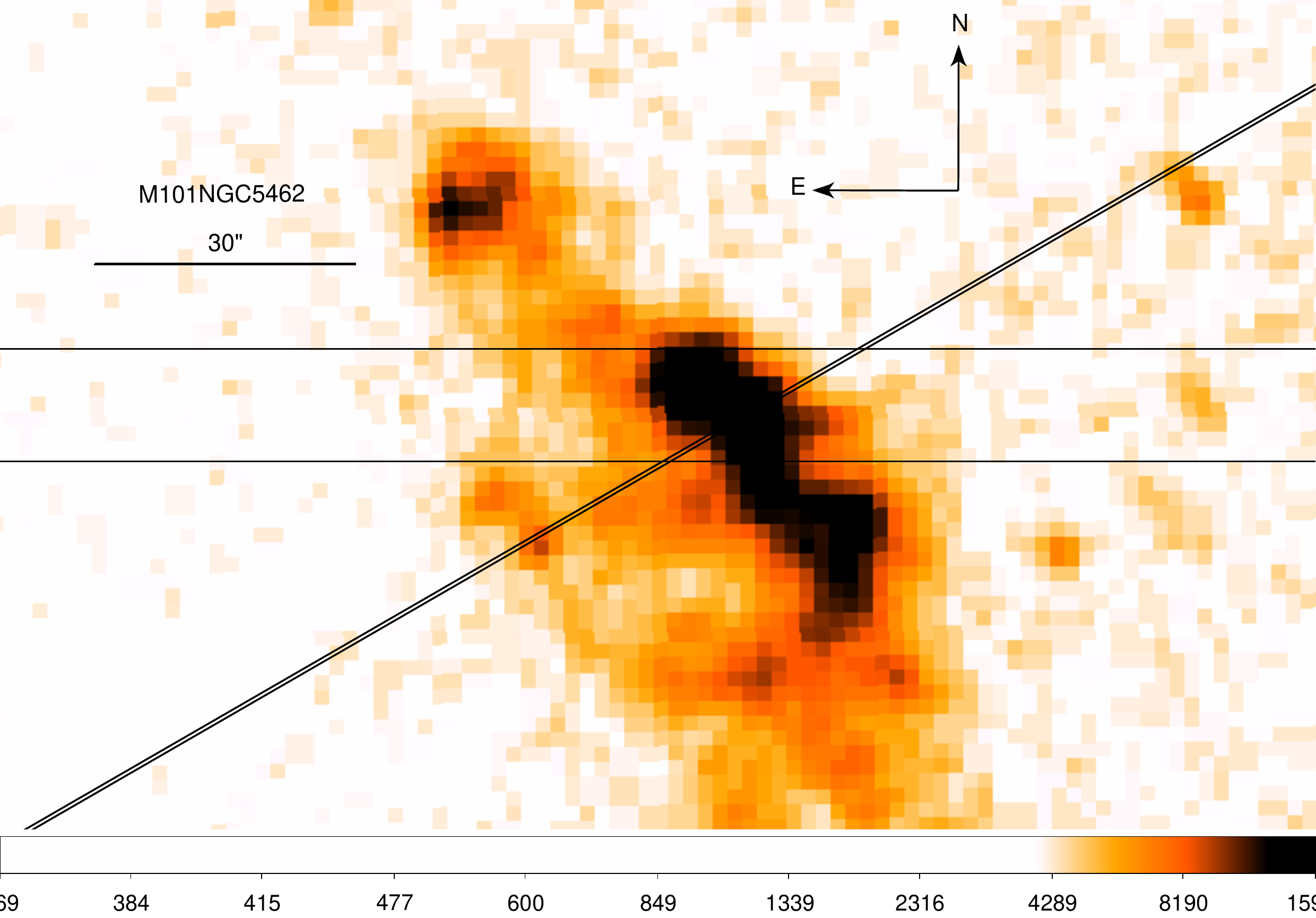}
 \includegraphics[scale=0.18]{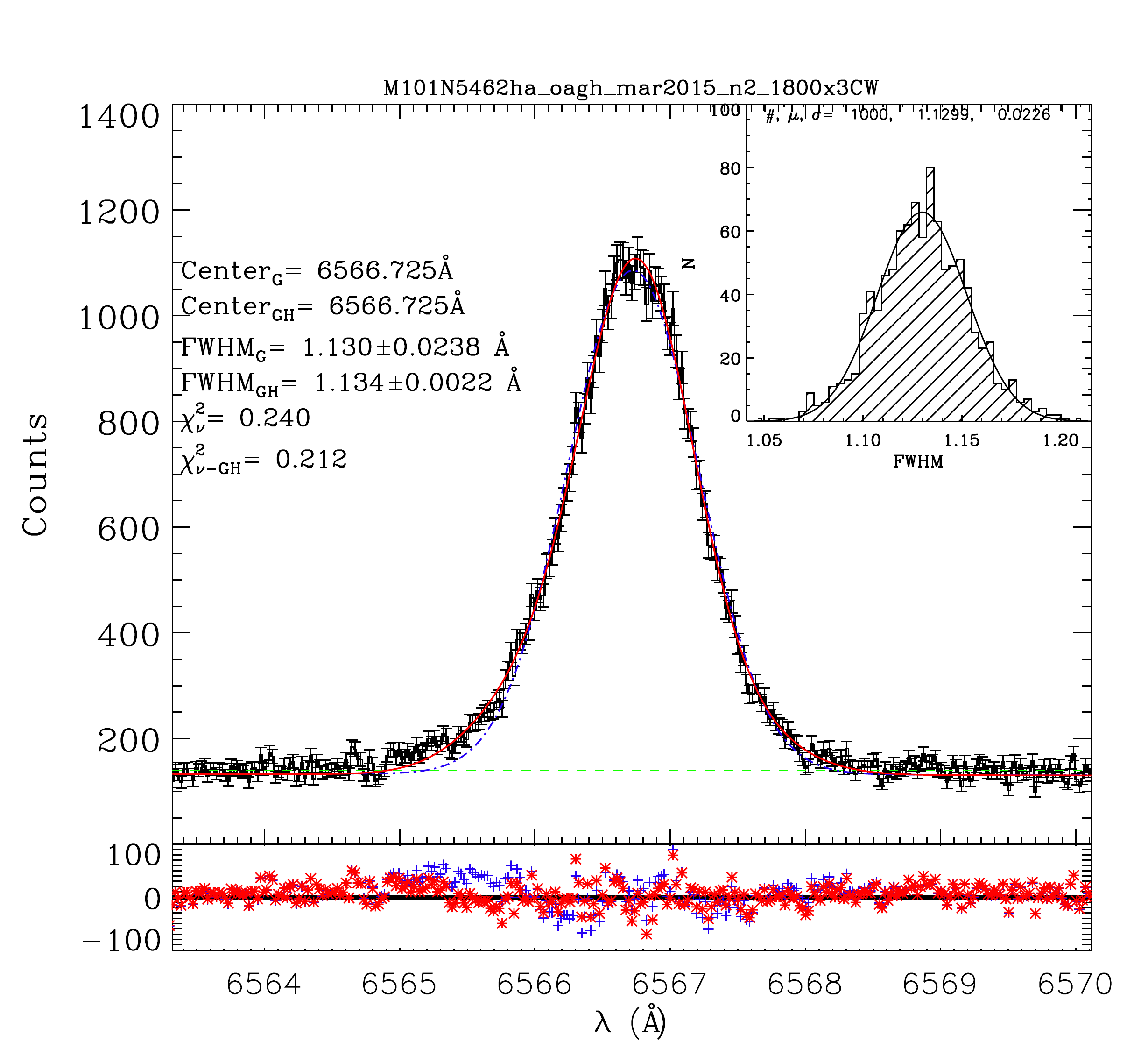}
 \includegraphics[scale=0.2]{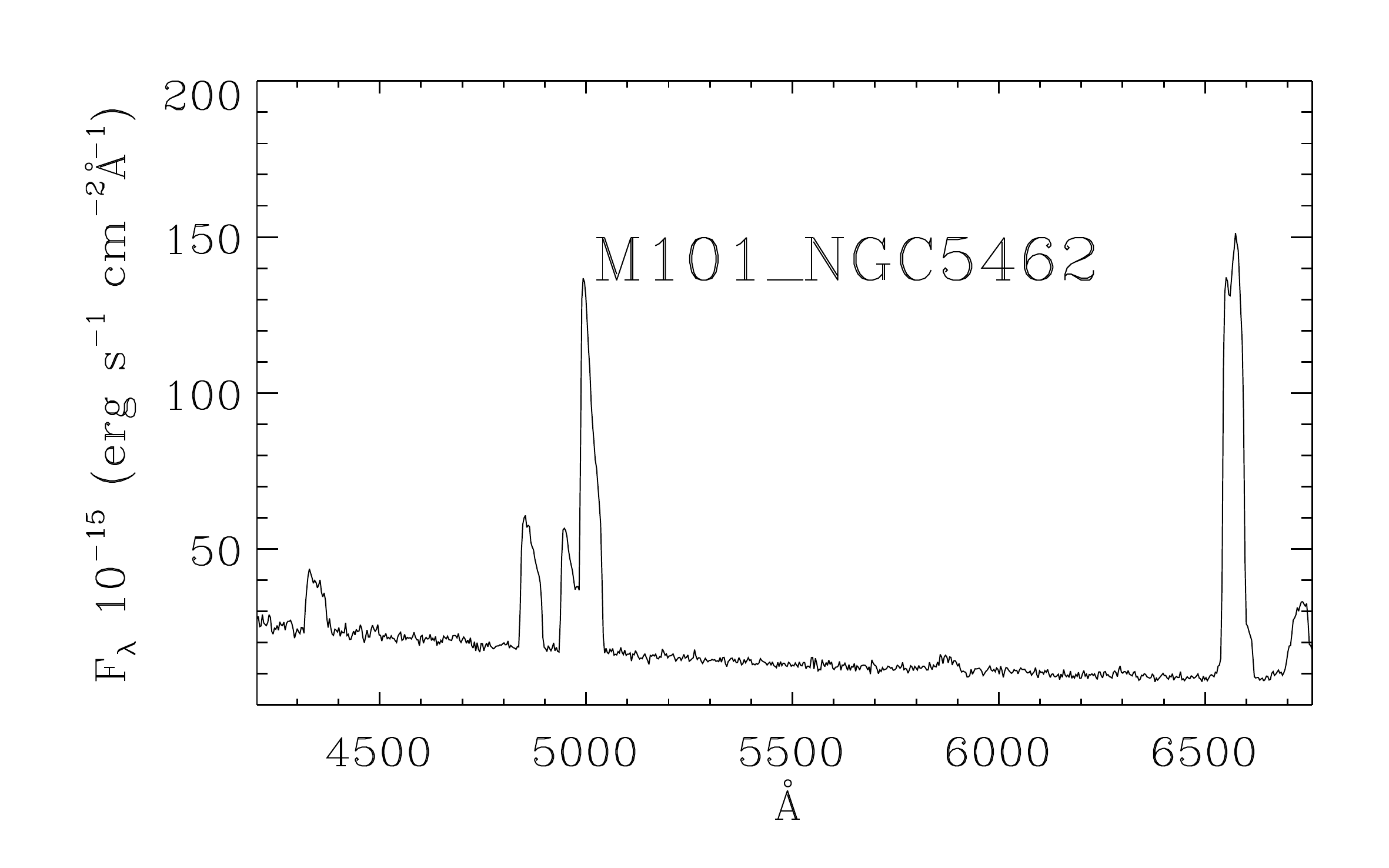}
 
 \includegraphics[scale=0.2]{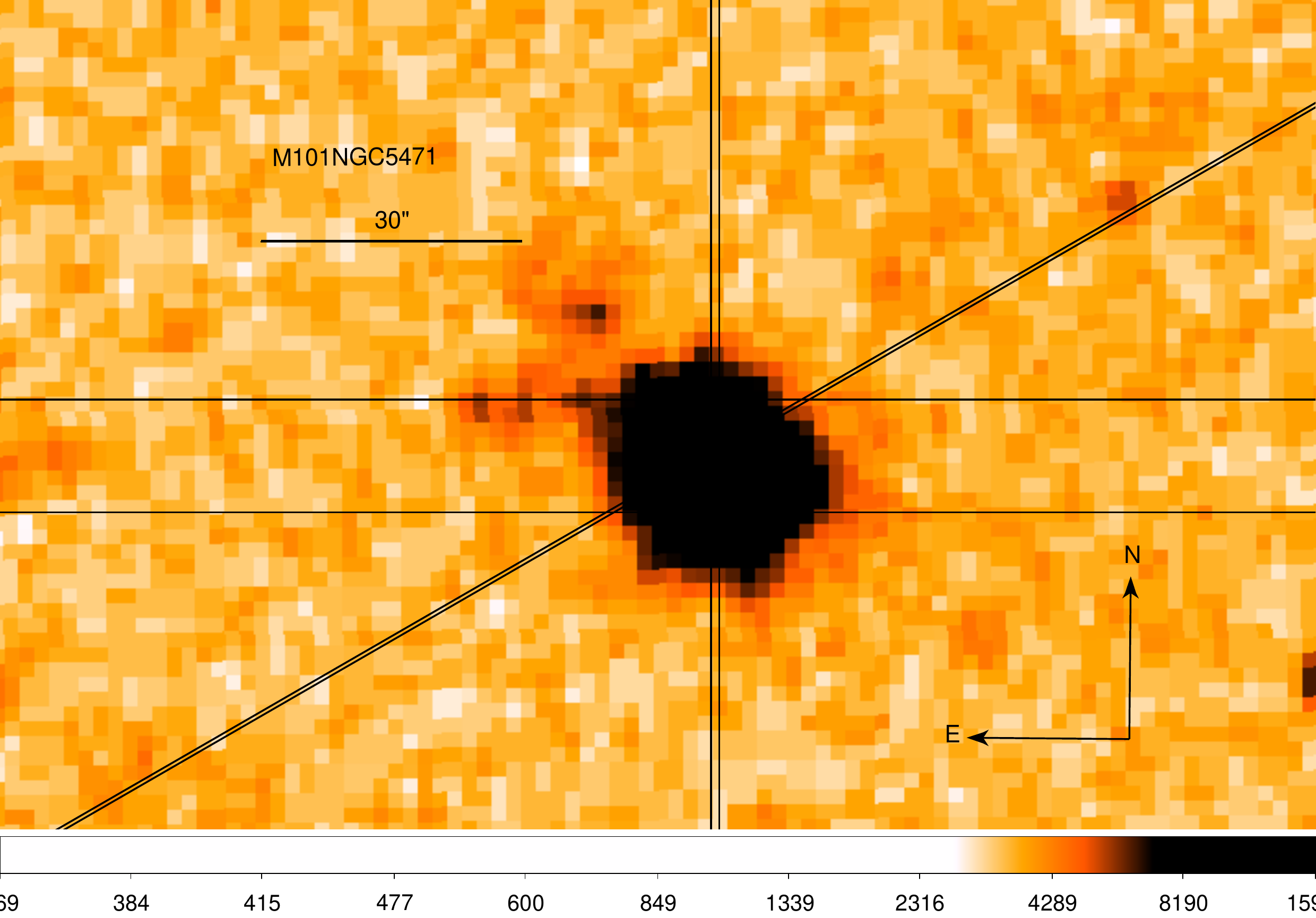}
 \includegraphics[scale=0.18]{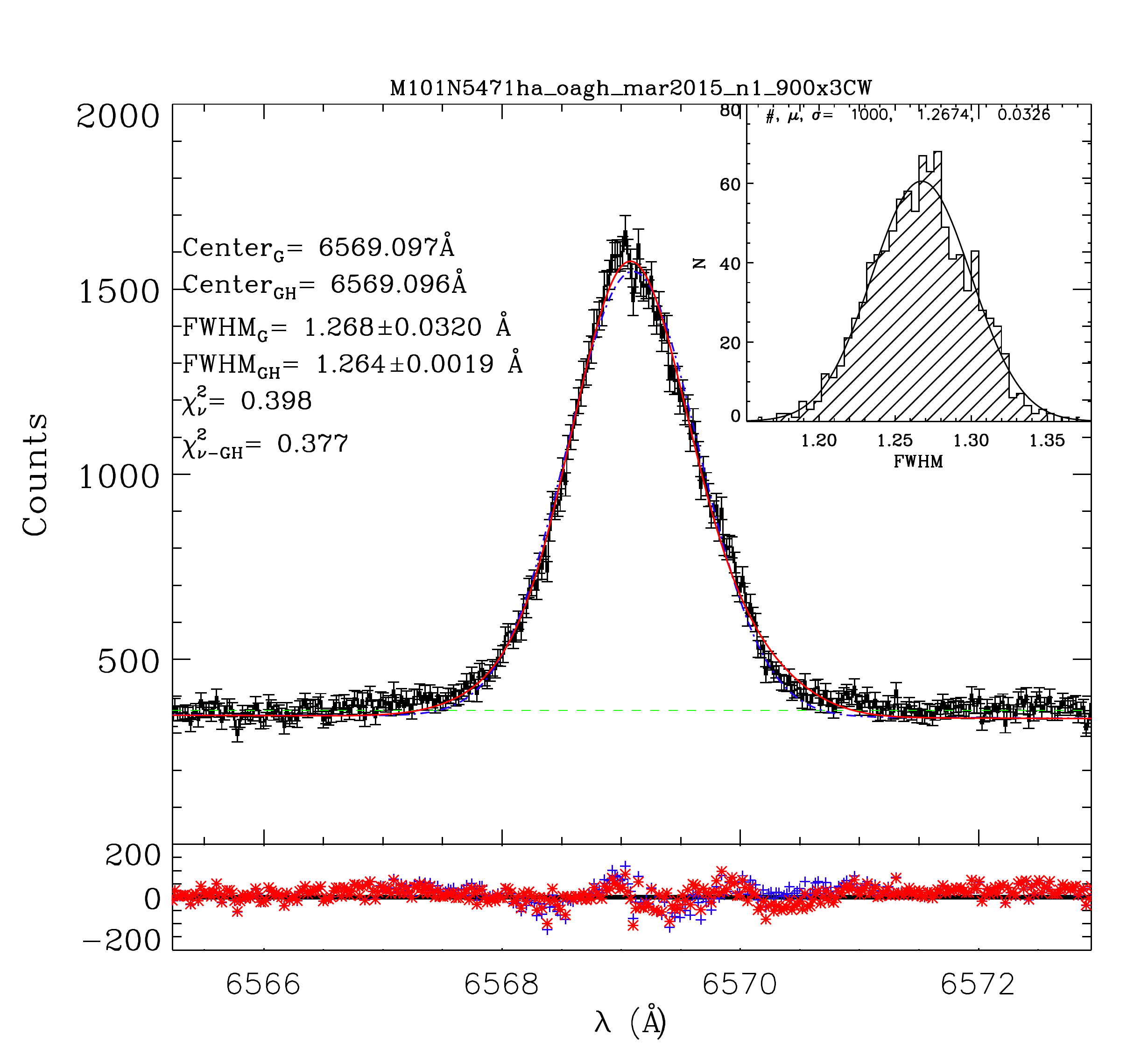}
 \includegraphics[scale=0.2]{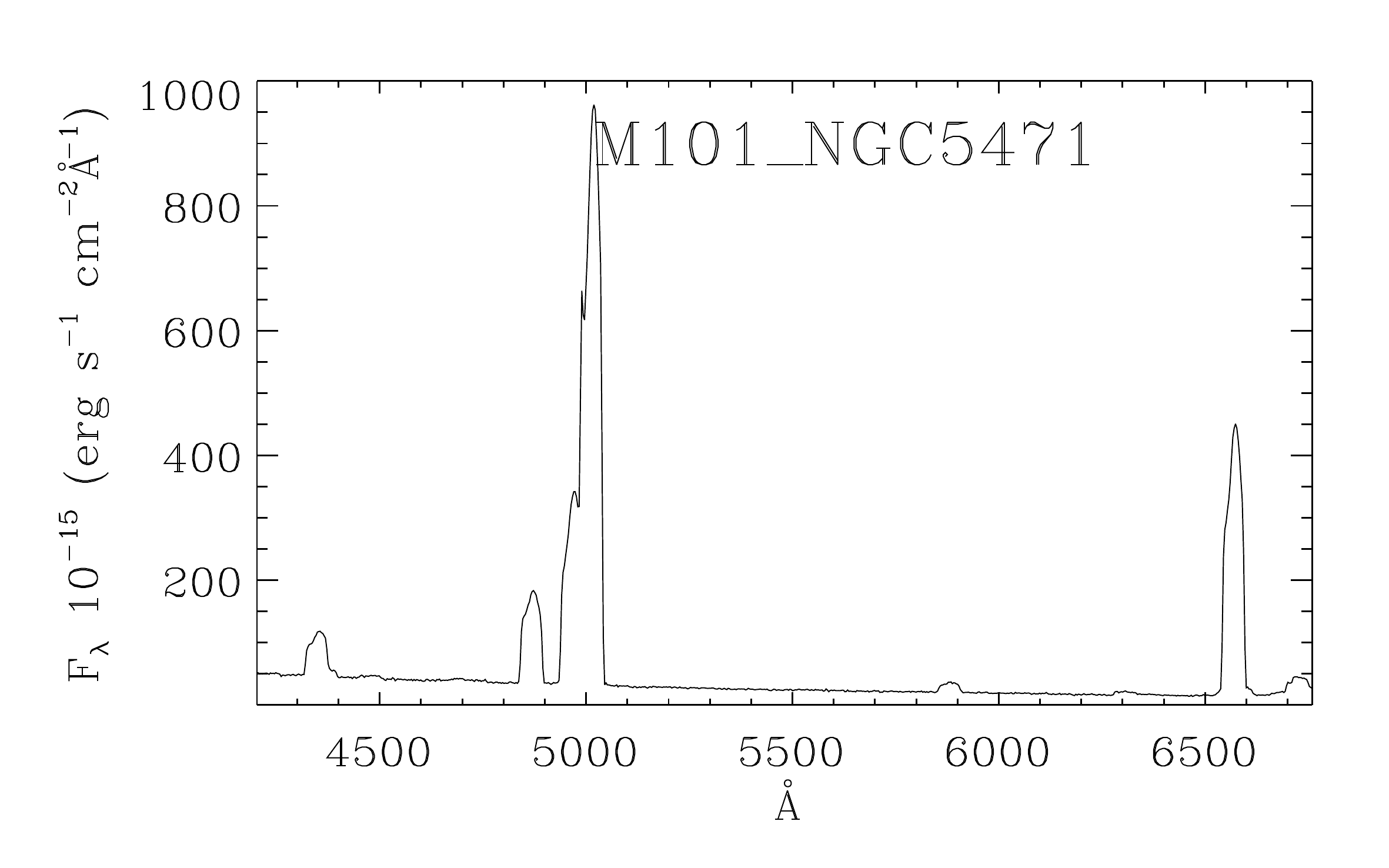}
 
 \includegraphics[scale=0.2]{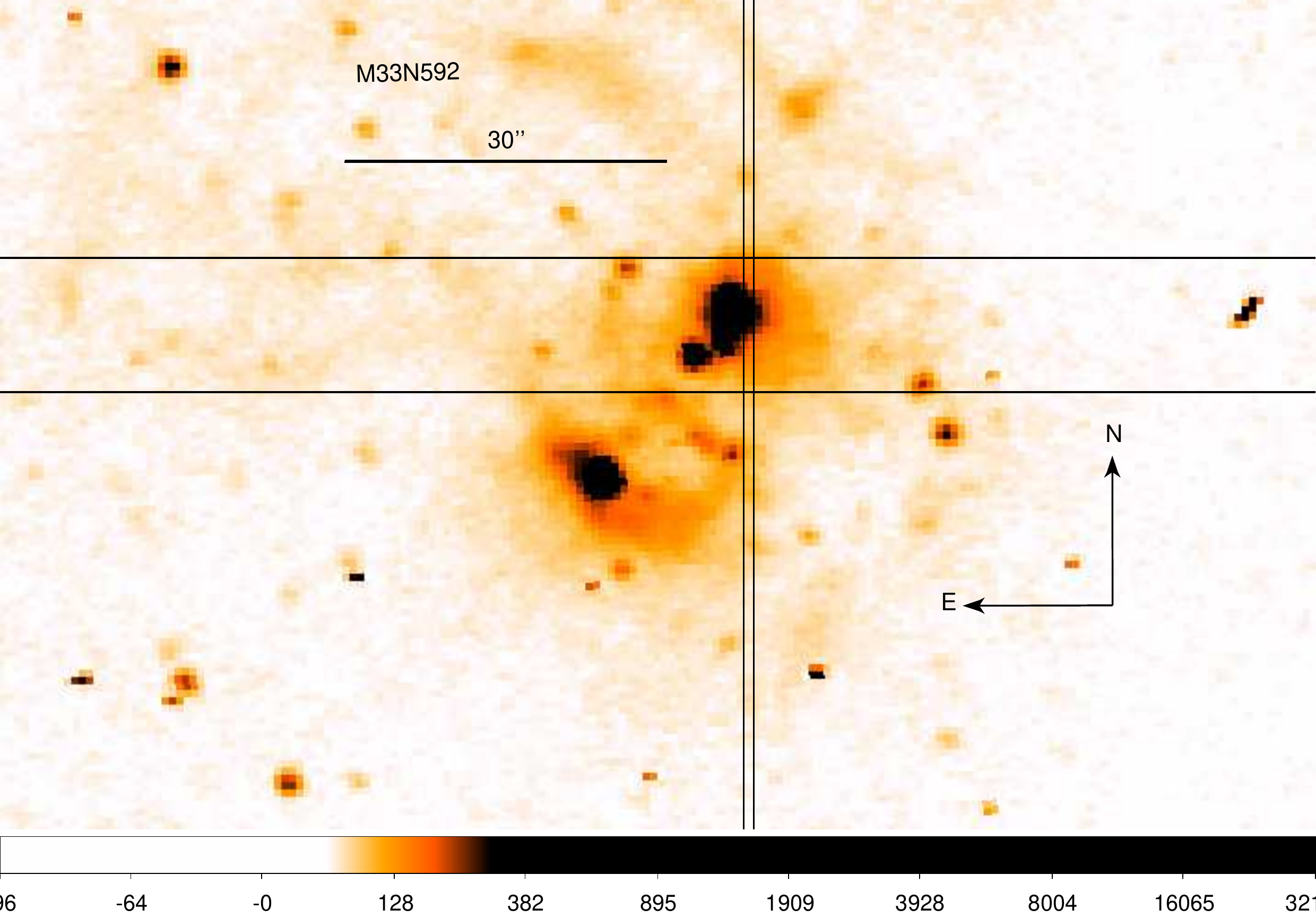}
 \includegraphics[scale=0.18]{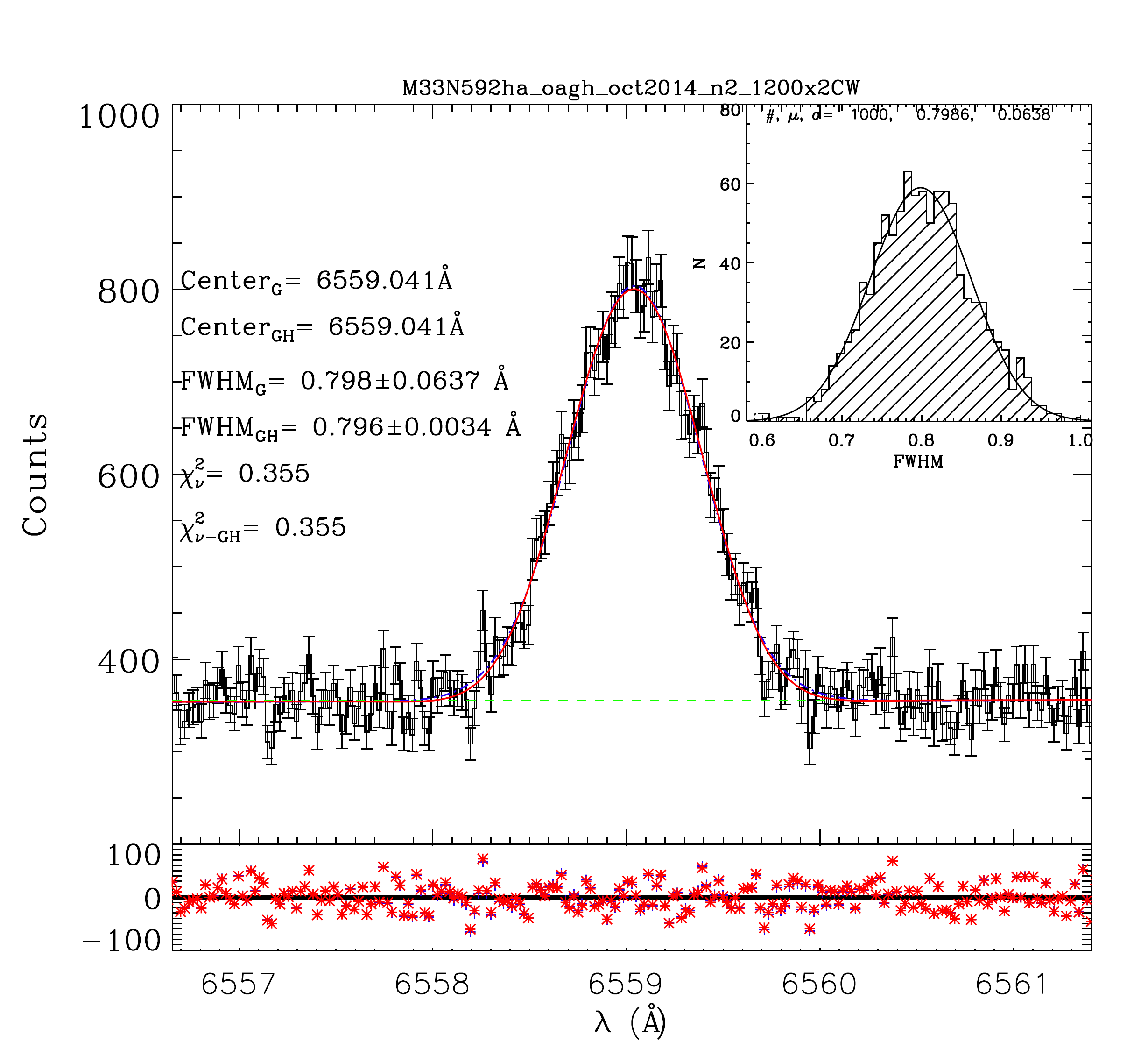}
 \includegraphics[scale=0.2]{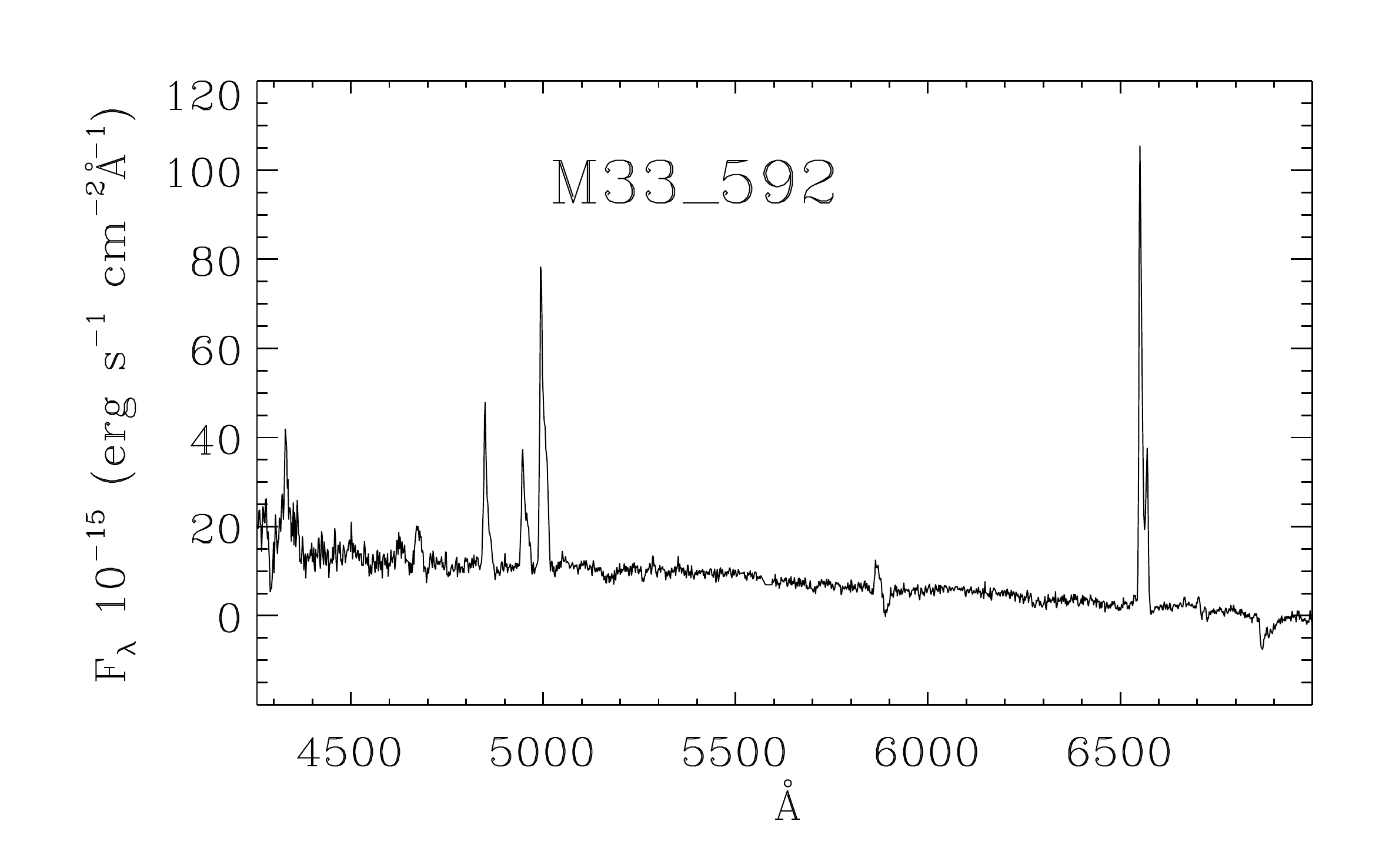}
 
 \includegraphics[scale=0.2]{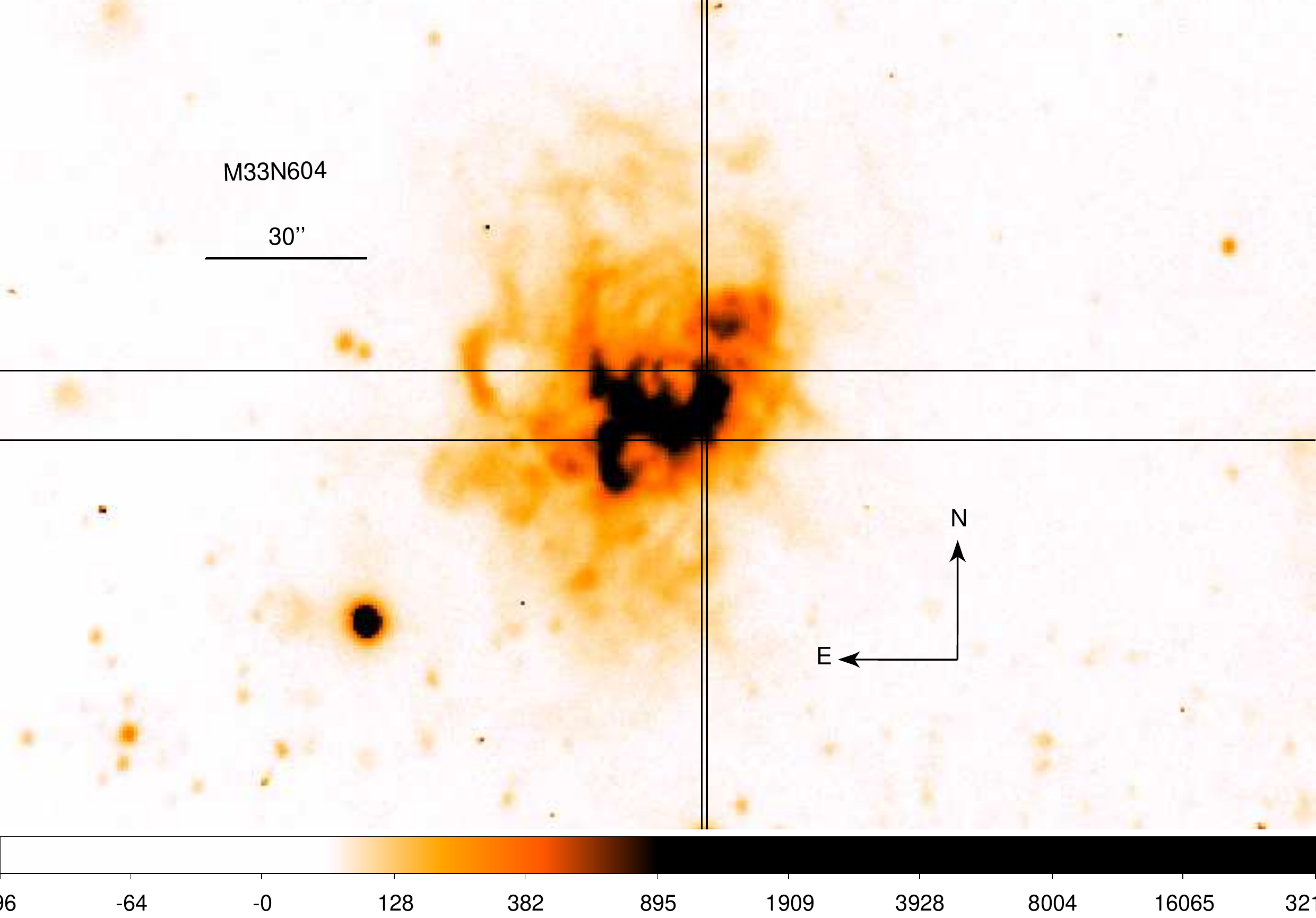}
 \includegraphics[scale=0.18]{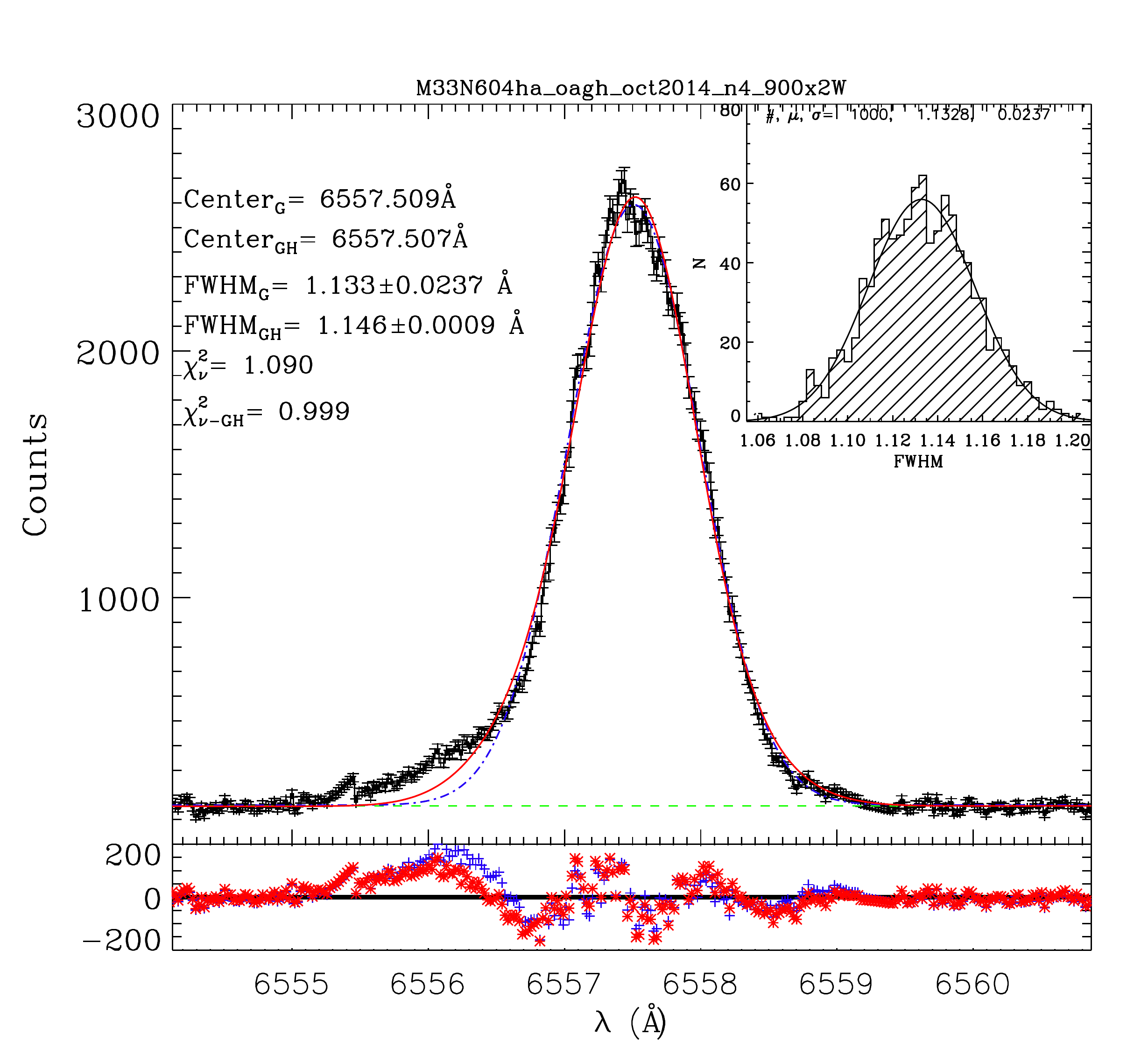}
 \includegraphics[scale=0.2]{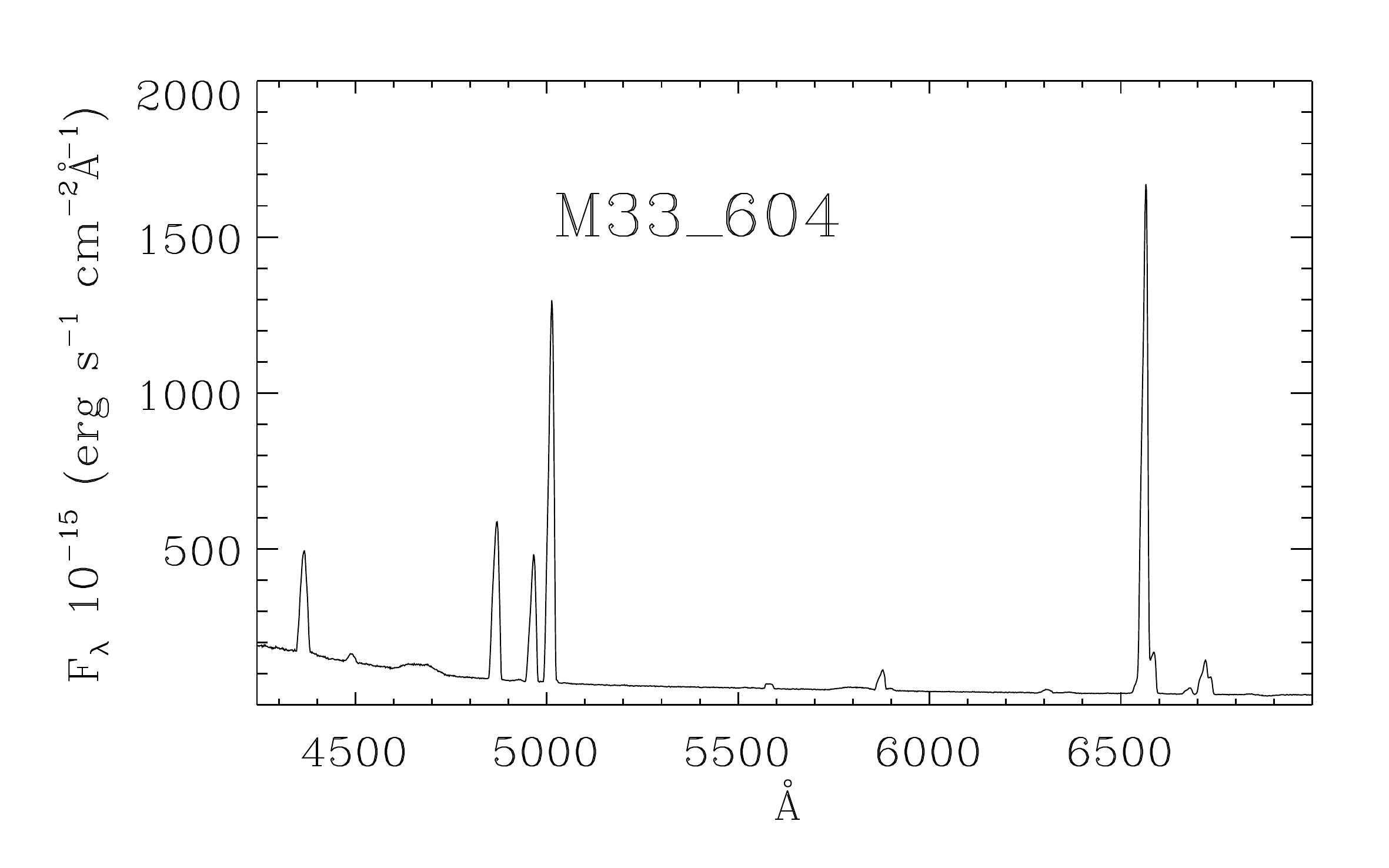}

 \caption[]{ (continued)}
\end{figure*}

  \begin{figure*}
  
  \includegraphics[scale=0.2]{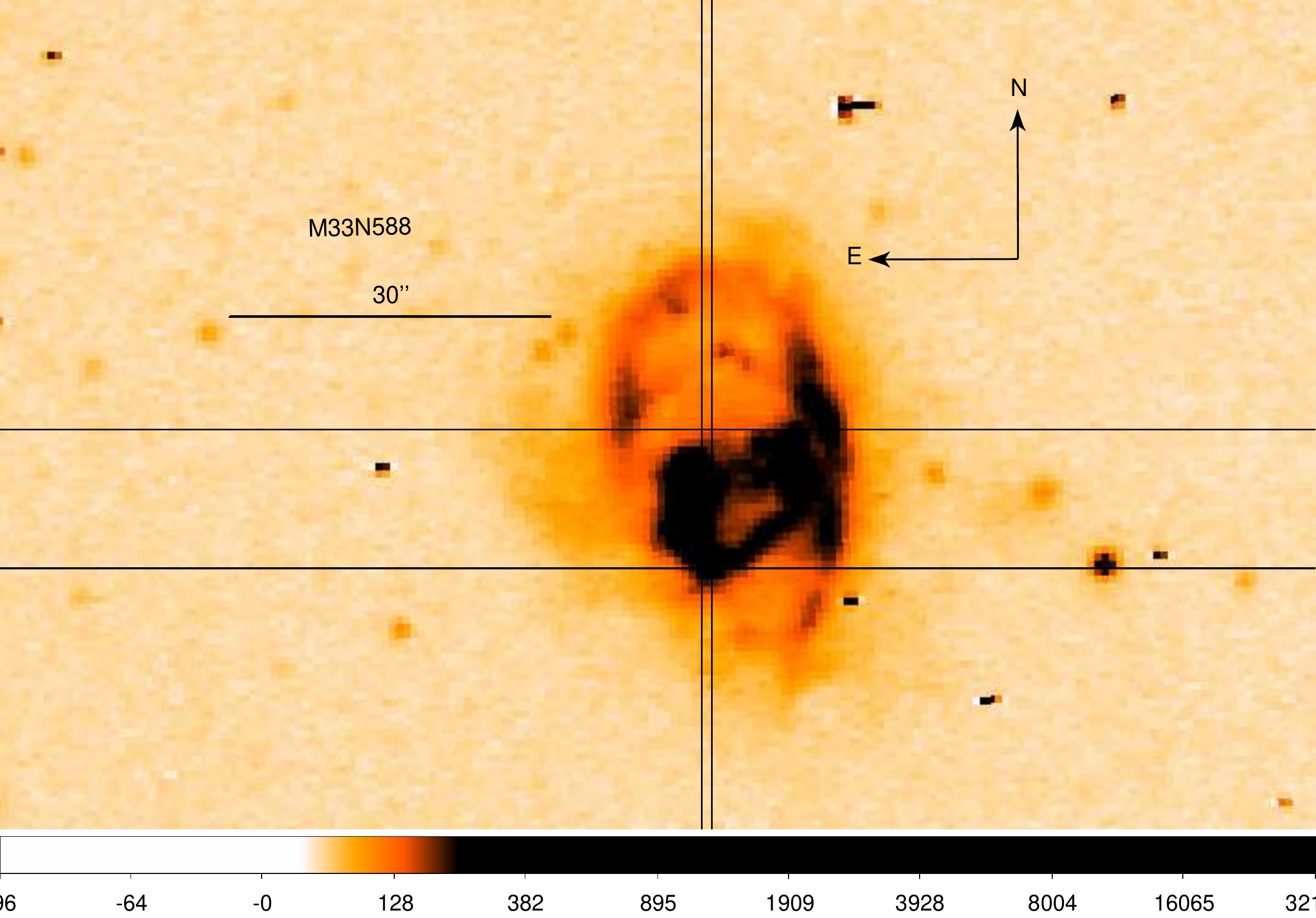}
 \includegraphics[scale=0.18]{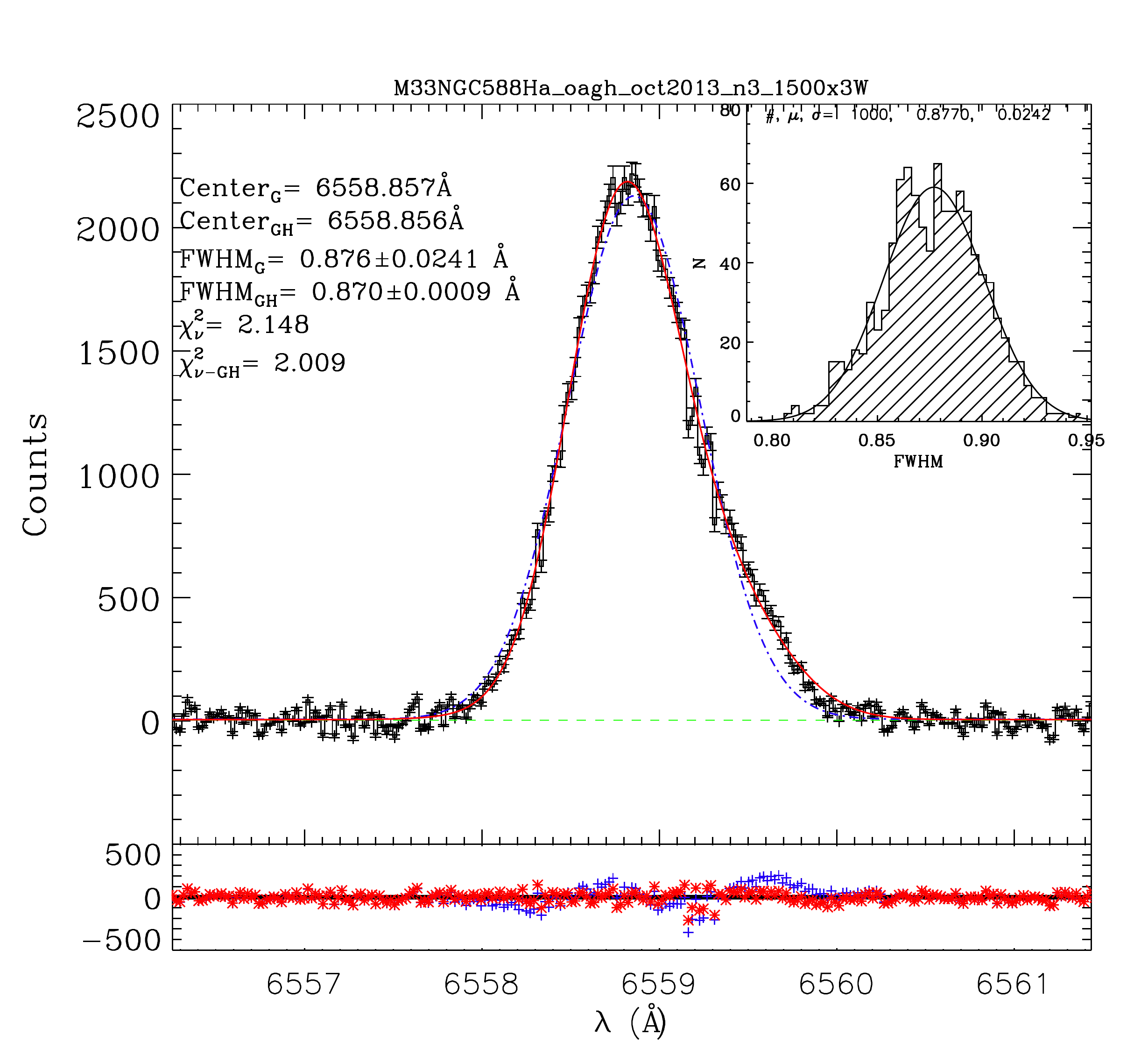}
  \includegraphics[scale=0.2]{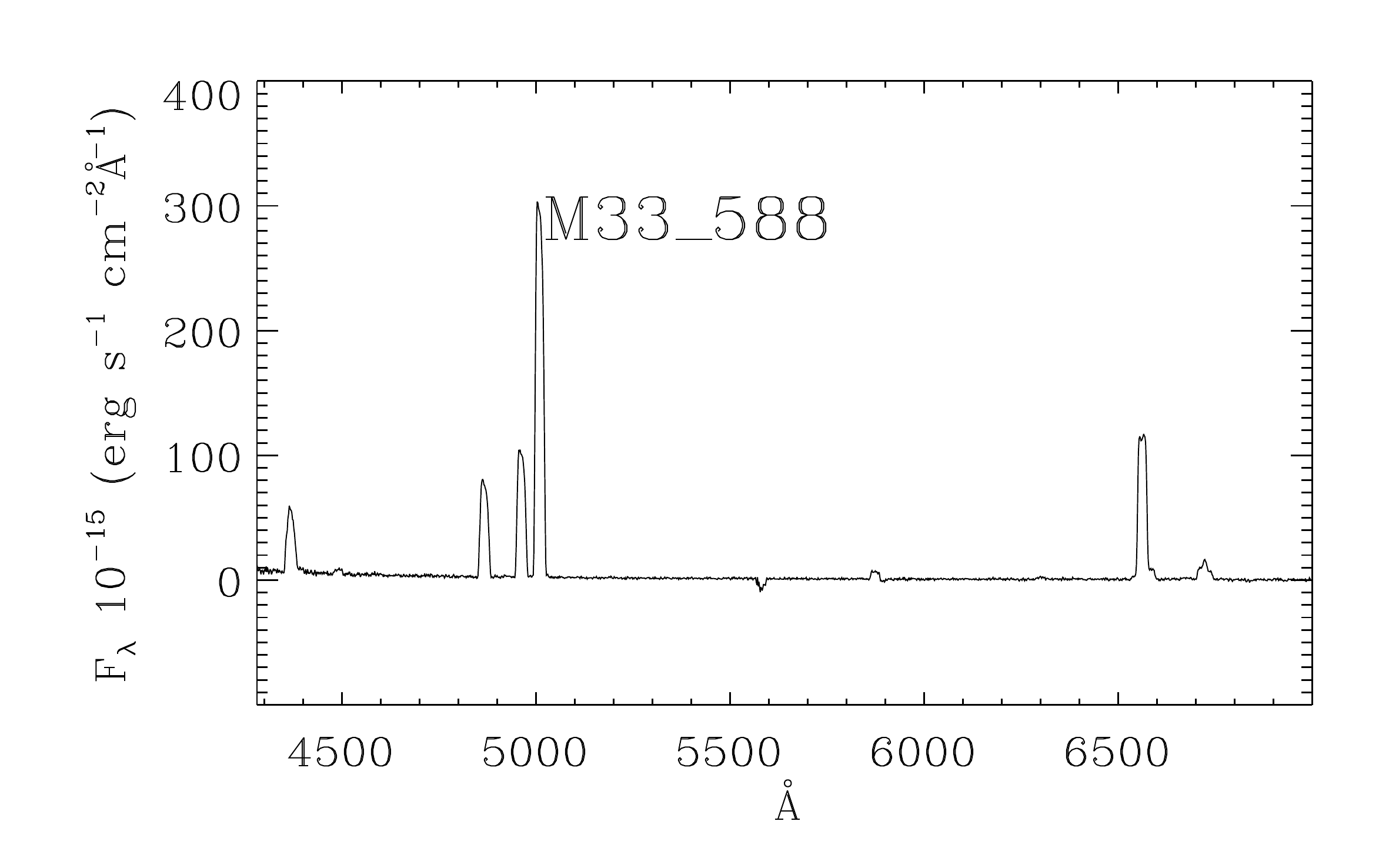}
  
  \includegraphics[scale=0.2]{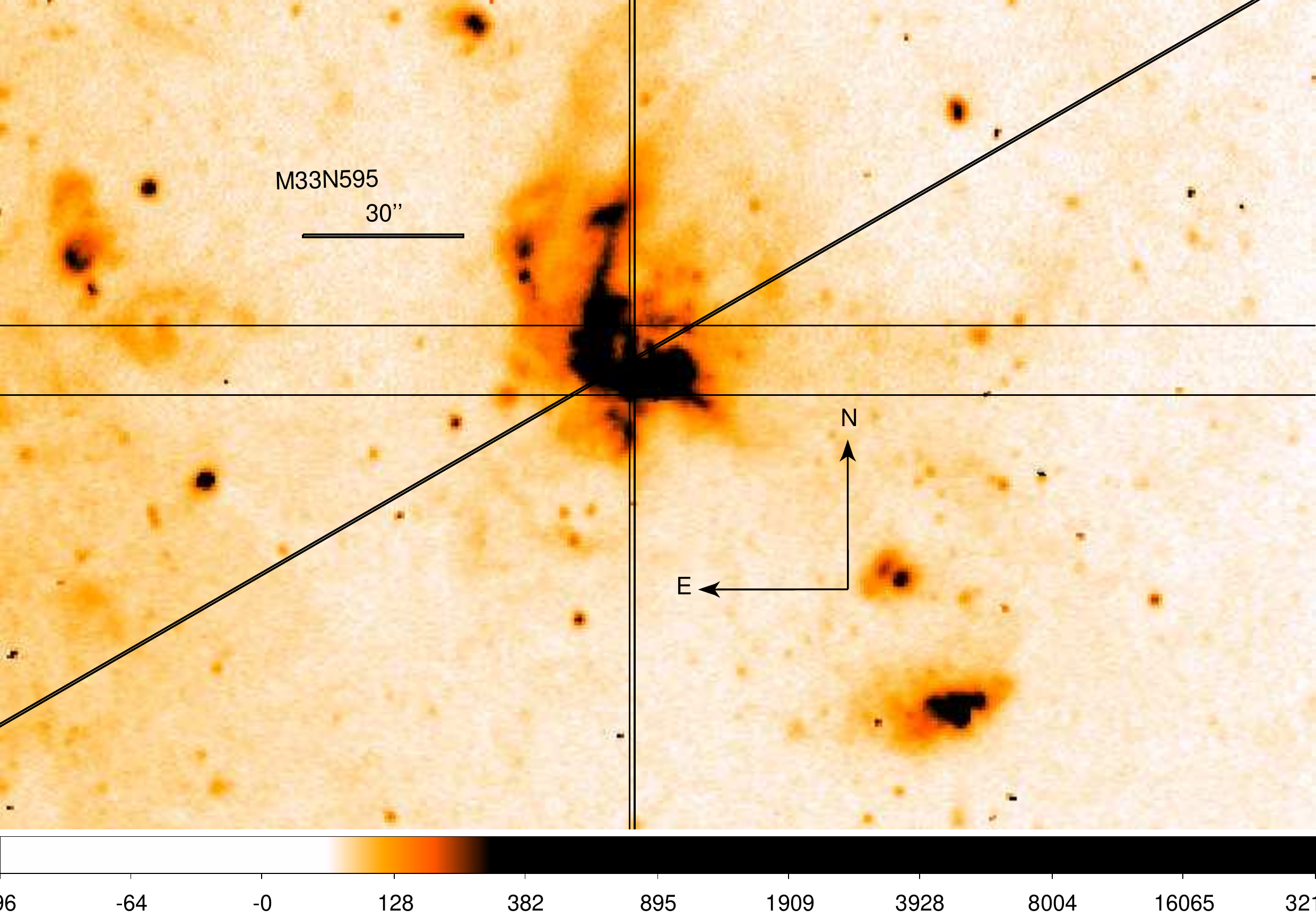}
 \includegraphics[scale=0.18]{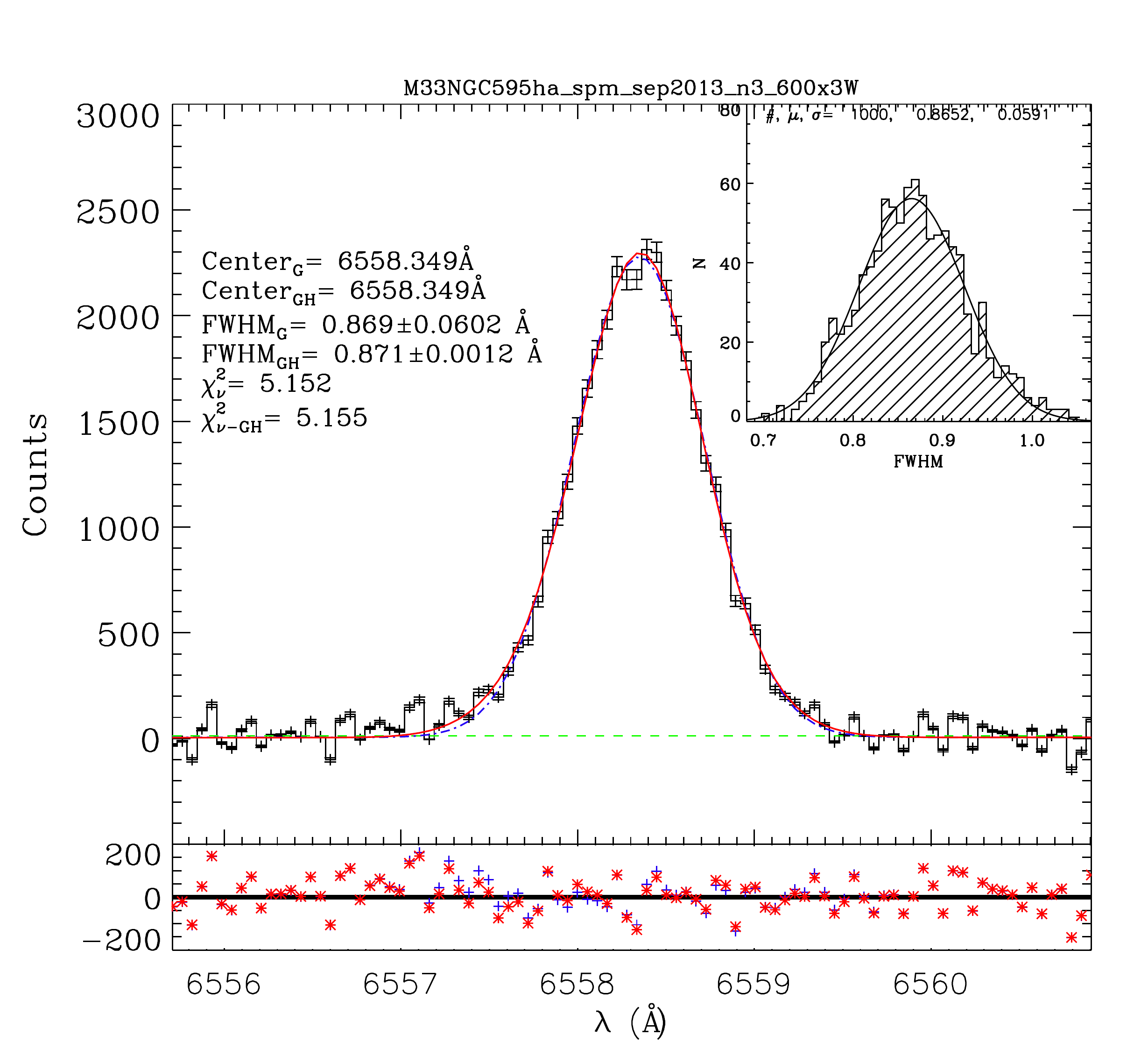}
 \includegraphics[scale=0.2]{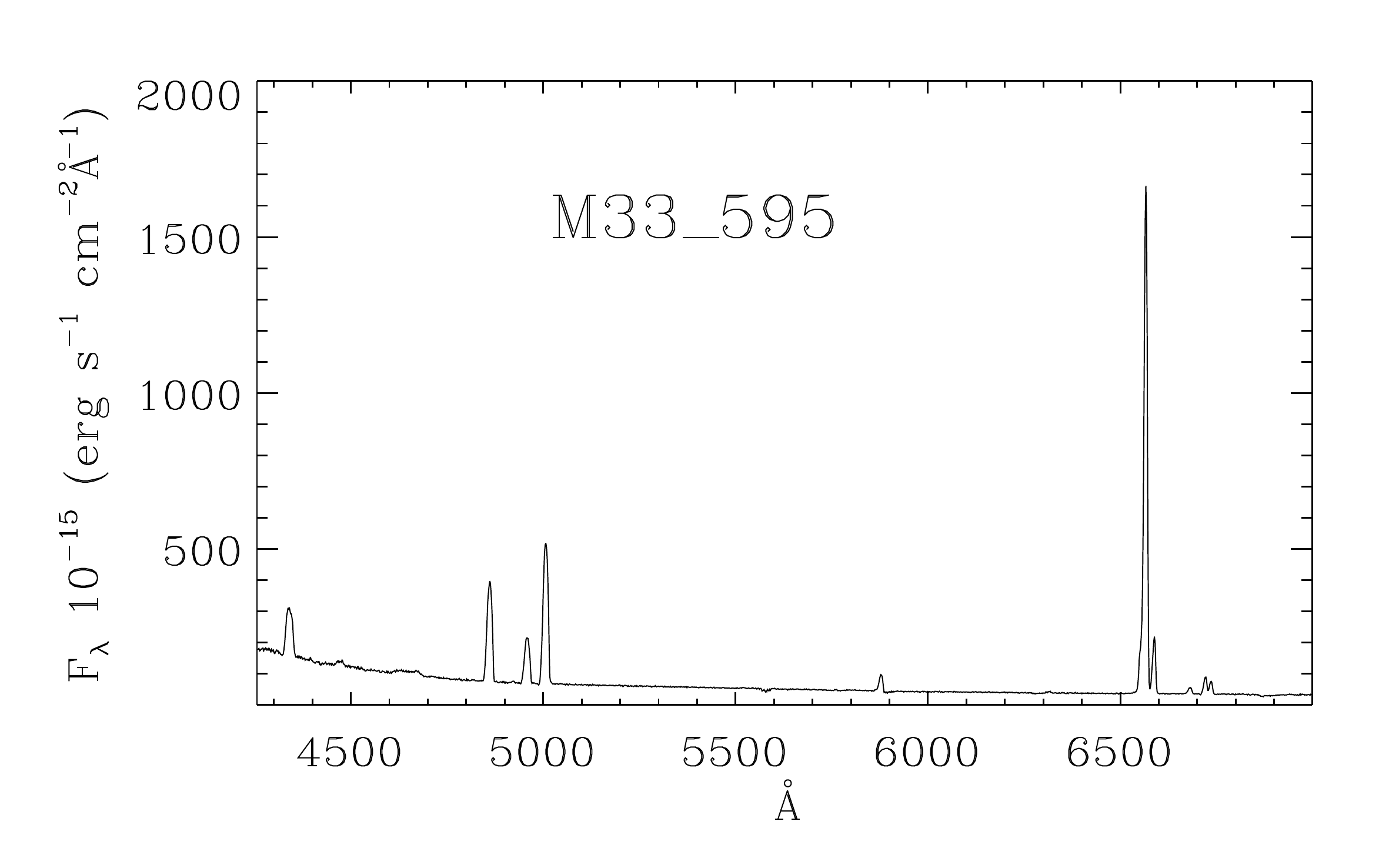}
  
  \includegraphics[scale=0.2]{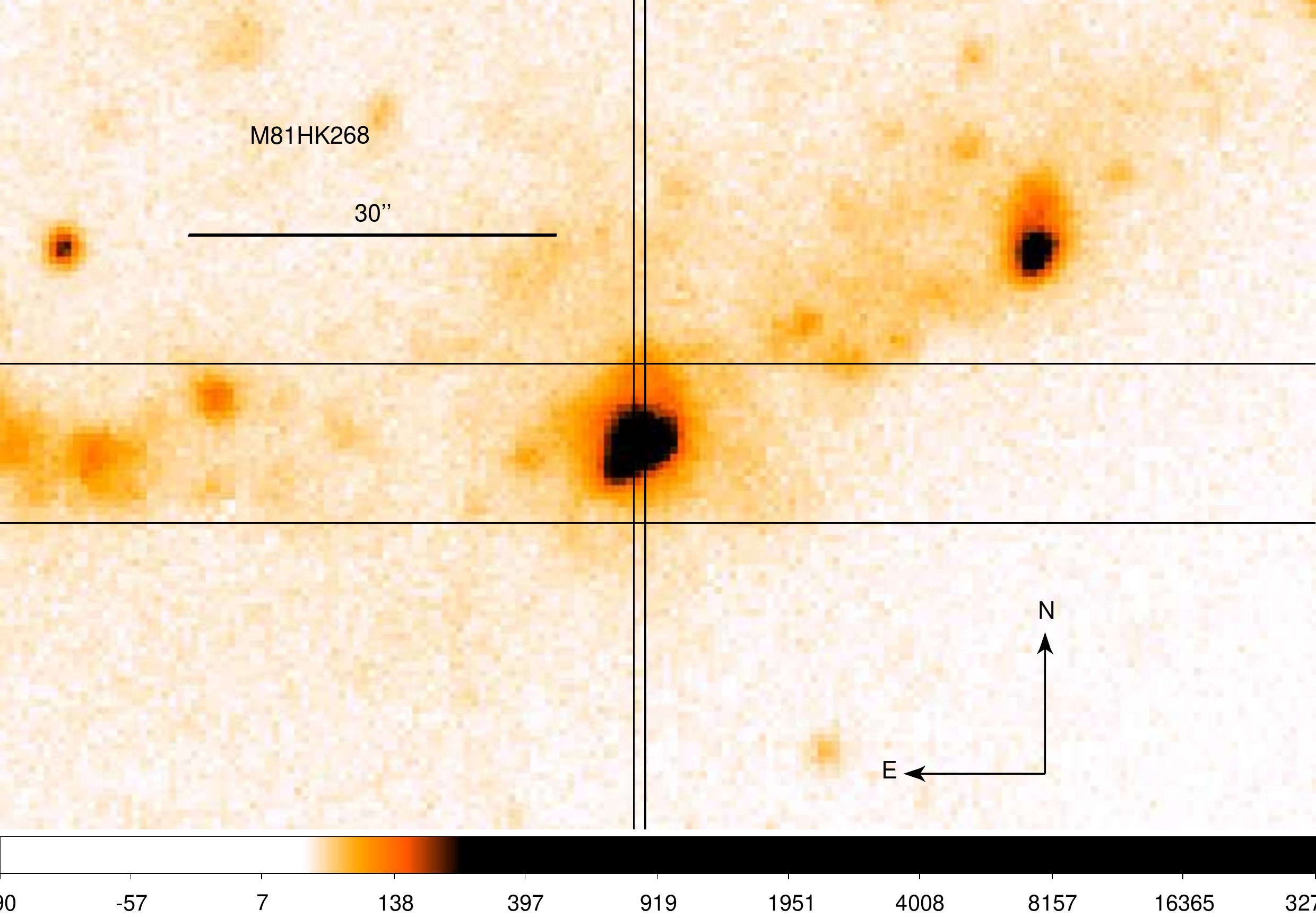}
  \includegraphics[scale=0.18]{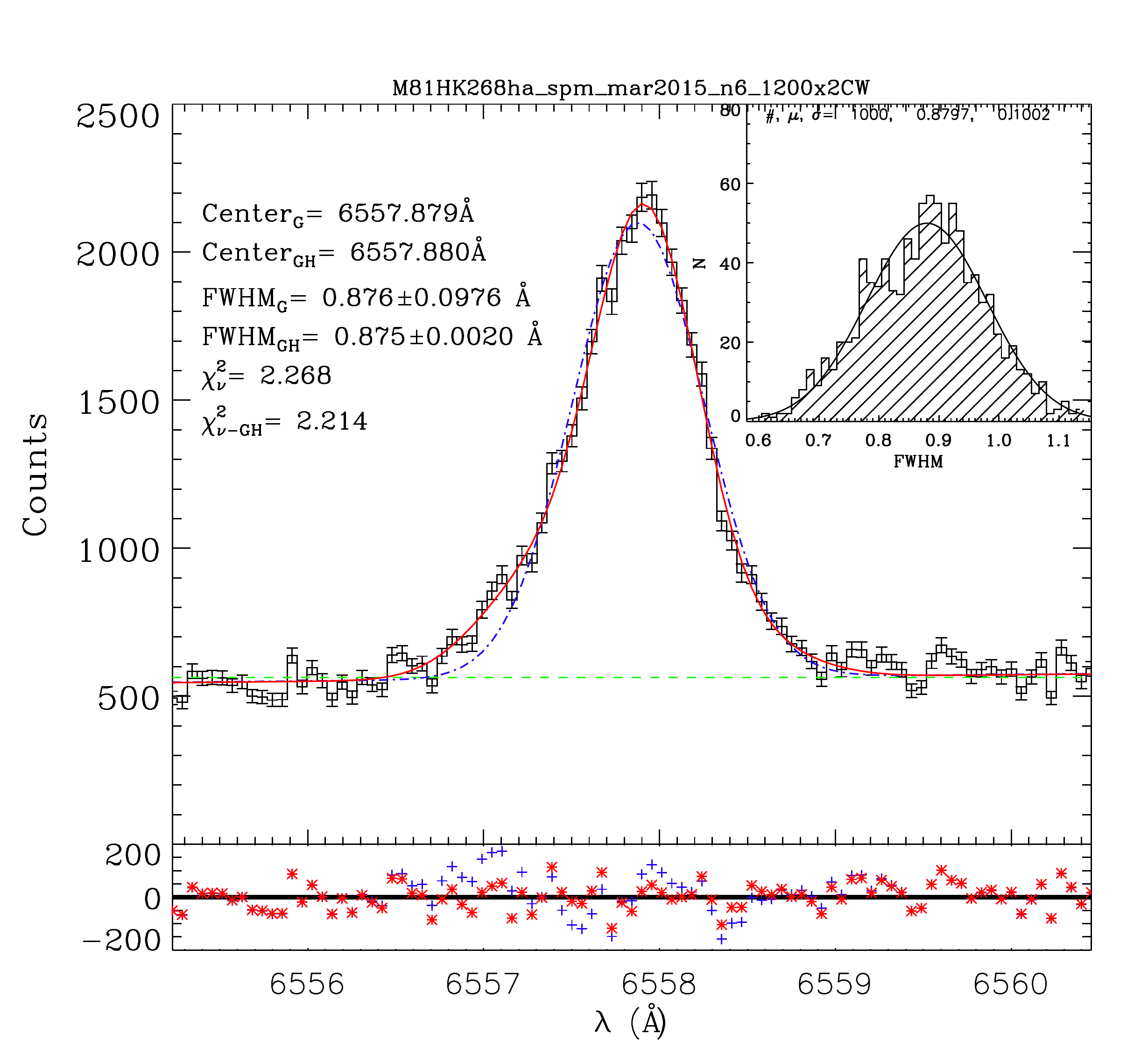}
  \includegraphics[scale=0.2]{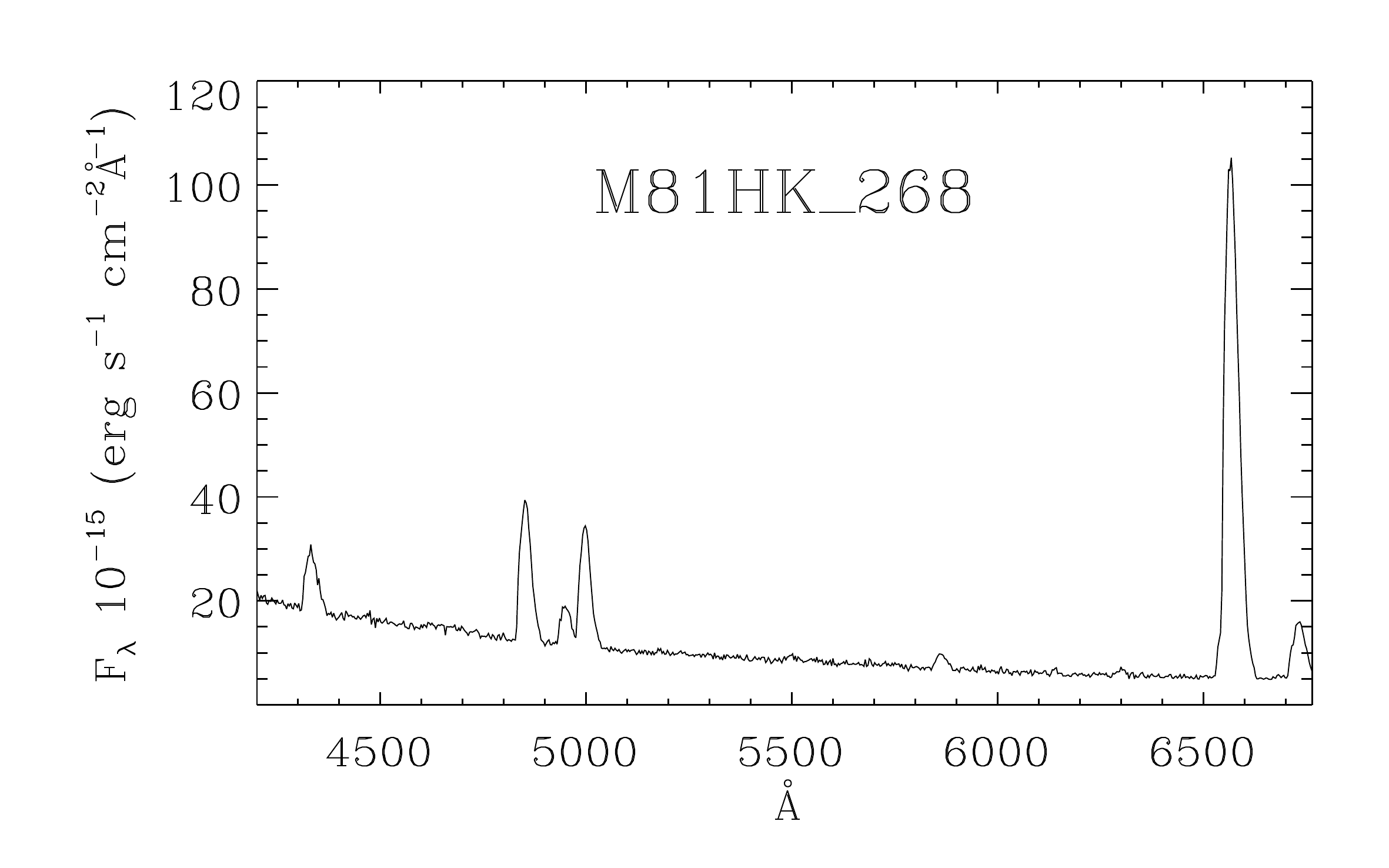}
  
  \includegraphics[scale=0.2]{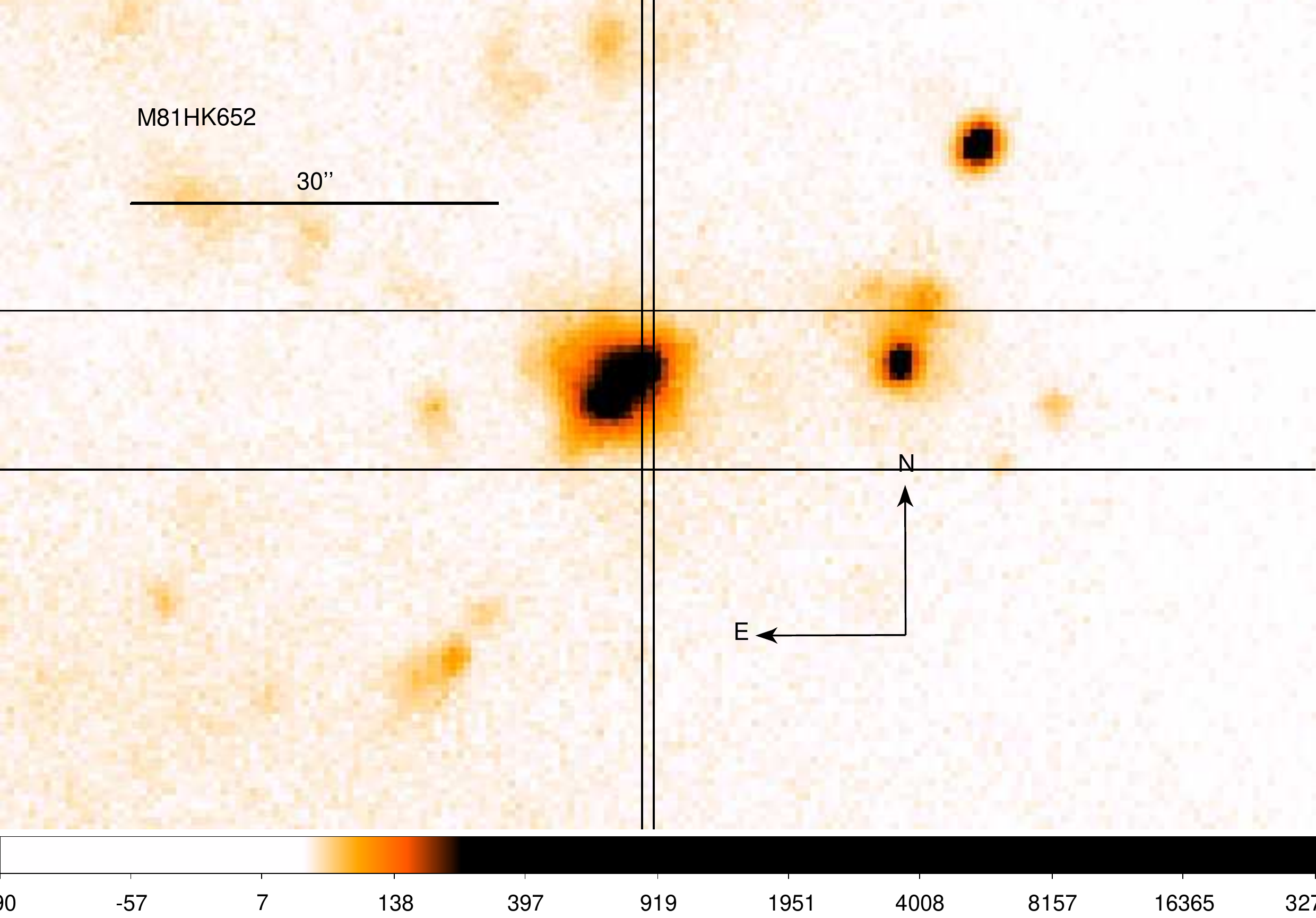}
  \includegraphics[scale=0.18]{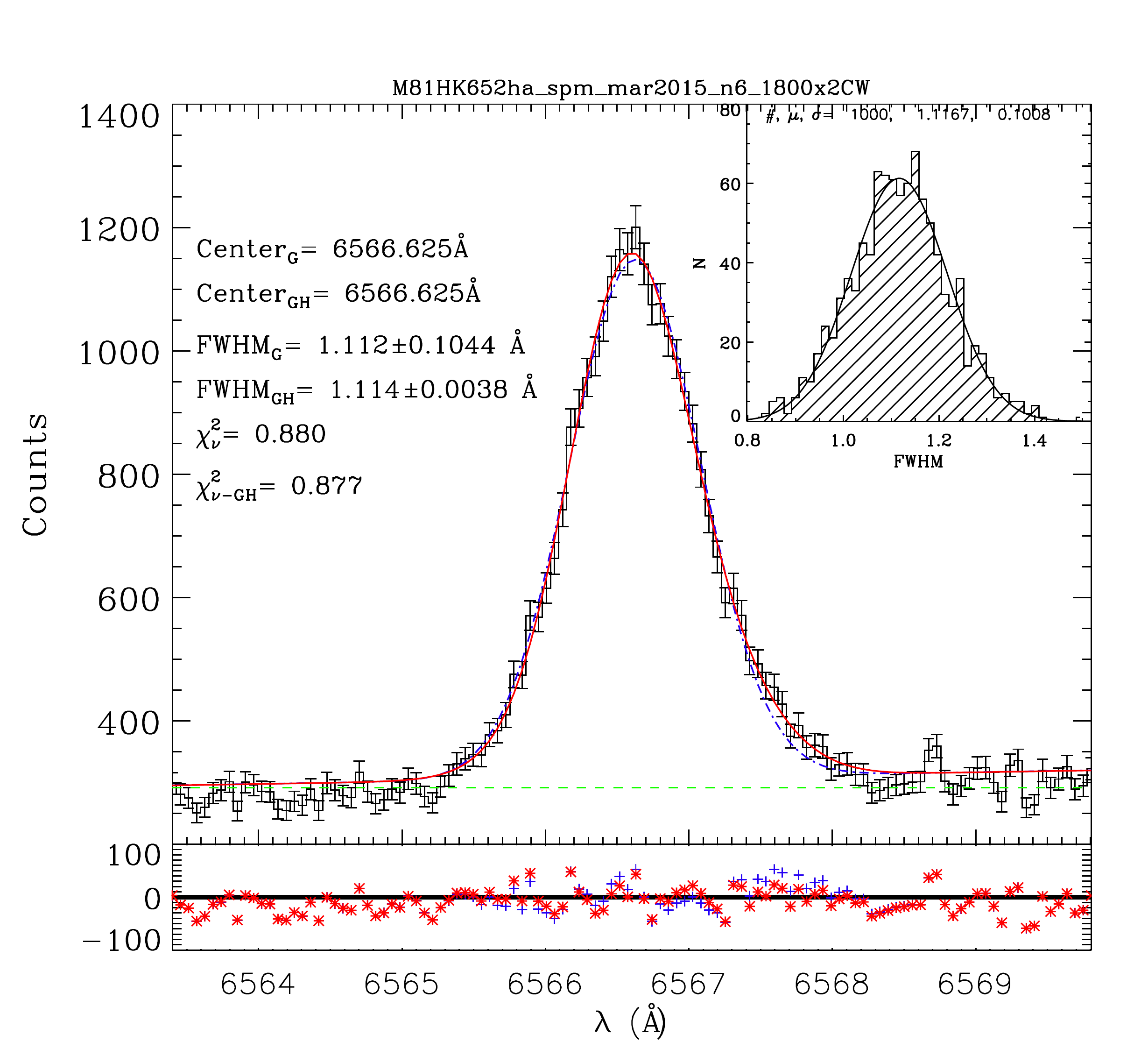}
  \includegraphics[scale=0.2]{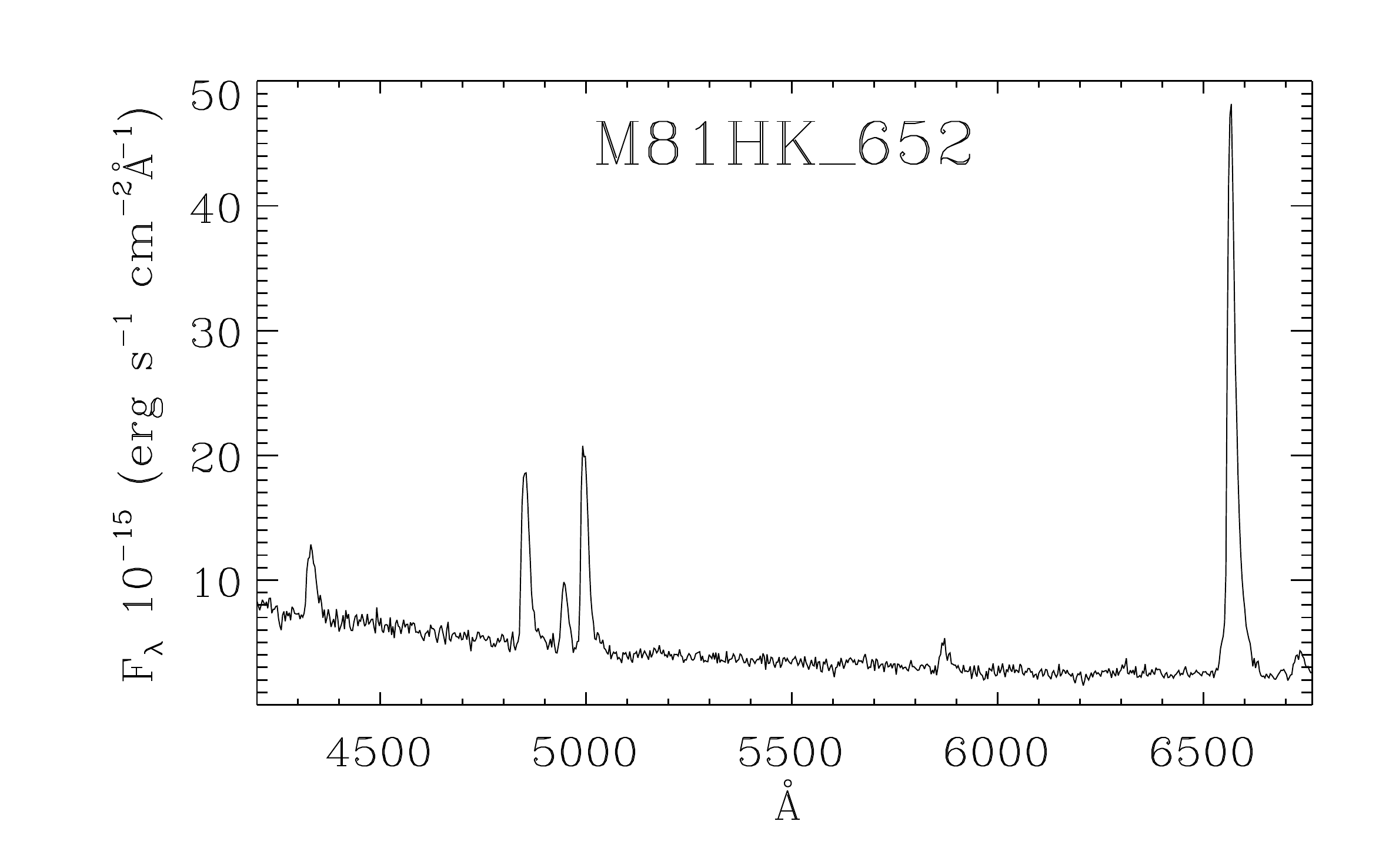}
  
  \includegraphics[scale=0.2]{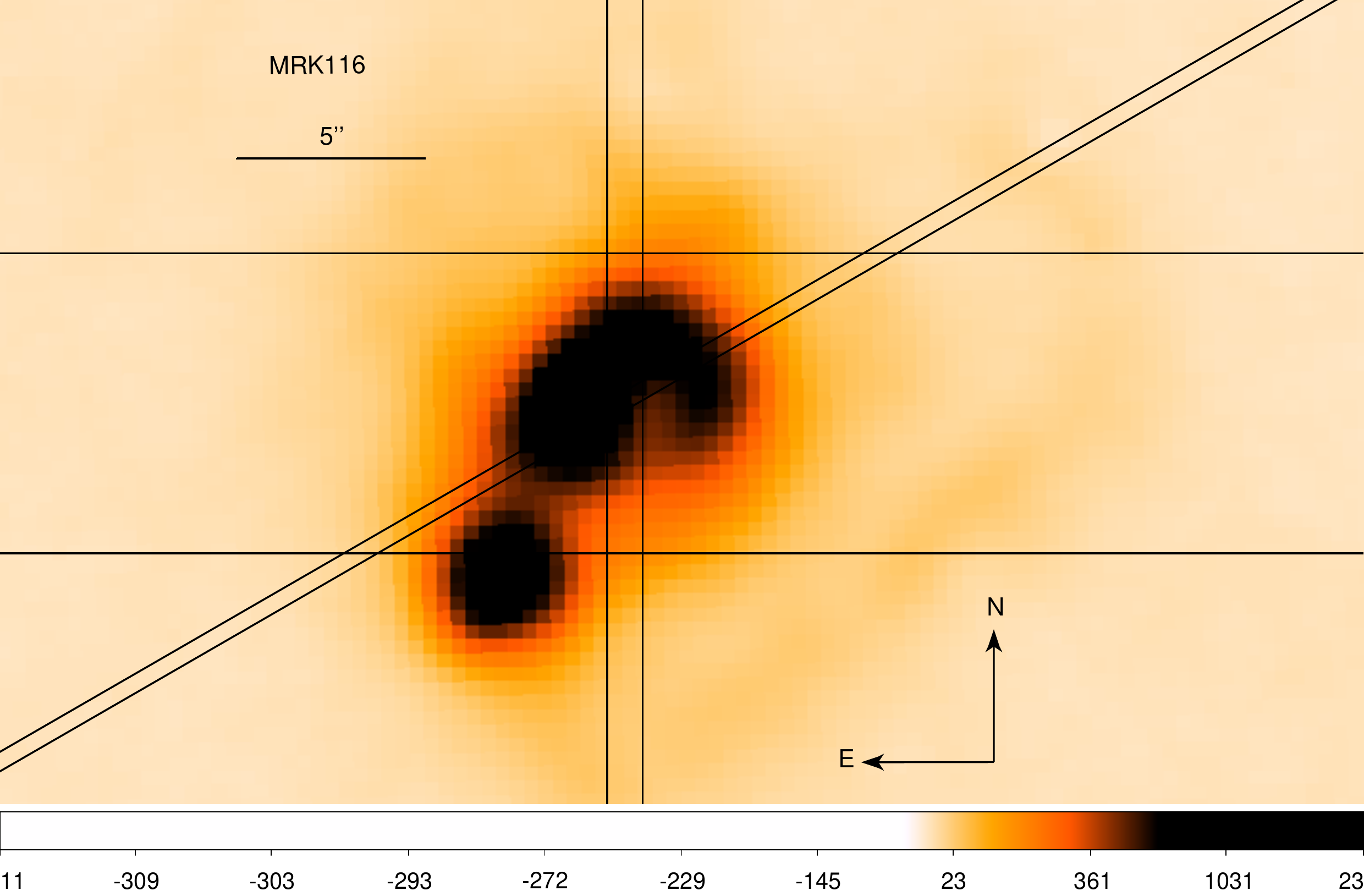}
  \includegraphics[scale=0.18]{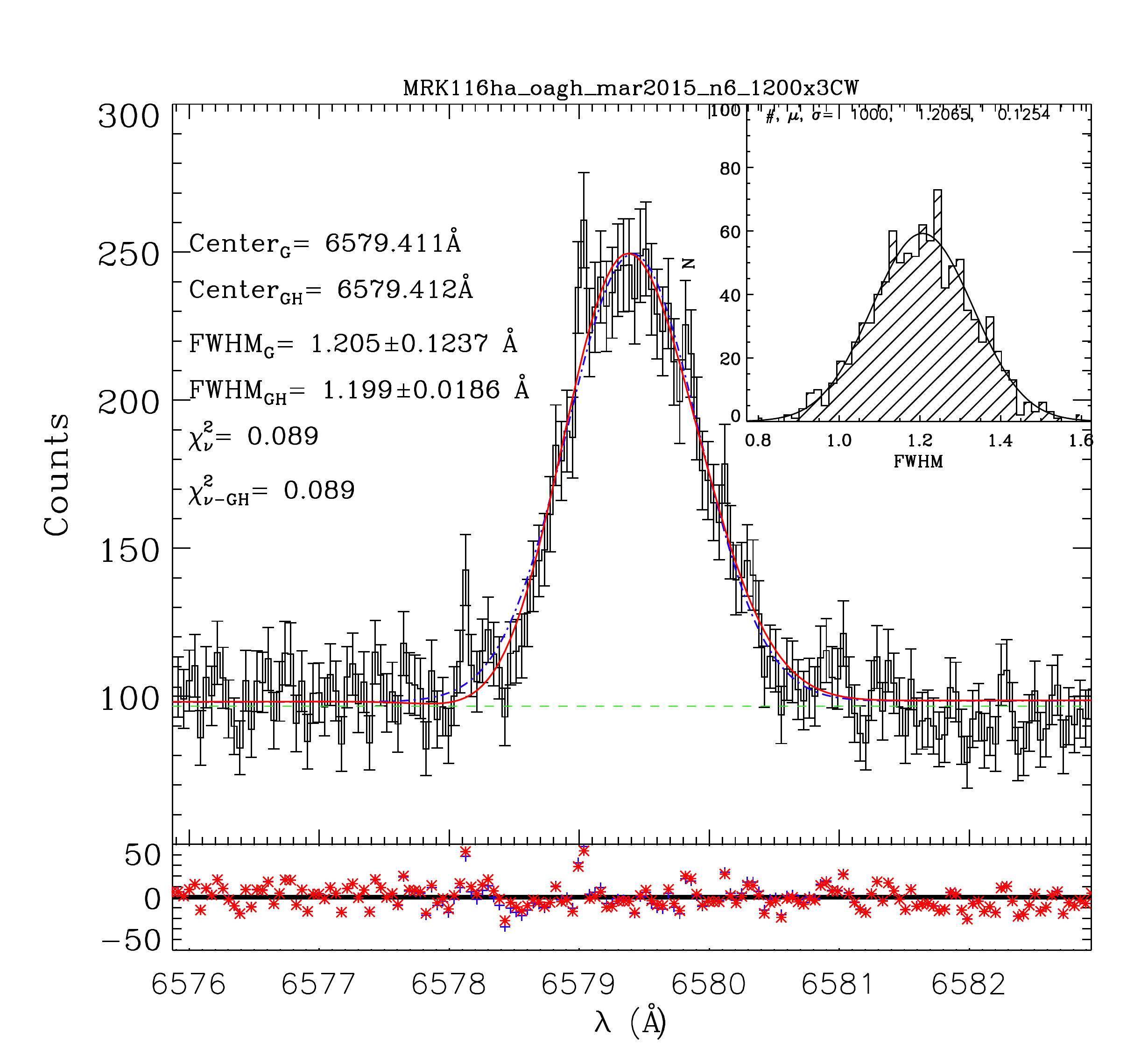}
  \includegraphics[scale=0.2]{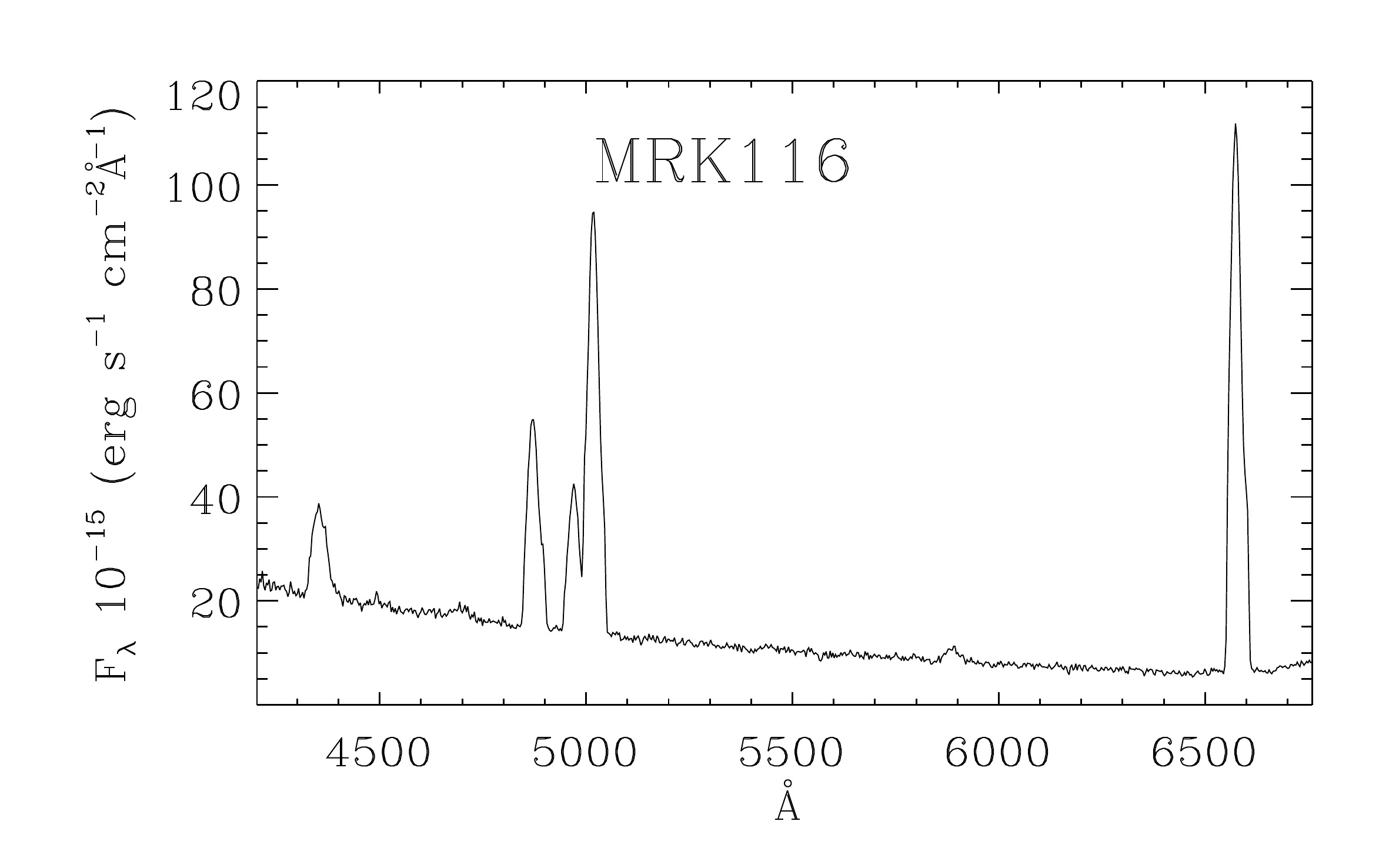}

  \caption[]{ (continued)}
 \end{figure*}

  \begin{figure*}
  
   \includegraphics[scale=0.2]{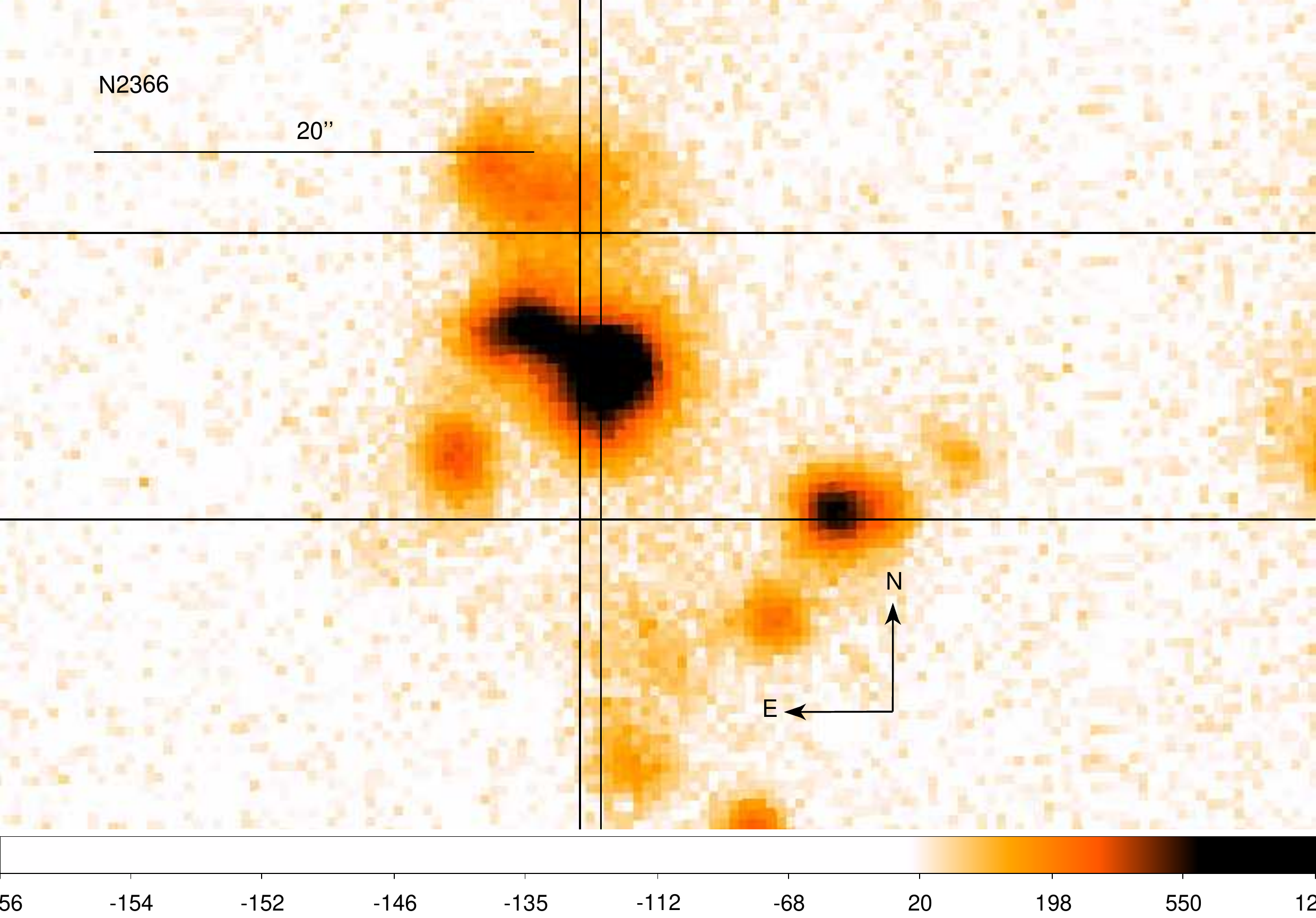}
  \includegraphics[scale=0.18]{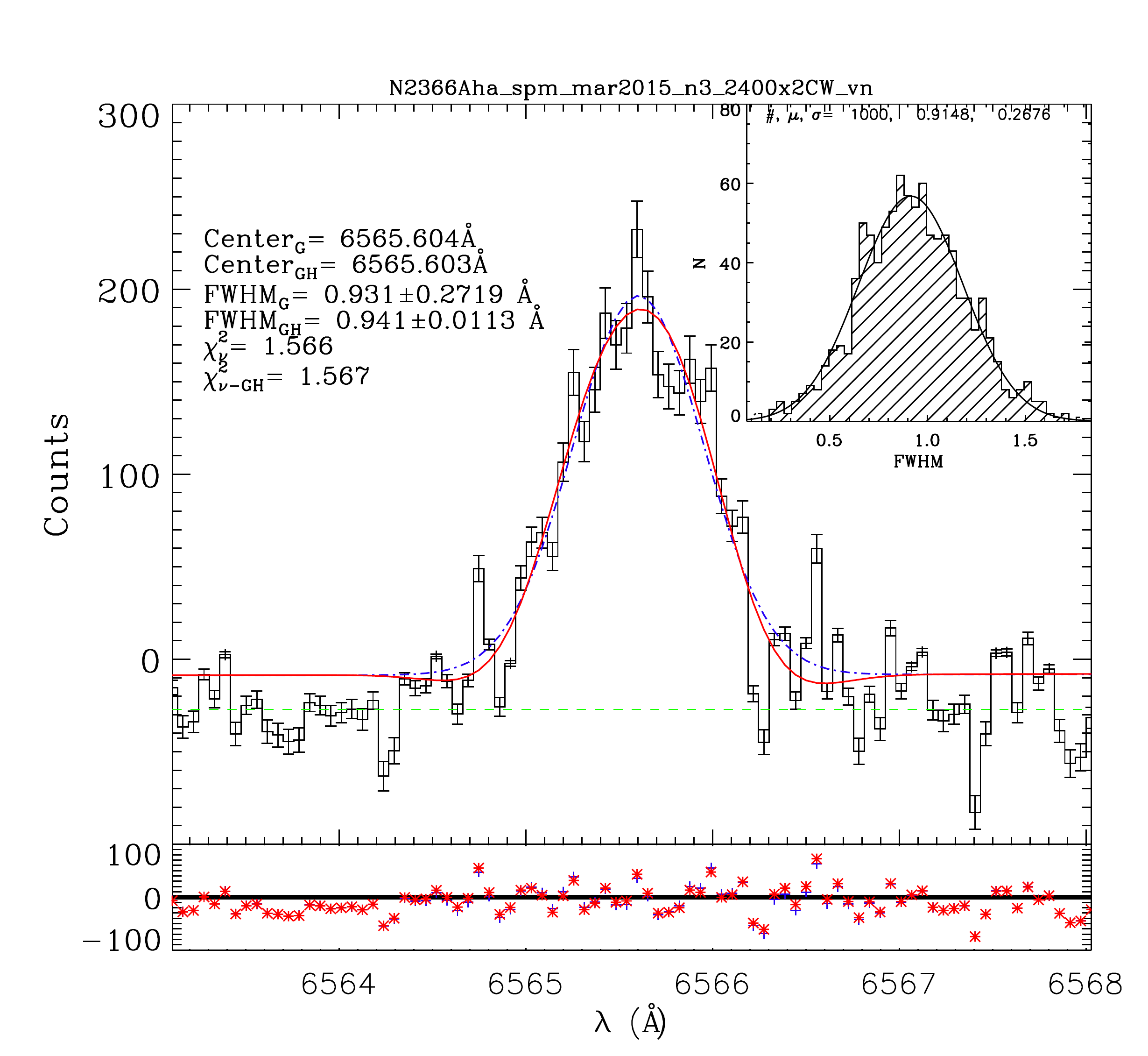}
  \includegraphics[scale=0.2]{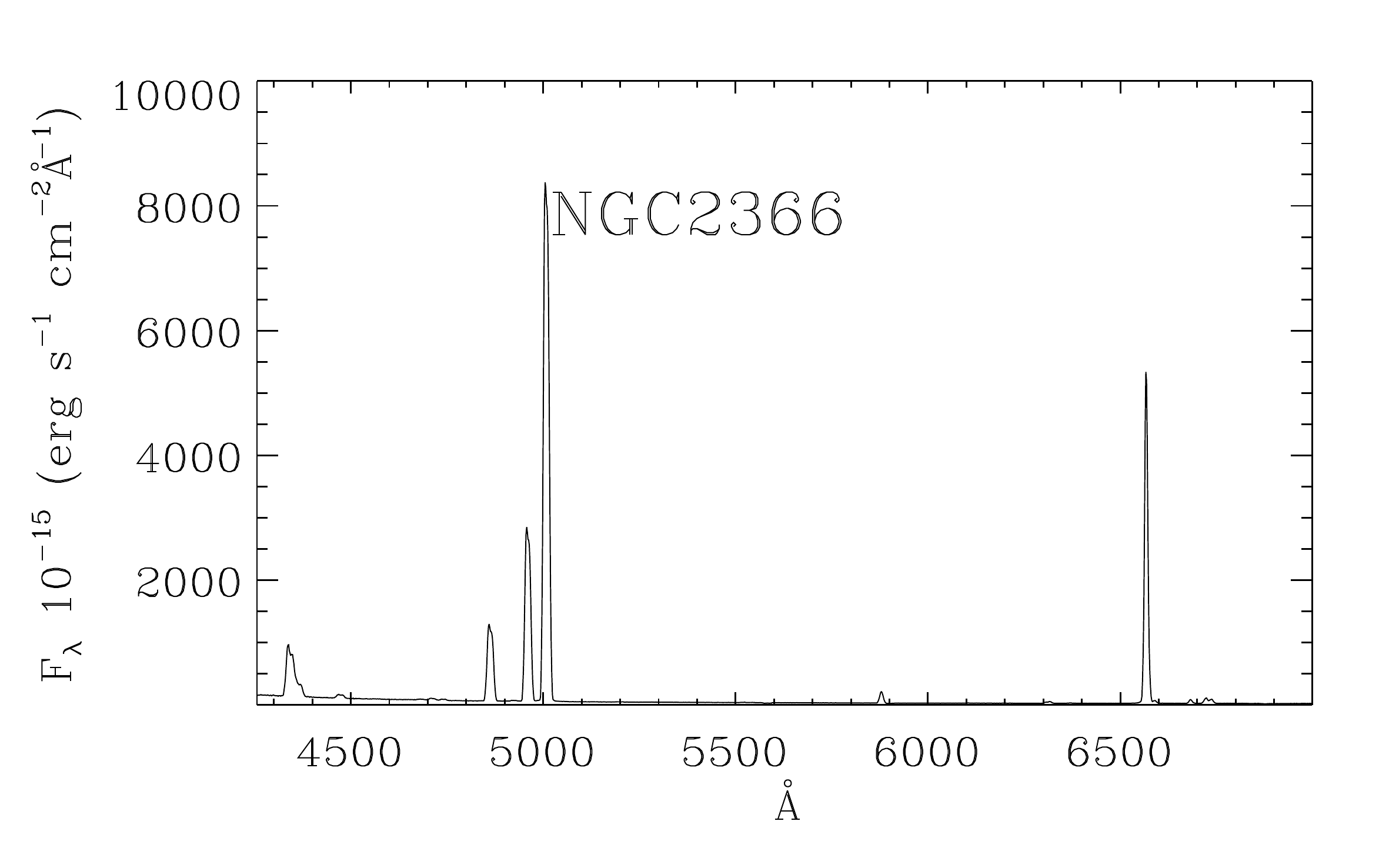}
  
  \includegraphics[scale=0.2]{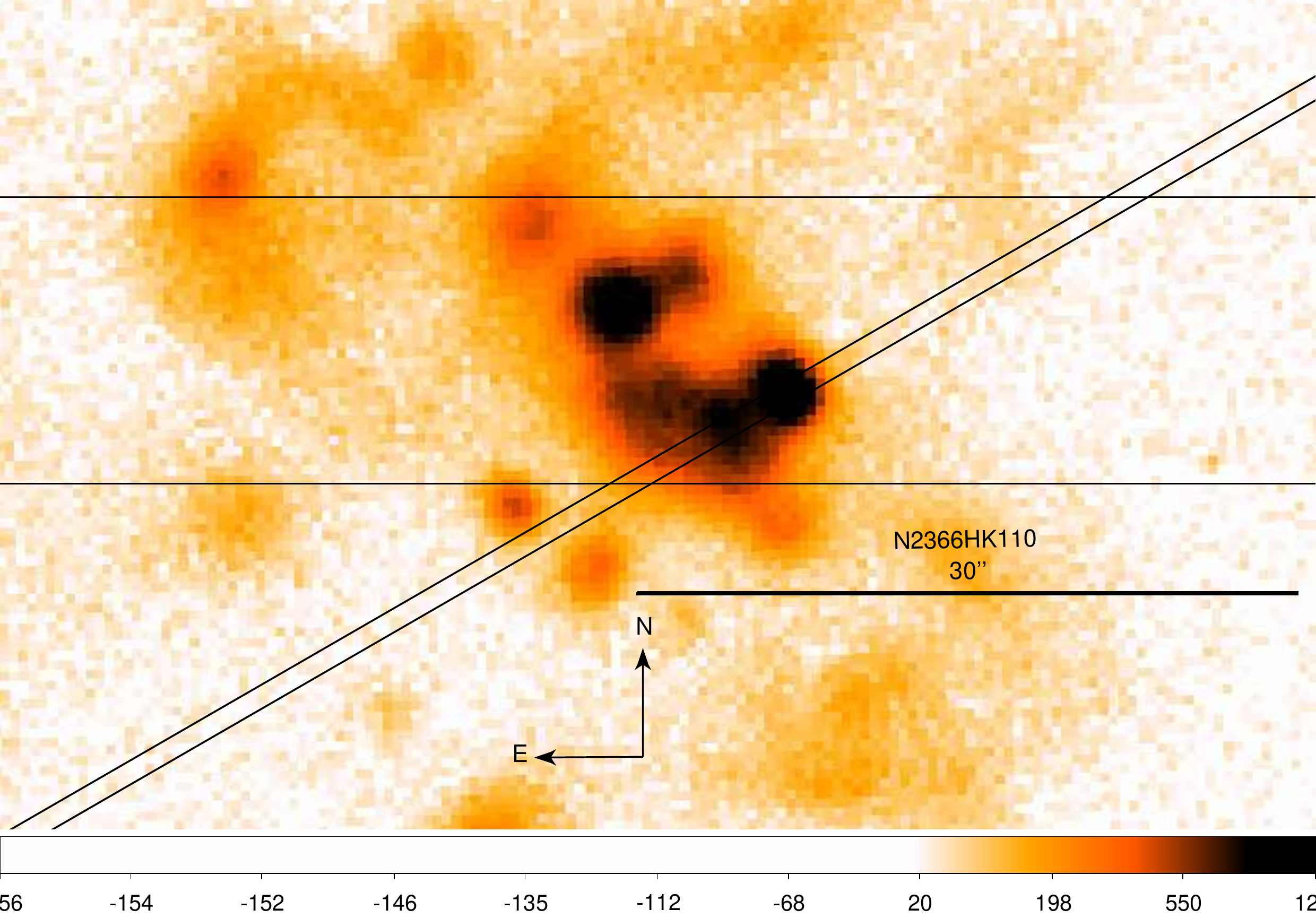}
  \includegraphics[scale=0.18]{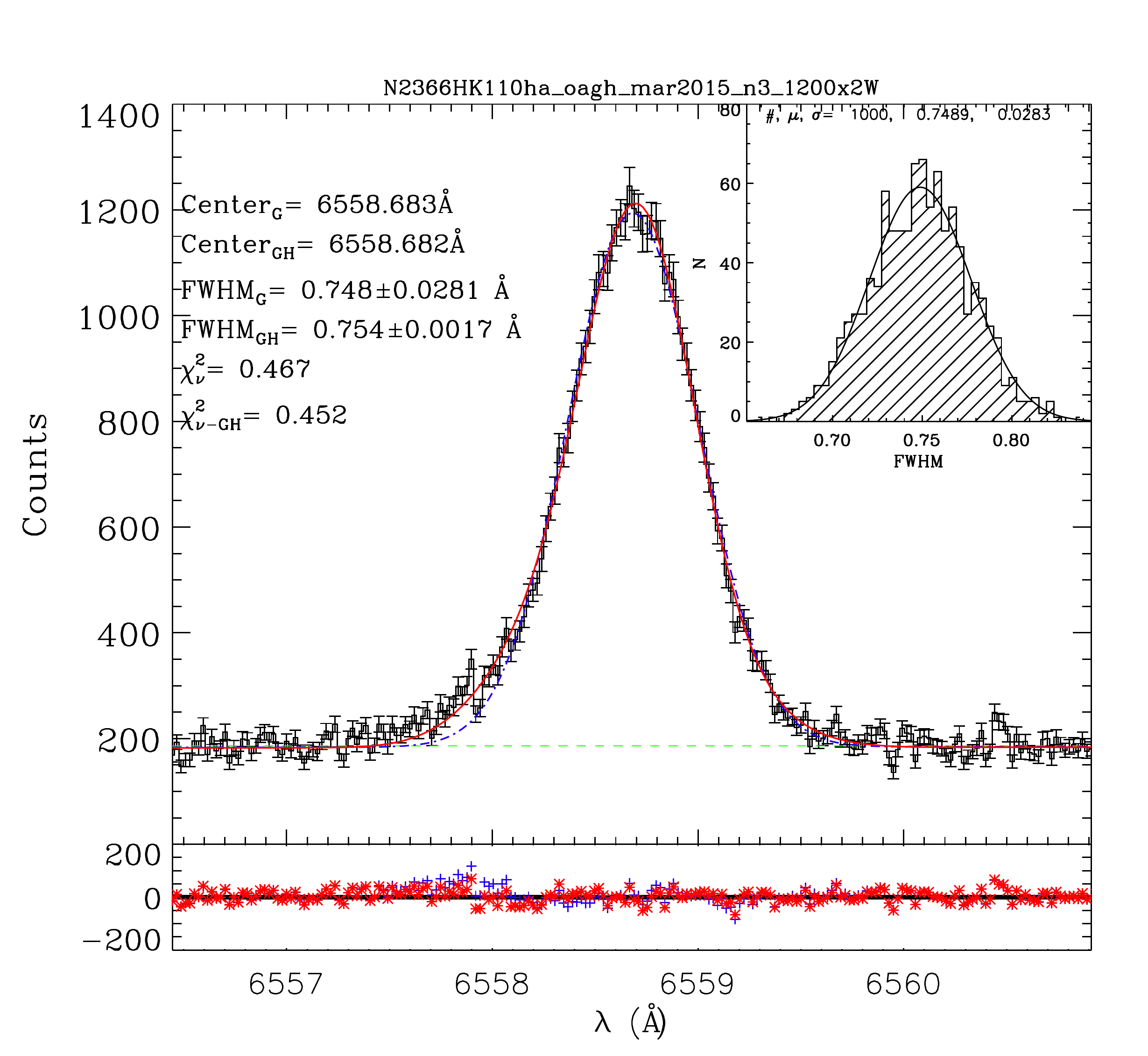}
  \includegraphics[scale=0.2]{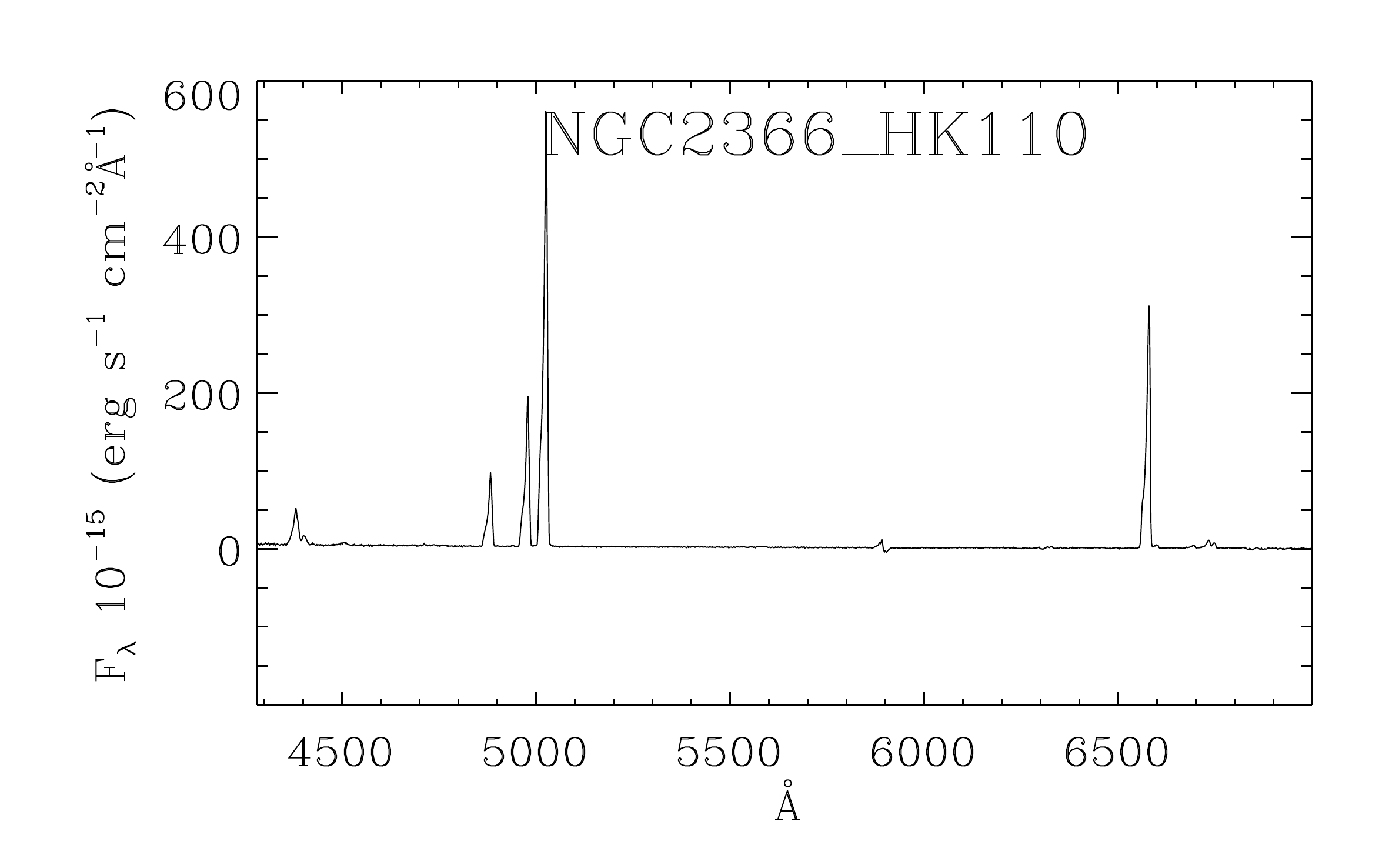}
  
  \includegraphics[scale=0.2]{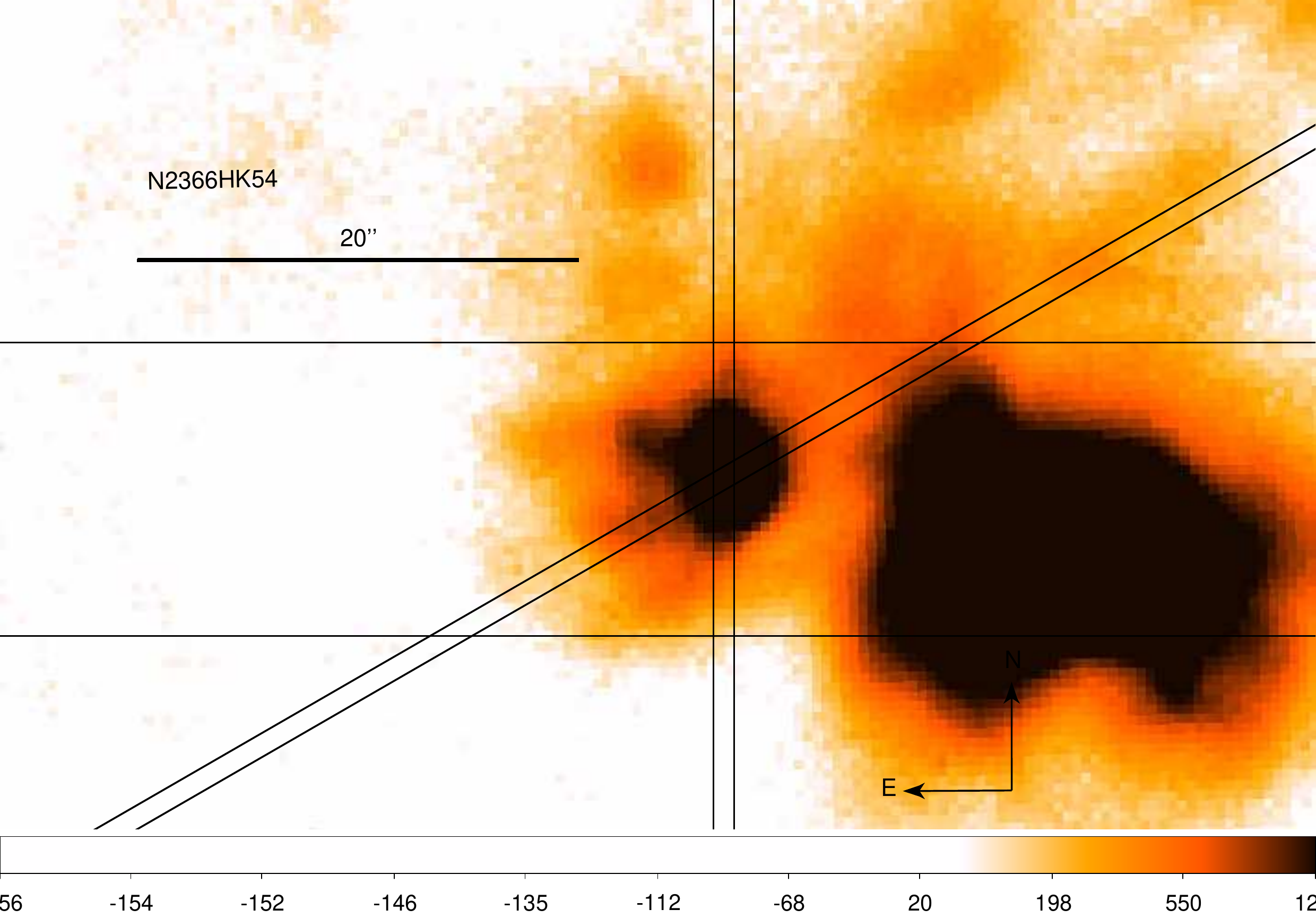}
  \includegraphics[scale=0.18]{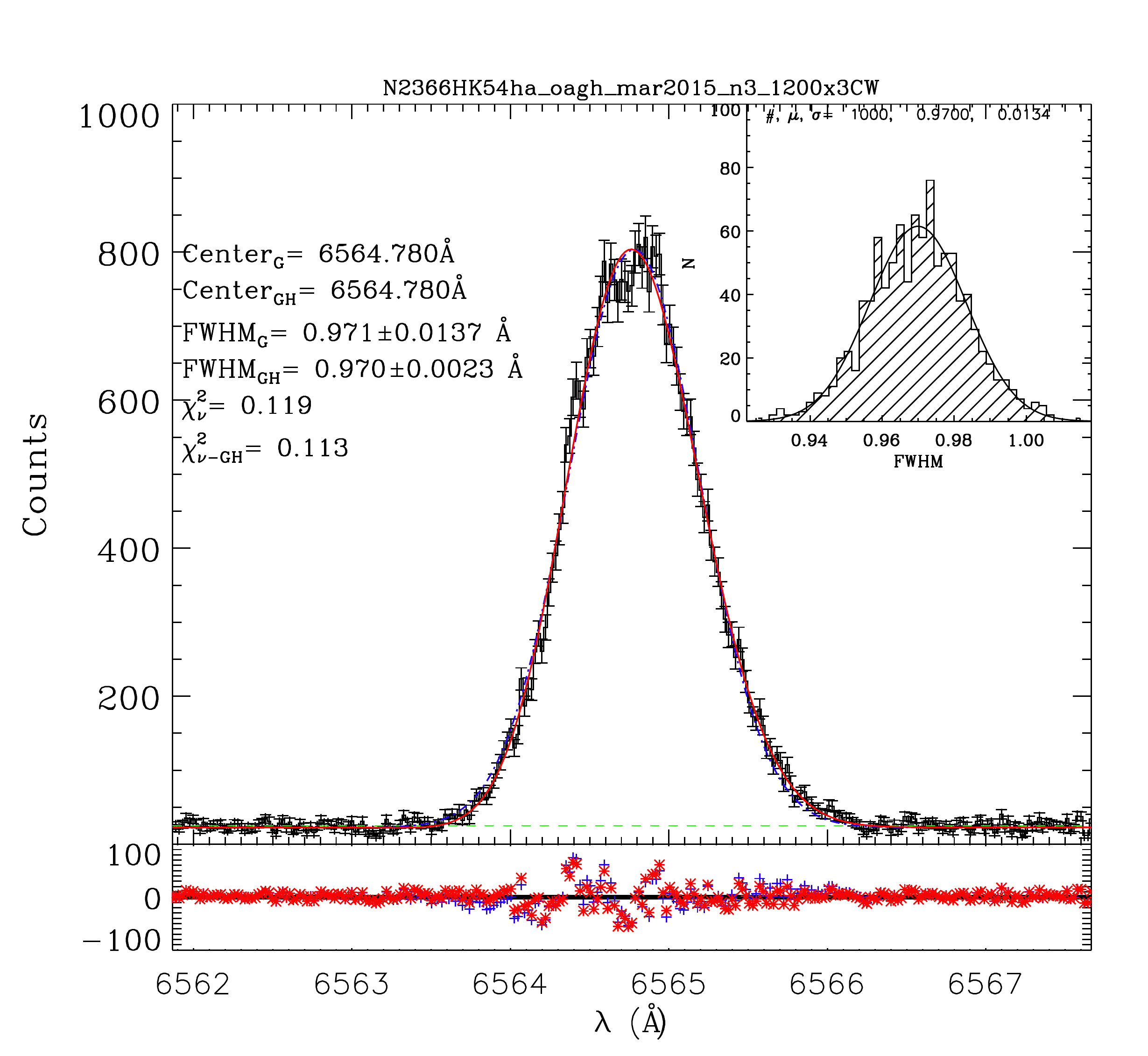}
   \includegraphics[scale=0.2]{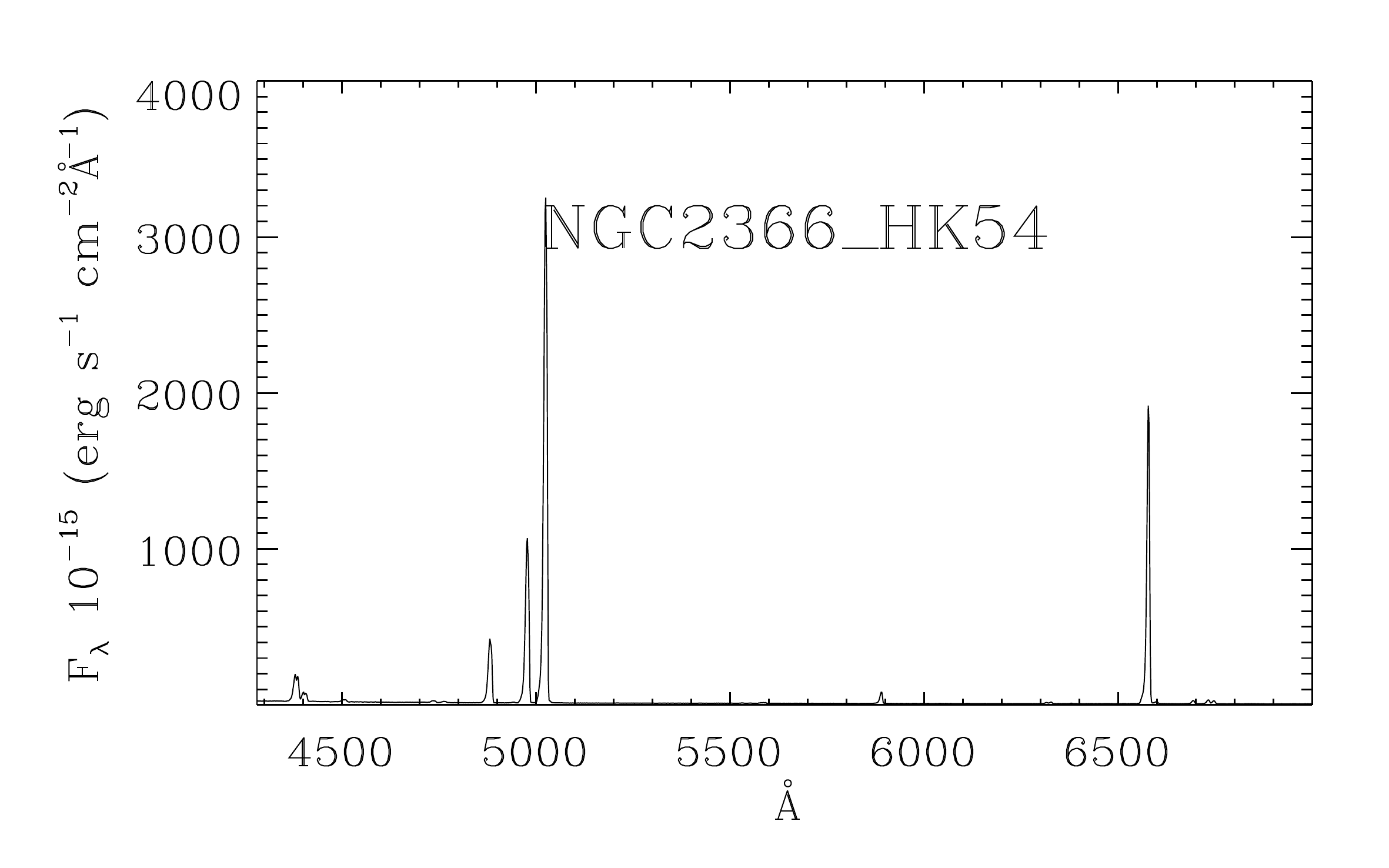}

\includegraphics[scale=0.2]{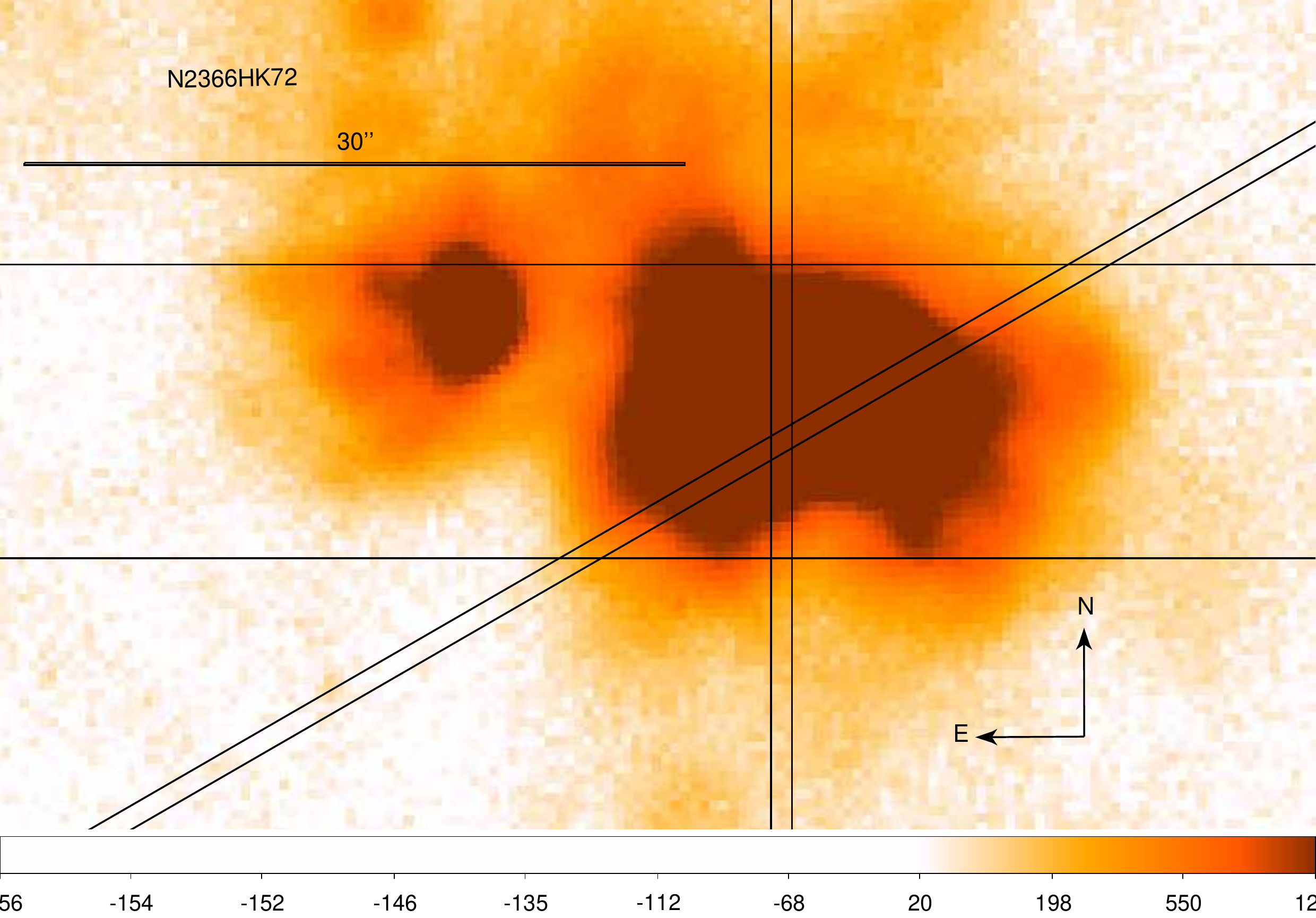}
  \includegraphics[scale=0.18]{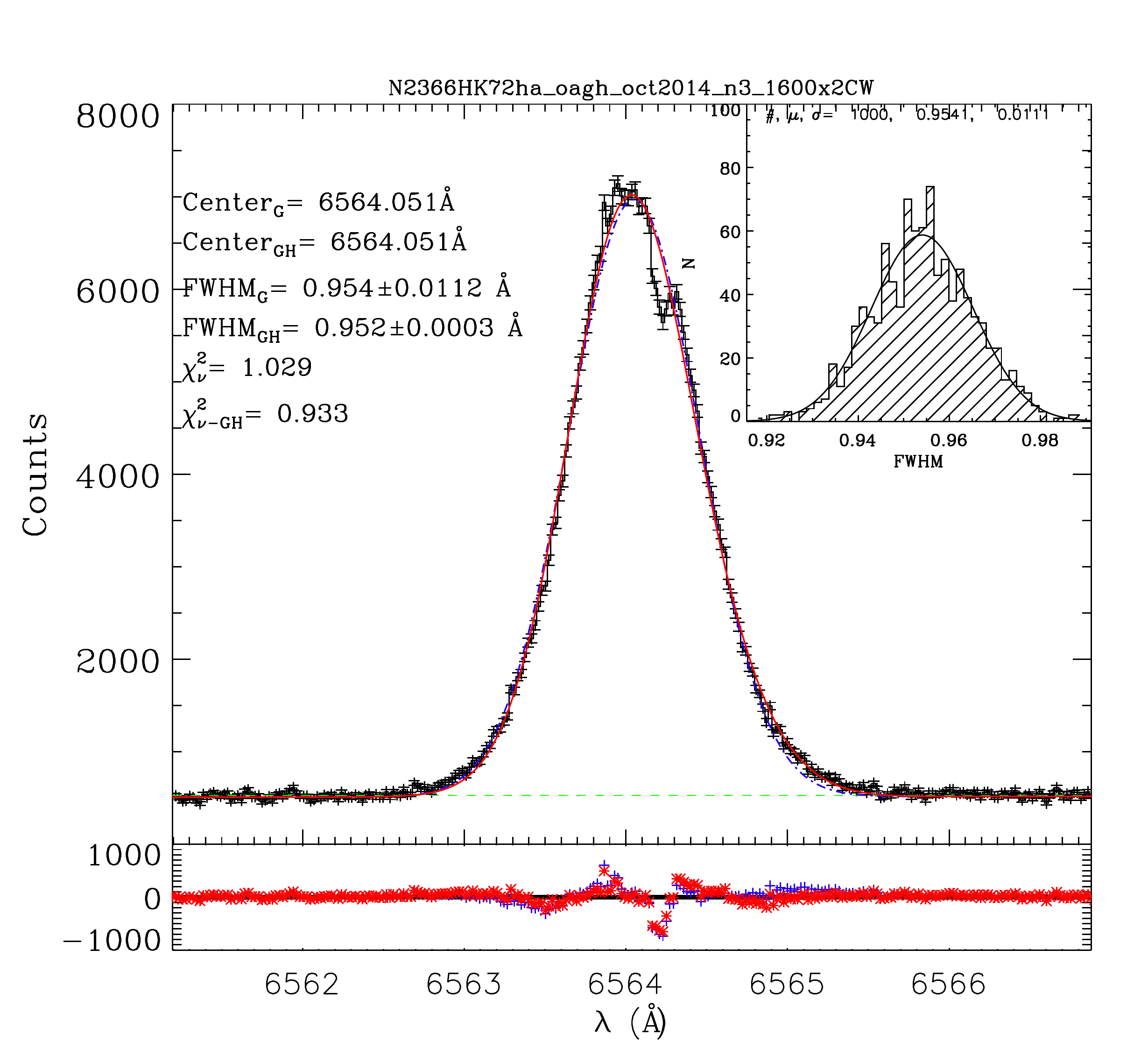}
  \includegraphics[scale=0.2]{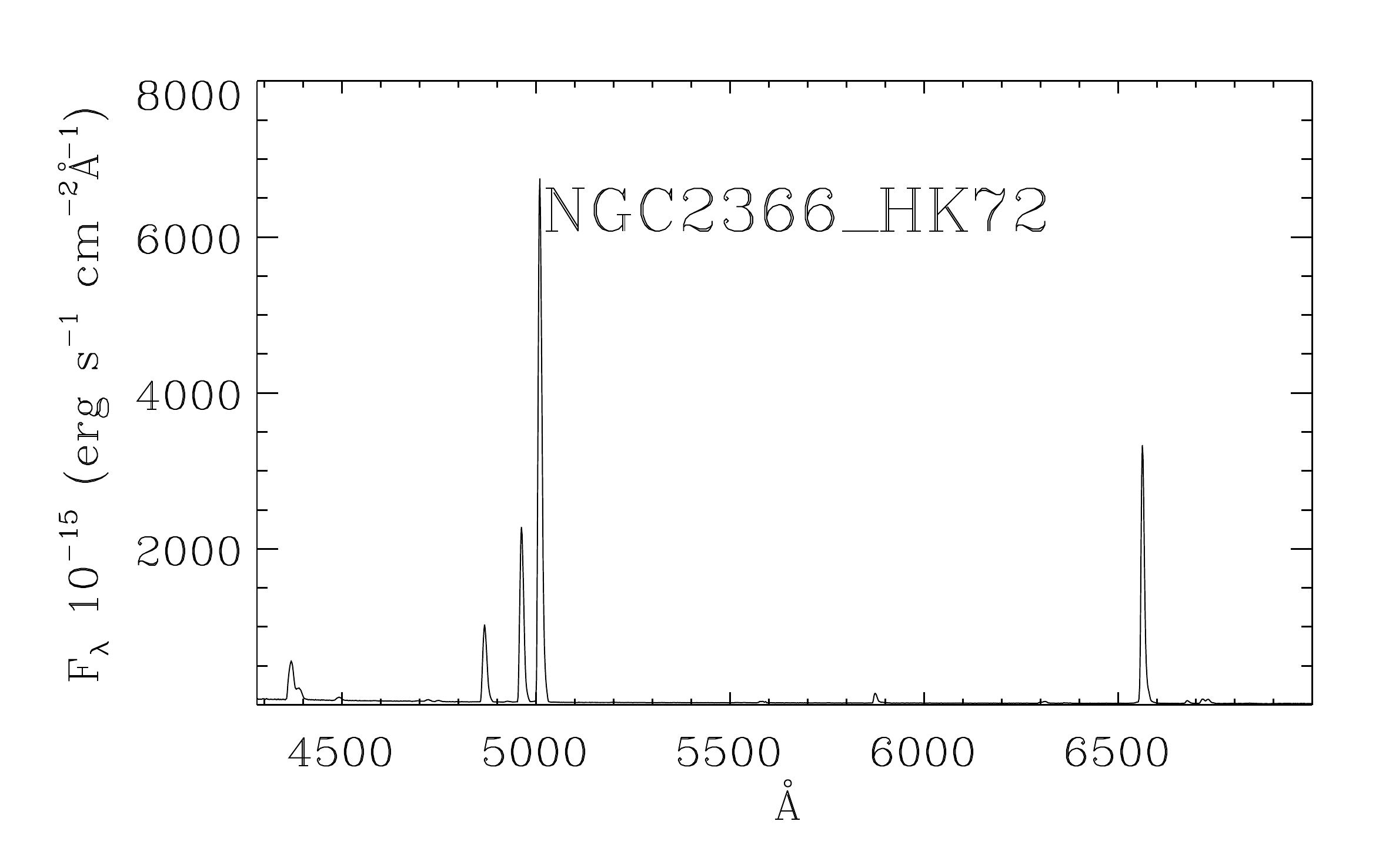}
  
  \includegraphics[scale=0.2]{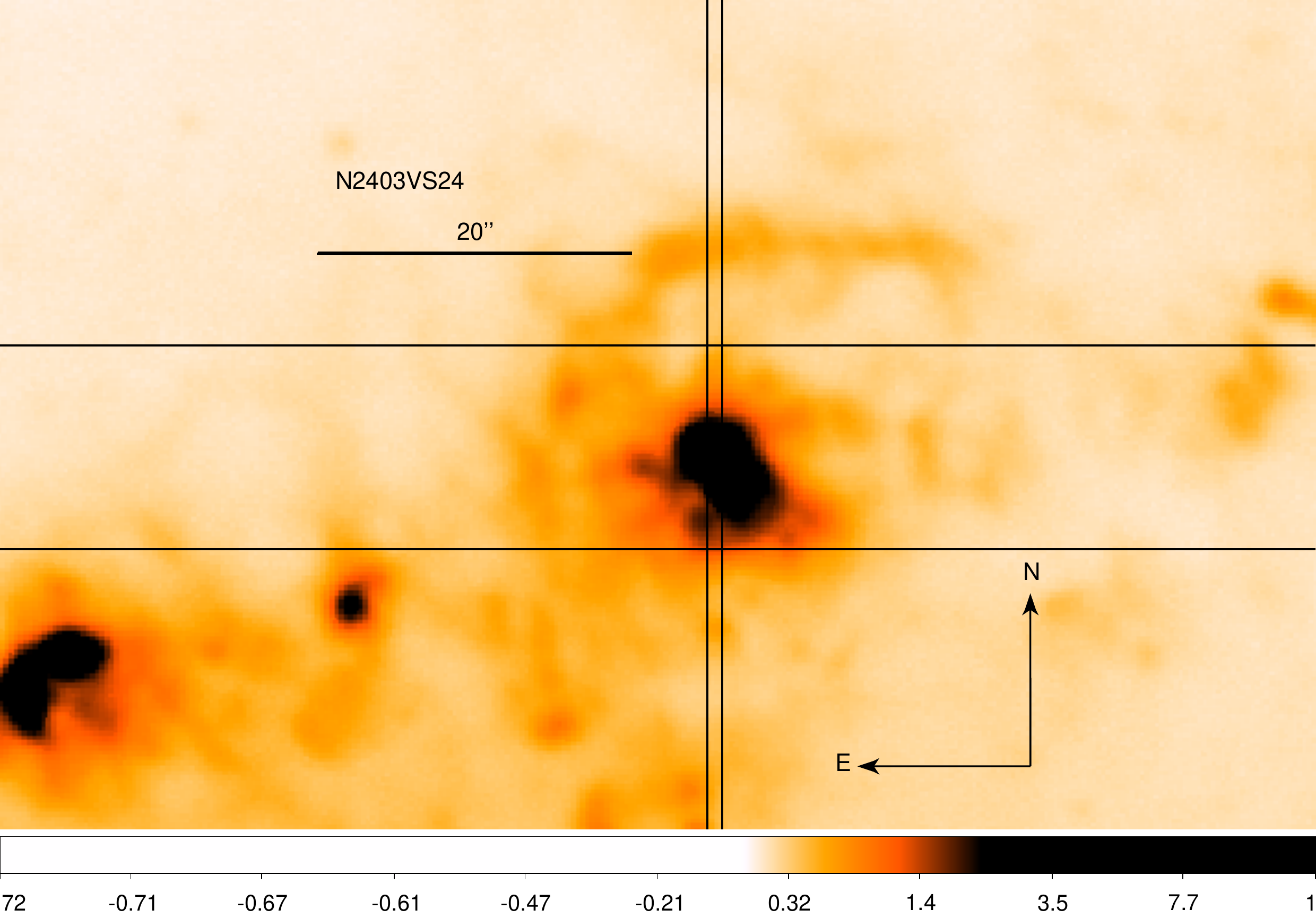}
  \includegraphics[scale=0.18]{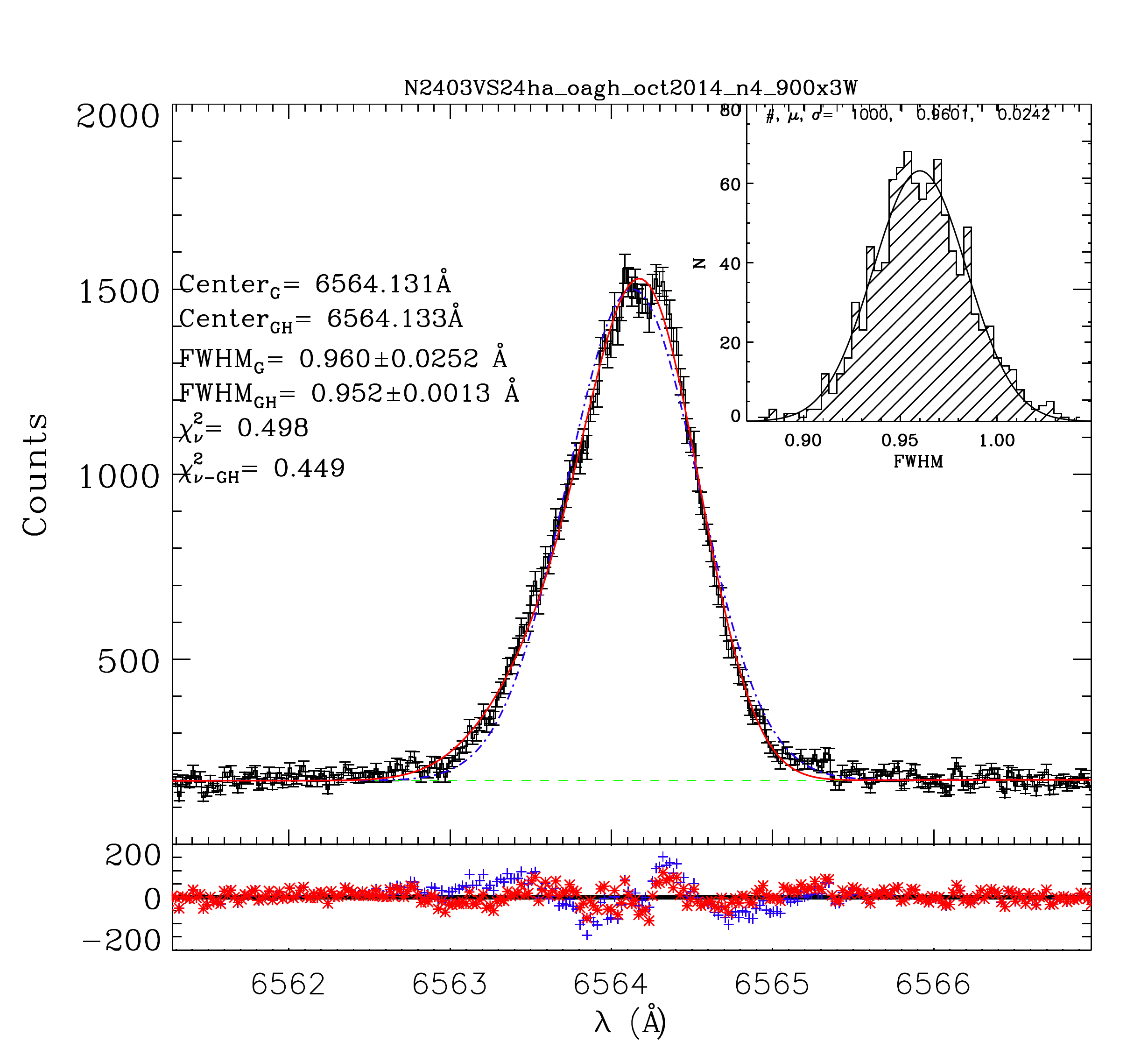}
   \includegraphics[scale=0.2]{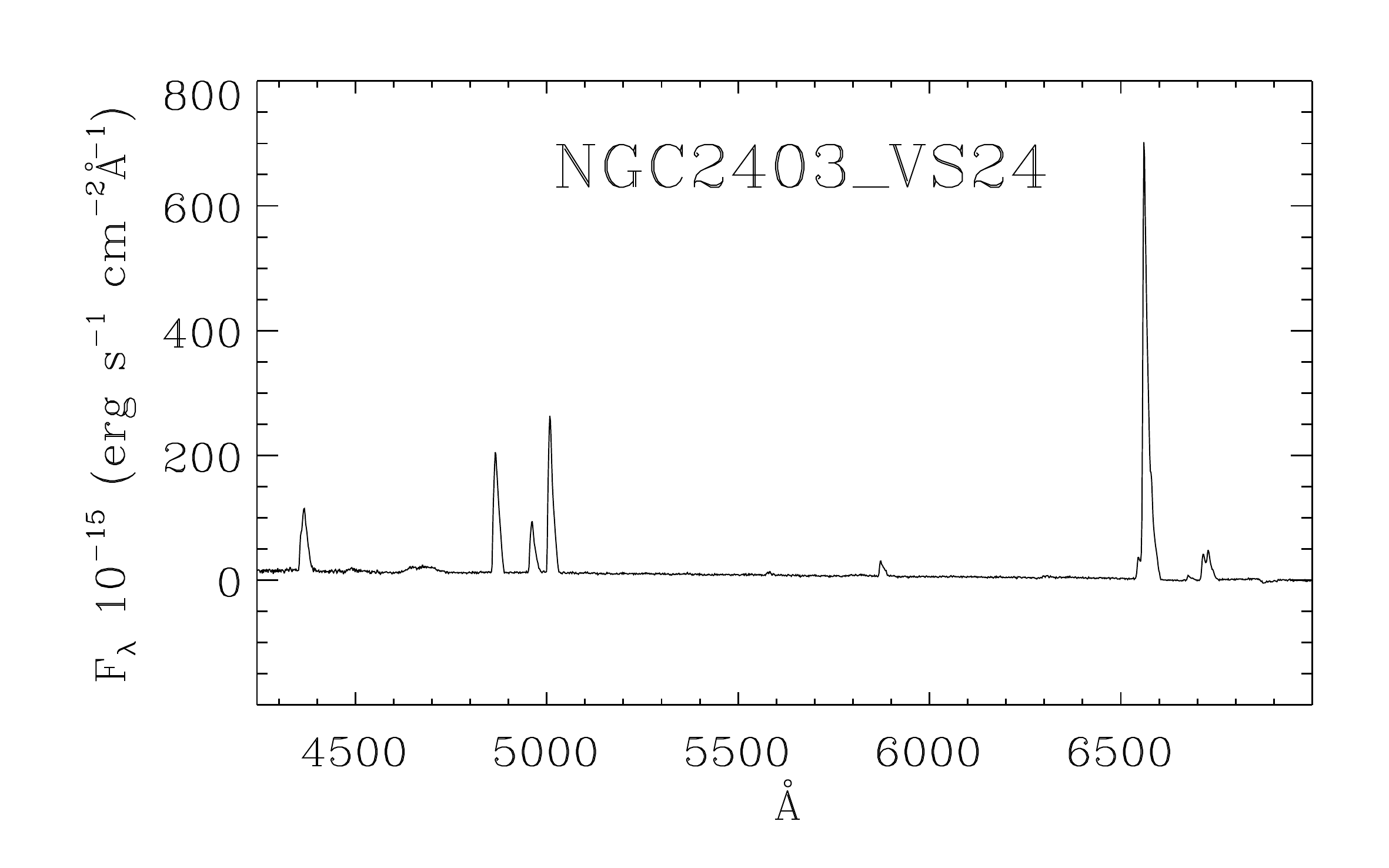}
         \caption[]{ (continued)}
 \end{figure*}

  \begin{figure*}       
        \includegraphics[scale=0.2]{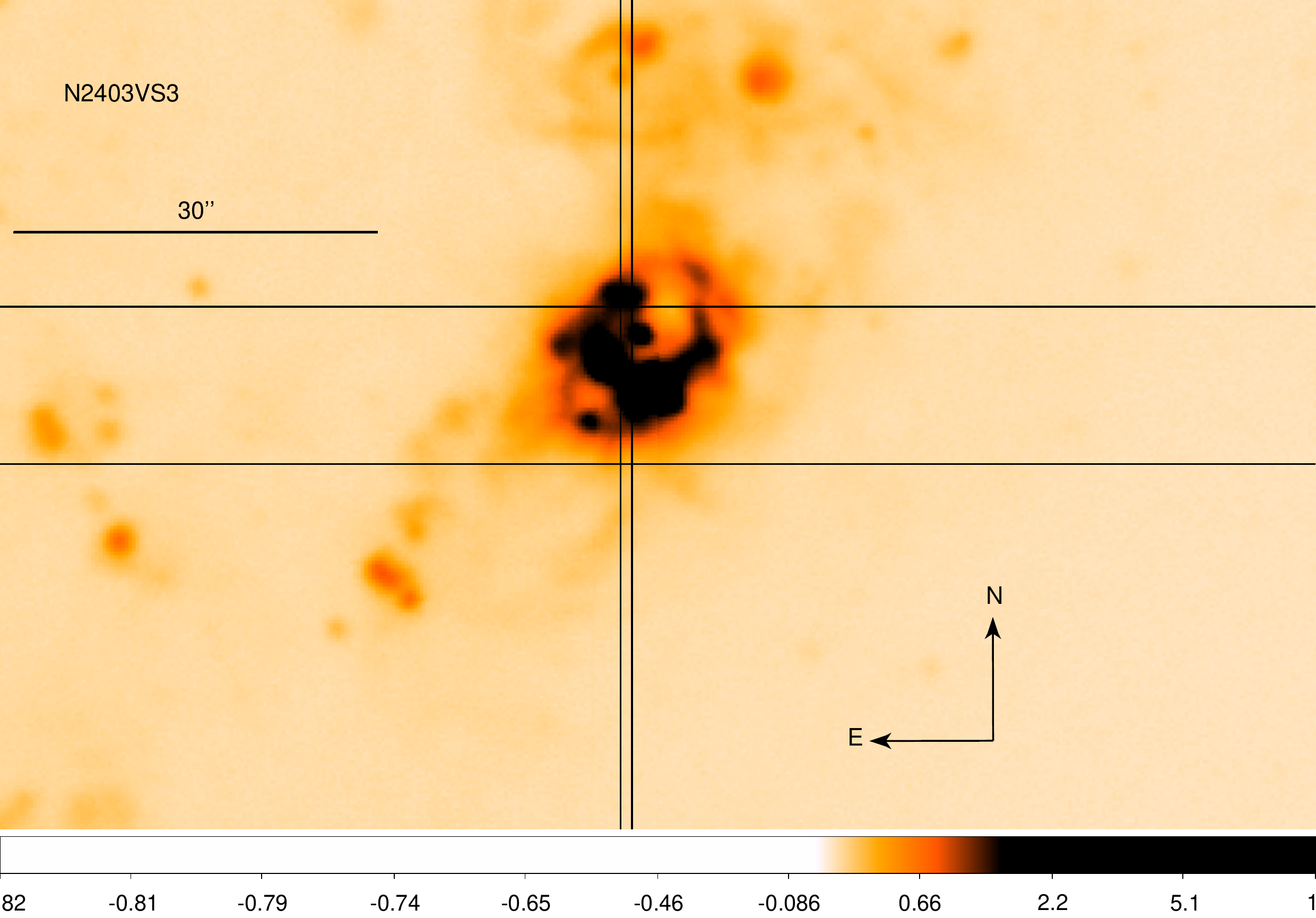}
  \includegraphics[scale=0.18]{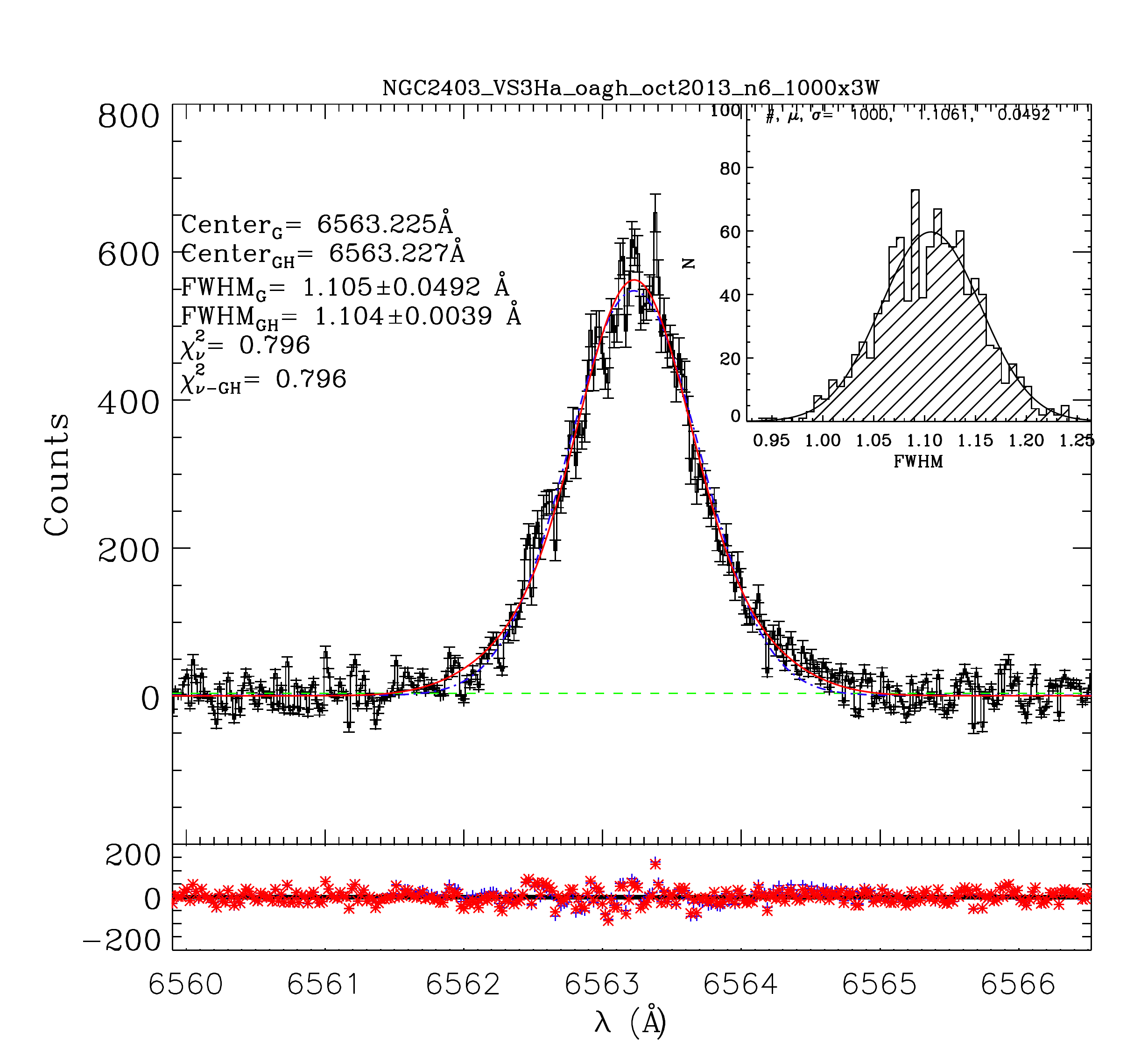}
 \includegraphics[scale=0.2]{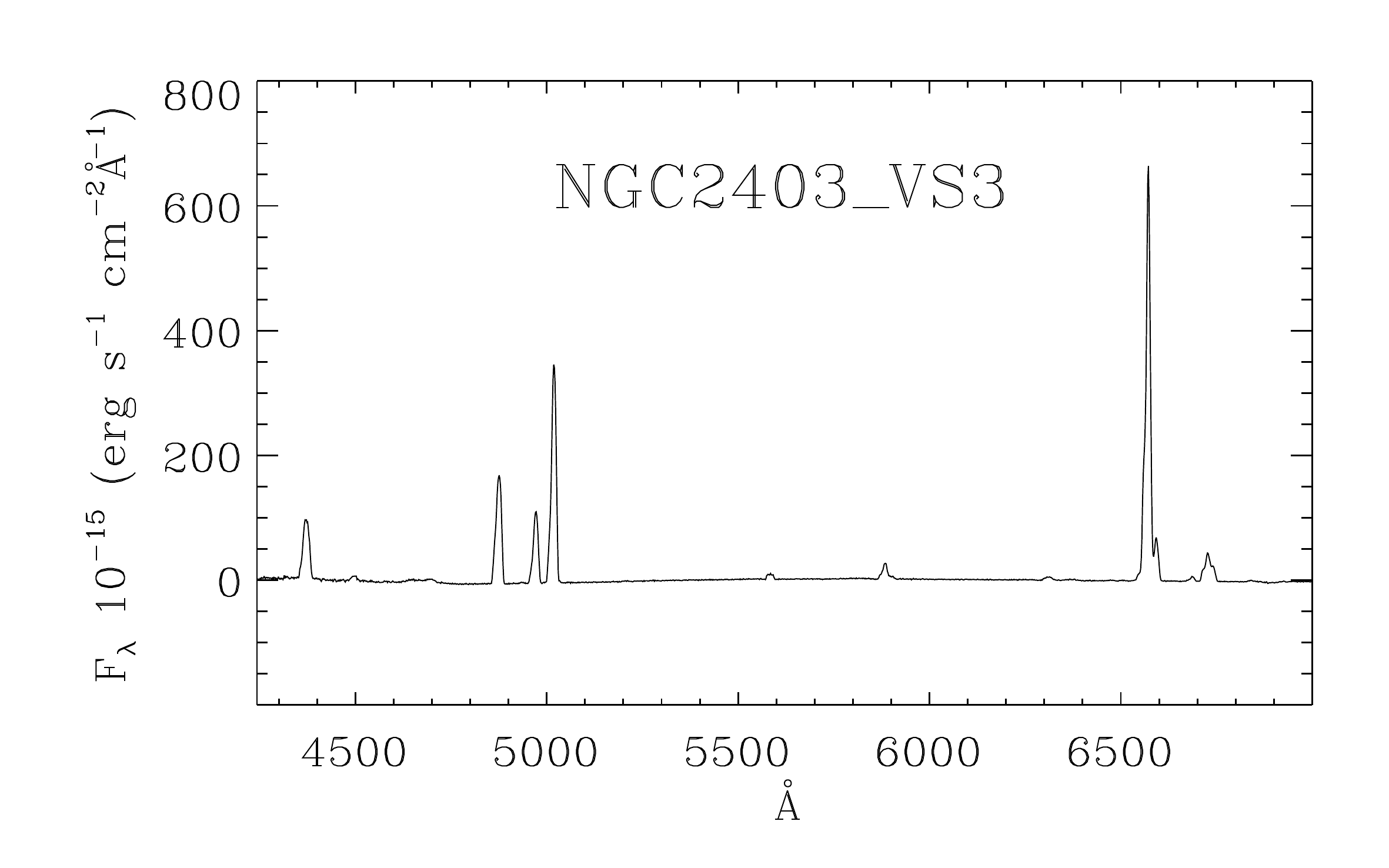}
 
 \includegraphics[scale=0.2]{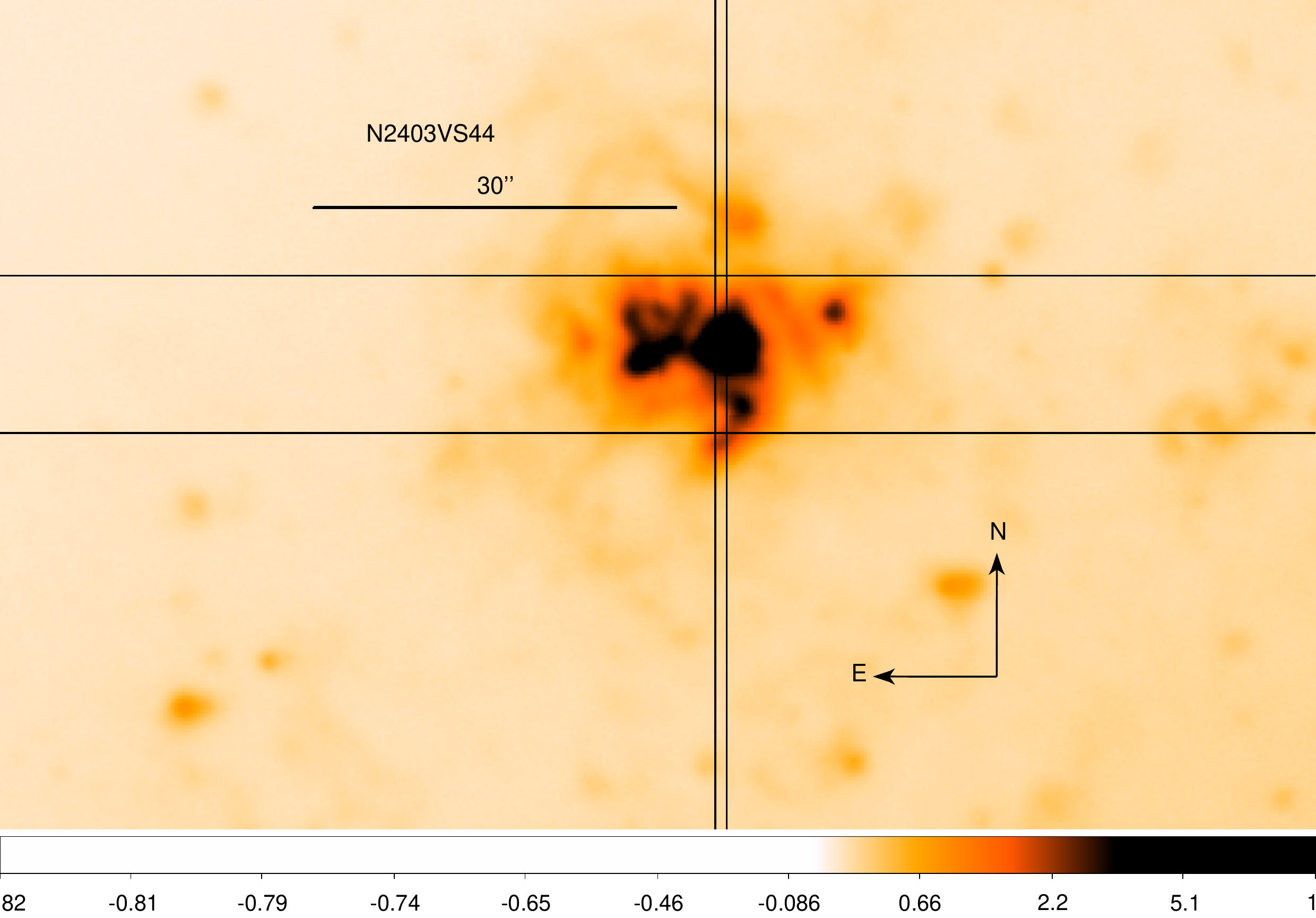}
  \includegraphics[scale=0.18]{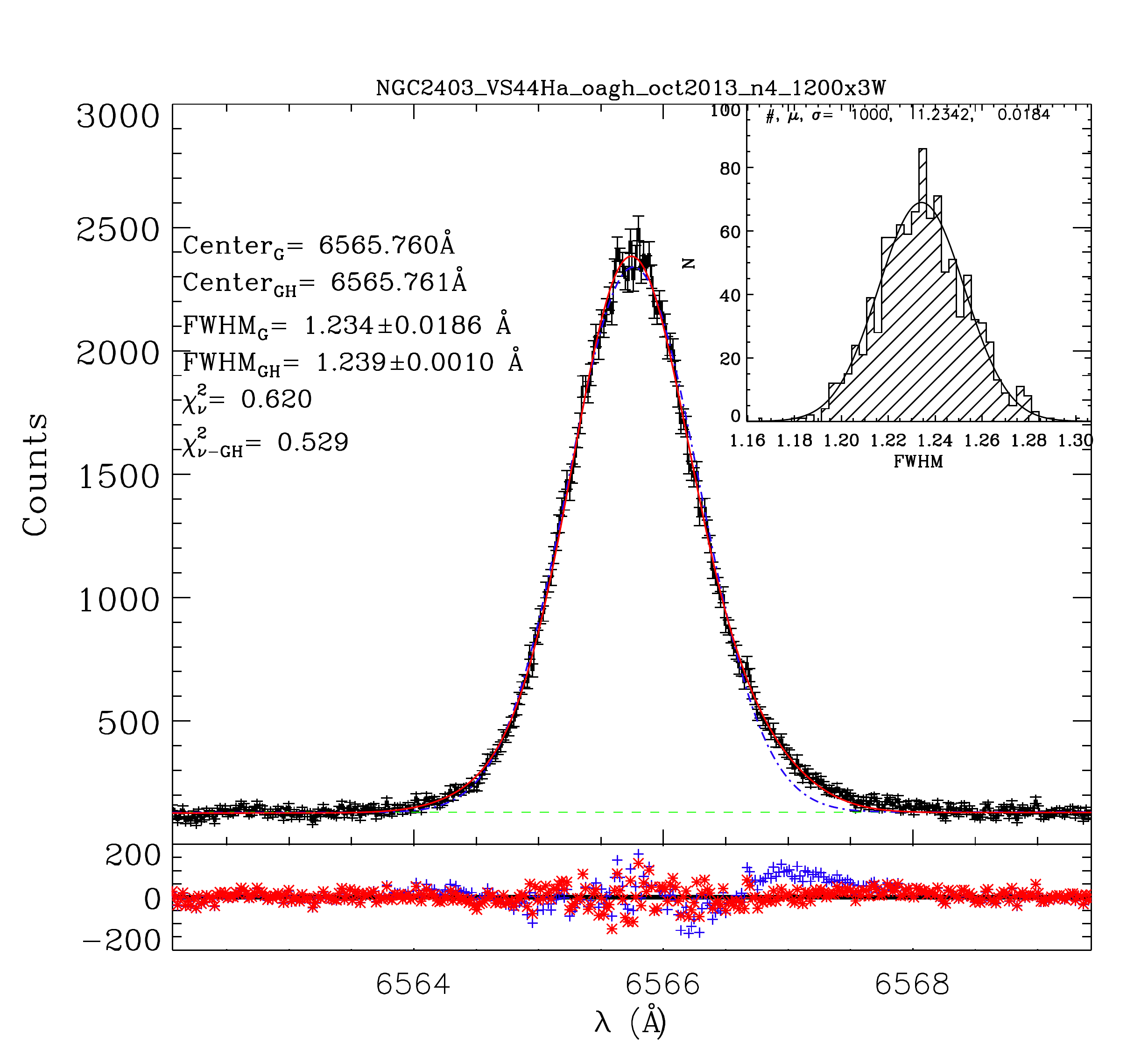}
\includegraphics[scale=0.2]{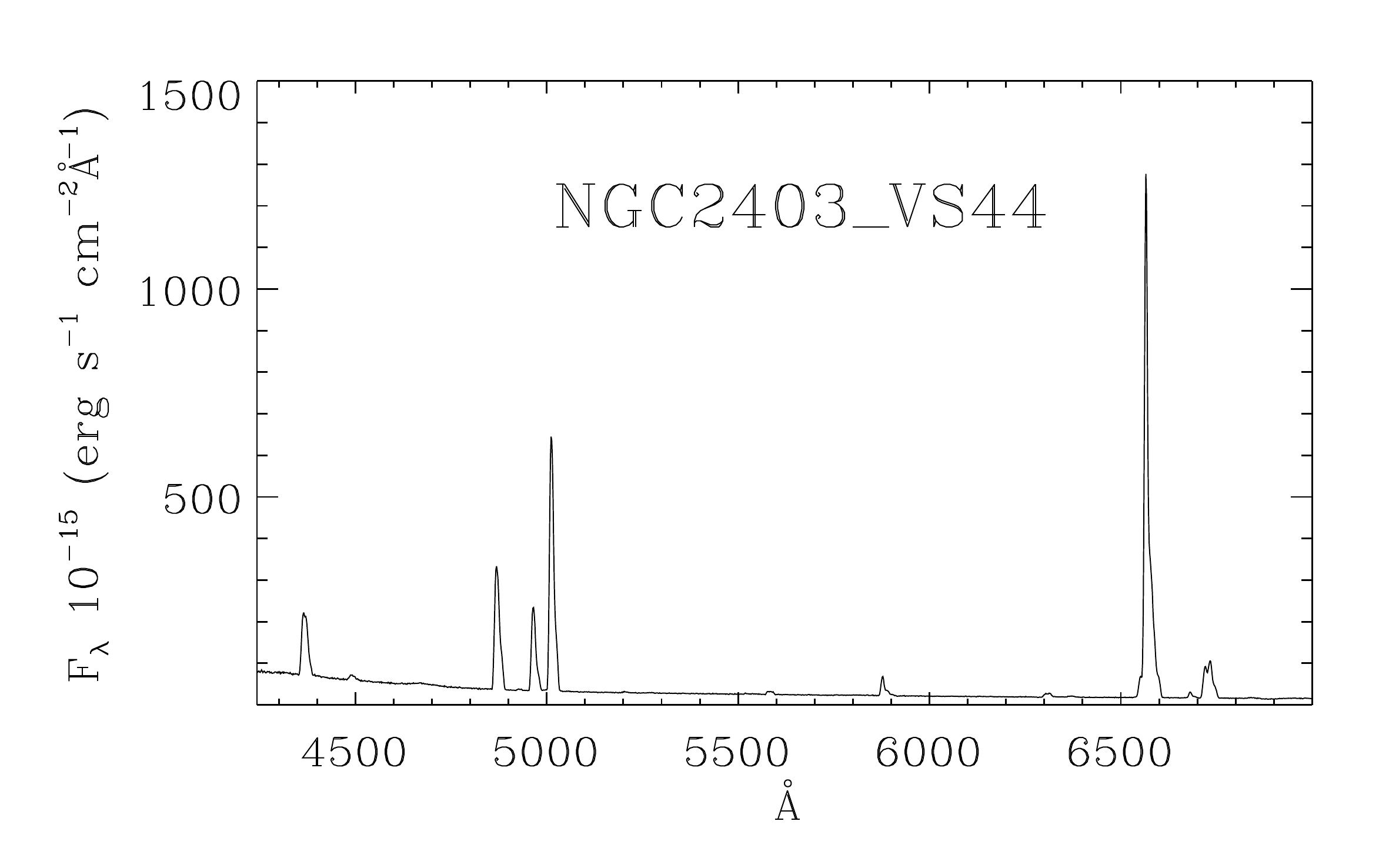}

\includegraphics[scale=0.2]{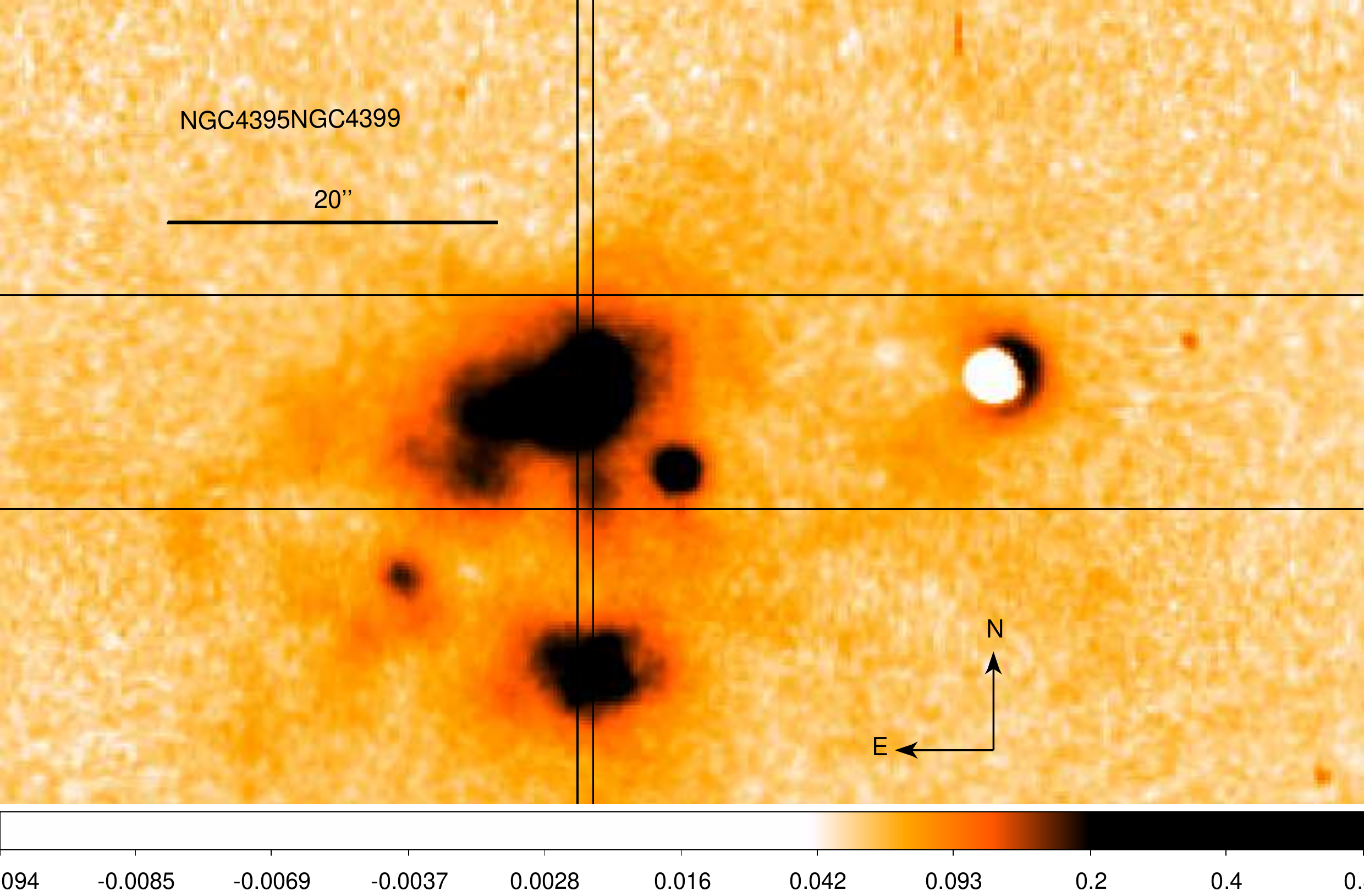}
 \includegraphics[scale=0.18]{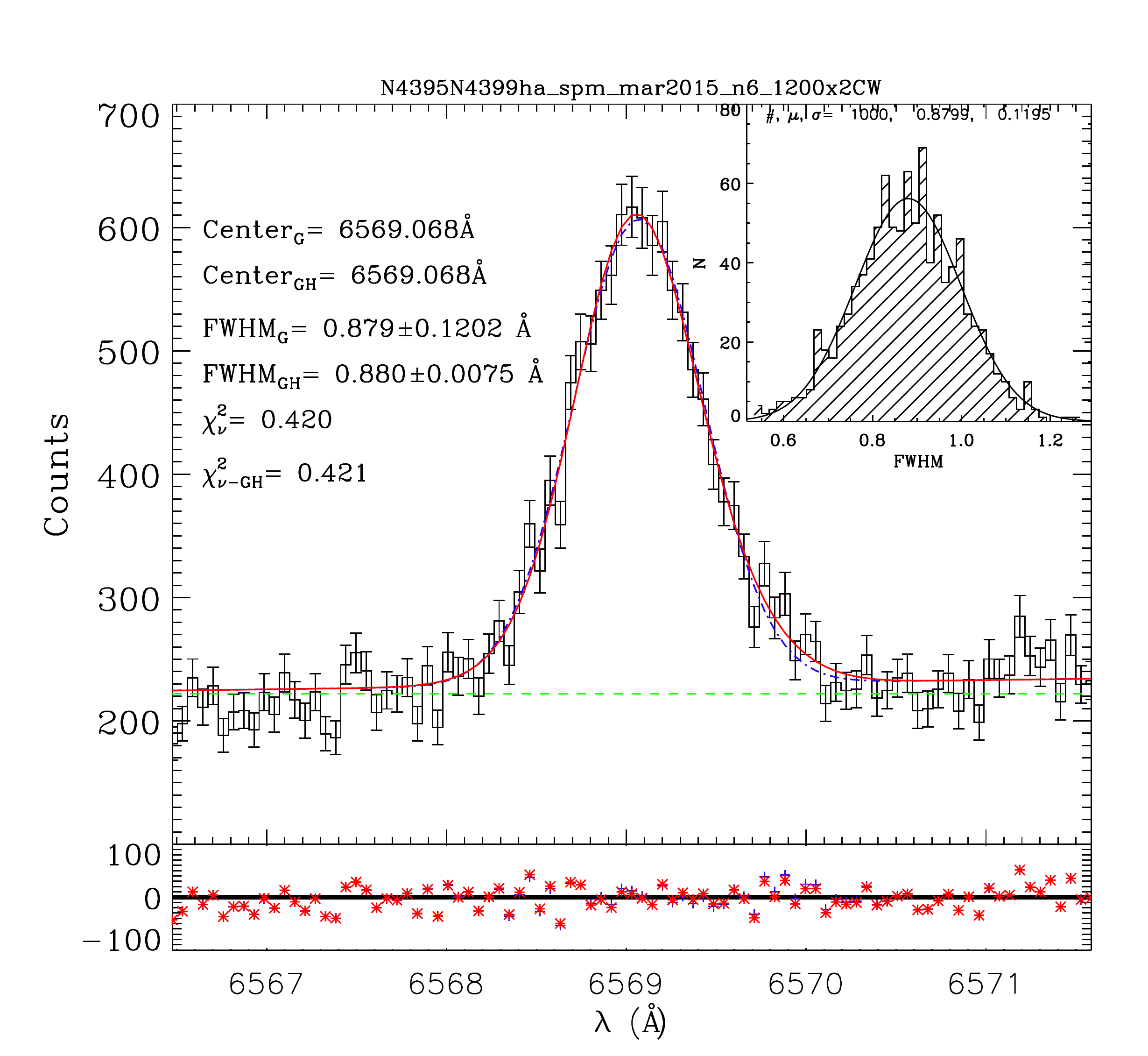}
  \includegraphics[scale=0.2]{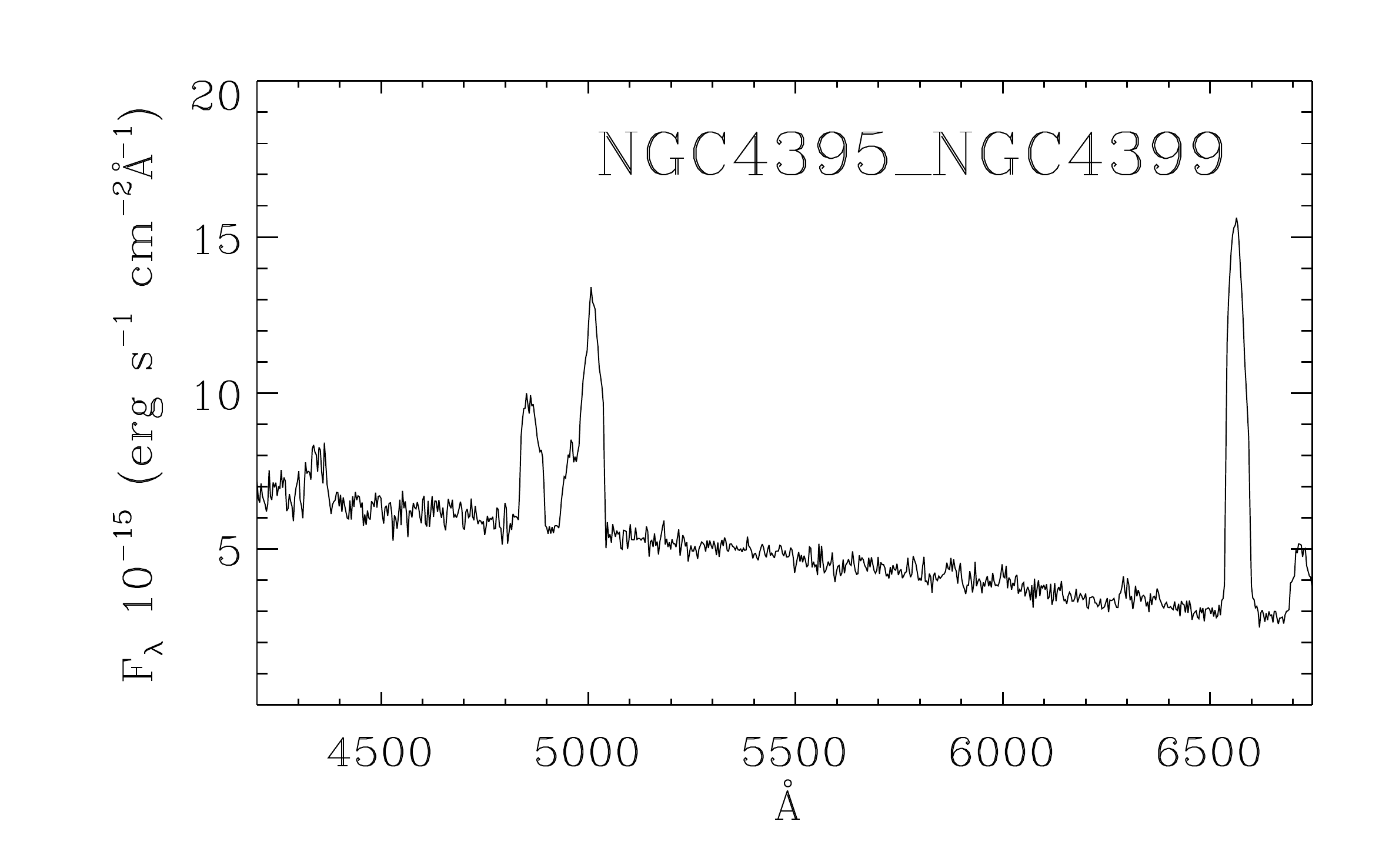}

\includegraphics[scale=0.2]{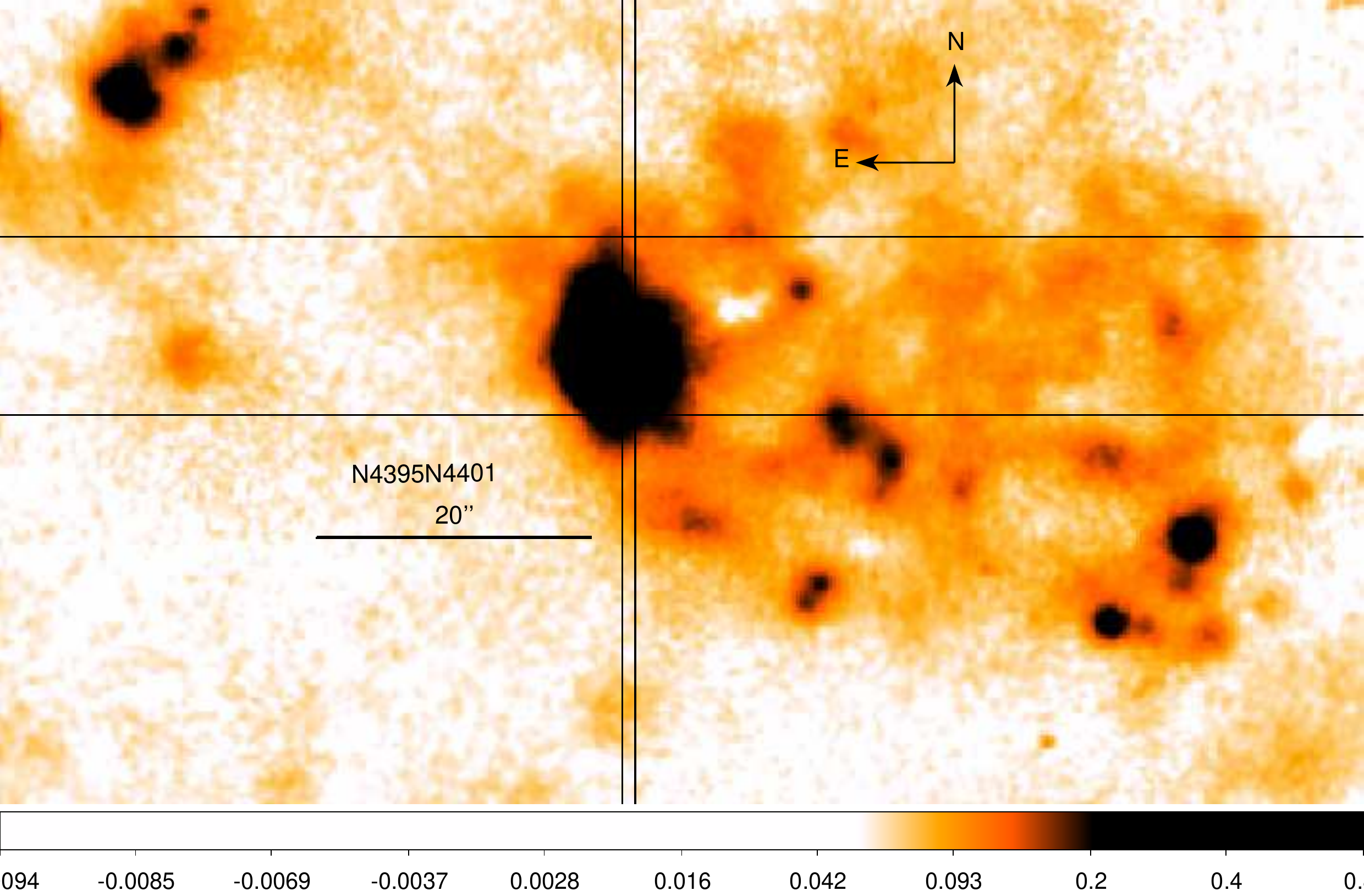}
  \includegraphics[scale=0.18]{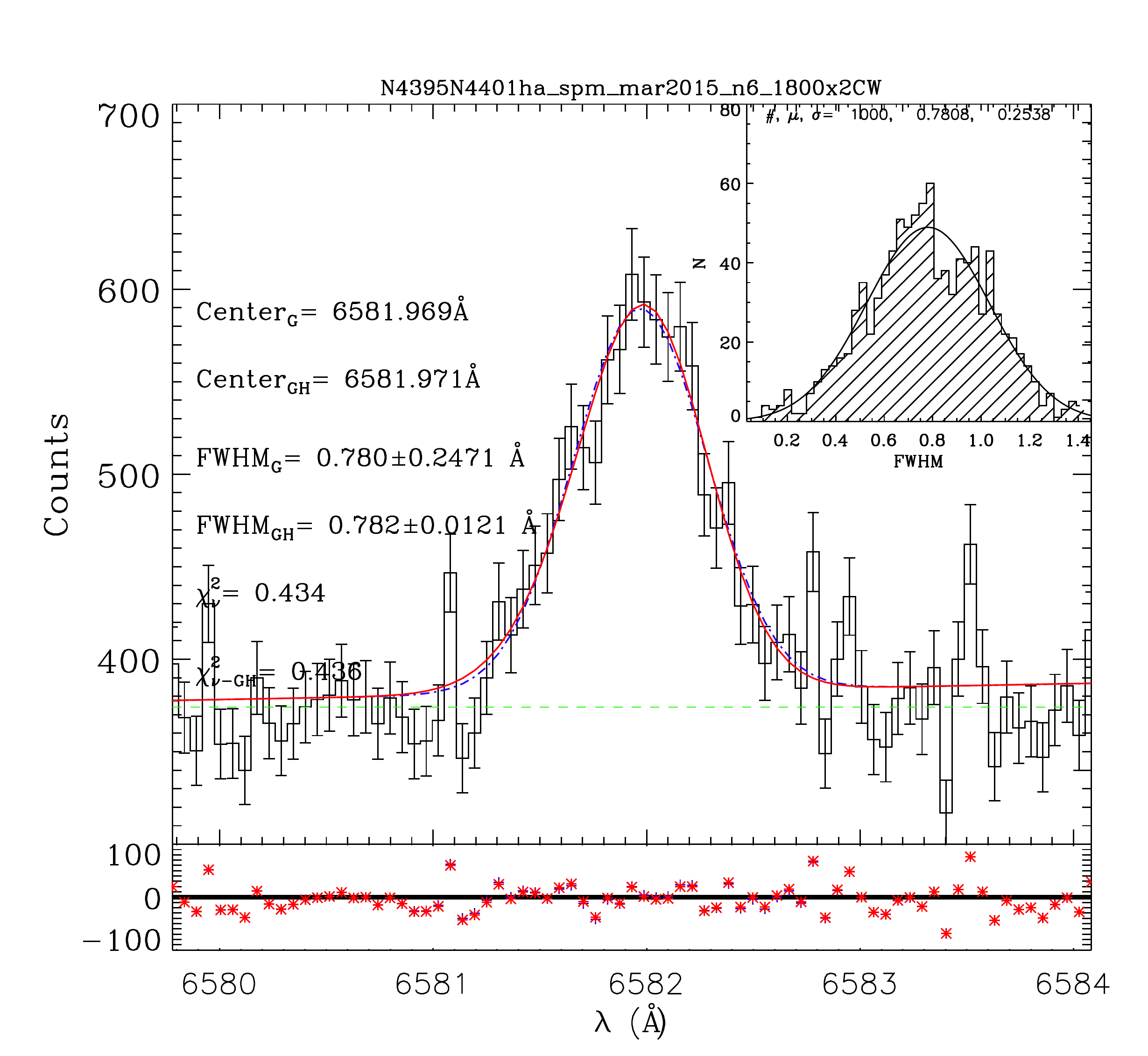}
  \includegraphics[scale=0.2]{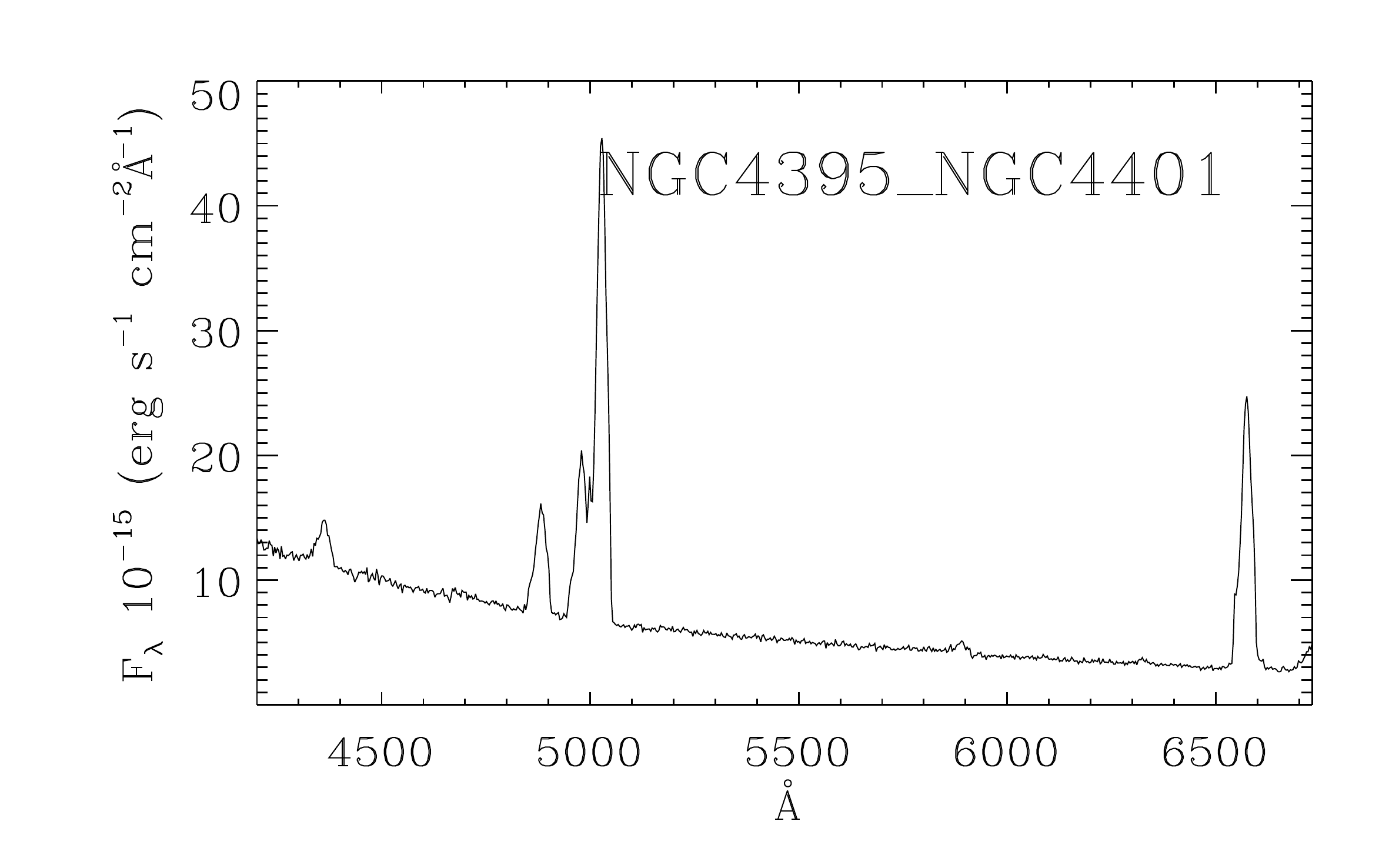}

  \includegraphics[scale=0.2]{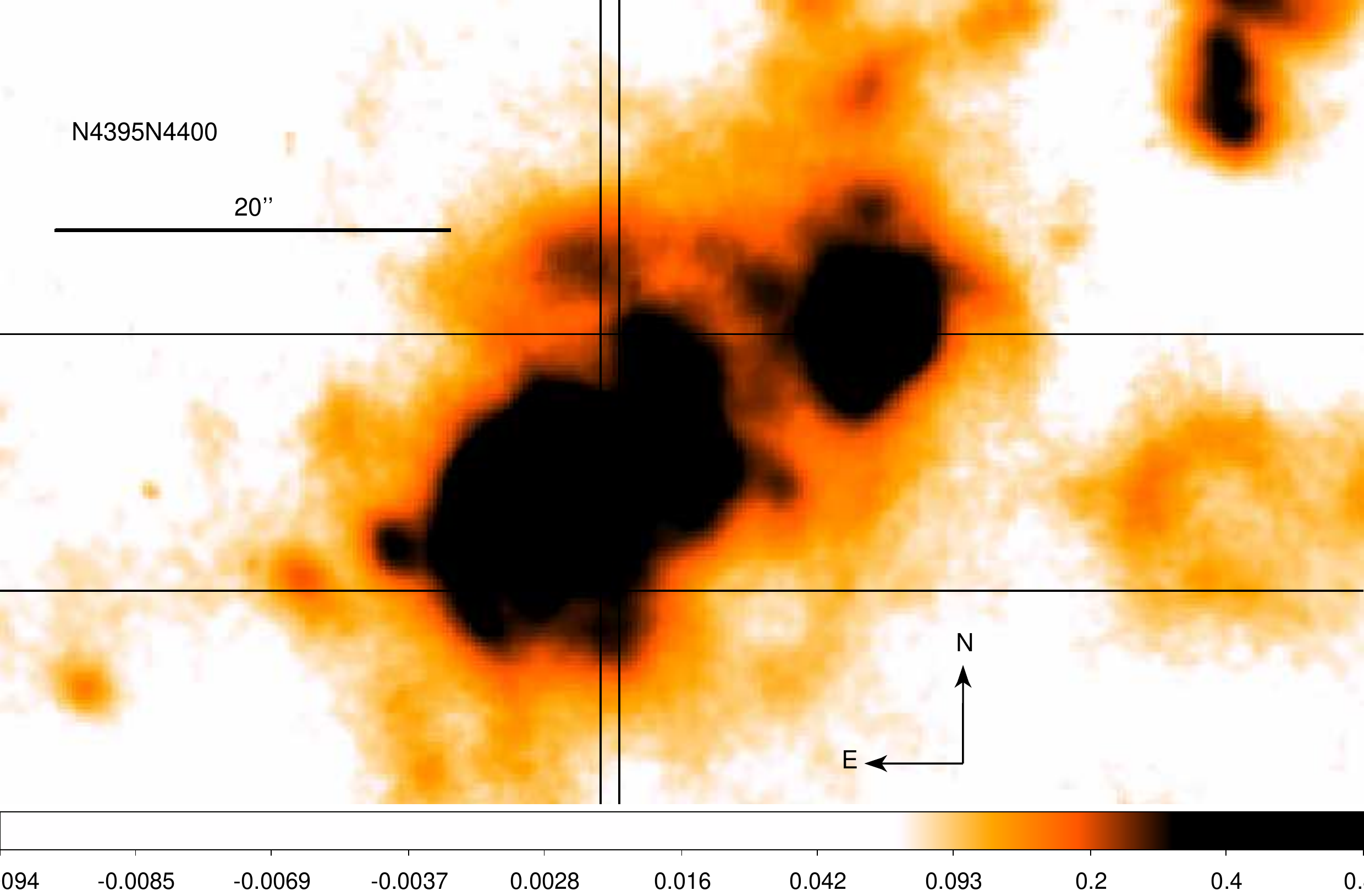}
  \includegraphics[scale=0.18]{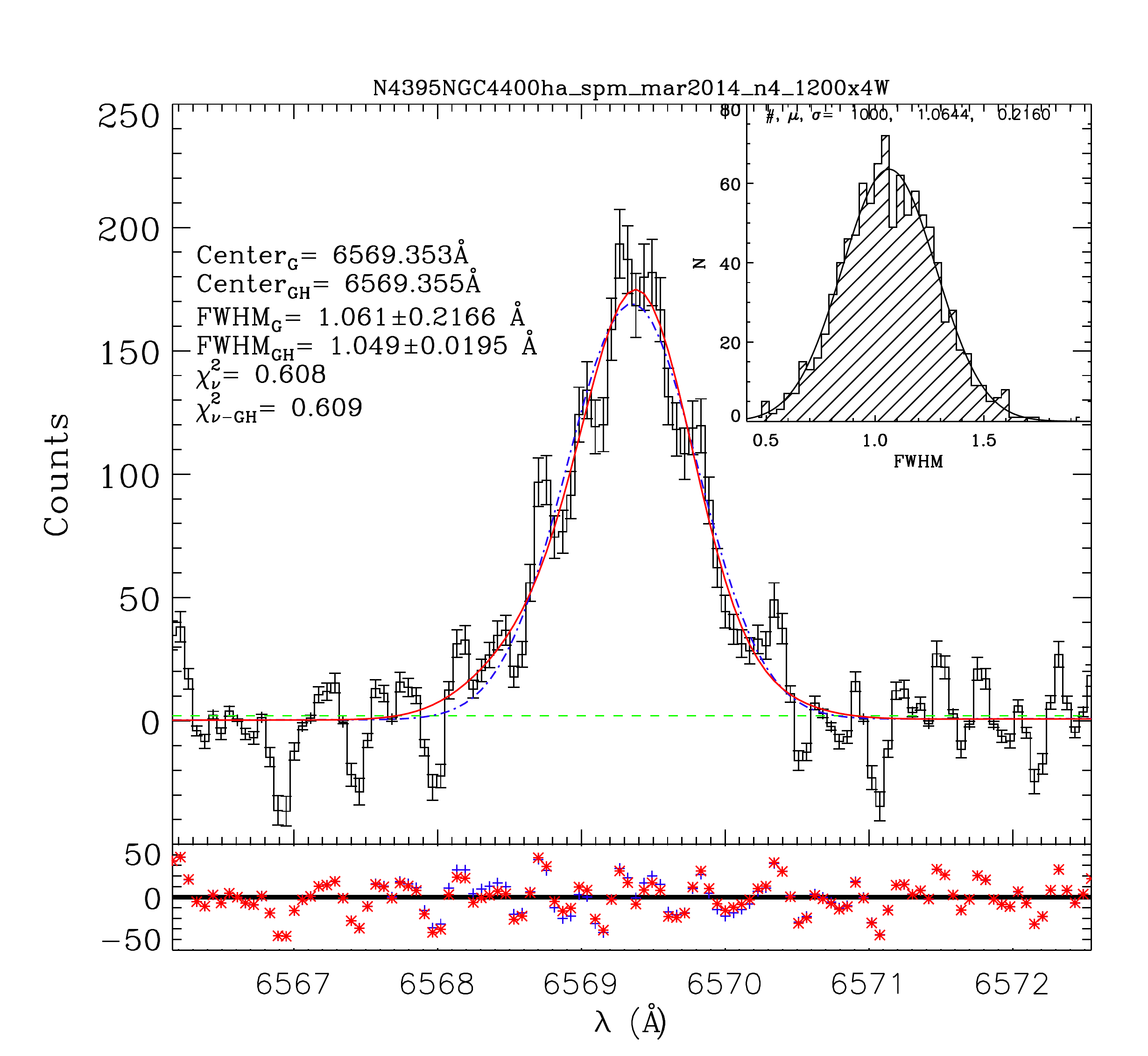}
  \includegraphics[scale=0.2]{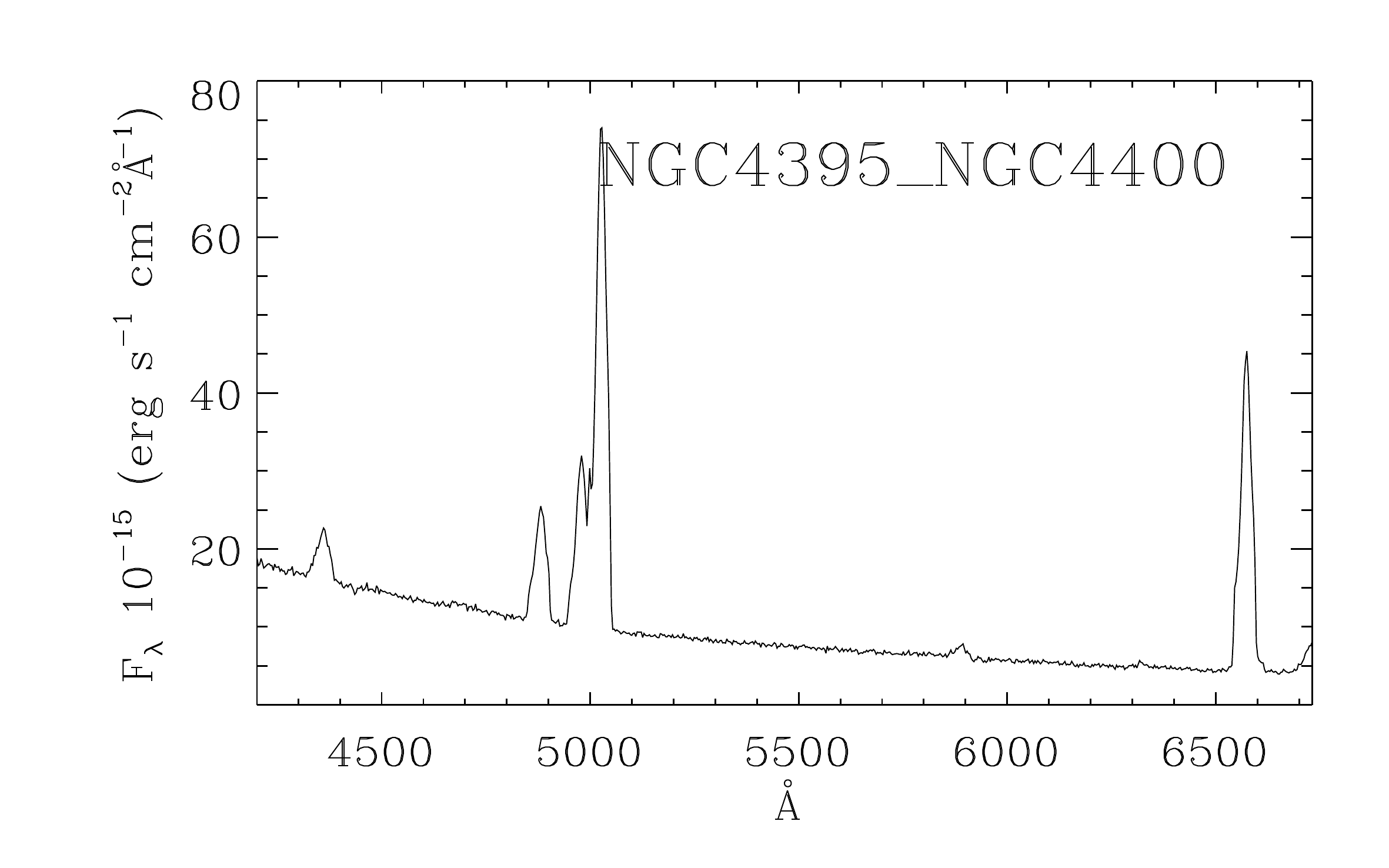}
  \caption[]{ (continued)}
 \end{figure*}

  \begin{figure*}
\includegraphics[scale=0.2]{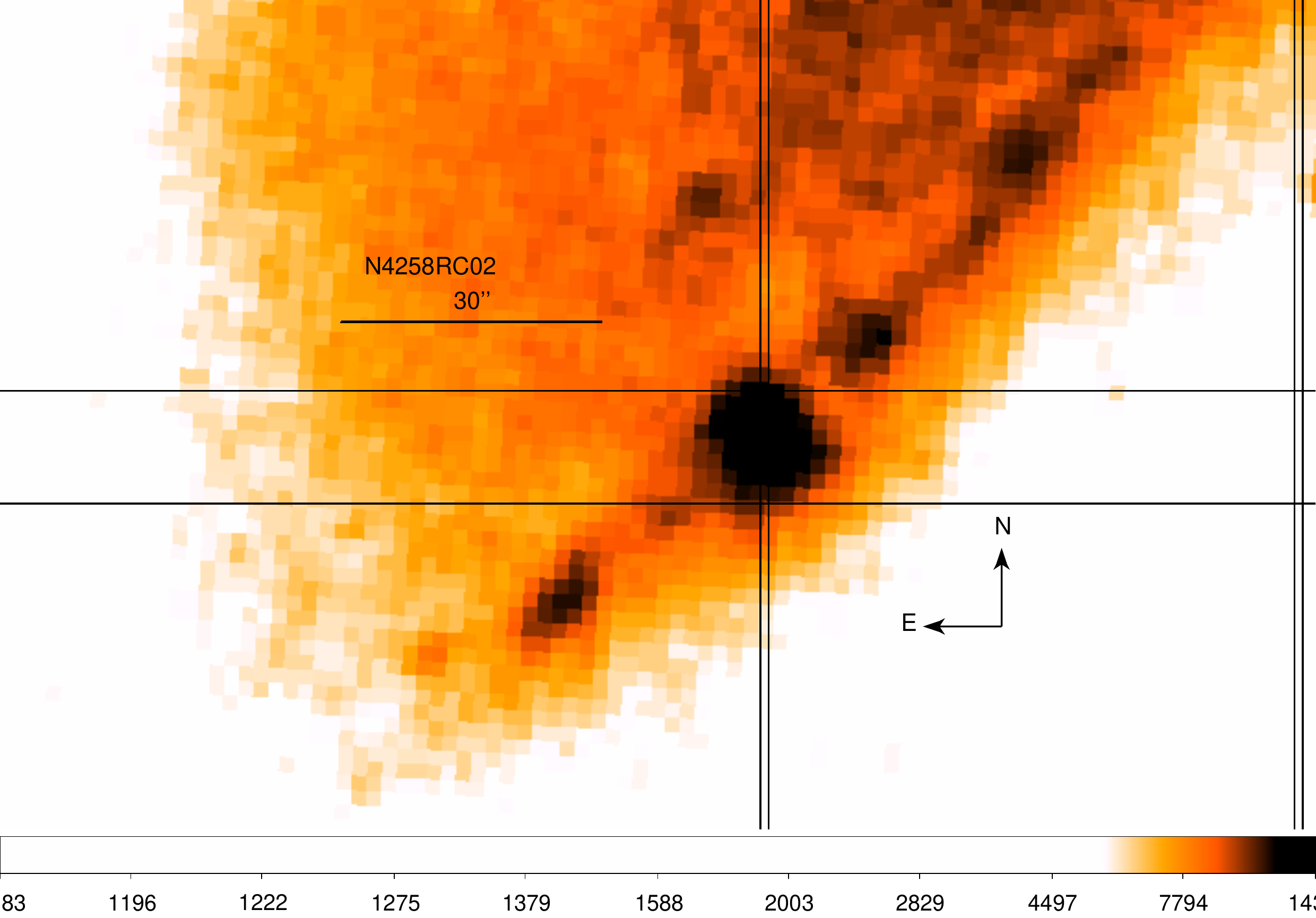}
  \includegraphics[scale=0.18]{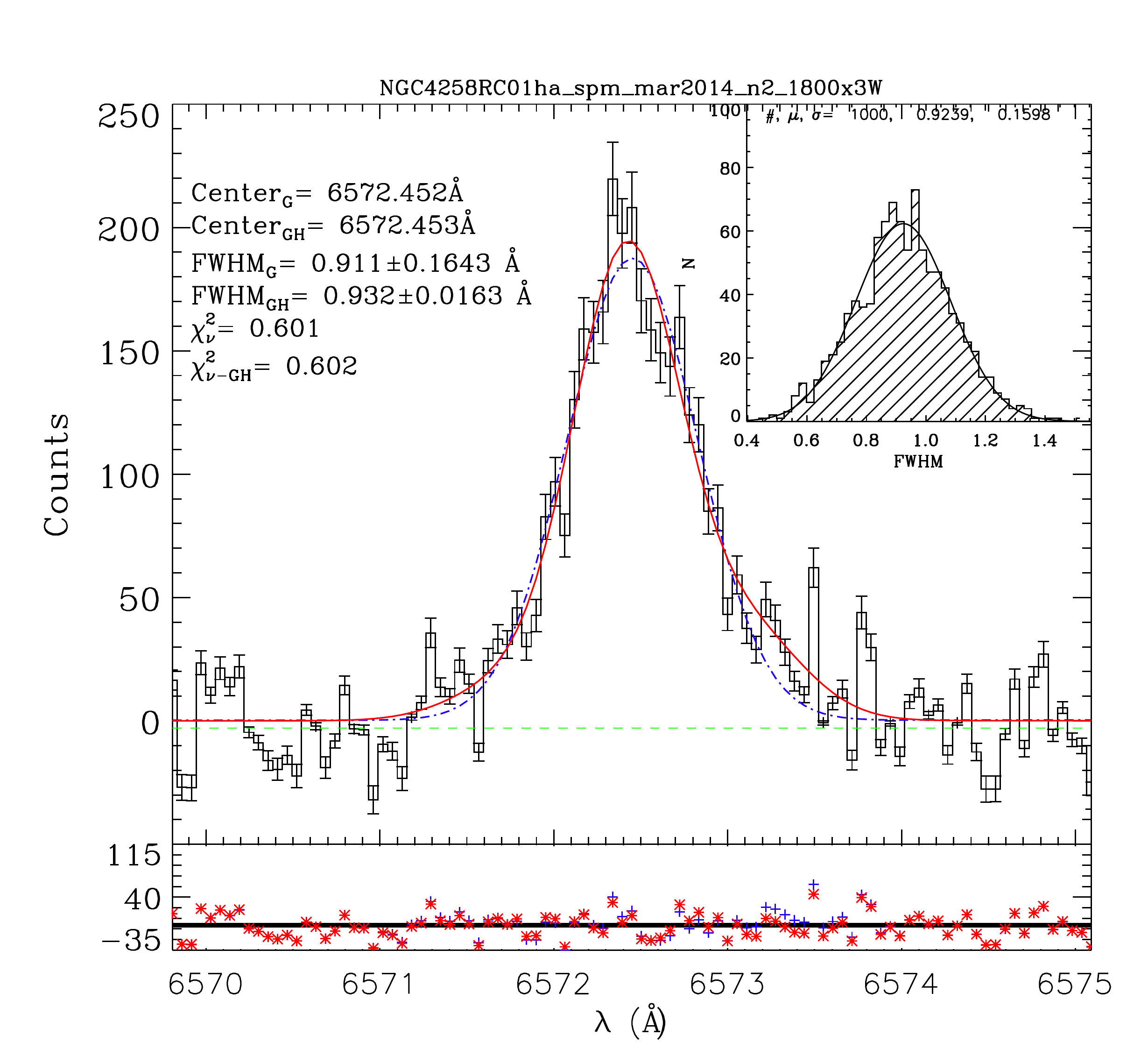}
\includegraphics[scale=0.2]{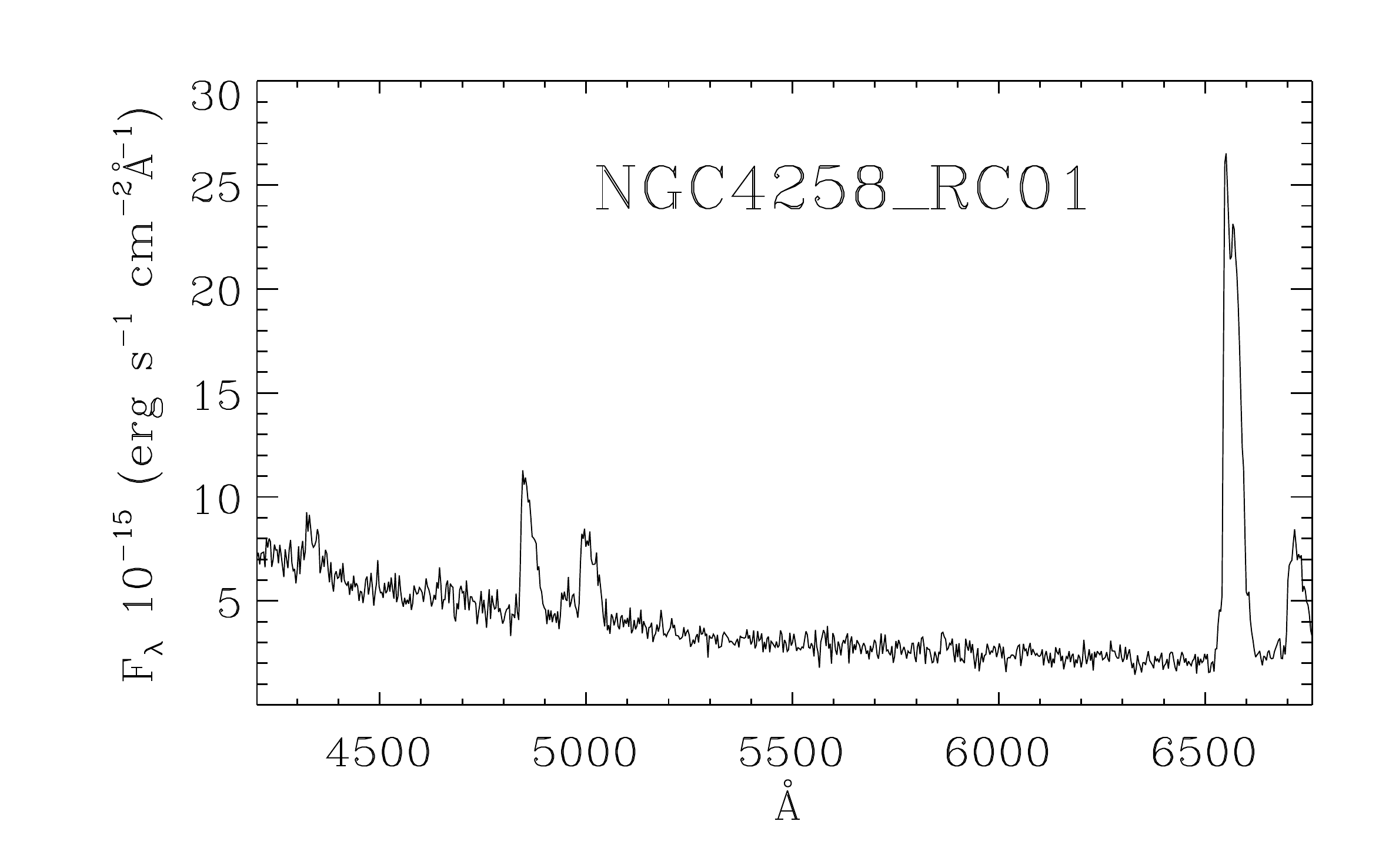}

\includegraphics[scale=0.2]{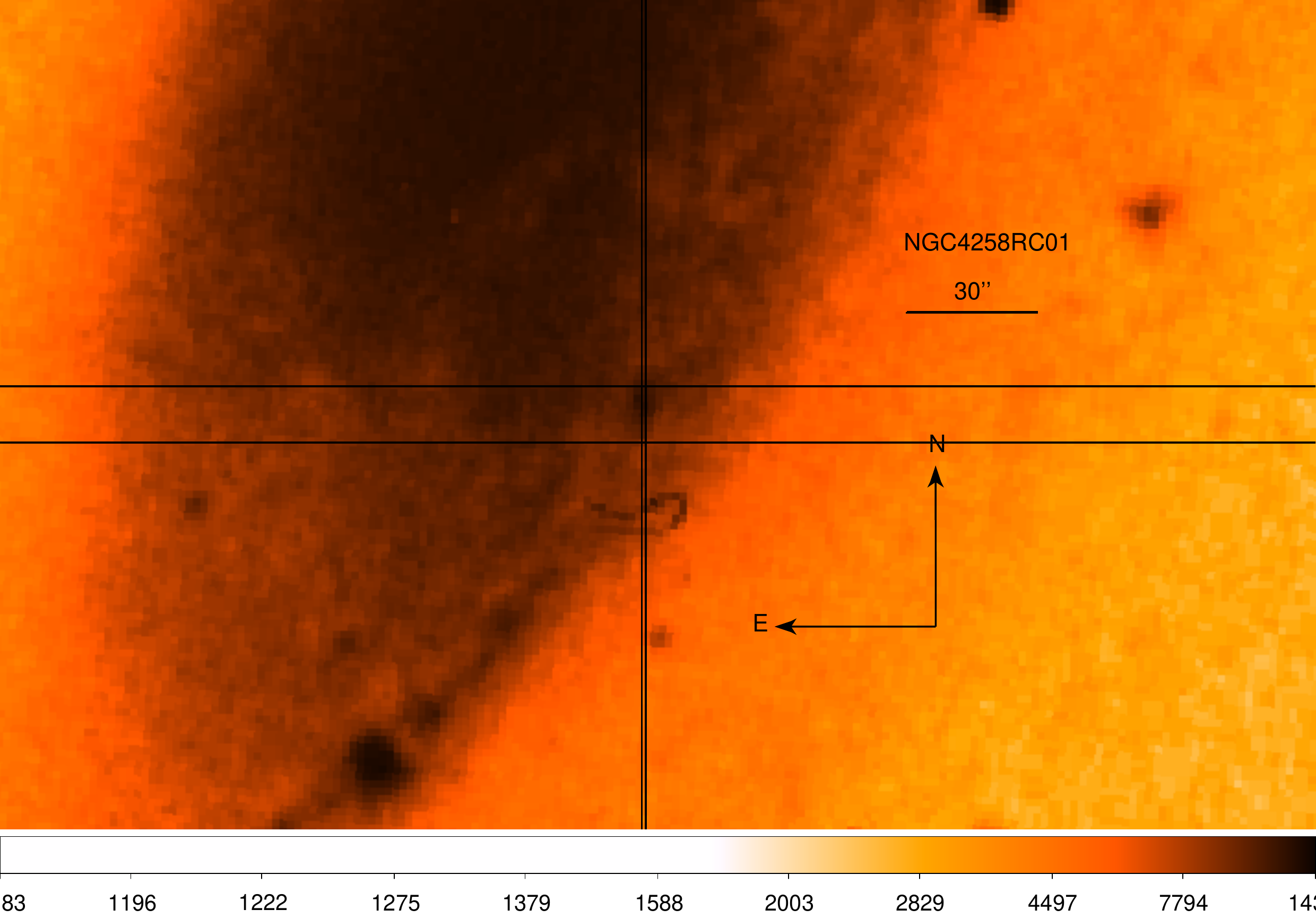}
  \includegraphics[scale=0.18]{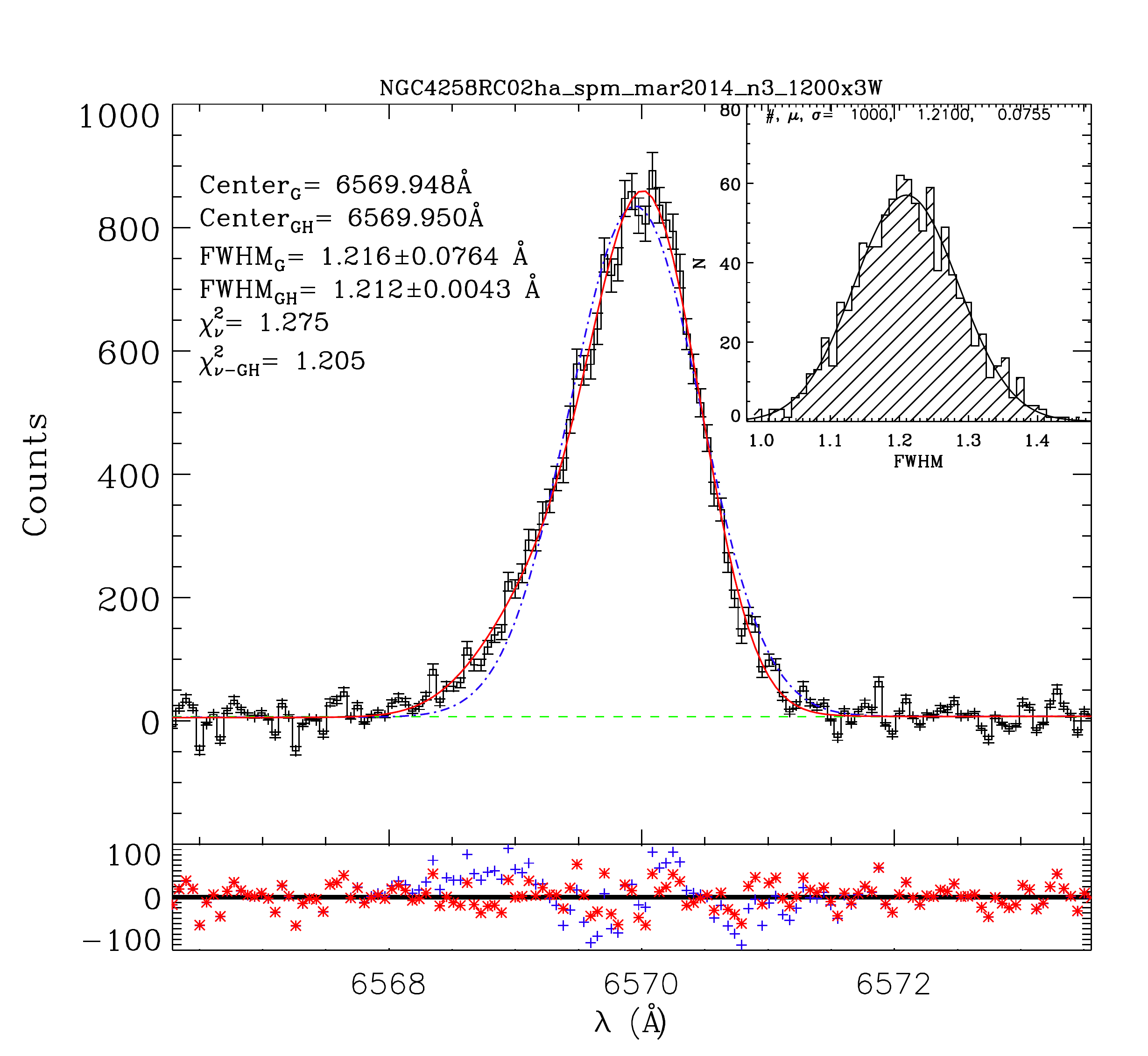}
   \includegraphics[scale=0.2]{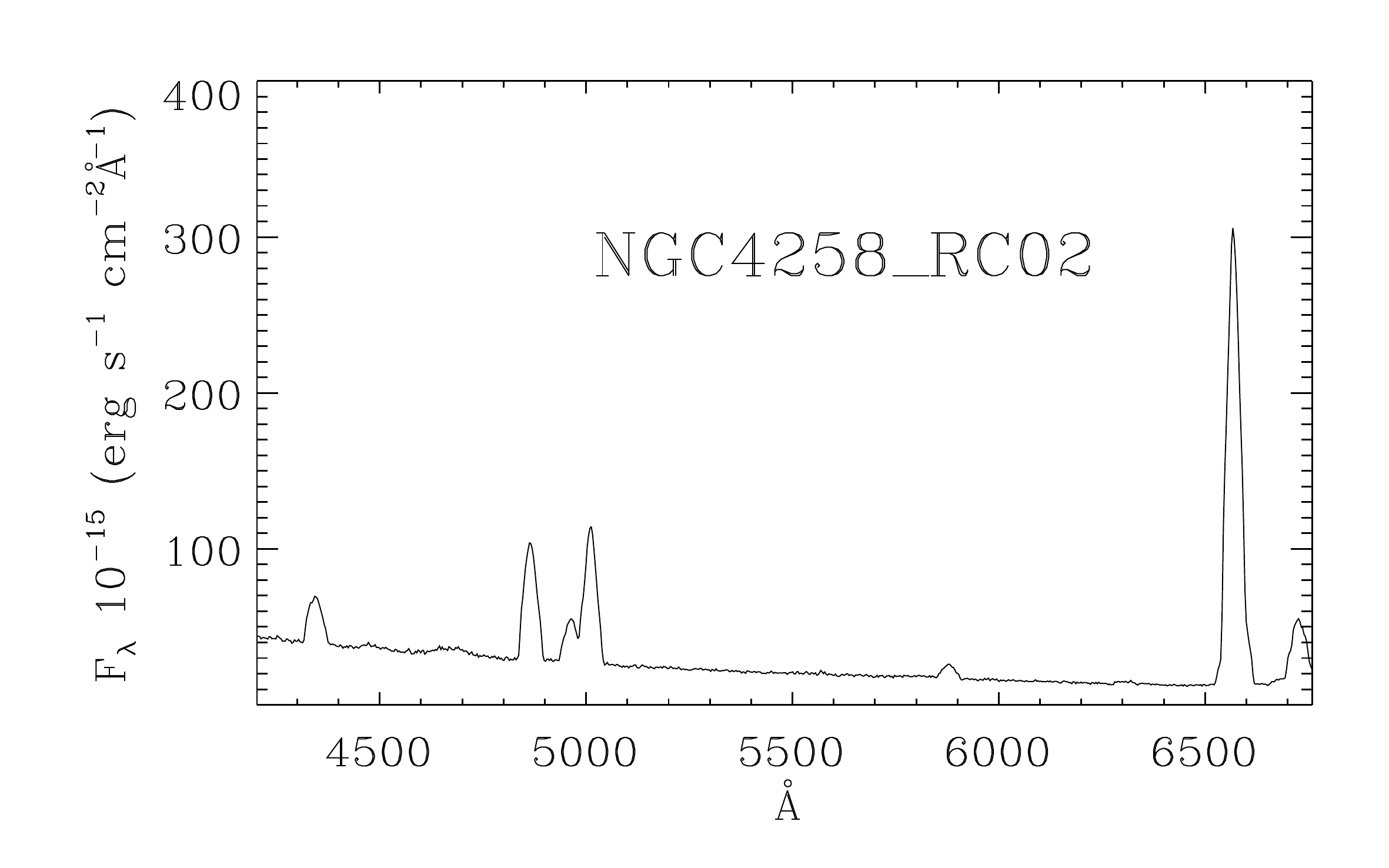}
  
\includegraphics[scale=0.2]{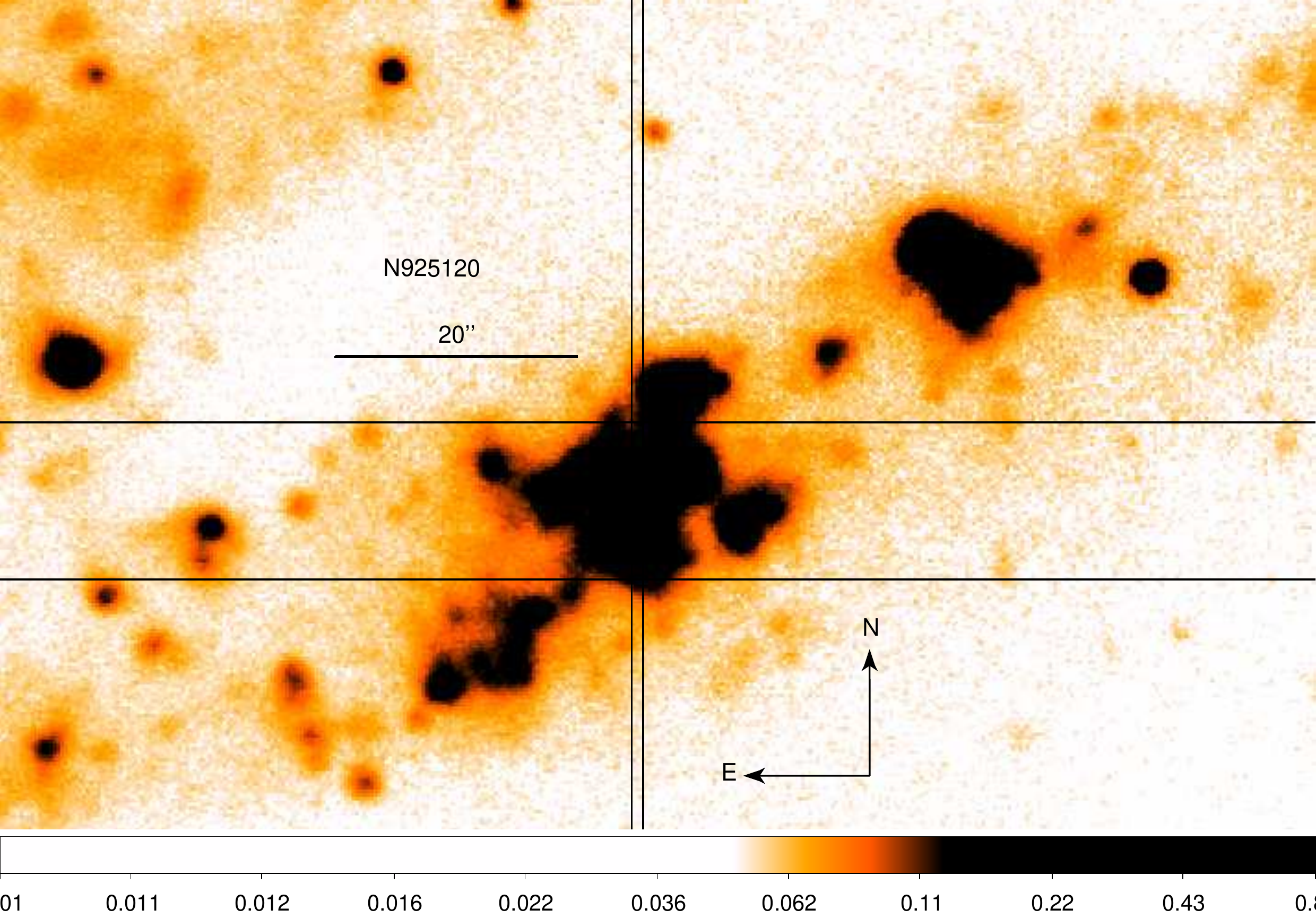}
  \includegraphics[scale=0.18]{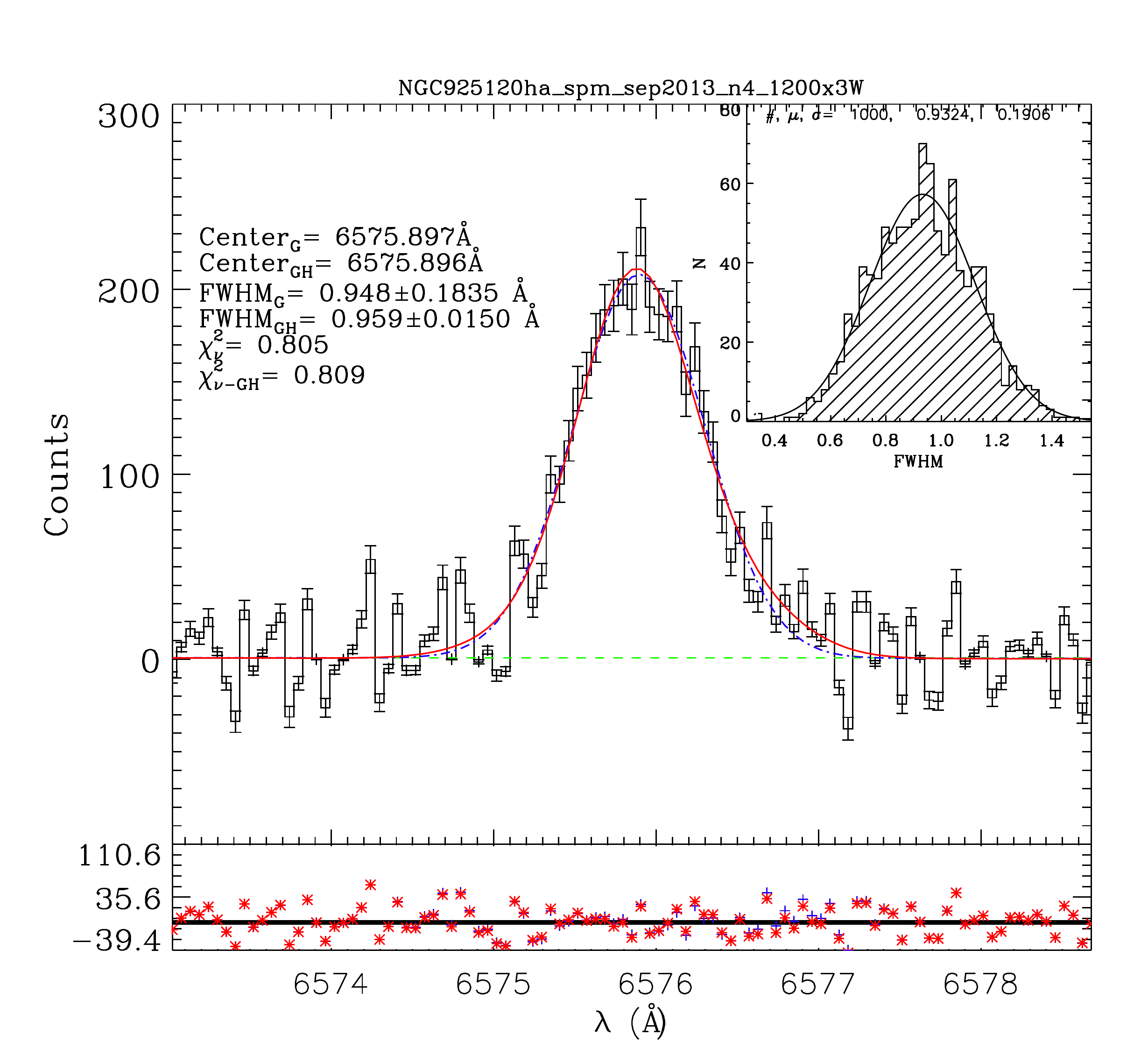}
   \includegraphics[scale=0.2]{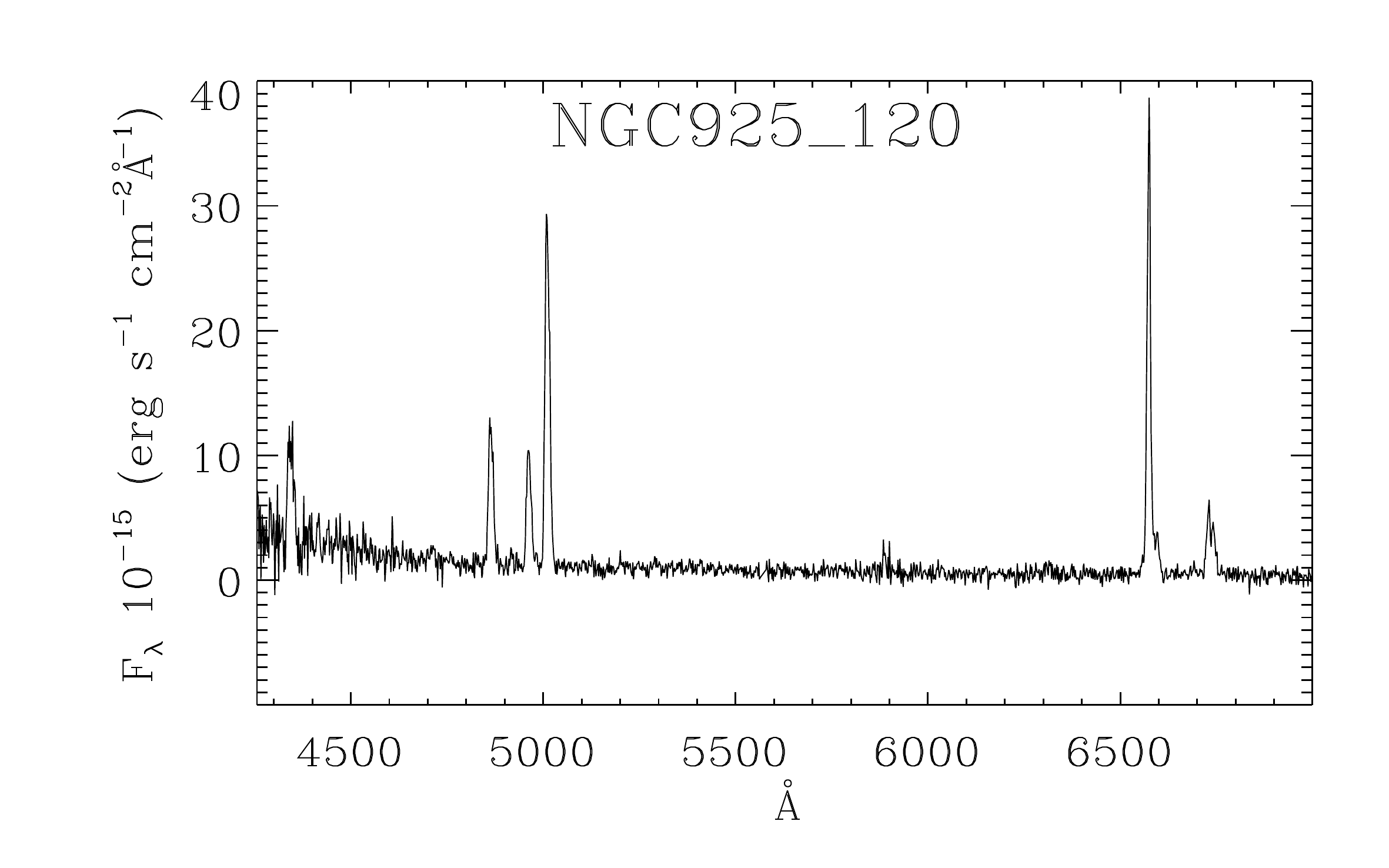}

   \includegraphics[scale=0.2]{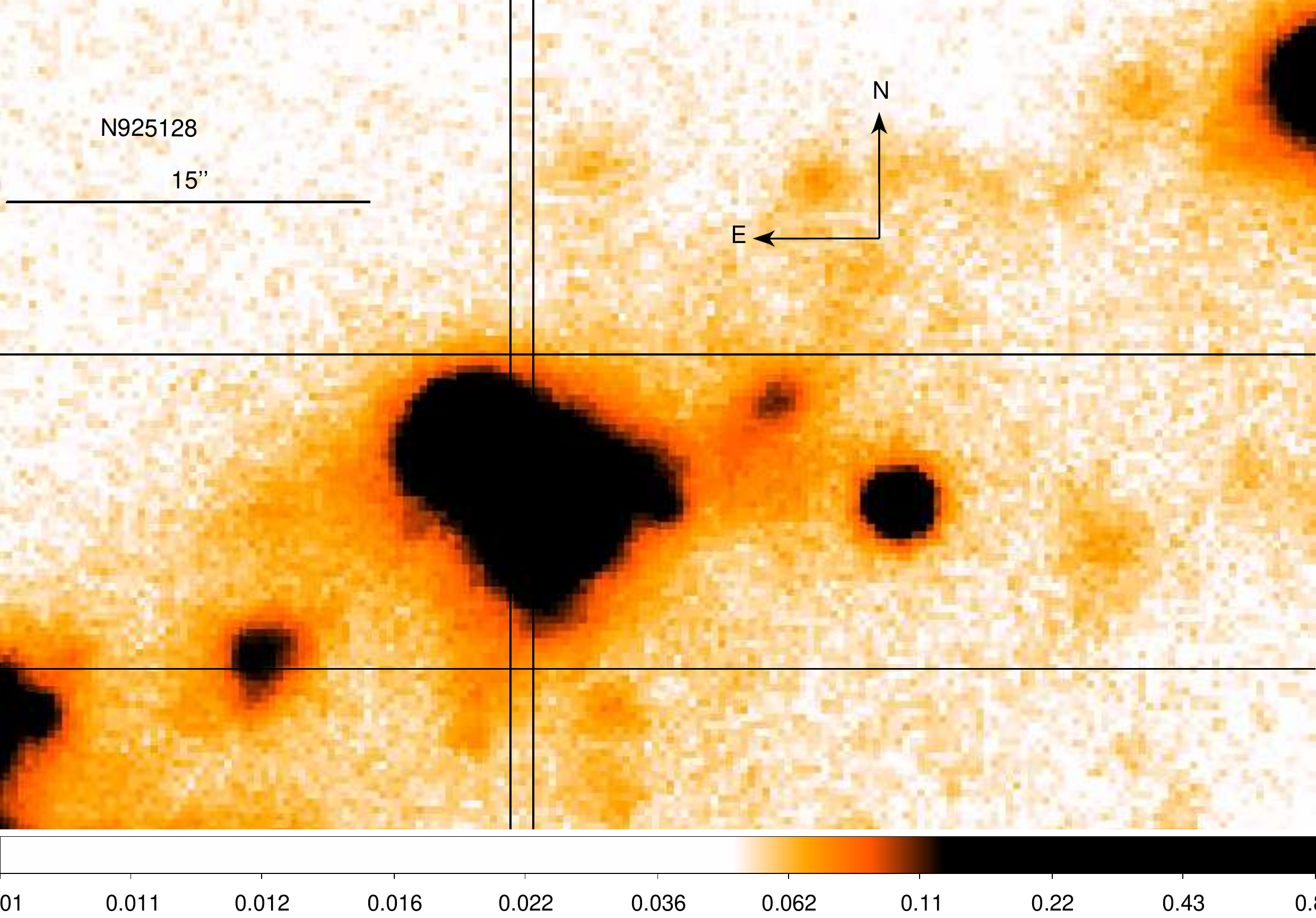}
  \includegraphics[scale=0.18]{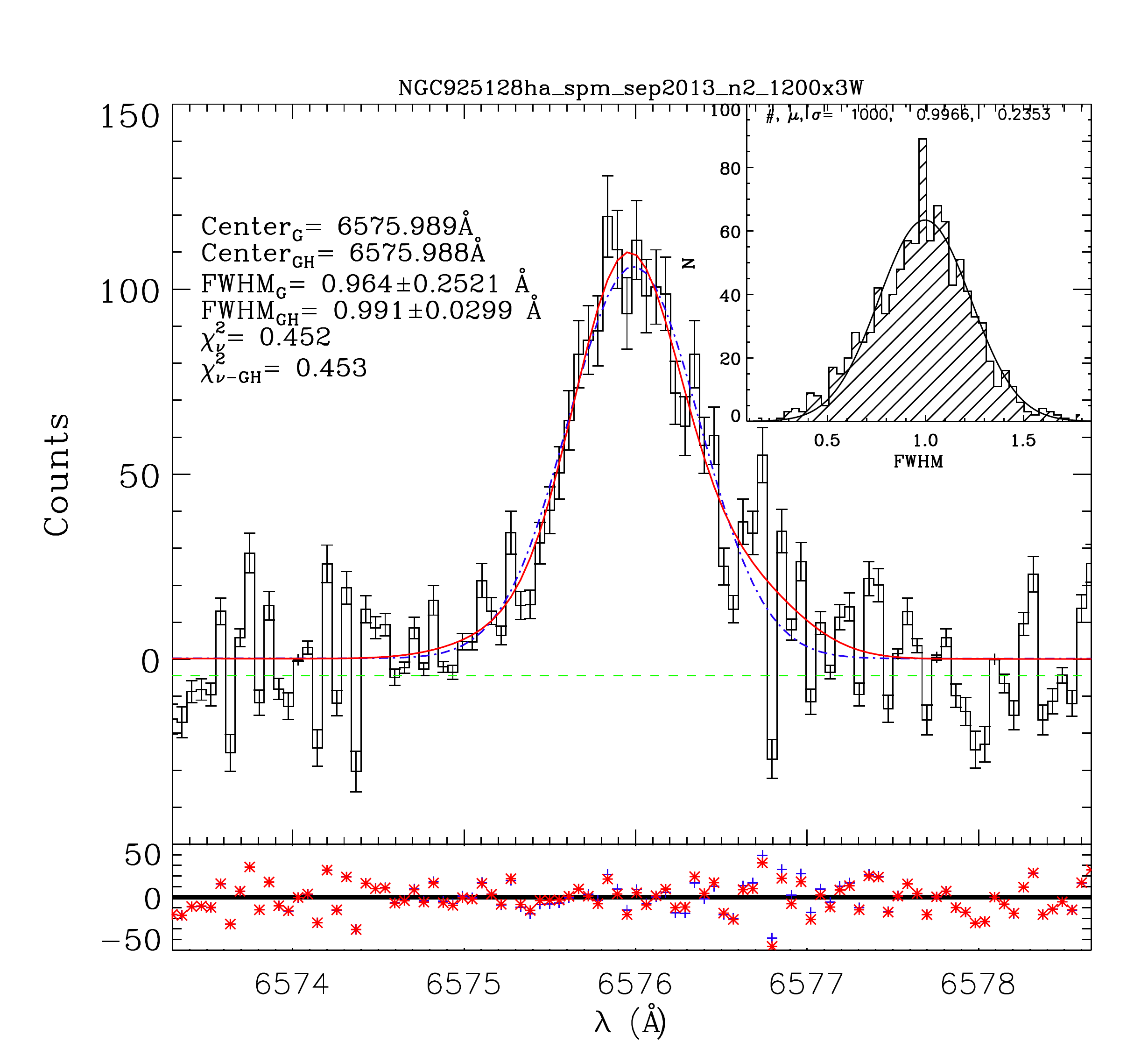}
  \includegraphics[scale=0.2]{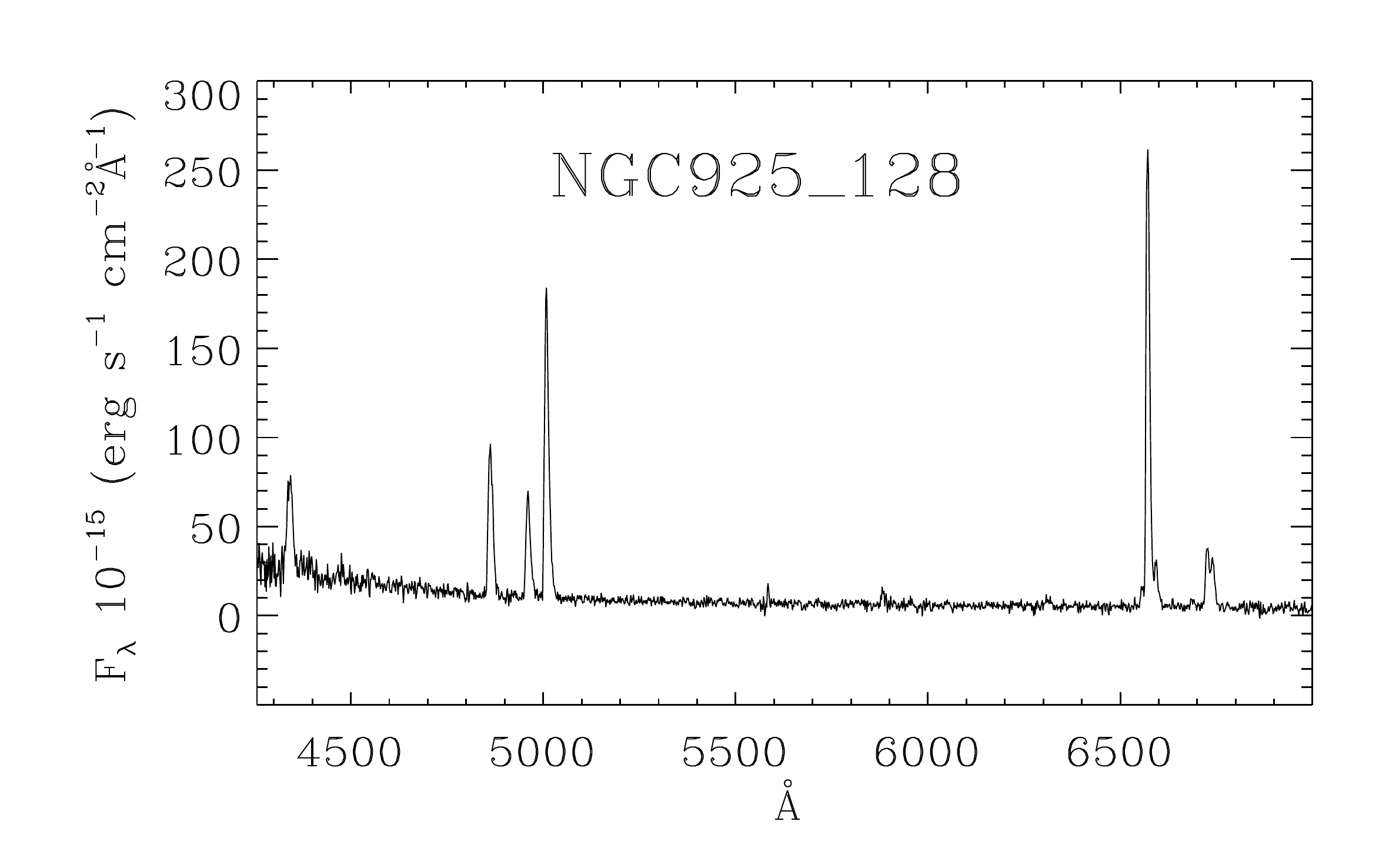}
  
   \includegraphics[scale=0.2]{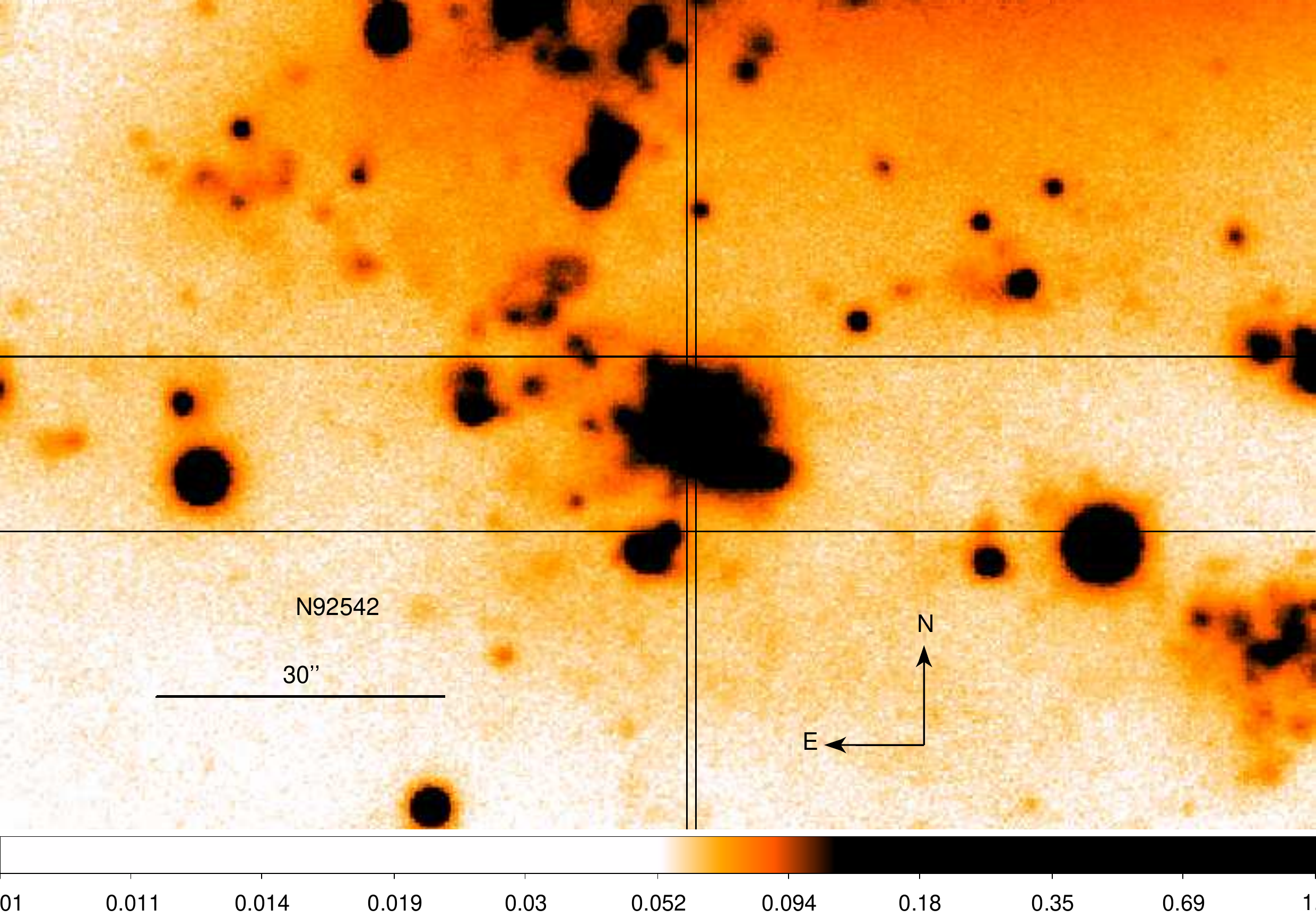}
  \includegraphics[scale=0.18]{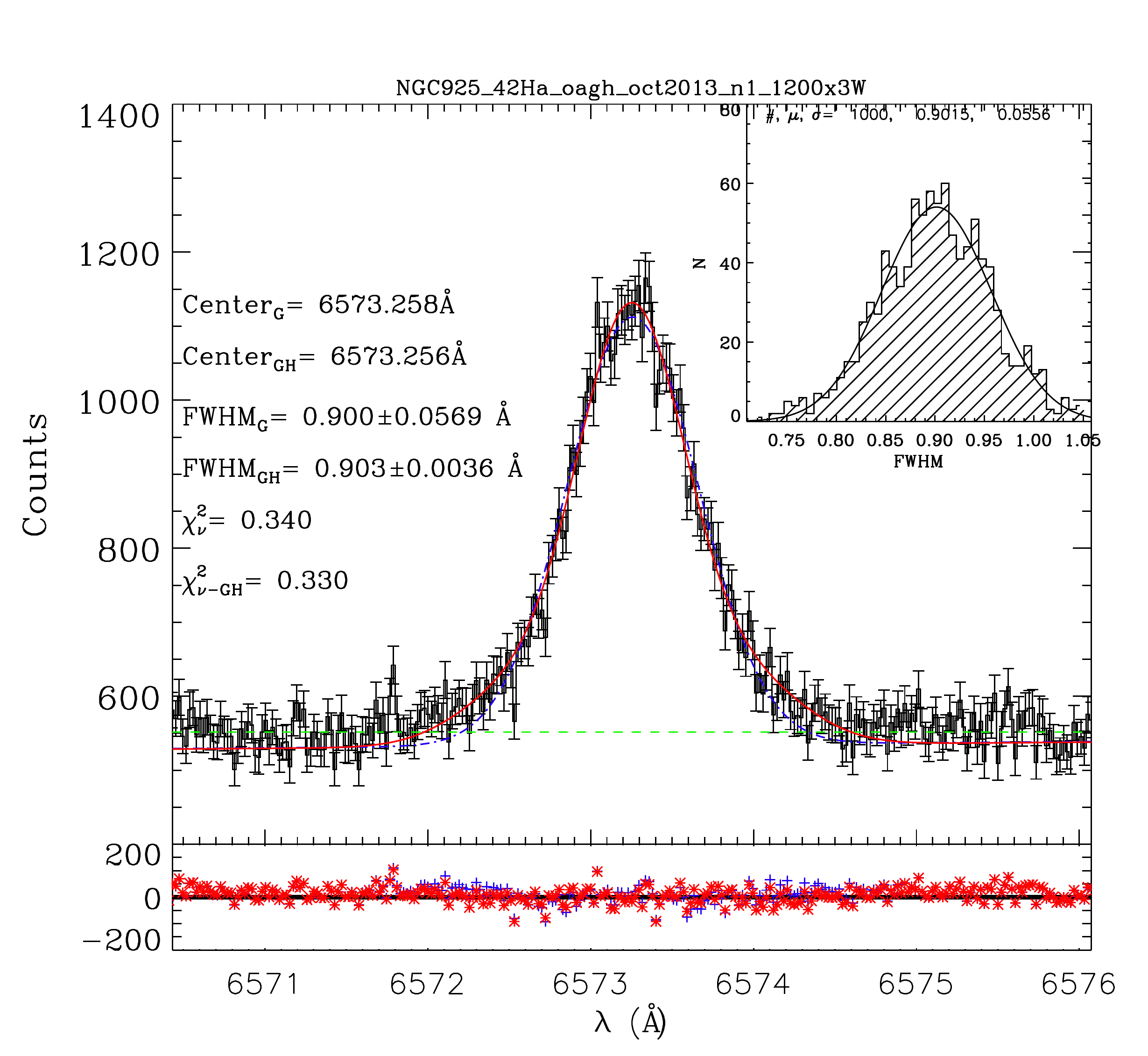}
  \includegraphics[scale=0.2]{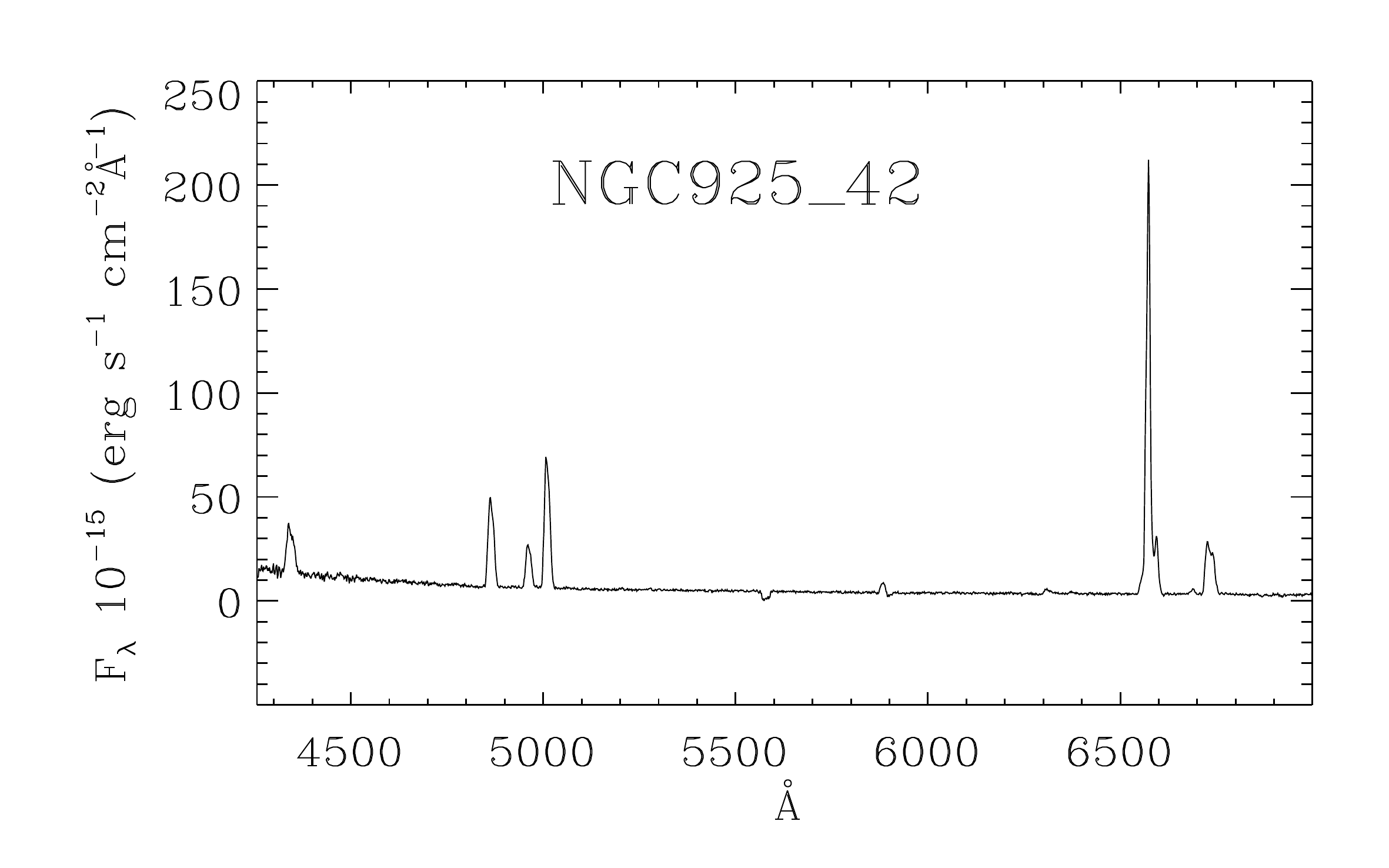}
  
  \caption[]{ (continued)}
 
  \end{figure*}

  \begin{figure*}

  \includegraphics[scale=0.2]{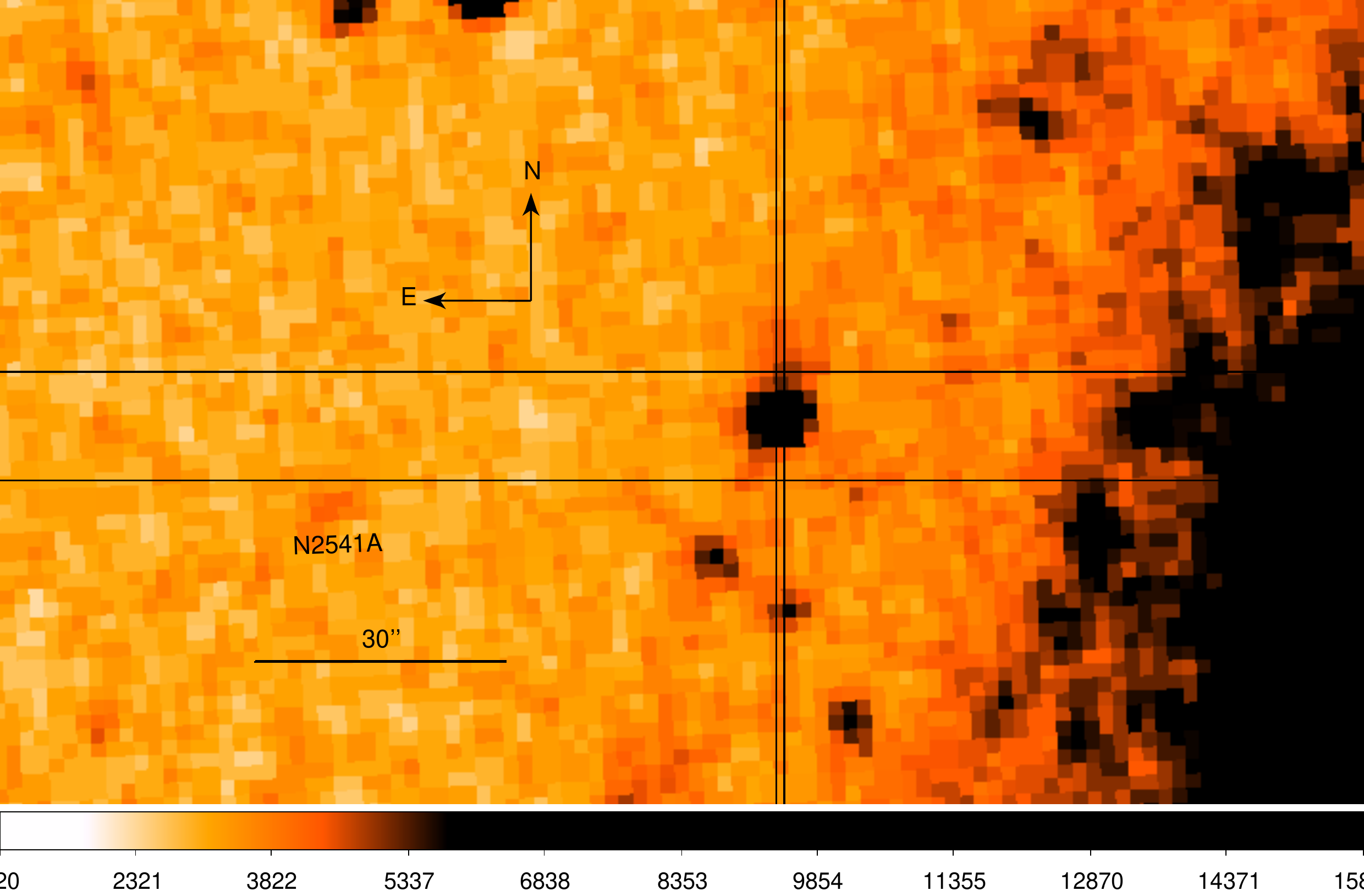}
  \includegraphics[scale=0.18]{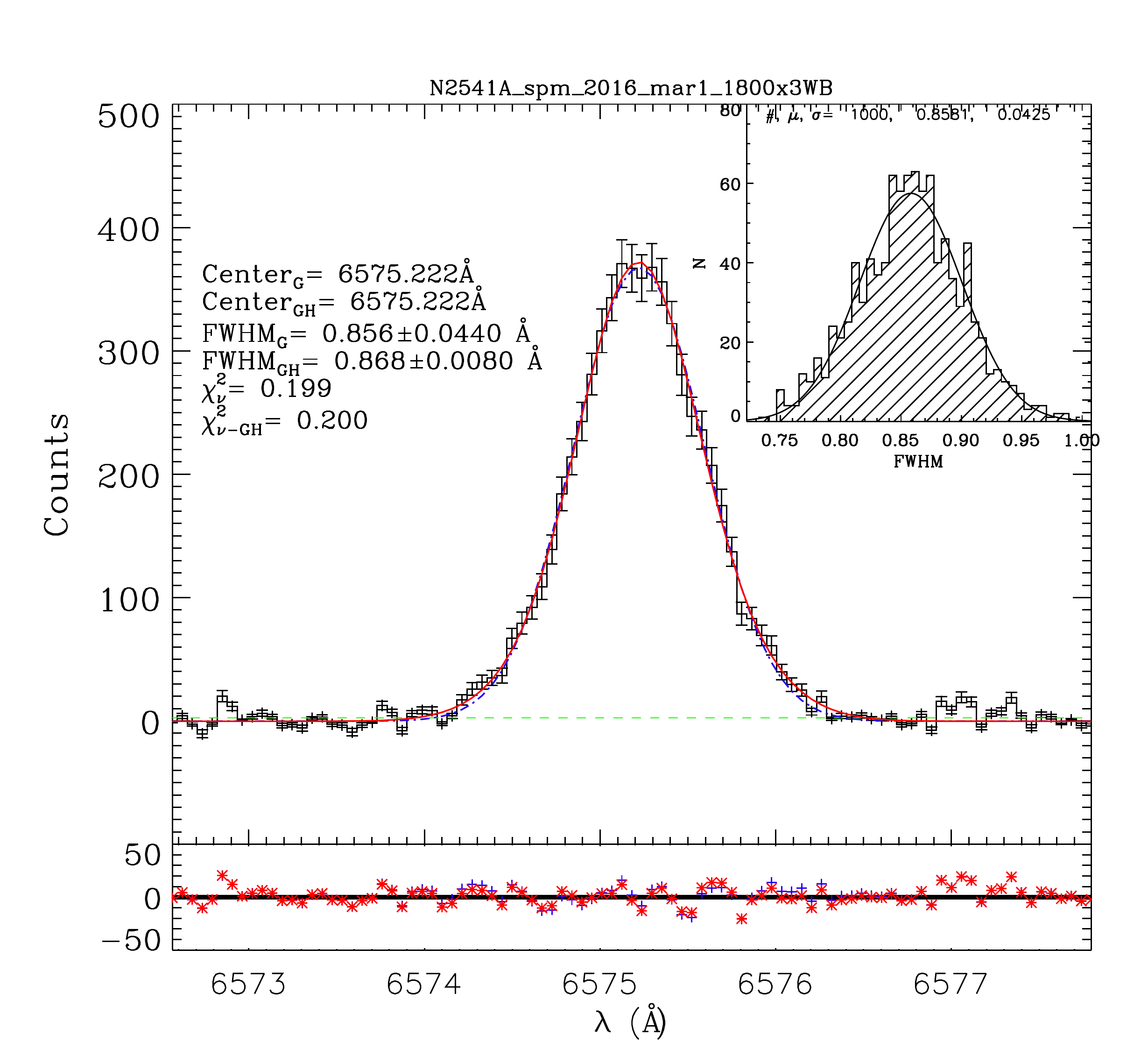}
  \includegraphics[scale=0.2]{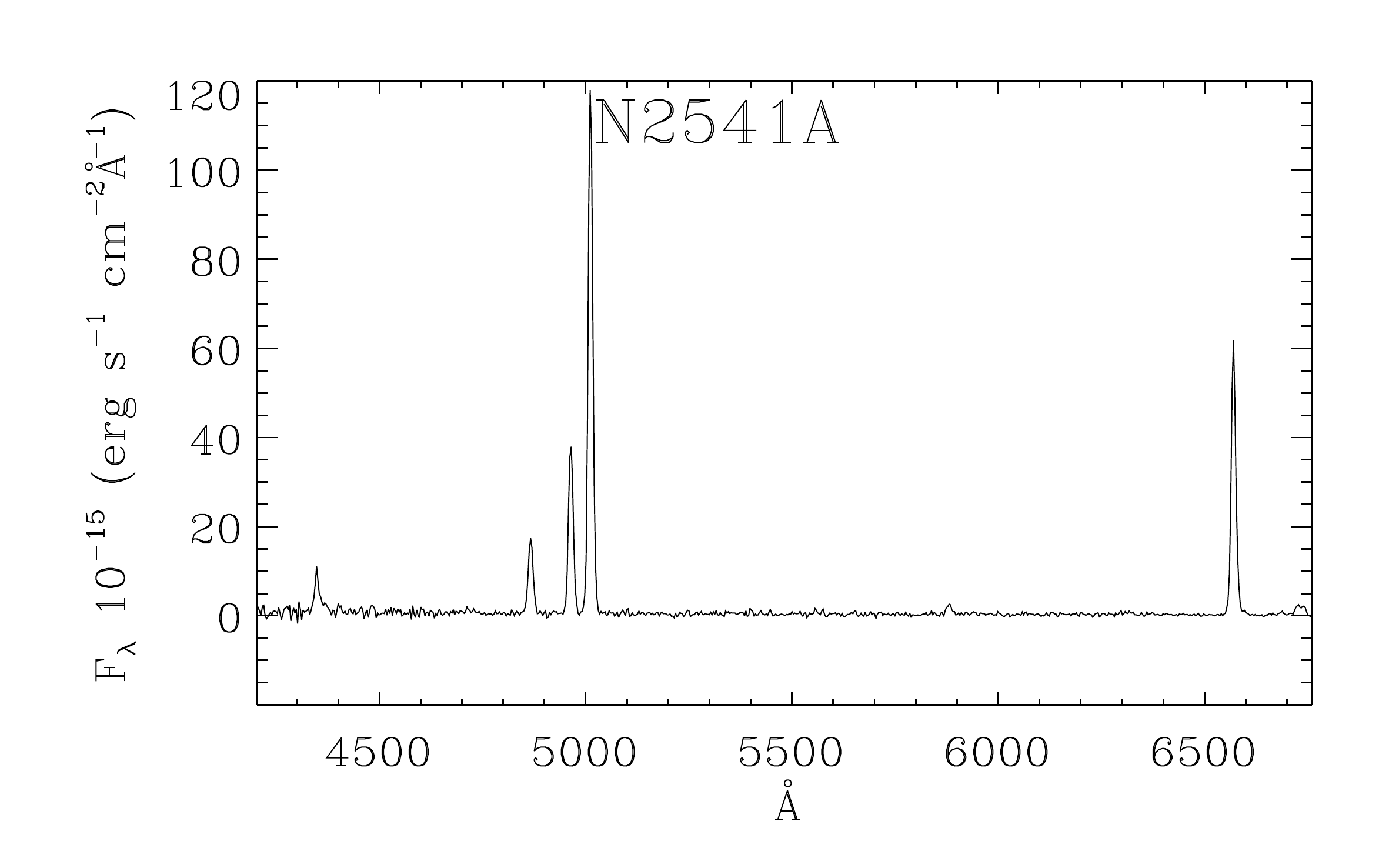}

  \includegraphics[scale=0.2]{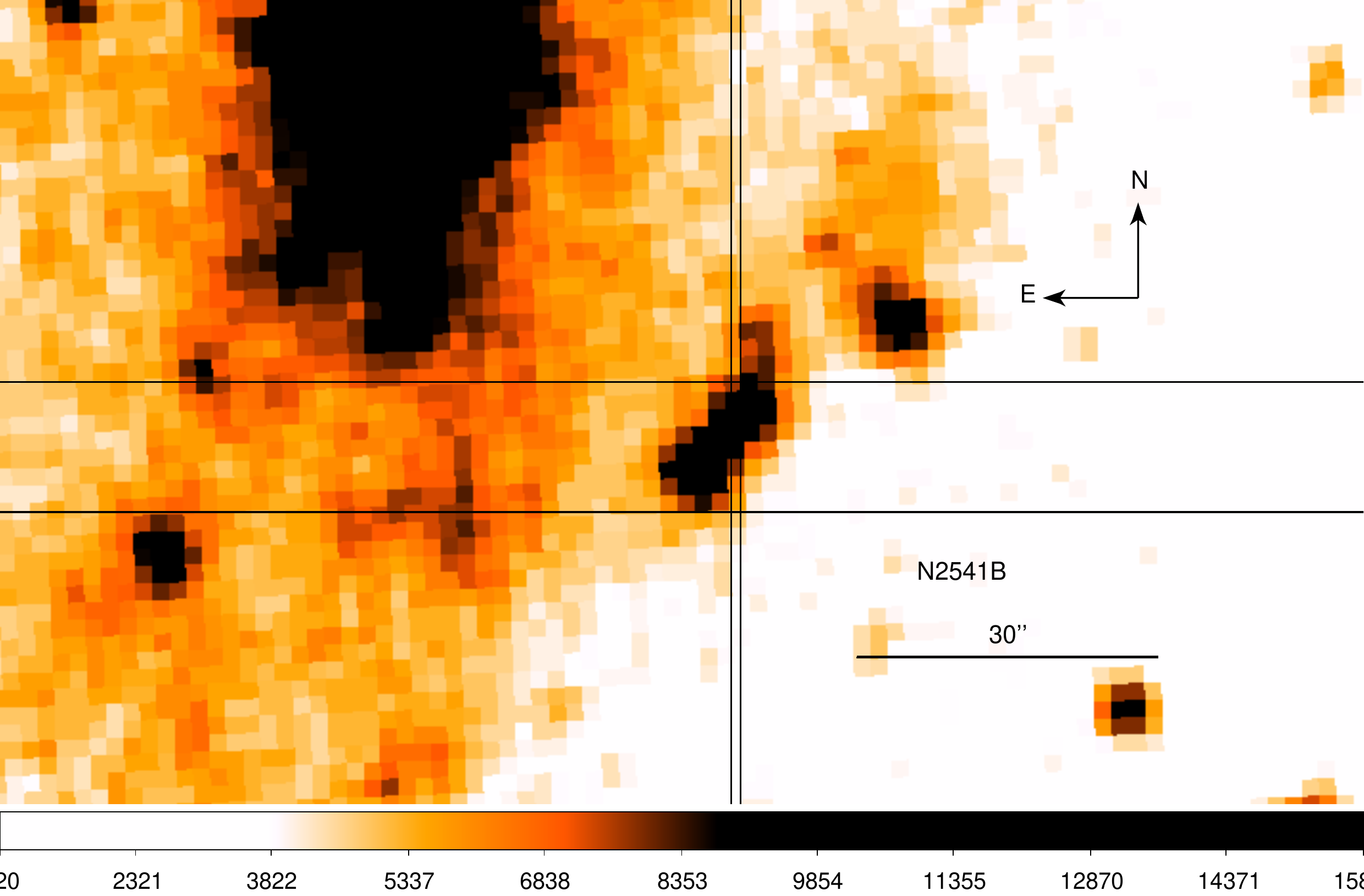}
  \includegraphics[scale=0.18]{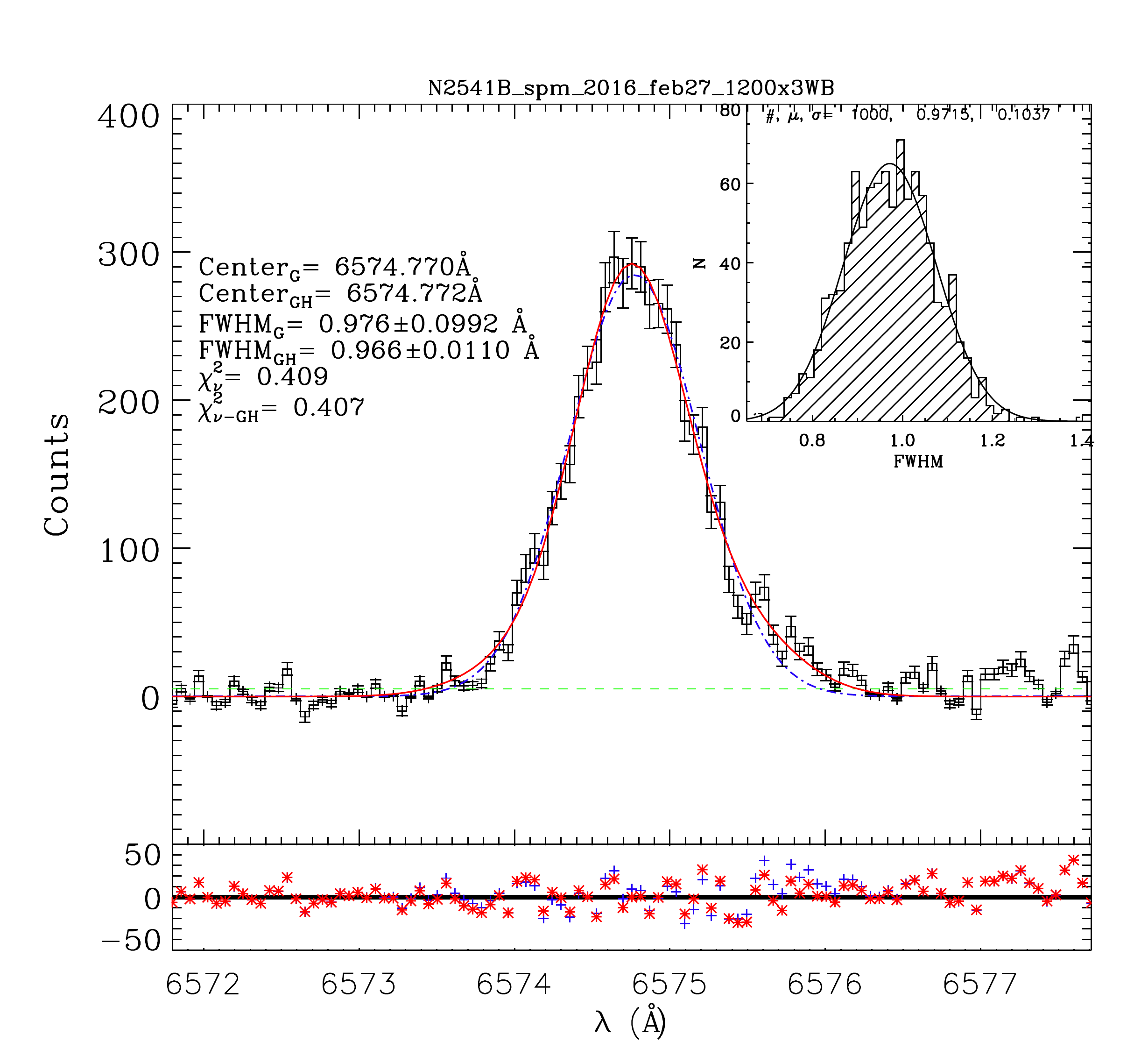}
  \includegraphics[scale=0.2]{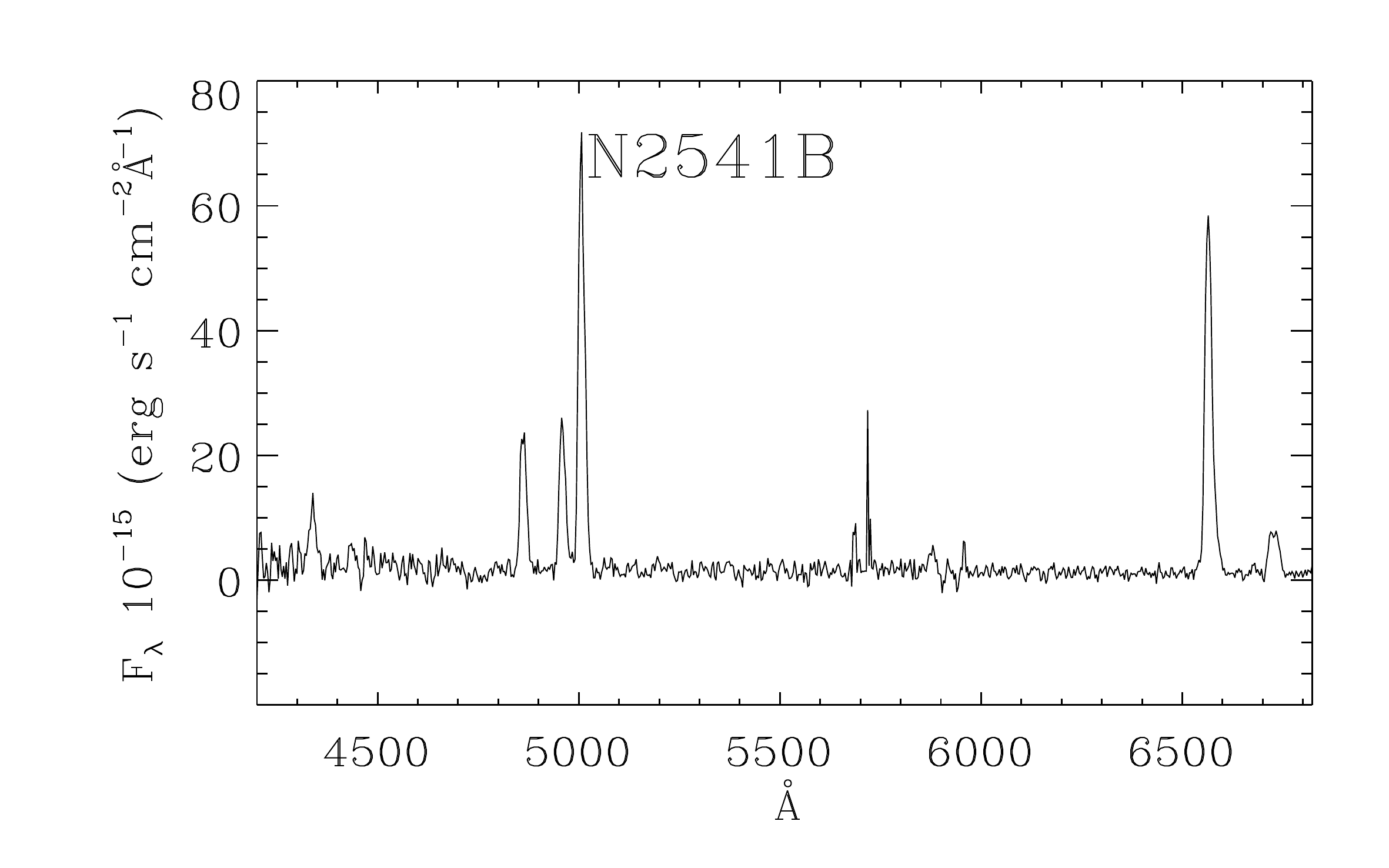}
  
  \includegraphics[scale=0.2]{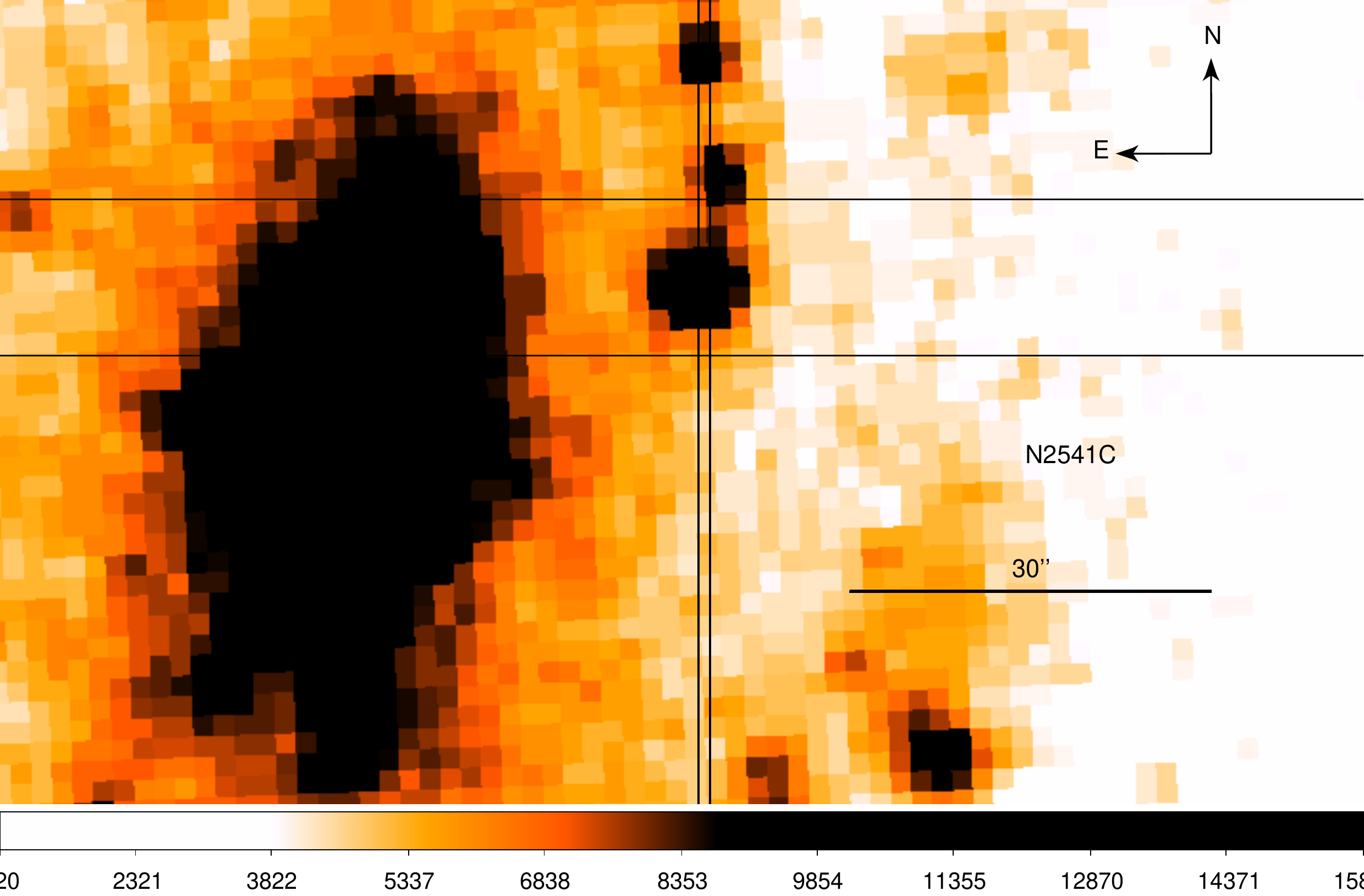}
  \includegraphics[scale=0.18]{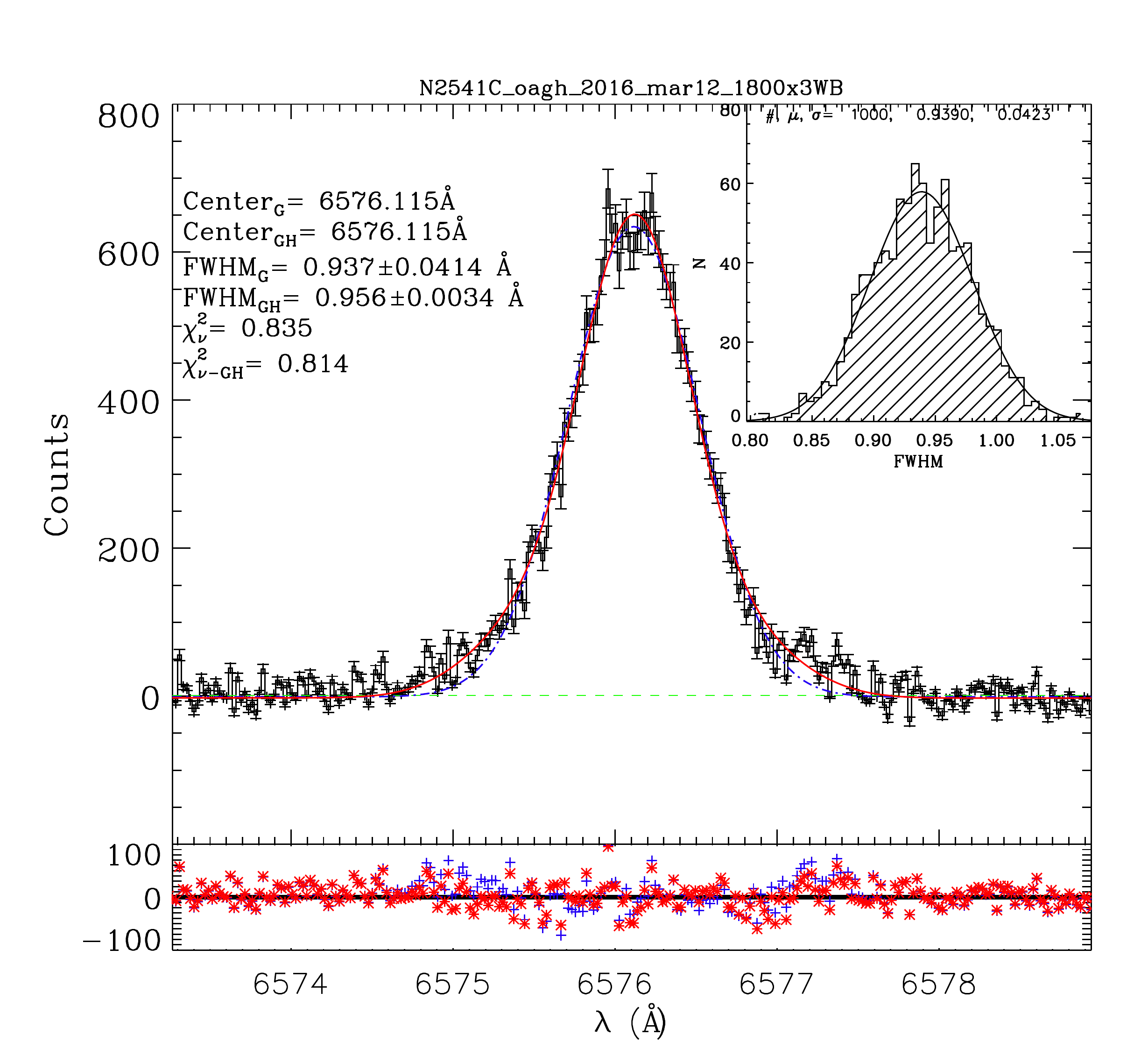}
  \includegraphics[scale=0.2]{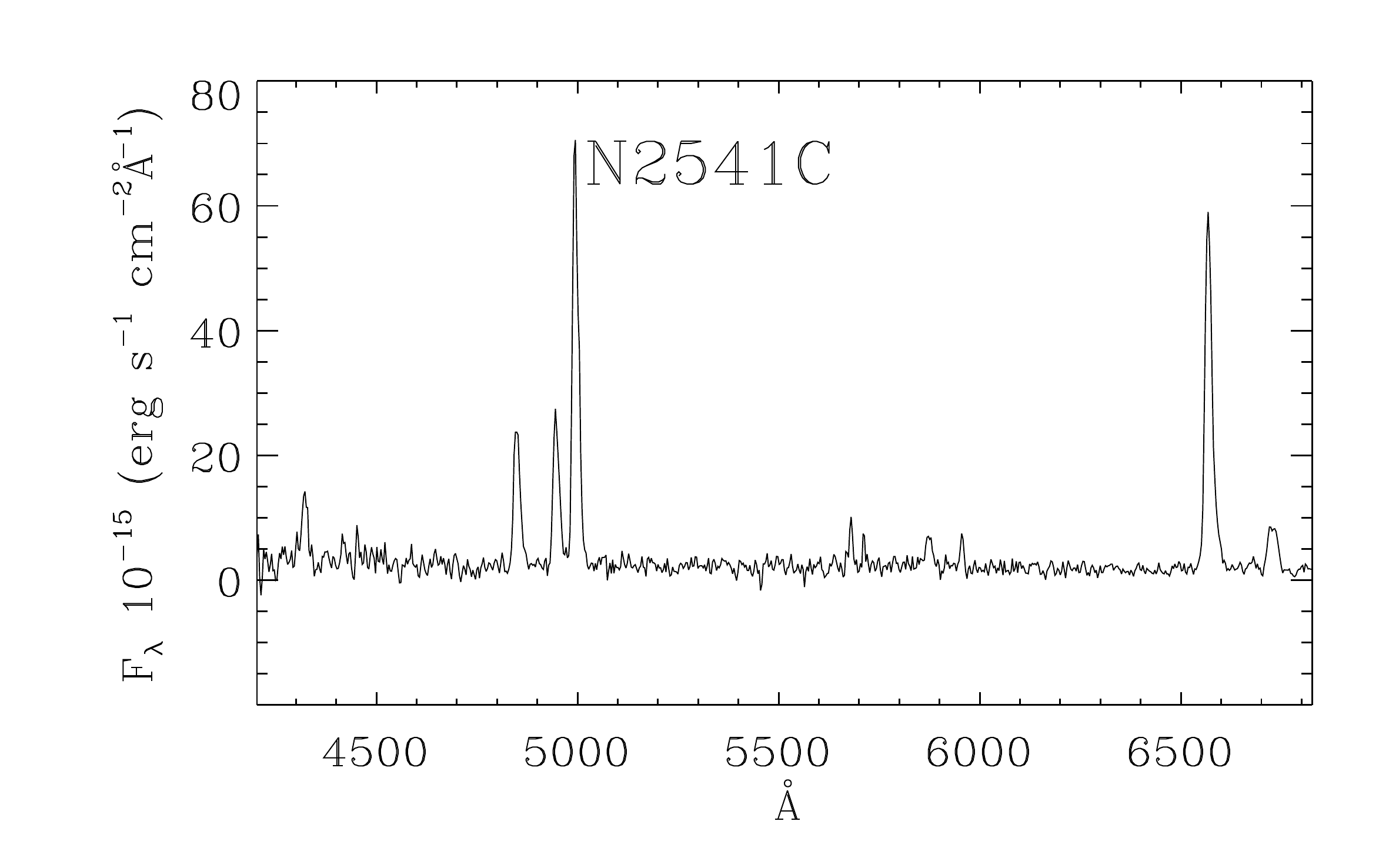}

\includegraphics[scale=0.2]{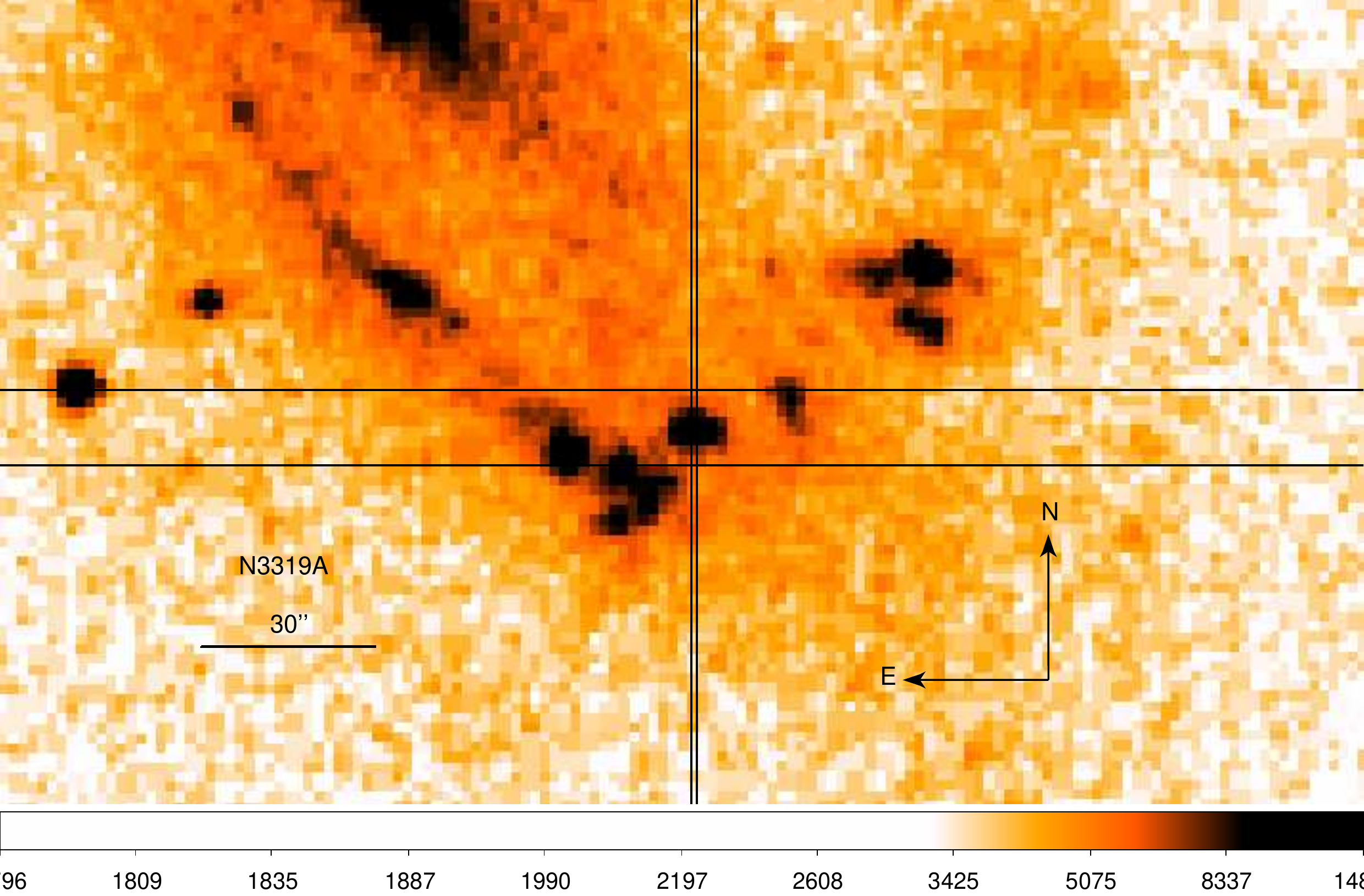}
  \includegraphics[scale=0.18]{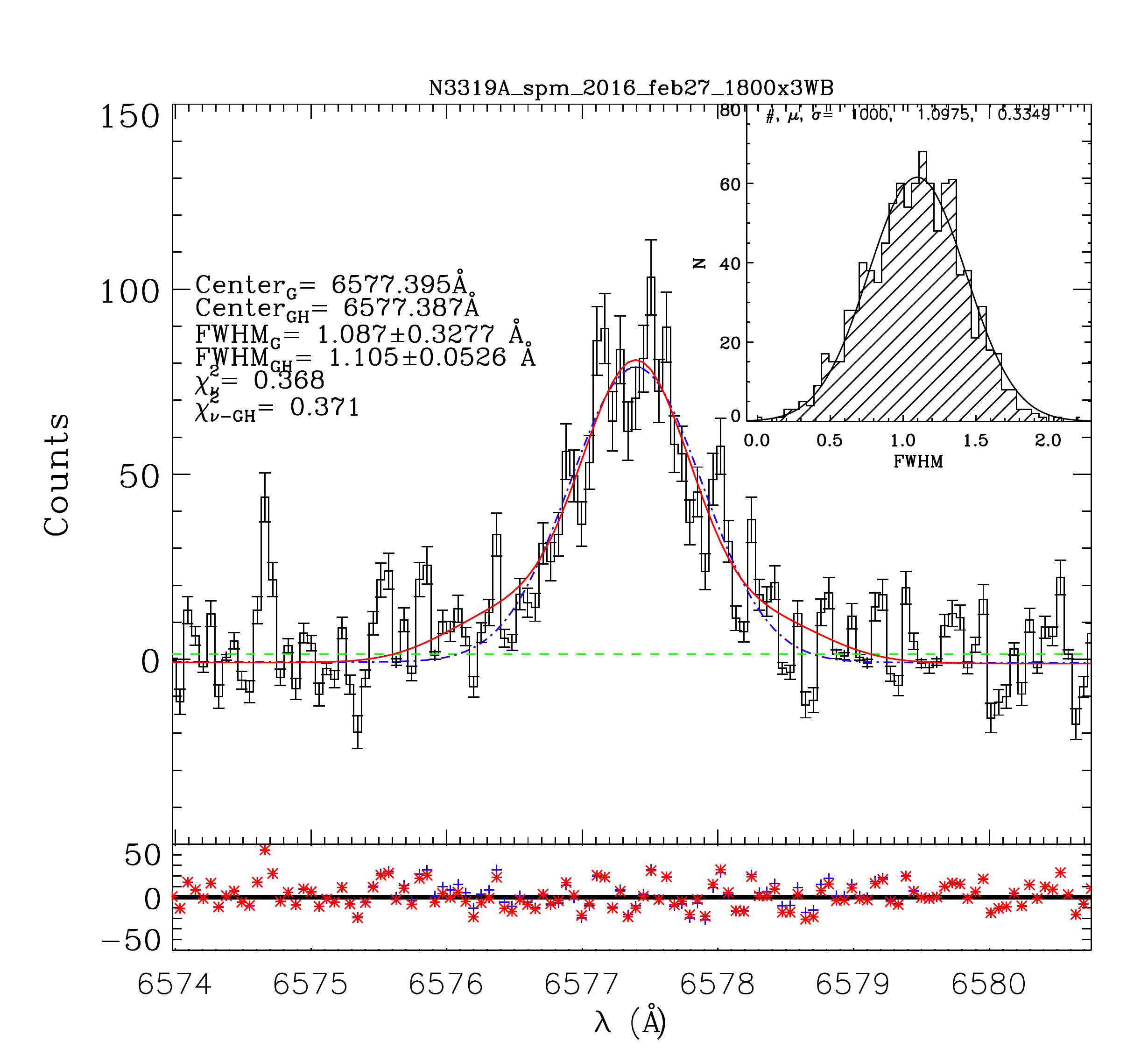}
  \includegraphics[scale=0.2]{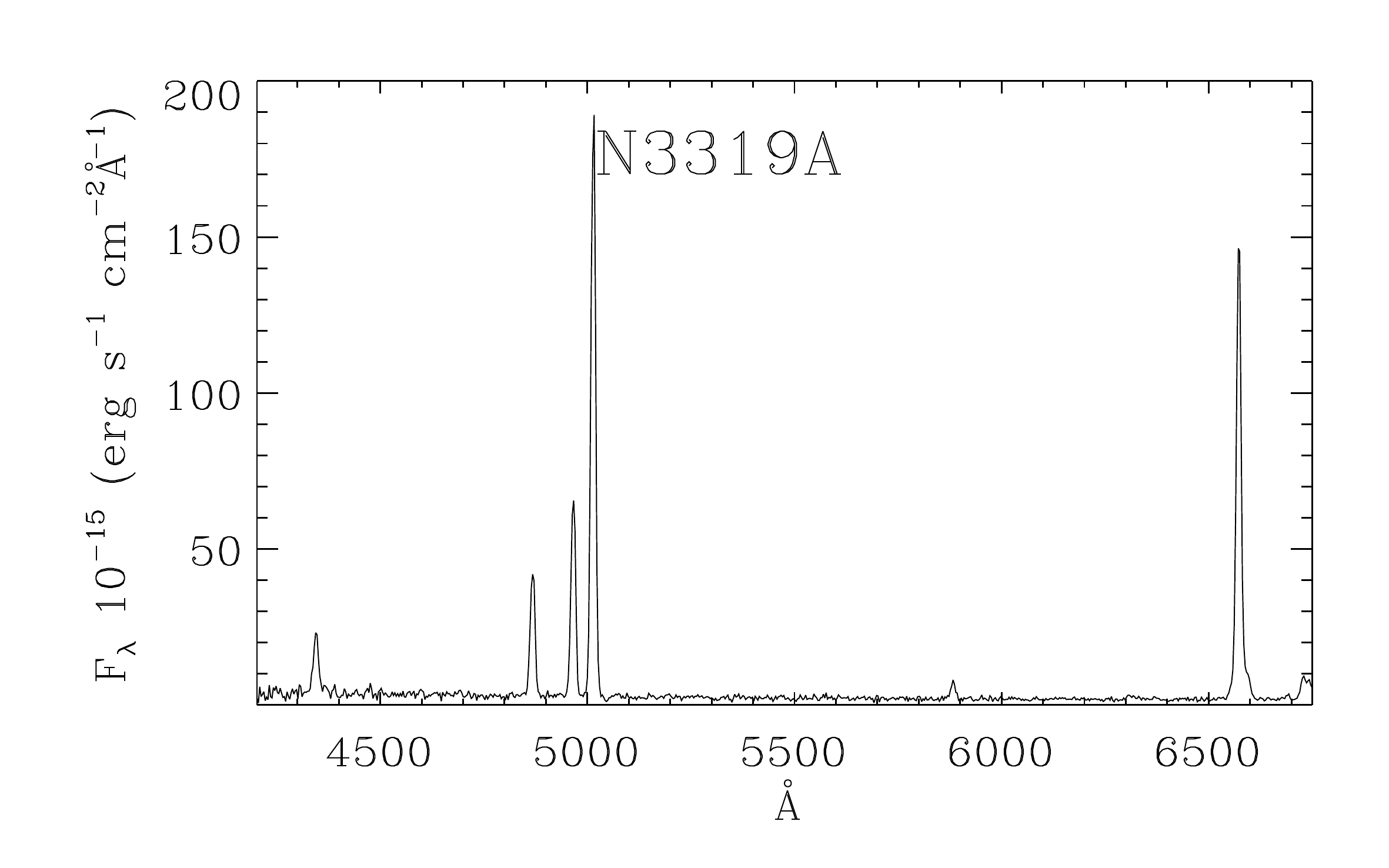}
  
  \includegraphics[scale=0.2]{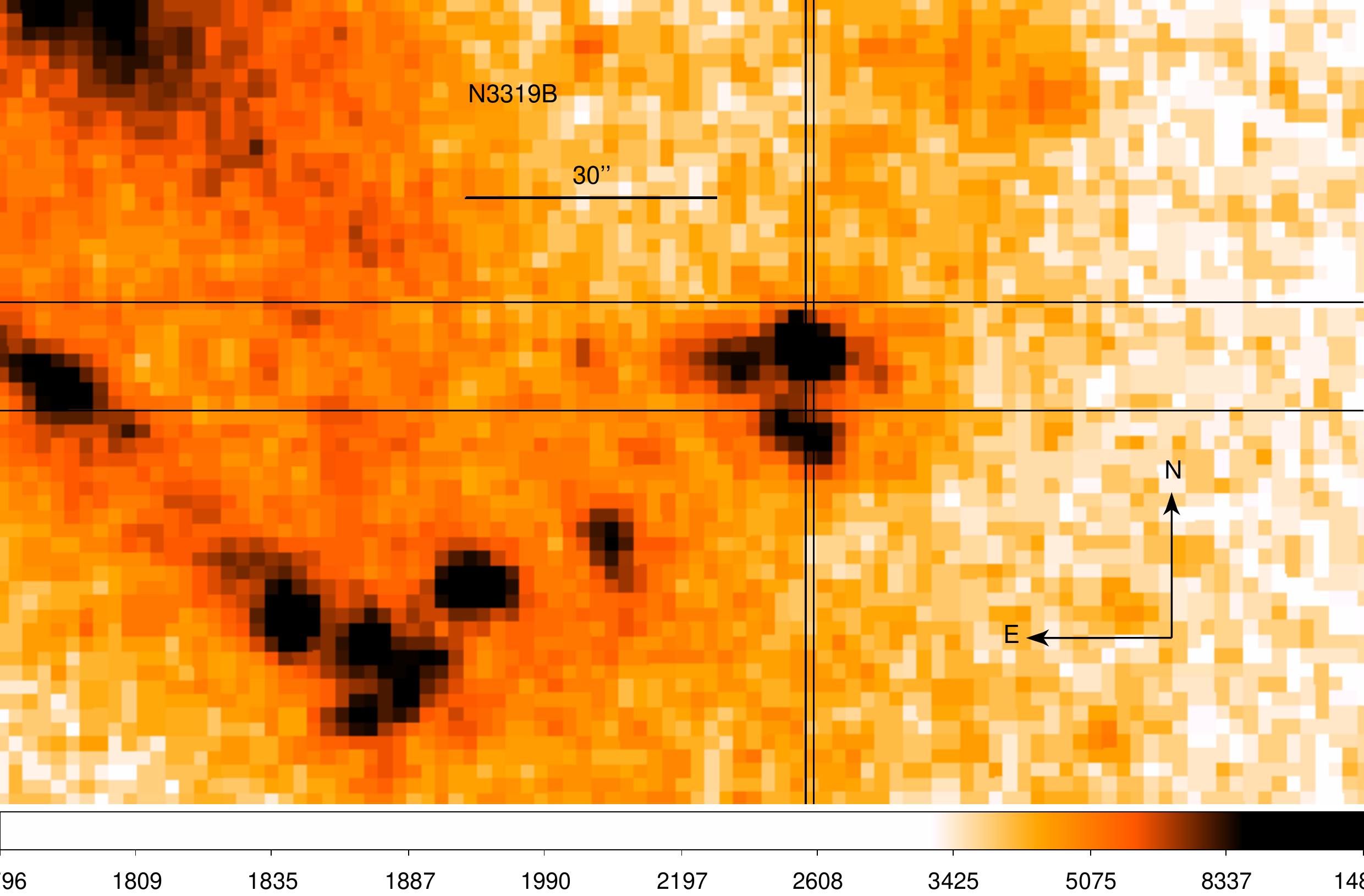}
  \includegraphics[scale=0.18]{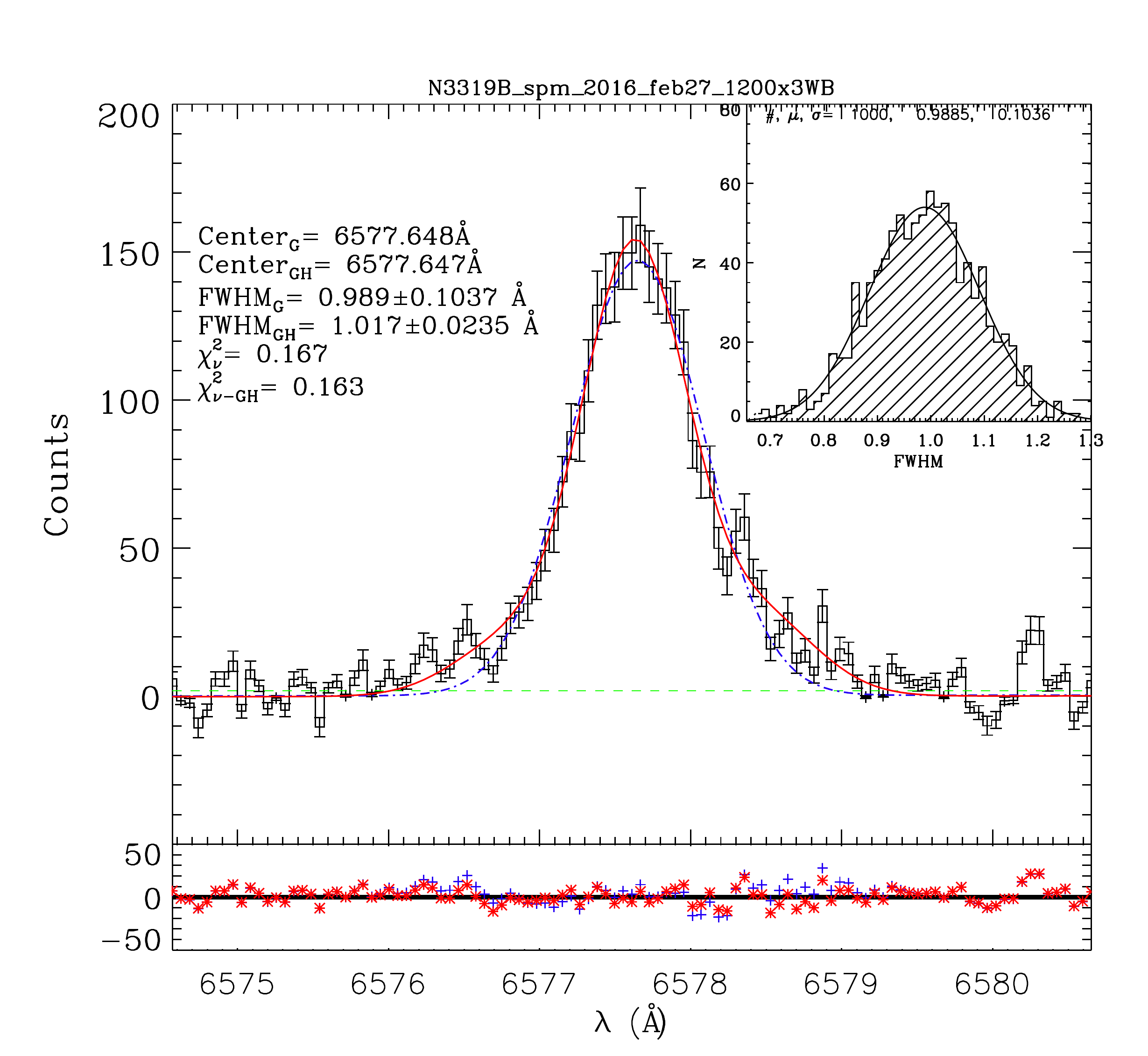}
  \includegraphics[scale=0.2]{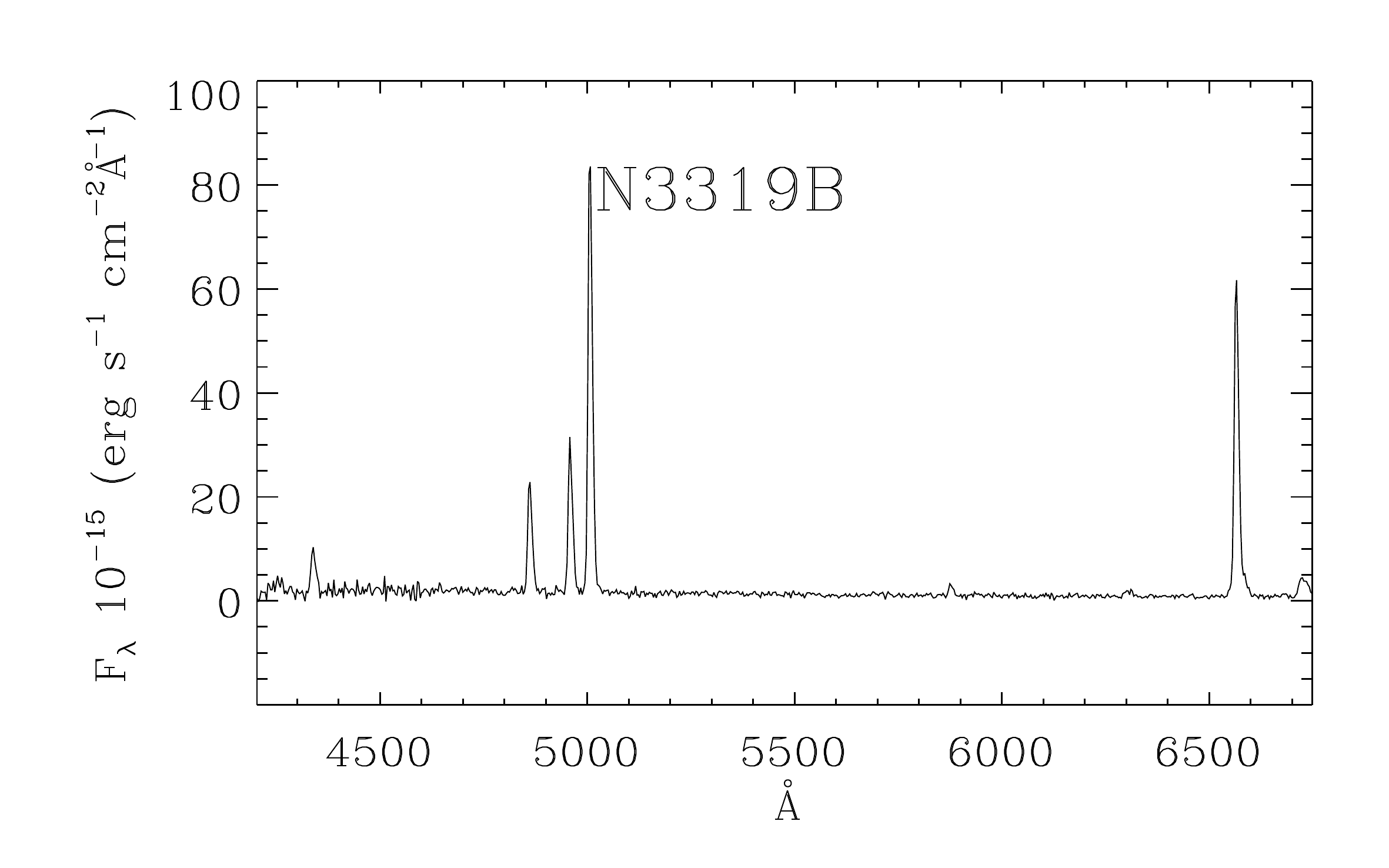}
  
  \caption[]{ (continued)}
 
  \end{figure*}  
  
   \begin{figure*}
  \includegraphics[scale=0.2]{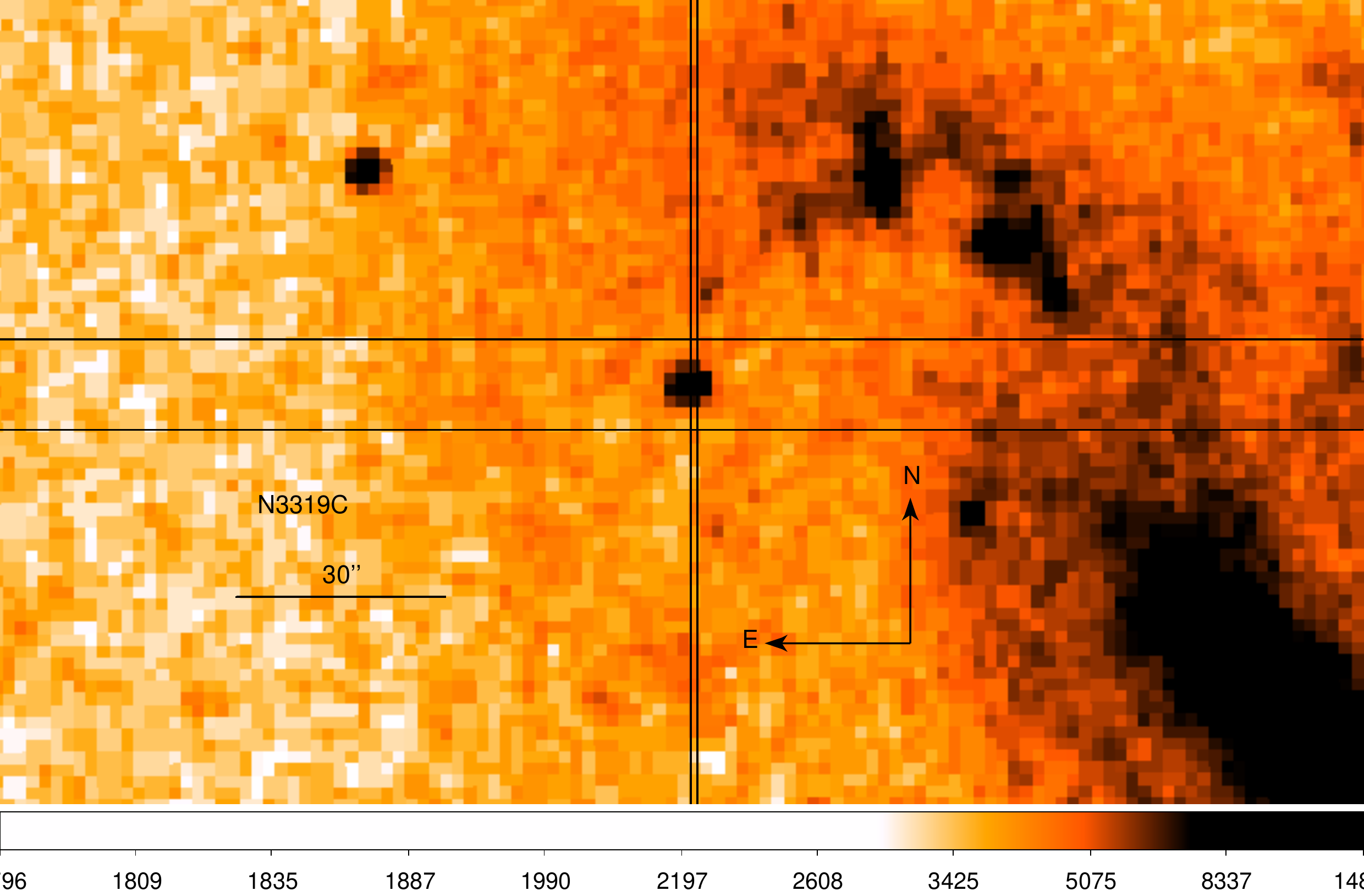}
  \includegraphics[scale=0.18]{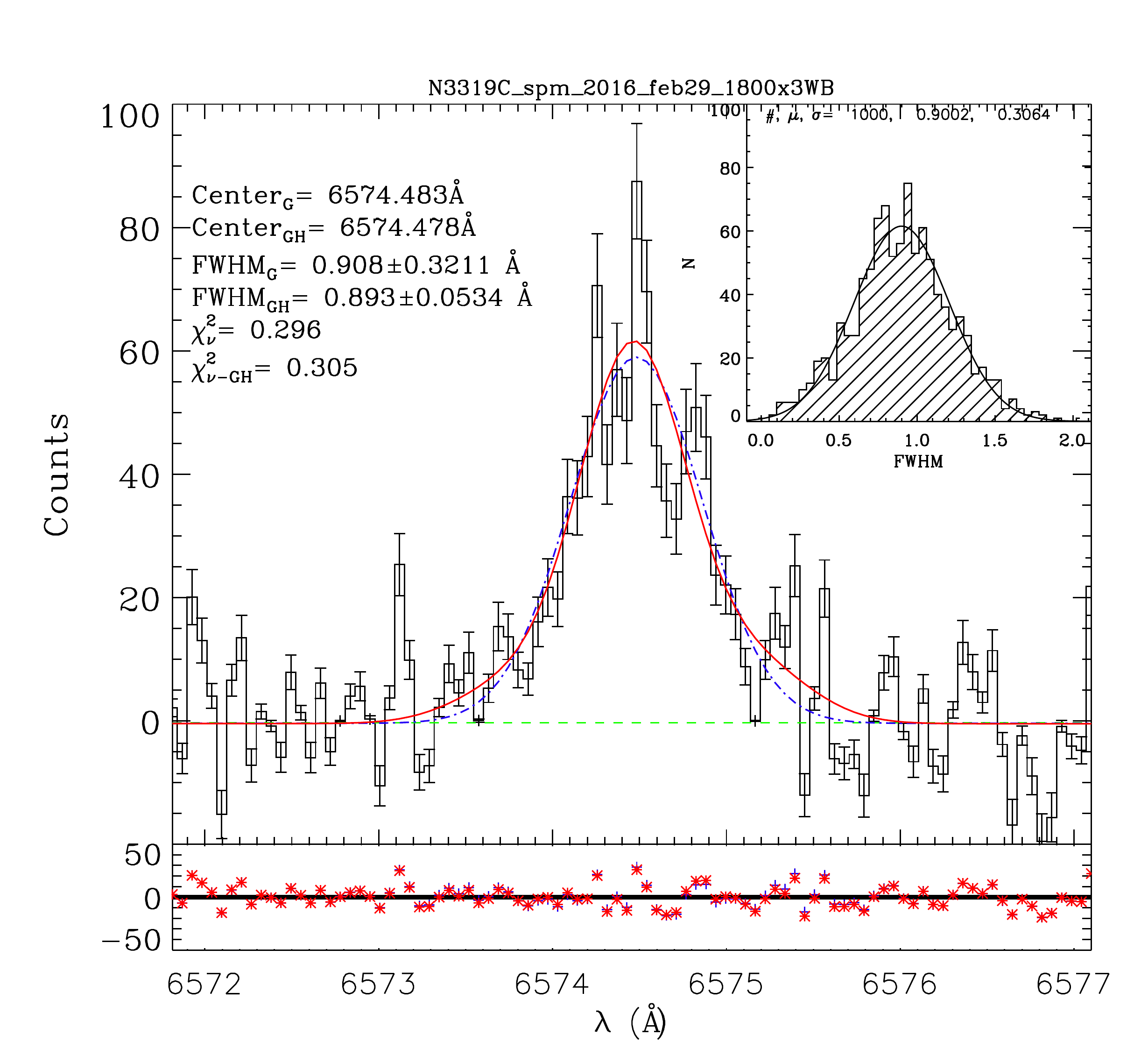}
  \includegraphics[scale=0.2]{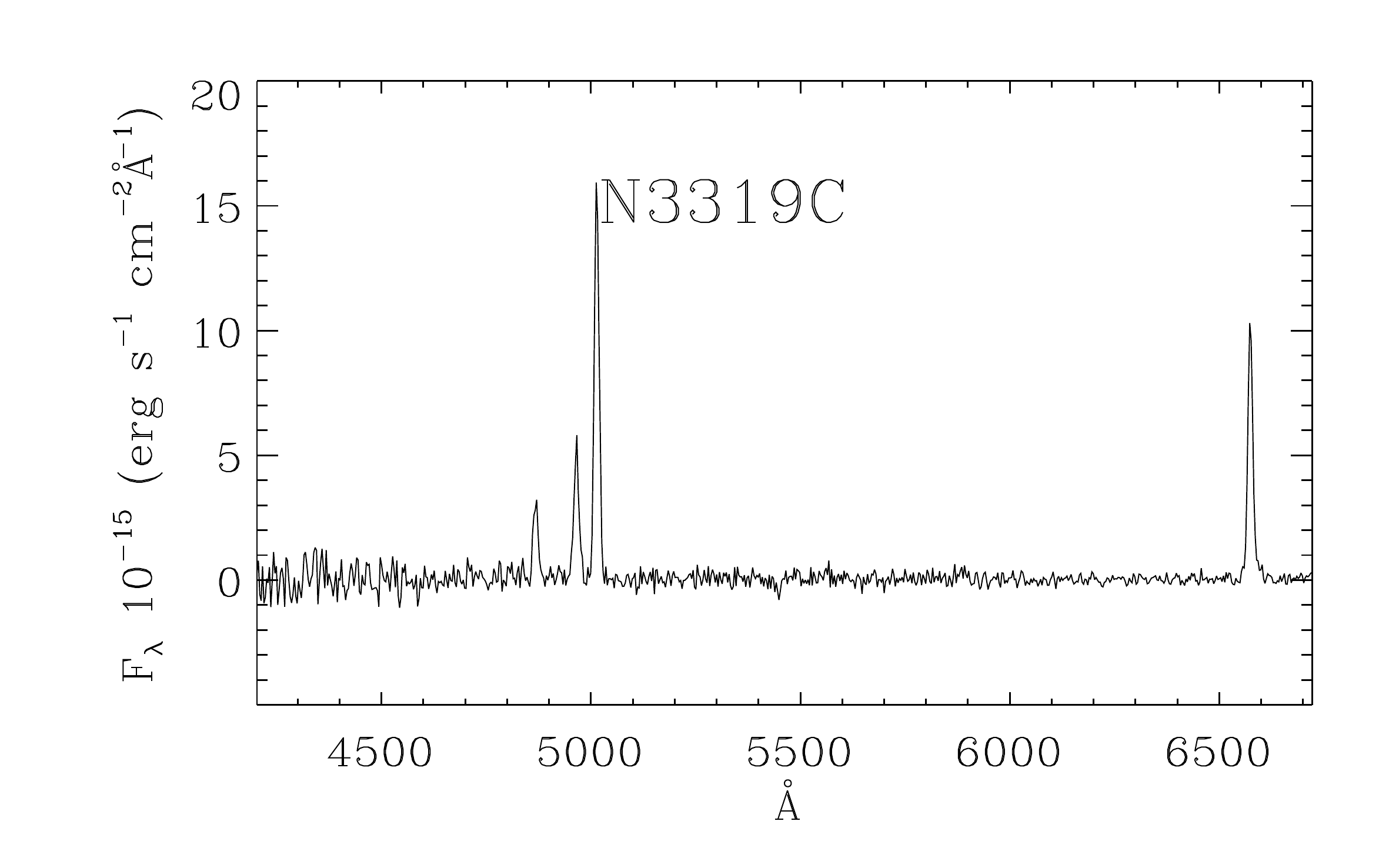}
  
  \includegraphics[scale=0.2]{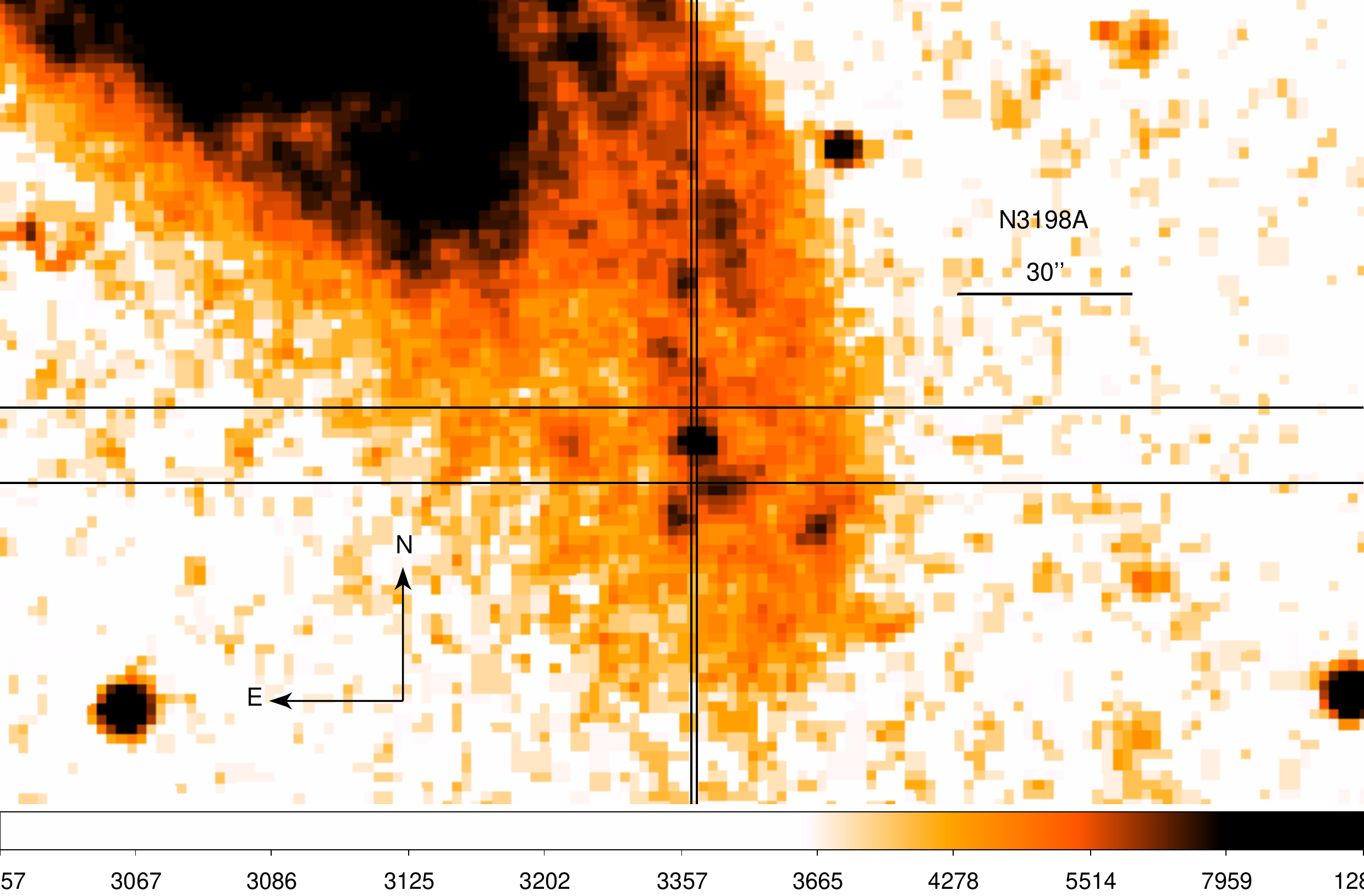}
  \includegraphics[scale=0.18]{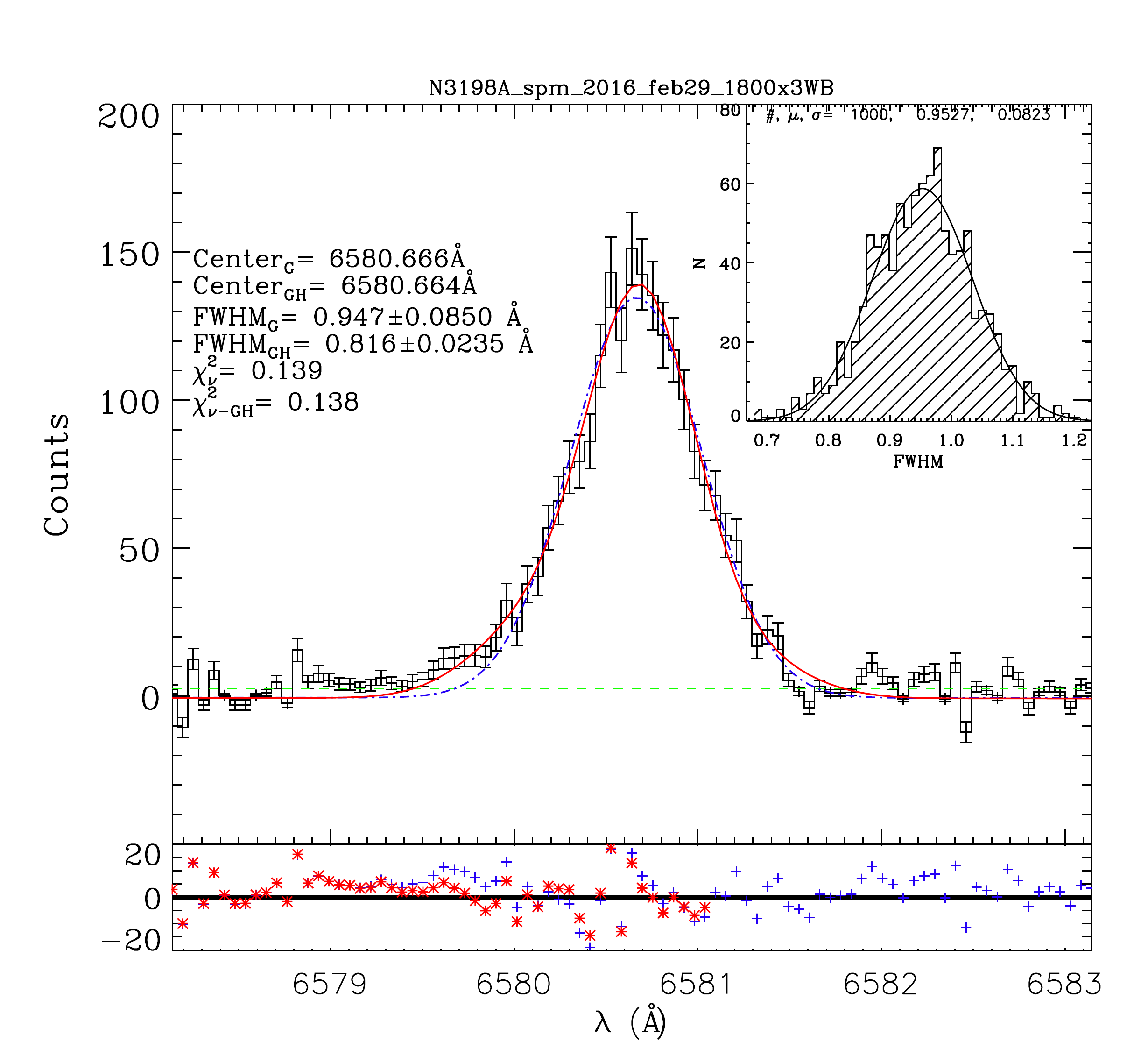}
\includegraphics[scale=0.2]{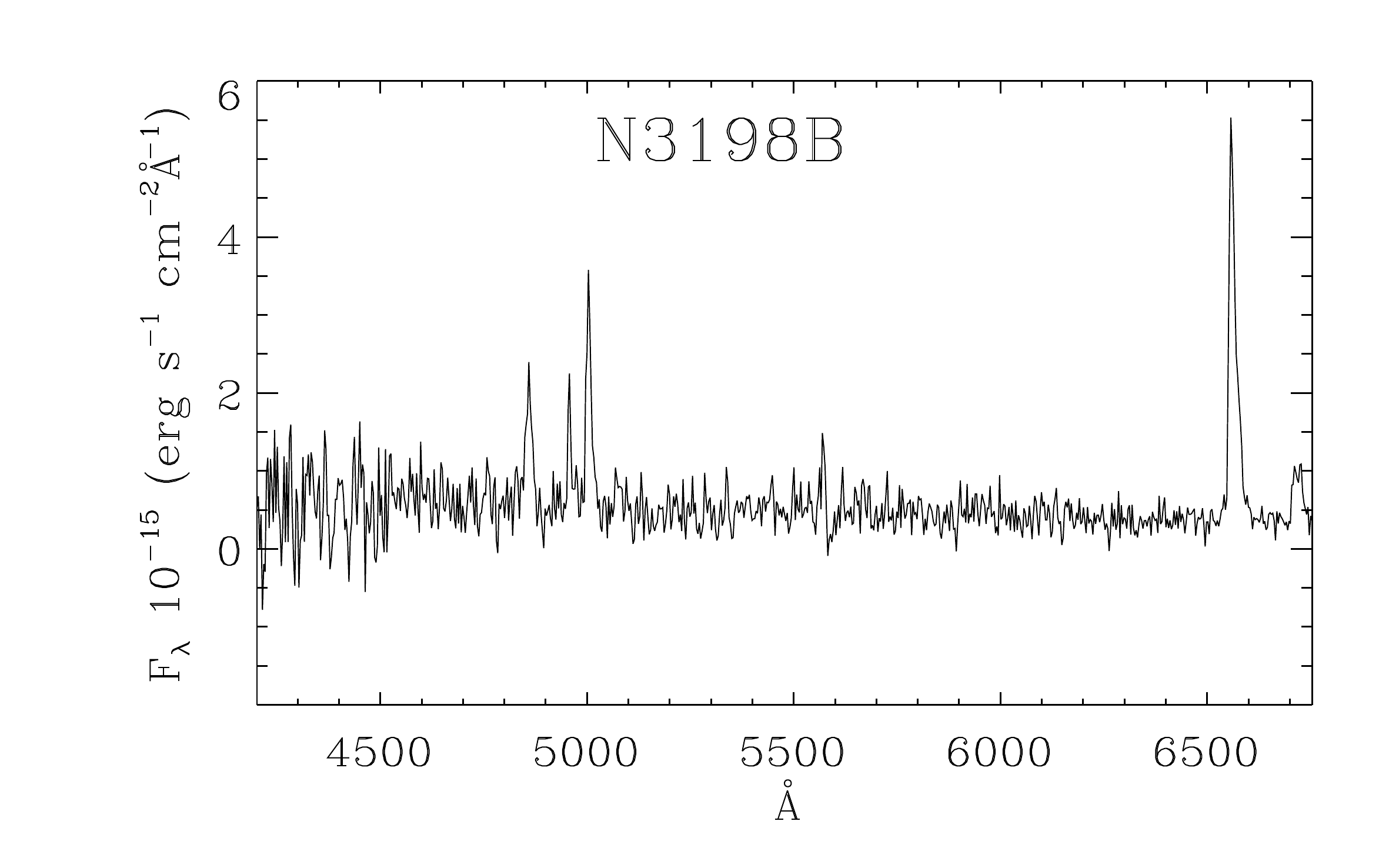}

  \caption[]{ (continued)}
 
  \end{figure*}

\bsp

\end{document}